\documentclass[11pt]{article}
\pdfoutput=1

\usepackage[letterpaper,margin=1in,bottom=1.4in]{geometry} % alt size
\usepackage[format=plain,font=small,labelfont=bf,textfont=up]{caption}

\usepackage[utf8]{inputenc} % allow utf-8 input
\usepackage[T1]{fontenc}    % use 8-bit T1 fonts
\usepackage[hidelinks]{hyperref}       % hyperlinks
\usepackage{url}            % simple URL typesetting
\usepackage{booktabs}       % professional-quality tables
\usepackage{amsfonts}       % blackboard math symbols
\usepackage{nicefrac}       % compact symbols for 1/2, etc.
\usepackage{microtype}      % microtypography
\usepackage[affil-it]{authblk} % author affiliations
\usepackage{tikz}
\usepackage{lmodern}
\usepackage{float}

\usepackage[linesnumbered,ruled,vlined]{algorithm2e}

\SetCommentSty{mycommfont}

\usepackage{amsmath}
\usepackage{amssymb}
\usepackage{amsthm}
\usepackage{mathtools}
\mathtoolsset{showonlyrefs,showmanualtags}
\allowdisplaybreaks
\usepackage{bbm}
\usepackage{graphicx}
\usepackage{subcaption}
\usepackage{thmtools, thm-restate}  %restatable
\usepackage[toc,page,header]{appendix}
\usepackage{minitoc}

\newcommand{\figpath}{./figs}

\newcounter{spacesave}
\usepackage[style=authoryear,sorting=nyt,sortcites=false, natbib=true, backend=bibtex]{biblatex}
\DeclarePairedDelimiter\paren\lparen\rparen
\DeclarePairedDelimiter\bracket\lbrack\rbrack
\DeclarePairedDelimiter\braces\lbrace\rbrace

\DeclarePairedDelimiter\abs\lvert\rvert

\providecommand{\bbone}{\mathbf{1}}
\DeclarePairedDelimiterXPP\indicator[1]{\bbone}{\lbrack}{\rbrack}{}{#1}

\DeclarePairedDelimiterXPP\expf[1]{\exp}{\lparen}{\rparen}{}{#1}
\DeclarePairedDelimiterXPP\logf[1]{\log}{\lparen}{\rparen}{}{#1}
\DeclarePairedDelimiterXPP\maxf[1]{\max}{\lparen}{\rparen}{}{#1}
\DeclarePairedDelimiterXPP\minf[1]{\min}{\lparen}{\rparen}{}{#1}

\DeclareMathOperator*{\sgn}{sgn}
\DeclarePairedDelimiterXPP\sgnf[1]{\sgn}{\lparen}{\rparen}{}{#1}

\DeclarePairedDelimiterXPP\func[2]{#1}{\lparen}{\rparen}{}{#2}

\DeclareMathOperator*{\atan}{atan}
\DeclarePairedDelimiterXPP\atanf[1]{\atan}{\lparen}{\rparen}{}{#1}
\DeclarePairedDelimiterXPP\tanf[1]{\tan}{\lparen}{\rparen}{}{#1}

\DeclarePairedDelimiter\setb\lbrace\rbrace

\newcommand{\Reals}{\mathbb{R}}

\newcommand{\Naturals}{\mathbb{N}}

\DeclareMathOperator*{\argmax}{arg\,max}
\DeclareMathOperator*{\argmin}{arg\,min}

\newcommand{\mat}[1]{\boldsymbol{#1}}
\renewcommand{\vec}[1]{\boldsymbol{#1}}

\makeatletter
\newcommand*{\tran}{{\mathpalette\@tran{}}}
\newcommand*{\@tran}[2]{\raisebox{\depth}{$\m@th#1\intercal$}}
\makeatother

\DeclarePairedDelimiter\norm\lVert\rVert
\DeclarePairedDelimiterXPP\tnorm[1]{}{\lVert}{\rVert_{1}}{}{#1}
\DeclarePairedDelimiterXPP\enorm[1]{}{\lVert}{\rVert_{2}}{}{#1}
\DeclarePairedDelimiterXPP\inorm[1]{}{\lVert}{\rVert_{\infty}}{}{#1}
\DeclarePairedDelimiterXPP\pnorm[2]{}{\lVert}{\rVert_{#1}}{}{#2}
\DeclarePairedDelimiterXPP\opnorm[1]{}{\lVert}{\rVert_{op}}{}{#1}

\DeclarePairedDelimiterXPP\detf[1]{\det}{\lparen}{\rparen}{}{#1}

\DeclarePairedDelimiterXPP\kerf[1]{\ker}{\lparen}{\rparen}{}{#1}

\DeclareMathOperator{\trsym}{tr}
\DeclarePairedDelimiterXPP\tr[1]{\trsym}{\lparen}{\rparen}{}{#1}

\DeclareMathOperator{\vspansym}{span}
\DeclarePairedDelimiterXPP\vspan[1]{\vspansym}{\lparen}{\rparen}{}{#1}

\DeclareMathOperator{\diagsym}{diag}
\DeclarePairedDelimiterXPP\diag[1]{\diagsym}{\lparen}{\rparen}{}{#1}

\DeclareMathOperator{\ranksym}{rank}
\DeclarePairedDelimiterXPP\rank[1]{\ranksym}{\lparen}{\rparen}{}{#1}

\DeclareMathOperator{\vectorizesym}{vec}
\DeclarePairedDelimiterXPP\vectorize[1]{\vectorizesym}{\lparen}{\rparen}{}{#1}

\DeclareMathOperator*{\esssup}{ess\,sup}
\DeclarePairedDelimiterXPP\esssupf[1]{\esssup}{\lparen}{\rparen}{}{#1}

\let\Prsym\Pr
\let\Pr\relax
\DeclarePairedDelimiterXPP\Pr[1]{\Prsym}{\lparen}{\rparen}{}{%
	#1}

\DeclarePairedDelimiterXPP\Prsub[2]{\Prsym_{#1}}{\lparen}{\rparen}{}{%
	#2}

\DeclareMathOperator{\Esym}{E}
\DeclarePairedDelimiterXPP\E[1]{\Esym}{\lbrack}{\rbrack}{}{%
	#1}

\DeclarePairedDelimiterXPP\Esub[2]{\Esym_{#1}}{\lbrack}{\rbrack}{}{%
	#2}

\DeclareMathOperator{\Varsym}{Var}
\DeclarePairedDelimiterXPP\Var[1]{\Varsym}{\lparen}{\rparen}{}{%
	#1}

\DeclarePairedDelimiterXPP\Varsub[2]{\Varsym_{#1}}{\lparen}{\rparen}{}{%
	#2}

\DeclarePairedDelimiterXPP\EstVar[1]{\widehat{\Varsym}}{\lparen}{\rparen}{}{%
	#1}

\DeclareMathOperator{\Covsym}{Cov}
\DeclarePairedDelimiterXPP\Cov[1]{\Covsym}{\lparen}{\rparen}{}{%
	#1}

\DeclarePairedDelimiterXPP\Covsub[2]{\Covsym_{#1}}{\lparen}{\rparen}{}{%
	#2}

\DeclareMathOperator{\Corrsym}{Corr}
\DeclarePairedDelimiterXPP\Corr[1]{\Corrsym}{\lparen}{\rparen}{}{%
	#1}

\makeatletter
\newcommand{\indep}{\protect\mathpalette{\protect\@indep}{\perp}}
\newcommand*{\@indep}[2]{\mathrel{\rlap{$#1#2$}\mkern3mu{#1#2}}}
\makeatother

\newcommand{\bigOsym}{\mathcal{O}}
\DeclarePairedDelimiterXPP\bigO[1]{\bigOsym}{\lparen}{\rparen}{}{#1}
\DeclarePairedDelimiterXPP\bigOt[1]{\widetilde{\bigOsym}}{\lparen}{\rparen}{}{#1}

\newcommand{\littleOsym}{o}
\DeclarePairedDelimiterXPP\littleO[1]{\littleOsym}{\lparen}{\rparen}{}{#1}

\newcommand{\bigOpsym}{\bigOsym_p}
\DeclarePairedDelimiterXPP\bigOp[1]{\bigOpsym}{\lparen}{\rparen}{}{#1}

\newcommand{\littleOpsym}{\littleOsym_p}
\DeclarePairedDelimiterXPP\littleOp[1]{\littleOpsym}{\lparen}{\rparen}{}{#1}

\newcommand{\bigOmegasym}{\Omega}
\DeclarePairedDelimiterXPP\bigOmega[1]{\bigOmegasym}{\lparen}{\rparen}{}{#1}

\newcommand{\littleOmegasym}{\omega}
\DeclarePairedDelimiterXPP\littleOmega[1]{\littleOmegasym}{\lparen}{\rparen}{}{#1}

\newcommand{\bigThetasym}{\Theta}
\DeclarePairedDelimiterXPP\bigTheta[1]{\bigThetasym}{\lparen}{\rparen}{}{#1}

\DeclarePairedDelimiterXPP\cosf[1]{\cos}{\lparen}{\rparen}{}{#1}
\DeclarePairedDelimiterXPP\sinf[1]{\sin}{\lparen}{\rparen}{}{#1}

\newcommand{\quadtext}[1]{\quad\text{#1}\quad}

\newcommand{\quadand}{\quadtext{and}}

\newcommand{\quadwhere}{\quadtext{where}}

\newcommand{\newvar}[2]{
	\expandafter\newcommand\csname #1\endcsname{#2}
}

\newcommand{\newvars}[2]{
	\expandafter\newcommand\csname #1\endcsname[1]{#2_{##1}}
}

\newcommand{\newvarss}[2]{
	\expandafter\newcommand\csname #1\endcsname[2]{#2_{##1,##2}}
}

\newcommand{\newvarsss}[2]{
	\expandafter\newcommand\csname #1\endcsname[3]{#2_{##1,##2,##3}}
}

\newcommand{\newvarssss}[2]{
	\expandafter\newcommand\csname #1\endcsname[4]{#2_{##1,##2,##3,##4}}
}

\newcommand{\newfunconly}[2]{
	\expandafter\DeclarePairedDelimiterXPP\csname #1\endcsname[1]{#2}{\lparen}{\rparen}{}{##1}
}

\newcommand{\newfuncs}[2]{
	\expandafter\DeclarePairedDelimiterXPP\csname #1\endcsname[2]{#2_{##1}}{\lparen}{\rparen}{}{##2}
}

\newcommand{\newfuncss}[2]{
	\expandafter\DeclarePairedDelimiterXPP\csname #1\endcsname[3]{#2_{##1,##2}}{\lparen}{\rparen}{}{##3}
}

\newcommand{\newfuncsss}[2]{
	\expandafter\DeclarePairedDelimiterXPP\csname #1\endcsname[4]{#2_{##1,##2,##3}}{\lparen}{\rparen}{}{##4}
}

\newcommand{\newfuncssss}[2]{
	\expandafter\DeclarePairedDelimiterXPP\csname #1\endcsname[5]{#2_{##1,##2,##3,##4}}{\lparen}{\rparen}{}{##5}
}

\newcommand{\varpartialeval}[3]{
	\expandafter\newcommand\csname #2\endcsname{\csname #1\endcsname{#3}}
}

\newcommand{\funcpartialeval}[3]{
	\expandafter\newcommand\csname #2\endcsname[1][]{\csname #1\endcsname[##1]{#3}}
}

\newcommand{\varevali}[2][]{
	\varpartialeval{#2#1}{#2i}{i}
	\varpartialeval{#2#1}{#2j}{j}
}

\newcommand{\varevalk}[2][]{
	\varpartialeval{#2#1}{#2k}{k}
	\varpartialeval{#2#1}{#2l}{\ell}
}

\newcommand{\varevals}[2][]{
	\varpartialeval{#2#1}{#2s}{s}
	\varpartialeval{#2#1}{#2r}{r}
}

\newcommand{\funcevali}[2][]{
	\funcpartialeval{#2#1}{#2i}{i}
	\funcpartialeval{#2#1}{#2j}{j}
}

\newcommand{\funcevalk}[2][]{
	\funcpartialeval{#2#1}{#2k}{k}
	\funcpartialeval{#2#1}{#2l}{\ell}
}

\newcommand{\varevalii}[2][]{
	\varevali[#1]{#2}
	\varevali{#2i}
	\varevali{#2j}
}

\newcommand{\varevalik}[2][]{
	\varevali[#1]{#2}
	\varevalk{#2i}
	\varevalk{#2j}
}

\newcommand{\varevaliik}[2][]{
	\varevalii[#1]{#2}
	\varevalk{#2ii}
	\varevalk{#2ij}
	\varevalk{#2ji}
	\varevalk{#2jj}
}

\newcommand{\varevaliis}[2][]{
	\varevalii[#1]{#2}
	\varevals{#2ii}
	\varevals{#2ij}
	\varevals{#2ji}
	\varevals{#2jj}
}

\newcommand{\varevaliikk}[2][]{
	\varevaliik[#1]{#2}
	\varevalk{#2iik}
	\varevalk{#2ijk}
	\varevalk{#2jik}
	\varevalk{#2jjk}
	\varevalk{#2iil}
	\varevalk{#2ijl}
	\varevalk{#2jil}
	\varevalk{#2jjl}
}

\newcommand{\funcevalii}[2][]{
	\funcevali[#1]{#2}
	\funcevali{#2i}
	\funcevali{#2j}
}

\newcommand{\funcevalik}[2][]{
	\funcevali[#1]{#2}
	\funcevalk{#2i}
	\funcevalk{#2j}
}

\newcommand{\funcevaliik}[2][]{
	\funcevalii[#1]{#2}
	\funcevalk{#2ii}
	\funcevalk{#2ij}
	\funcevalk{#2ji}
	\funcevalk{#2jj}
}

\newcommand{\funcevaliikk}[2][]{
	\funcevaliik[#1]{#2}
	\funcevalk{#2iik}
	\funcevalk{#2ijk}
	\funcevalk{#2jik}
	\funcevalk{#2jjk}
	\funcevalk{#2iil}
	\funcevalk{#2ijl}
	\funcevalk{#2jil}
	\funcevalk{#2jjl}
}

\newcommand{\funcevalX}[3]{
	\expandafter\newcommand\csname #1v#2\endcsname{\csname #1v\endcsname{#3}}
}

\newcommand{\funcevaliX}[3]{
	\expandafter\newcommand\csname #1v#2\endcsname{\csname #1v\endcsname{#3}}
	\expandafter\newcommand\csname #1v#2e\endcsname[1]{\csname #1ve\endcsname{##1}{#3}}
	\expandafter\newcommand\csname #1v#2i\endcsname{\csname #1vi\endcsname{#3}}
	\expandafter\newcommand\csname #1v#2j\endcsname{\csname #1vj\endcsname{#3}}
}

\newcommand{\funcevalikX}[3]{
	\expandafter\newcommand\csname #1v#2\endcsname{\csname #1v\endcsname{#3}}
	\expandafter\newcommand\csname #1v#2e\endcsname[2]{\csname #1ve\endcsname{##1}{##2}{#3}}
	\expandafter\newcommand\csname #1v#2i\endcsname[1]{\csname #1vi\endcsname{##1}{#3}}
	\expandafter\newcommand\csname #1v#2j\endcsname[1]{\csname #1vj\endcsname{##1}{#3}}
	\expandafter\newcommand\csname #1v#2ik\endcsname{\csname #1vik\endcsname{#3}}
	\expandafter\newcommand\csname #1v#2il\endcsname{\csname #1vil\endcsname{#3}}
	\expandafter\newcommand\csname #1v#2jk\endcsname{\csname #1vik\endcsname{#3}}
	\expandafter\newcommand\csname #1v#2jl\endcsname{\csname #1vjl\endcsname{#3}}
}

\newcommand{\newvari}[2]{
	\newvar{#1}{#2}
	\newvars{#1e}{\csname #1\endcsname}
	\varevali[e]{#1}
}

\newcommand{\newvark}[2]{
	\newvar{#1}{#2}
	\newvars{#1e}{\csname #1\endcsname}
	\varevalk[e]{#1}
}

\newcommand{\newvarik}[2]{
	\newvar{#1}{#2}
	\newvarss{#1e}{\csname #1\endcsname}
	\varevalik[e]{#1}
}

\newcommand{\newvarii}[2]{
	\newvar{#1}{#2}
	\newvarss{#1e}{\csname #1\endcsname}
	\varevalii[e]{#1}
}

\newcommand{\newvariik}[2]{
	\newvar{#1}{#2}
	\newvarsss{#1e}{\csname #1\endcsname}
	\varevaliik[e]{#1}
}

\newcommand{\newvariis}[2]{
	\newvar{#1}{#2}
	\newvarsss{#1e}{\csname #1\endcsname}
	\varevaliis[e]{#1}
}

\newcommand{\newvariikk}[2]{
	\newvar{#1}{#2}
	\newvarssss{#1e}{\csname #1\endcsname}
	\varevaliikk[e]{#1}
}

\newcommand{\newfunc}[2]{
	\newvar{#1}{#2}
	\newfunconly{#1v}{\csname #1\endcsname}
}

\newcommand{\newfunci}[2]{
	\newvari{#1}{#2}
	\newfunconly{#1v}{\csname #1\endcsname}
	\newfuncs{#1ve}{\csname #1\endcsname}
	\funcevali[e]{#1v}
}

\newcommand{\newfunck}[2]{
	\newvark{#1}{#2}
	\newfunconly{#1v}{\csname #1\endcsname}
	\newfuncs{#1ve}{\csname #1\endcsname}
	\funcevalk[e]{#1v}
}

\newcommand{\newfuncik}[2]{
	\newvarik{#1}{#2}
	\newfunconly{#1v}{\csname #1\endcsname}
	\newfuncss{#1ve}{\csname #1\endcsname}
	\funcevalik[e]{#1v}
}

\newcommand{\newfuncii}[2]{
	\newvarii{#1}{#2}
	\newfunconly{#1v}{\csname #1\endcsname}
	\newfuncss{#1ve}{\csname #1\endcsname}
	\funcevalii[e]{#1v}
}

\newcommand{\newfunciik}[2]{
	\newvariik{#1}{#2}
	\newfunconly{#1v}{\csname #1\endcsname}
	\newfuncsss{#1ve}{\csname #1\endcsname}
	\funcevaliik[e]{#1v}
}

\newcommand{\newfunciikk}[2]{
	\newvariikk{#1}{#2}
	\newfunconly{#1v}{\csname #1\endcsname}
	\newfuncssss{#1ve}{\csname #1\endcsname}
	\funcevaliikk[e]{#1v}
}

\theoremstyle{plain}
\newtheorem{theorem}{Theorem}[section]

\newtheorem{corollary}[theorem]{Corollary}
\newtheorem{lemma}[theorem]{Lemma}
\newtheorem{proposition}[theorem]{Proposition}

\newtheorem*{mainresult}{Main Result}

\newtheorem{innerrefcorollary}{Corollary}
\newenvironment{refcorollary}[1]
{\renewcommand\theinnerrefcorollary{#1}\innerrefcorollary}
{\endinnerrefcorollary}

\newtheorem{innerreflemma}{Lemma}
\newenvironment{reflemma}[1]
{\renewcommand\theinnerreflemma{#1}\innerreflemma}
{\endinnerreflemma}

\newtheorem{innerrefproposition}{Proposition}
\newenvironment{refproposition}[1]
{\renewcommand\theinnerrefproposition{#1}\innerrefproposition}
{\endinnerrefproposition}

\newtheorem{innerreftheorem}{Theorem}
\newenvironment{reftheorem}[1]
{\renewcommand\theinnerreftheorem{#1}\innerreftheorem}
{\endinnerreftheorem}

\theoremstyle{definition}
\newtheorem{assumption}{Assumption}

\newtheorem{definition}{Definition}

\theoremstyle{remark}

\newcommand{\contrastz}[2]{\vec{z}_{#1,(#2)}}
\newcommand{\zonei}{\contrastz{i}{1}}
\newcommand{\zzeroi}{\contrastz{i}{0}}

\newcommand{\ate}{\tau}

\newcommand{\eate}{\hat{\ate}}

\newcommand{\eatetrue}{\eate_{\textrm{standard}}}

\newcommand{\evb}{\widehat{\mathrm{VB}}}
\newcommand{\vb}{\mathrm{VB}}

\newcommand{\ourdesign}{\textsc{Conflict-Graph-Design}}

\newcommand{\cH}{\mathcal{H}}
\newcommand{\cD}{\mathcal{D}}
\newcommand{\ev}{\vec{v}}
\newcommand{\adjH}{\mat{A}_{\cH}}
\newcommand{\lamH}{\lambda(\cH)}
\newcommand{\lamG}{\lambda(G)}

\newcommand{\eigpi}{\pi_{\textup{eig}}}
\newcommand{\degpi}{\pi_{\textup{deg}}}

\newcommand{\lp}{\left}
\newcommand{\rp}{\right}

\newcommand{\R}{\mathbb{R}}

\newcommand{\constvar}{\frac{2r}{(1- 1/(2r))^2}}
\newcommand{\constcovar}{ \frac{-2 \ln (1- 1/(2r))}{(1 - 1/ (2r))^{2}}}
\newcommand{\lammax}{\lambda_{\textrm{max}}}
\newcommand{\finalconst}{12.08}
\newcommand{\cleanconst}{12.5}

\newcommand{\vx}{\mathbf{x}}
\newcommand{\vy}{\mathbf{y}}

\newcommand{\varM}{\mat{V}}
\newcommand{\lamV}{\lambda(\varM)}

\newcommand{\dmax}{d_{\textrm{max}}}

\newcommand{\Pp}[2][]{\Prsym_{#1}\left(#2\right)}

\newcommand{\kc}[1]{#1^{(c)}}
\newcommand{\kzeroc}{k_0^{(c)}}

\newcommand{\alphaind}[2]{\alpha_{#1}^{(#2)}}
\newcommand{\betaind}[2]{\beta_{#1}^{(#2)}}

\newcommand{\Nb}[2]{N_b^{(#2)}(#1)}
\newcommand{\Na}[2]{N_a^{(#2)}(#1)}

\newcommand{\Aone}{A^{(1)}}
\newcommand{\Bone}{B^{(1)}}
\newcommand{\Gone}{G^{(1)}}
\newcommand{\Gtwo}{G^{(2)}}
\newcommand{\Gthree}{G^{(3)}}
\newcommand{\Vone}{V^{(1)}}
\newcommand{\Vtwo}{V^{(2)}}
\newcommand{\Vthree}{V^{(3)}}
\newcommand{\Eone}{E^{(1)}}
\newcommand{\Etwo}{E^{(2)}}
\newcommand{\Ethree}{E^{(3)}}

\newcommand{\Path}[3]{P^{(#3)}_{#1,#2}}

\newcommand{\predcover}{\textup{Pr}_{\textup{theory}}(\alpha)}

\usepackage{expcmd}
\newcommand\mainref\ref
\newcommand\suppref\ref


\title{The Conflict Graph Design: Estimating Causal Effects \\ under Arbitrary Neighborhood Interference}
\author[1]{Vardis Kandiros}
\author[2]{Charilaos Pipis}
\author[2]{Constantinos Daskalakis}
\author[1]{Christopher Harshaw}
\affil[1]{Columbia University}
\affil[2]{Massachusetts Institute of Technology}
\date{\today}

\begin{document}
	
	% acknowledgements
	\makeatletter%
	
	% this is so we can re-state algorithms
	% https://tex.stackexchange.com/questions/690161/how-to-restate-an-algorithm-with-the-same-algorithm-number
	\def\thmt@innercounters{equation,algocf}
	
	% this create acknowledgements
	\begin{NoHyper}\gdef\@thefnmark{}\@footnotetext{\hspace{-1em}We thank 
P.M. Aronow,
Arun Chandrasekhar,
Dean Eckels,
Dan Malinsky,
Caleb Miles,
Betsy Ogburn,
Phevos Paschalidis,
Fredrik S{\"a}vje,
Michael Sobel,
Panos Toulis,
Johan Ugander,
and 
Gernot Z{\"o}cklein
for insightful discussions which helped to shape this work.
We similarly thank participants in the 
BIRS ``Causal Inference and Prediction for Network Data'' workshop, 
the Columbia Causal Inference Learning Group,
and
the MIT Stochastics and Statistics Seminar.
Vardis Kandiros gratefully acknowledges support from the Eric and Wendy Schmidt Center at the Broad Institute of MIT and Harvard.
Vardis Kandiros and Charilaos Pipis gratefully acknowledge support from the Onassis Foundation.
Christopher Harshaw gratefully acknowledges support from Foundations of Data Science Institute (FODSI) NSF grant DMS2023505.
All authors gratefully acknowledge support from NSF Awards CCF-1901292, DMS-2022448 and DMS-2134108, a Simons Investigator Award, and the Simons Collaboration on the Theory of Algorithmic Fairness.}\end{NoHyper}%
	\makeatother%
	
	\maketitle
	\thispagestyle{empty}
	
	% abstract
	\begin{abstract}
		A fundamental problem in network experiments is selecting an appropriate experimental design in order to precisely estimate a given causal effect of interest.
In this work, we propose the Conflict Graph Design, a general approach for constructing experiment designs under network interference with the goal of precisely estimating a pre-specified causal effect.
A central aspect of our approach is the notion of a \emph{conflict graph}, which captures the fundamental unobservability associated with the causal effect and the underlying network.
In order to estimate effects, we propose a modified Horvitz--Thompson estimator.
We show that its variance under the Conflict Graph Design is bounded as $\bigO{\lamH/n}$, where $\lamH$ is the largest eigenvalue of the adjacency matrix of the conflict graph. 
These rates depend on both the underlying network and the particular causal effect under investigation.
Not only does this yield the best known rates of estimation for several well-studied causal effects (e.g. the global and direct effects) but it also provides new methods for effects which have received less attention from the perspective of experiment design (e.g. spill-over effects).
Finally, we construct conservative variance estimators which facilitate asymptotically valid confidence intervals for the causal effect of interest.

	\end{abstract}

	\newpage
	
	% front matter
	\pagenumbering{roman}
	
	% table of contents
	\doparttoc % Tell to minitoc to generate a toc for the parts
	\faketableofcontents % Run a fake tableofcontents command for the partocs
	\part{} % Start the document part
	\parttoc % Insert the document TOC
	
	\newpage

	% begin main body
	\pagenumbering{arabic}
	
	\section{Introduction} \label{sec:intro}

% PARAGRAPH 1 (interference is great, but hard)
Randomized experiments are among the most reliable methodological tools for investigating causal effects.
A key assumption underpinning the design and analysis of classical randomized experiments is the Stable Unit Treatment Value Assumption (SUTVA), which includes the assumption of no interference, i.e. that treatment given to one subject does not affect the outcome of other subjects.
The past two decades have seen a growing interest in \emph{network experiments}, where subjects are assumed to interact through a network.
Network experiments have been used across a variety of disciplines to investigate causal effects in cash transfer programs \citep{Haushofer2016Short}, education policy \citep{Duflo2011Peer}, policing policy \citep{Blattman2021Place}, conflict prevention \citep{Paluck2016Changing}, and viral marketing campaigns \citep{Aral2011Creating,Eckles2016Estimating}, to name a few.
Part of what makes these network experiments so attractive is their ability to investigate new types of causal questions regarding the nature of these interactions, including global effects, direct effects, spill-over effects, and more.
The major methodological challenge in designing and analyzing network experiments is the violation of SUTVA, which renders classical methods invalid in these settings.

% PARAGRAPH 2 (what has been going on so far) 
A fundamental problem in the design and analysis of network experiments is the construction of appropriate mechanisms for randomizing treatment assignments across the network.
The randomized assignment mechanism is referred to as the \emph{experimental design}.
In this paper, we focus on selecting an experiment design in order to facilitate precise estimation of a single experimenter-specified causal effect.
Under SUTVA, independent treatment assignment is often an appropriate experimental design, as it facilitates estimation of the average treatment effect at the parametric rate.
However, the variance of typical network effect estimators under independent treatment can be prohibitively large, growing exponentially in the degrees.
The recent literature has highlighted that the best choices of experiment design should depend not only on the underlying network, but also on the causal effect under investigation.
For example, experimental designs for the global average treatment effect often involve clustering units and assigning treatments to clusters \citep[see e.g.,][]{Ugander2013Graph,Eckles2017Design,Harshaw2023Design,Ugander2023Randomized} while experimental designs for direct effects involve assigning treatment to an independent set in the network \citep{Karwa2018Systematic, Jagadeesan2020Designs, Fatemi2020Network}.

% PARAGRAPH 3 (what are the questions)
The work in this paper is motivated by two central questions in the design and analysis of network experiments.
The first question is of interest from methodological and theoretical perspectives:
\begin{center}
	\textit{
		Given a causal effect and an underlying network, what are the optimal rates of estimation of the causal effect from experimental data?
	}
\end{center}
It has been implicitly understood in the literature that the answer to this question depends not only on the underlying network but also on the causal effect under investigation.
For example, effects which are ``more global'' are believed to be harder to estimate than effects which are ``more local.''
The experimental design problem plays a central role in answering this question because it determines the correlation structure of observed outcomes.
Indeed, a sophisticated estimation procedure is unlikely to fix the problems of a poorly designed experiment.
%We remark that the model of interference plays an additional role in the answer to this question, but we keep the interference model fixed in this paper.

In the current literature on network experiments, the experimental designs are constructed using different kinds of graph theoretic ideas which are specific to a given causal effect.
As discussed above, graph clustering is typically used for global effects while independent sets are used for direct effect.
Our second motivating question is primarily of methodological interest:

\begin{center}
	\textit{
		Is there a unifying graph theoretic principle for designing network experiments for general causal effects?
	}
\end{center}
The benefit of a unifying approach for network experiment design is that it may be applied to all causal effects, including those for which specific experimental designs have not yet been developed. 
It stands to reason that the questions above are intertwined: understanding the optimal rates of estimation requires constructing optimal experiment designs which itself requires understanding the underlying principles of the experimental design problem.

% PARAGRAPH 4 (what are the main contributions)
In this work, we propose a general method for constructing experimental designs in order to facilitate precise estimation of a single experimenter-specified effect.
A central aspect of our methodology and theoretical analysis in the notion of a \emph{conflict graph}, which describes the fundamental unobservability in the causal estimation problem.
Specifically, the conflict graph encodes which pairs of units have estimand-relevant potential outcomes which cannot be simultaneously observed.
In this way, the conflict graph depends on both the underlying network and the causal effect.
We use the spectral properties of the conflict graph to guide the experiment design, which we call the Conflict Graph Design.
We propose a modified Horvitz—Thompson estimator and provide theoretical analysis of its variance under the Conflict Graph Design.
The main contribution of these methods is summarized in the following informally stated theorem (see Section~\ref{sec:estimation-analysis} for formal statements):

\begin{mainresult}[Informal] \label{thm:informal-main-result}
	For any network $G$ and aggregated contrastive causal estimand $\ate$, the Conflict Graph Design and unbiased effect estimator $\eate$ attain the following rate under mild regularity conditions on the potential outcomes:
	\[
	\Var{\eate} \leq \bigO[\Big]{ \frac{\lamH}{n} } \enspace,
	\]
	where $\cH$ is the conflict graph corresponding to $G$ and $\ate$ and $\lamH$ is the largest eigenvalue of its adjacency matrix.
\end{mainresult}
% PARAGRAPH  5 (brief interpretation of the rates)
One can view $\lamH$ as a global measure of connectivity of the conflict graph $\cH$.
In this way, our derived rates depend on global aspects of the fundamental unobservability constraints, as dictated by both the causal estimand and the underlying network.
Our results corroborate two implicitly understood points in the literature: (1) that in order to increase precision, experiment designs should be tailored to specific causal effects of interest and (2) that ``more local’’ effects are easier to estimate than ``more global’’ effects.

Perhaps surprisingly, this general approach to experiment design achieves the best known rates for the most well-studied causal effects in the literature.
We refer the reader to Section~\ref{sec:main-application} for a discussion of these improvements.
Although they constitute an improvement over prior work, we do not claim that these rates are optimal;
however, we believe that the methodology and theory in this work serves as a promising initial step towards answering the two questions raised above.

% PARAGRAPH 6 (technical contribution)
From a technical perspective, we are able to achieve these improved rates through an \emph{operator norm} analysis.
In fact, focusing on an operator norm analysis led to the development of the Conflict Graph Design and estimator, so the methods and analysis are intertwined in this sense.
The operator norm analysis stands in contrast to the dependency graph analysis which has been used throughout the literature on causal inference under interference.
The dependency graph technique produces rates which incur additional dependence on the maximum degree.
Our operator norm analysis is a secondary contribution of this work which is likely to be of independent interest.

% PARAGRAPH 7 (inference)
The final contribution of this work is the corresponding development of methods for inference of causal effects.
Our proposed confidence intervals are obtained by constructing a conservative estimator of the variance and using either Chebyshev or Wald-type intervals centered at the effect estimator.
We show that the expected width of these intervals enjoys the favorable $\sqrt{ \lamH / n }$ rates and provide sufficient conditions for their asymptotic coverage.

\subsection{Application of Main Result} \label{sec:main-application}

The main result in this paper is the methodological development of the Conflict Graph Design and modified Horvitz—Thompson estimator, together with the theoretical analysis of the resulting variance. 
In this section, we illustrate these main results by focusing on three estimands: global, direct, and spillover effects.
We highlight the improvement of the rates obtained in this work in comparison to existing literature.

\paragraph{Global Average Treatment Effect:}
The Global Average Treatment Effect (GATE) is perhaps the most well-studied causal effect under interference.
The estimand seeks to capture the causal effect of treating everyone versus treating no one.
Experiment designs for this problem have aimed to create clusters of units and apply treatments at the cluster level, so that it appears from a local point of view that either everyone received treatment or control.
To our knowledge, the previously best known rates of estimation for this causal effect are for the Randomized Graph Cluster Randomization design of \citet{Ugander2023Randomized}, which builds upon the Graph Cluster Randomization approach of \citet{Ugander2013Graph}.
The rates of $n \cdot \Var{\eate}$ in this work are $\bigO{\lambda(G^2) d_{\max} \kappa^3}$, where $\lambda(G^2)$ is the largest eigenvalue of the adjacency of the two-hop connected graph $G^2$, $d_{\max}$ is the maximum degree of $G$, and $\kappa$ is the restricted growth coefficient, which characterizes the growth of neighborhoods in the graph.
The rates hold under a uniform bound on all potential outcomes.
Moreover, the authors remark that $\kappa$ is typically on the order of $d_{\max}$.

In this setting, we show that the conflict graph $\cH$ is the two-hop connected graph $G^2$ and thus the Conflict Graph Design achieves the improved rate of $n \cdot \Var{\eate} = \bigO{\lambda(G^2)}$.
This improves the previously best known rates by a factor of $d_{\max}^4$ in typical settings where $\kappa = \bigTheta{d_{\max}}$.
Moreover, these rates hold under the weaker condition of bounded second moments of the potential outcomes.

\paragraph{Direct Treatment Effect:}
The direct treatment effect seeks to capture the (aggregate) causal effect of treatment on a particular unit when all other units are given the control.
While several works have focused on estimation of the direct treatment effect under a fixed design \citep[see e.g.,][]{Aronow2017Estimating}, there have been considerably fewer works exploring the experimental design problem.
A natural idea is to consider experiment designs based on random independent sets so that treated units have untreated neighbors.
This idea has been explored in various ways \citep{Karwa2018Systematic,Fatemi2020Network} but those methods lack formal statistical analysis.
To the best of our knowledge, the previously best known rates of estimation are due to the quasi-coloring approach of \citet{Jagadeesan2020Designs} which attains $n \cdot \Var{\eate} = \bigO{d_{\max}^2}$.

In this setting, we show that the conflict graph $\cH$ is the original graph $G$ with additional self-loops and thus the Conflict Graph Design achieves the improved rate of $n \cdot \Var{\eate} = \bigO{\lambda(G)}$.
The rate is an improvement by at least a factor of $d_{\max}$, as $\lambda(G) \leq d_{\max}$.
In fact, $\lambda(G)$ can sometimes be substantially smaller than $d_{\max}$.
This can best be seen when $G$ is a star graph where $d_{\max} = \bigTheta{n}$ while $\lambda(G) = \bigTheta{\sqrt{n}}$.
In this case, the Conflict Graph Design guarantees consistent estimation as $\Var{\eate} = \bigO{n^{-1/2}}$.
On the other hand, the analysis of the quasi-coloring design provided by \citet{Jagadeesan2020Designs} shows only that $\Var{\eate} = \bigO{n}$, which is an increasing function of the sample size.

\paragraph{Spill-over Effects:}
Spill-over effects aim to capture the (aggregate) causal effect on a particular unit of treating other nearby units.
While GATE and direct effects have fairly standardized definitions, spill-over effects may be defined in a variety of different ways.
We are not aware of prior work which specifically considers the experimental design problem to increase precision for spill-over effects.
We consider a type of spill-over effect where each unit has a set of neighboring ``seed subjects'' and we seek to understand the effect of treating the seed subjects on the original subject.
This formalization recovers the spill-over effects studied by \citet{Paluck2016Changing} in conflict mitigation experiments.
In Section~\suppref{sec:extention-to-marginal-effects} of the supplement, we consider extensions to ``anonymous'' spill-over effects, e.g. the effect of treating a random half of neighbors.

Our proposed approach to experiment design yields rates of estimation which depend on the complexity of the spill-over effect to be estimated.
In particular, the conflict graph $\cH$ depends on the overlapping structure of the seed subjects and the relevant quantity $\lamH$ becomes larger as the collection of seed subjects features more overlapping.
For spill-over effects defined by a collection of seeds which are small and minimally overlapping, our methods attain parametric rates, i.e. $n \cdot \Var{\eate} \leq \bigO{1}$.
For spill-over effect defined by a collection of seeds which are maximally overlapping, our methods attain  $n \cdot \Var{\eate} \leq \bigO{ \lambda(G^2) }$.
In this way, our experimental design and the corresponding rates adapt to the underlying complexity of the estimation problem, as measured by the overlap in seed subjects. 

\subsection{Related Work} \label{sec:related-work}

% Paragraph 1: General Potential Outcomes
We work in the potential outcomes framework of causal inference \citep{Neyman1923,Rubin1980Randomization,Holland1986Statistics}.
For textbook treatments of the subject, we refer readers to the books of \citet{ImbensRubin2015}, \citet{Hernan2020What}, and \citet{Ding2024First}.
We work more specifically in the design-based framework, where treatment assignment is the only source of randomness and serves as the sole basis for our inference.
For this reason, we focus our literature review primarily on causal inference within this framework.

% Paragraph 2: Early Interference
To the best of our knowledge, \citet{Cox1958} was the first to discuss the no-interference assumption while \citet{Rubin1980Randomization} provided its first formalization.
The Stable Unit Treatment Value Assumption (SUTVA) states that (among other things) treatment given to one unit does not affect the outcome of another unit.
Violations of this aspect of SUTVA are referred to as \emph{interference}.
Early investigations into causal inference under interference began with \emph{partial interference}, where it was assumed that interference occurs within, but not between, pre-specified clusters \citep{Sobel2006What, Hudgens2008, Liu2014, Liu2016, Rigdon2015, Tchetgen2012, VanderWeele2011}.

% Paragraph 3: Network Interference
More complex types of \emph{network interference} were introduced in the exposure mappings framework of \citet{Aronow2017Estimating} which has served as the basis for much of the interference literature since, though similar ideas may be found in \citet{Manski2013Identification}.
\citet{Aronow2017Estimating} propose to model interference between units through an exposure mapping function which is typically defined through the underlying network.
Causal effects are then defined as contrasts of outcomes under different exposures.
Their approach has been extended in a variety of ways including to observational settings \citep{Forastiere2021}, settings where the exposure mapping is only approximately correct \citep{Leung2022Causal}, as well as settings with more severely misspecified exposure mappings \citep{Saevje2021Average}.
The papers above do not consider the experimental design problem in the sense that they focus on effect estimation when the experimental design to considered to be fixed.
When considering contrastive treatment effects, the conditions for consistency in these works place strong assumptions on the underlying network (e.g. constant degree) which are typically invoked only implicitly (i.e. through assumptions on exposure probabilities).

%% Paragraph 4: Different Approaches to Experiment Design
%Broadly speaking, there are two approaches to the problem of experiment design.
%The first approach assumes a generative model on the potential outcomes or the network while keeping the experimental design relatively simple.
%For example, \citet{Li2022Random} consider estimation of direct and indirect effects when the network is drawn from a graphon model and treatment is assigned independently to each subject.
%By leveraging the network model, they show that improved rates of estimation are possible.
%\citet{Viviano2020Experimental} considers a two-stage network experiment where the potential outcomes are assumed to follow a network regression model and treatment assignment in the second round is deterministic.
%In this first approach, the generative assumptions were crucial to obtaining improved estimation procedures.
%The second approach to experiment design treats the potential outcomes and the network as being deterministic so that any precision improvement must come from the choice of experiment design.
%Our paper falls into the second approach, which we review in the remainder of the section.

% Paragraph 5: Experimental Designs for Estimating GATE
The experimental design problem has been considered extensively for the global average treatment effect, or variants thereof.
\citet{Ugander2013Graph} proposed Graph Cluster Randomization with the goal of reducing the variance of standard effect estimators.
This line of work was improved upon by \citet{Ugander2023Randomized} who gave the previously best known rates for this problem, as described in Section~\ref{sec:main-application}.
\citet{Eckles2017Design} described experimental designs which aim to minimize the bias of estimators under network interference.
Cluster designs have been also proposed in the bipartite experiment setting to estimate corresponding variants of the global average treatment effect \citep{Pouget2019Variance, Harshaw2023Design, Brennan2022Cluster} as well as low-degree interaction models \citep{Eichhorn2024Low}.

 % Paragraph 6: Experimental Designs for Estimating Direct Effect
As discussed in Section~\ref{sec:main-application}, the experimental design problem for the direct effect has received less attention.
In a survey of network experiments, \citet{Karwa2018Systematic} propose an experimental design based on the random greedy algorithm for constructing maximal independent sets.
In an appendix, \citet{Eckles2017Design} describe a ``hole--punching'' design which modifies Graph Cluster Randomization by inverting treatment on a small fraction of vertices. 
\citet{Fatemi2020Network} propose using a single deterministically chosen maximal independent set.
None of these works provides formal analysis of the variance and the positivity violations in last design imply that no unbiased estimator exists without further strong assumptions (e.g. constant treatment effects).
To the best of our knowledge, \citet{Jagadeesan2020Designs} propose the only experimental design for direct effect that includes an analysis of the variance, which is reviewed in Section~\ref{sec:main-application}.

% Paragraph 7: Model assisted
A complimentary literature has investigated model-assisted approaches to experimental design, where the experimenter posits a working statistical model on the potential outcomes in order to guide the choice of experimental design \citep{Basse2018Model,Toulis2013Estimation}.
Such an approach aims to ensure precision gains when the model is correct while still guaranteeing unbiased estimation if the model is wrong.
With varying degrees of model assistance / reliance, this approach has been applied to randomized saturation designs \citep{Baird2018Optimal}, cluster designs \citep{Viviano2025Causal}, and single-assignment alphabetical optimal design \citep{Zhang2022Locally, Zhang2023Asymptotics}.
While the model-based approach improves precision when the statistical model is correct, precision may be worsened otherwise.
In this paper, we focus on rates of estimation which only require the assumption of arbitrary neighborhood interference.

% Paragraph 8: Operator Norm Analysis
The operator norm analysis, which plays a central role in this work, is related to two strands in the literature.
The first strand uses the operator norm primarily as a means for evaluating experimental designs.
In the no-interference setting, \citet{Efron1971Forcing} refers to the largest eigenvalue of $\Cov{Z}$ as ``accidental bias'', which serves as a metric for the performance of an experiment design, a perspective which was later extended by \citet{Kapelner2021Harmonizing}.
\citet{Harshaw2022Design} present a general design-based framework which captures estimands and interference models beyond exposure mappings in which the operator norm was proposed as the relevant measure of consistency.
This work contributes to an emerging second strand of literature where the experimental design is carefully constructed around the operator norm.
The covariate balancing design of  \citet{Harshaw2024Balancing} is derived by attempting to simultaneously minimize two operator norms which they refer to as ``the balance-robustness trade-off''.
To the best of our knowledge, our paper is the first to make use of the operator norm analysis to construct experimental designs under interference and consequently attain rates which do not depend on the maximum degree.

\section{Design-Based Inference Framework} \label{sec:preliminaries}

%We work within the potential outcomes framework of causal inference \citep{Neyman1923,Rubin1980Randomization,ImbensRubin2015}, with a particular focus on the design-based setting.
%In this section, we review the preliminary concepts of network experiments in this setting.
%

\subsection{Interventions, Experiment Designs, and Potential Outcomes} \label{sec:preliminaries-design-based}

There are $n$ experimental subjects which are denoted by integers $i \in [n]$.
For each unit $i \in [n]$, we denote its individual treatment assignment as $Z_i \in \setb{0,1}$.
The \emph{intervention} $Z \in \setb{0,1}^n$ is the vector that collects all individual treatment assignments, 
\ifnum \value{spacesave}=1 {
	$Z = (Z_1, Z_2, \dots Z_n )$.
	
} \else {
	\[
	Z = (Z_1, Z_2, \dots Z_n ) \enspace.
	\]
} \fi
Informally, an experimental design is a randomized treatment mechanism which is chosen by (and thus completely known to) the experimenter.
Formally, we consider the measurable space $(\mathcal{Z}, \mathcal{F})$ whose sample space is the set of interventions $\mathcal{Z} = \setb{0,1}^n$ and the $\sigma$-algebra $\mathcal{F}$ is simply the power set.
The \emph{experimental design} is a probability measure $P : \mathcal{F} \to [0,1]$ which formally represents the treatment assignment mechanism.
In the design-based framework, treatment assignment is the only source of randomness and serves as the sole basis for inference (i.e. units and their outcomes are not assumed to be sampled from a super-population).
Thus, all probability statements are made with respect to the probability space induced by the experimental design.
We emphasize that only one intervention is selected according to the experimental design $P$.
Our goal in this work is to construct an experimental design to facilitate improved rates for causal inference under interference.

Each unit $i \in [n]$ has a potential outcome function $y_i : \mathcal{Z} \to \Reals$ which maps interventions to real-valued observable outcomes.
Because only one intervention is performed, only one outcome is actually observed.
However, all outcomes have the \emph{potential} to be observed, hence the name ``potential outcomes''.
We emphasize that in the design-based framework, the potential outcome functions are considered to be deterministic, with treatment assignment being the only source of randomness.
For each unit $i \in [n]$, the observed outcome
\ifnum \value{spacesave}=1 {
	$Y_i = y_i(Z)$
} \else {
	\[
	Y_i = y_i(Z)
	\]
} \fi
is random, as it depends on the intervention which is randomly selected according to the design.
We also allow for heterogeneity in the sense that potential outcomes are not assumed to be related across units, except through mild moment conditions.
So far, no structural assumptions have been made on the potential outcome functions.
However, we must place some such assumptions in order to facilitate meaningful inference \citep{Basse2018Limitations}.

\subsection{Arbitrary Neighborhood Interference} \label{sec:preliminaries-inference-model}

We work under the exposure mapping framework of \citet{Aronow2017Estimating}.
An exposure mapping is typically defined with respect to a network structure and its purpose is to formalize the notion that the potential outcome function of a given unit depends only on the treatments of its neighbors in the network.
We focus on an arbitrary neighborhood interference model, where a unit's potential outcome function can depend arbitrarily on their own treatment and treatment of their neighbors.

We suppose that the experimenter has collected network data on the participants in the experiment, denoted as $G = (V,E)$ with vertex set $V = [n]$ and undirected edge set $E$.
We write $N(i)$ to denote the neighborhood of unit $i$ in $G$ and we define the extended neighborhood as $\widetilde{N}(i) = N(i) \cup \setb{i}$.
The exposure set of unit $i$ is defined as $\Delta_i = \mathcal{P}(\widetilde{N}(i))$, where $\mathcal{P}$ denotes the power set.
For each unit $i \in [n]$, we define the \emph{arbitrary neighborhood exposure mapping} $d_i : \mathcal{Z} \to \Delta_i$ as
\ifnum \value{spacesave}=1 {
	$d_i(z) = \setb{ j \in \widetilde{N}(i) : z_j = 1}$.
} \else {
	\[
	d_i(z) = \setb[\Big]{ j \in \widetilde{N}(i) : z_j = 1}
	\enspace.
	\]
} \fi
In words, this exposure mapping essentially maps the entire intervention vector down to the sub-vector corresponding to the unit's extended neighborhood.
Given intervention $z \in \mathcal{Z}$ and exposure $e \in \Delta_i$, we say that unit $i$ \emph{receives exposure} $e$ under intervention $z$ if $d_i(z) = e$. 
The following assumption states that the potential outcome function only depends on the treatments assigned to units in the extended neighborhood, i.e. on that sub-vector.

\begin{assumption}[Arbitrary Neighborhood Interference] \label{assumption:ani-model}
	For all units $i \in [n]$ and interventions $z, z' \in \mathcal{Z}$ it holds that if $d_i(z) = d_i(z')$, then $y_i(z) = y_i(z')$.
	Equivalently, for each unit $i \in [n]$ there exists coefficients $\alpha_{i,S}$ for $S \subseteq \widetilde{\mathcal{N}}(i)$ such that the potential outcome function $y_i$ may be expressed as
	\[
	y_i(z) 
	= \sum_{S \subseteq \widetilde{\mathcal{N}}(i)} \alpha_{i,S} \cdot \prod_{j \in S} z_i \prod_{j \notin S} (1-z_i) 
	\enspace.
	\]
\end{assumption}

Observe that when $d_i(z) = S$ (i.e. $S$ is the set of treated units in the extended neighborhood) then the observed outcome is $Y_i = y_i(Z) = \alpha_{i,S}$.
Because the coefficients $\alpha_{i,S}$ correspond to observed outcomes under the exposures, we sometimes refer to them directly as potential outcomes.
We do not place any homogeneity assumptions on these potential outcomes $\alpha_{i,S}$, i.e. they are unit-specific and are not necessarily related in any particular way across different units.
Likewise, the potential outcomes for a specific unit are not assumed to be related in any particular way, i.e. $\alpha_{i,S}$ is not presumed to be related to $\alpha_{i,S'}$ in any way.

When the graph $G$ has no edges, then each unit's extended neighborhood contains just itself, i.e. $\widetilde{N}(i) = \setb{i}$.
In this case, the standard SUTVA setting is recovered.

The arbitrary neighborhood interference stands in contrast to linear-in-means network models and other types of anonymous interference, which posit that outcomes depend only on the number of treated neighbors \citep{Hudgens2008}.
While these potential outcome models may facilitate improved rates of effect estimation, they also place stronger assumptions on the permissible types of interference.
The arbitrary neighborhood interference model posited by Assumption~\ref{assumption:ani-model} presumes that a unit's outcomes are affected only by their 1-hop neighbors.
The model can be accommodated to include arbitrary interference of more units (e.g. 2-hop or more generally $k$-hop neighbors) simply by re-defining the graph $G$ to include additional edges to these units.
For simplicity, we keep the interference focused to the 1-hop neighbors.

\subsection{Causal Estimands} \label{sec:causal-estimands} \label{sec:preliminaries-estimands}

The causal estimands considered in this paper are defined as the contrast of potential outcomes under two different interventions, aggregated over all subjects in the experiment.
The estimands are formally defined as follows: for each unit $i \in [n]$, fix two \emph{contrasting interventions} $\zonei$ and $\zzeroi \in \mathcal{Z} = \setb{0,1}^n$ and define the \emph{individual treatment effect} $\tau_i$ as
\ifnum \value{spacesave}=1 {
	$
	\tau_i = y_i(\zonei) - y_i(\zzeroi)
	$.
} \else {
	\[
	\tau_i = y_i(\zonei) - y_i(\zzeroi) \enspace.
	\]
} \fi
Given the individual effects, the \emph{aggregated contrastive treatment effect} is defined as
\[
\tau 
= \frac{1}{n} \sum_{i=1}^n \tau_i 
= \frac{1}{n} \sum_{i=1}^n \braces[\big]{ y_i(\zonei) - y_i(\zzeroi) } \enspace.
\]
We remark that the contrasting interventions $\zonei$ and $\zzeroi$ can be unit-specific, which allows experimenters to specify various types of spill-over or direct effects.
Without stronger assumptions such as constant treatment effect, individual treatment effects cannot be estimated with any reasonable precision.
On the other hand, aggregated treatment effects can typically be estimated to high precision with large sample sizes under appropriate conditions.

The contrastive causal estimands considered above capture the majority of causal effects considered in the literature\footnote{In mediation analysis, the terms ``direct effect'' and ``indirect effect'' refer to different causal effects. See Section~\suppref{sec:supp-mediation} of the supplement for a detailed comparison.}, including the following:

\begin{itemize}
	\item \textbf{Global Average Treatment Effect}: 
	When the contrasting interventions are selected as $\zonei = \mathbf{1}$ and $\zzeroi = \mathbf{0}$, we recover the Global Average Treatment Effect, also referred to as the All-or-Nothing Treatment Effect.
	This effects captures the contrast between the policy where everyone gets the treatment versus one where everyone gets the control.
	
	\item \textbf{Direct Effects}: 
	A standard definition of direct effect is recovered with contrasting \emph{direct} intervention $\zonei = \vec{a}_i$ and all control intervention $\zzeroi = \vec{0}$, where $\vec{a}_i$ is the vector whose $i$th entry is $1$ and remaining entries are $0$.
	This estimand captures the effect of treatment on a unit, while fixing all other units to receive the control.
	
	\item \textbf{Spill-Over Effects}:
	For each unit $i \in [n]$, the experimenter specifies a non-empty set of neighboring \emph{seed subjects} $M_i \subseteq N(i)$.
	The \emph{treatment} intervention $\zonei$ is the one where only the seed units are treated, i.e. $\zonei(j) = \indicator{j \in M_i}$.
	The \emph{control} intervention is $\zzeroi = \vec{0}$.
	The resulting estimand captures the effect on unit $i$ of treating the seed subjects $M_i$, keeping treatment to all subjects fixed at control.
	Note that the effect itself depends on the specification of the seed subjects.
	We remark that this is one formalization which captures many, but not all, notions of spill-over effects.
\end{itemize}

The contrastive causal estimands have been defined independently of the model of interference.
However, it will often be advantageous to refer to the estimand in terms of the coefficients of the interference model.
For each unit $i \in [n]$, given the two contrasting interventions $\zonei$ and $\zzeroi$, we use the shorthand $y_i(e_1)$ and $y_i(e_0)$ to refer to the potential outcomes $\alpha_{i,S_{(1)}}$ and $\alpha_{i,S_{(0)}}$ where $d_i(\zonei) = S_{(1)}$ and $d_i(\zzeroi) = S_{(0)}$.
In other words, $y_i(e_1)$ and $y_i(e_0)$ are the potential outcomes which are being contrasted in the estimand, i.e. $\tau = (1/n) \sum_{i=1}^n y_i(e_1) - y_i(e_0)$.
We acknowledge that using $y_i(e_1)$ and $y_i(e_0)$ to refer to the relevant coefficients of the potential outcome function $y_i$ is an overloading of notation, but we think this shorthand is especially appealing in what is to follow.

We assume that for each unit $i \in [n]$, the two contrastive interventions result in different exposures, i.e. $d_i(\zonei) \neq d_i(\zzeroi)$.
This assumption is essentially without loss of generality: for units satisfying  $d_i(\zonei) = d_i(\zzeroi)$, the individual treatment effect is zero under the arbitrary neighborhood interference model so they do not contribute to the aggregated treatment effect and thus no estimation is required for these units.
Indeed, all relevant causal estimands satisfy this assumption.

In this paper, we focus on constructing experimental designs which are specifically tailored to estimate a particular causal estimand which has the contrastive form given above.
In Section~\suppref{sec:extention-to-marginal-effects} of the supplement, we extend the methods proposed in this paper to the broader class of marginalized contrastive effects, where the individual effects are defined as an average with respect to two distinct auxiliary measure, i.e. $\ate = (1/n) \sum_{i=1}^n \Esub{e \sim \mathcal{D}_{i,1}}{y_i(e)} - \Esub{e \sim \mathcal{D}_{i,0}}{y_i(e)}$.
These types of estimands arise in policy learning \citep{Athey2021Policy}, where the effect to be estimated is an average outcome under a randomized policy, as well as causal inference under misspecified exposure mappings \citep{Saevje2023Misspecified}, where the estimand of interest is the individual effect marginalized over the experimental design.

\subsection{Asymptotic Framework}\label{sec:asymptotic_framework_intro}

We analyze the proposed methods using both finite sample and large sample analysis.
For large sample analysis, which can be found in Sections~\ref{sec:asymptotic-analysis} and \ref{sec:inference}, we work with triangular array asymptotics, which have become standard in the design-based causal inference literature \citep{Freedman2008Regression,Lin2013Agnostic,Saevje2021Average,Leung2022Causal}.
Formally, we consider a sequence of experiments indexed by $n \in \Naturals$.
For a fixed $n \in \Naturals$, there is a network $G^{(n)}$ of $n$ subjects, each of whom has a potential outcome function $y_i^{(n)} : \setb{0,1}^n \to \Reals$ from which we define the aggregated treatment effect $\ate^{(n)}$.
At each $n \in \Naturals$, there is an experimental design (i.e. probability measure over interventions $\setb{0,1}^n$) and a resulting effect estimator $\eate^{(n)}$.
Asymptotic statements are understood with respect to this triangular asymptotic framework.

The network and potential outcome functions are not considered to be related across experiments in the sequence except through mild moment bounds; in other words, we do not assume that the outcomes or the graphs are generated from an asymptotic model.
For positive sequences $x_n$ and $y_n$, we write $x_n = \bigO{y_n}$ if $\limsup x_n / y_n < \infty$, $x_n = \bigOmega{y_n}$ if $\liminf x_n / y_n > 0$, and $x_n = \bigTheta{y_n}$ if $x_n = \bigO{y_n}$ and $x_n = \bigOmega{y_n}$.
In keeping with the convention of the literature, we drop the superscript $n$ for notational clarity.

\section{The Conflict Graph Design and Effect Estimator} \label{sec:design-and-estimator}

In this section, we present the Conflict Graph Design and a corresponding Horvitz--Thompson style estimator for estimating aggregated causal effects under arbitrary neighborhood interference.
The experimental design and estimator are tailored to a specific choice of causal effect so that changing the causal estimand results in a different experimental design and estimator.

\subsection{Conflict Graph}\label{sec:conflict_graph}
	
One of the key constructions in the proposed experimental design is the conflict graph.
Intuitively, the conflict graph will encode information about which pairs of units have estimand-relevant potential outcomes which cannot be simultaneously observed.

\begin{definition}[Conflicting Units] \label{definition:conflict}
	Two units $i, j \in [n]$ are said to \emph{conflict} if there exists $k, \ell \in \setb{0,1}$ and contrastive interventions $\contrastz{i}{k}$ and $\contrastz{j}{\ell}$ such that there does not exist an intervention $z \in \mathcal{Z}$ satisfying $d_i(z) = d_i(\contrastz{i}{k})$ and $d_j(z) = d_j(\contrastz{j}{\ell})$.
\end{definition}

In other words, units $i$ and $j$ are conflicting if there is no intervention under which unit $i$ can receive exposure $d_i(\contrastz{i}{k})$ and unit $j$ can receive exposure $d_j(\contrastz{j}{\ell})$.
Under the arbitrary neighborhood exposure assumption, this means that potential outcomes $y_i(e_k)$ and $y_j(e_\ell)$ can never be observed simultaneously.
In this way, our definition of conflict encodes a fundamental unobservability between outcomes of units that are relevant for the causal estimand.
Whether two units are in conflict depends not only on the underlying network but also on the causal estimand under investigation.

The \emph{conflict graph} $\cH= (V_\cH, E_\cH)$ is defined as follows: the experimental subjects are the vertices $V_\cH = [n]$ and there is an edge between unit $i$ and $j$ if they are in conflict.
Let $\mathcal{N}(i)$ denote the experimental subjects $j \in [n]$ which are adjacent to $i$ in the conflict graph.
Note that the script symbols are reserved for the conflict graph, i.e. the original graph $G$ has neighborhoods $N(i)$ while the conflict graph $\cH$ has neighborhoods $\mathcal{N}(i)$.
Our notion of conflict permits that $i$ conflicts with itself, meaning that $\cH$ may contain self-edges.
Indeed our assumption that the contrastive interventions yield distinct exposures for each unit means that every node of $\cH$ contains a self-edge.

\begin{figure}[t]
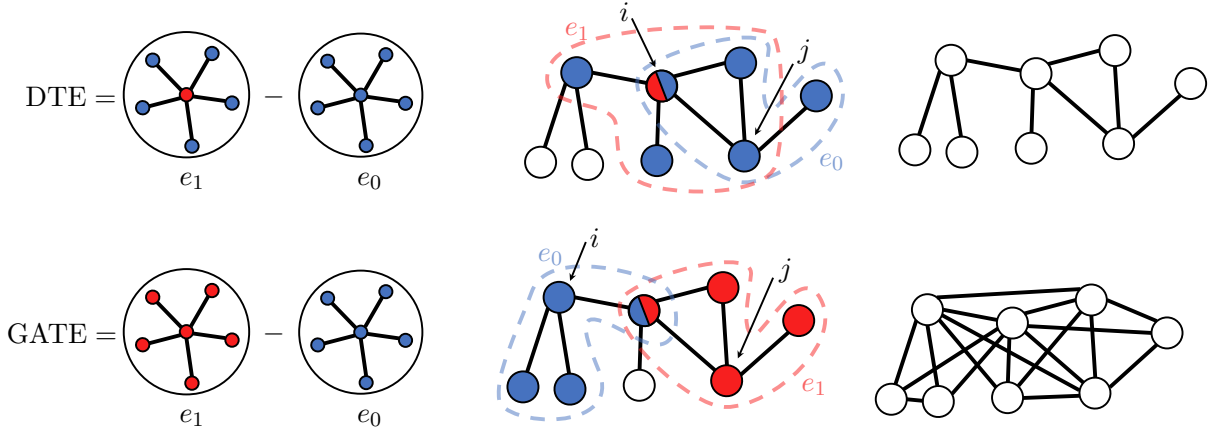
 
	\resizebox{\textwidth}{!}{
		\tikzset{every picture/.style={line width=0.75pt}} %set default line width to 0.75pt

\begin{tikzpicture}[x=0.75pt, y=0.75pt, yscale=-0.8,xscale=0.8]
% , show background rectangle

\begin{scope}[xshift=-110, yshift=15, xscale=0.9, yscale=0.9]
    \input{\figpath/conflict-graph/dte-desc}
\end{scope}

\begin{scope}[xshift=80, yscale=0.6, xscale=0.6]
    \input{\figpath/conflict-graph/dte-conflict-1hop}
\end{scope}

\begin{scope}[xshift=270, yscale=0.6, xscale=0.6]
    \input{\figpath/conflict-graph/dte-conflict-graph}
\end{scope}

\begin{scope}[xshift=-110, yshift=130, xscale=0.9, yscale=0.9]
    \input{\figpath/conflict-graph/gate-desc-v2}
\end{scope}

\begin{scope}[xshift=80, yshift=115, yscale=0.6, xscale=0.6]
    \input{\figpath/conflict-graph/gate-conflict-2hop}
\end{scope}

\begin{scope}[xshift=250, yshift=115, yscale=0.6, xscale=0.6]
    \input{\figpath/conflict-graph/gate-conflict-graph}
\end{scope}

% \draw [black!70]   (257,215.8) .. controls (302.54,217.38) and (321.03,117.04) .. (389.7,111.93) ;
% \draw [shift={(391.8,111.8)}, rotate = 177.4] [fill=black!70  ][line width=0.08]  [draw opacity=0] (10.72,-5.15) -- (0,0) -- (10.72,5.15) -- (7.12,0) -- cycle    ;
% %Curve Lines [id:da17867640169249244] 
% \draw [black!70]   (257,223.8) .. controls (304.14,225.38) and (324.98,343.8) .. (393.7,343.06) ;
% \draw [shift={(395.8,343)}, rotate = 177.4] [fill=black!70  ][line width=0.08]  [draw opacity=0] (10.72,-5.15) -- (0,0) -- (10.72,5.15) -- (7.12,0) -- cycle    ;

% \node[black!80,text width=6.3cm] at (280, 80) {\small \textbf{Case (I)}. \shellelpsd{}$(\cF)$ returns a transformation $\phi' \in \Phi_\text{FP} \cap \cF$. We terminate and return $\phi'$.};

% \node[black!80,text width=6.4cm] at (270, 360) {\small \textbf{Case (II)}. \shellelpsd{}$(\cF)$\\ returns a separating frontier. We curtail $\tilde\Phi$ and increase the radius $q$.};

\end{tikzpicture}
	}
	\caption{
		Example conflict graphs for the Direct Treatment Effect (DTE) and the Global Average Treatment Effect (GATE).
		For each effect, the left panel shows the contrastive exposures, the middle panel shows examples of conflicting units, and the right panel shows the full conflict graph.
	}
	\label{fig:conflict-graph-examples}
\end{figure}

In general, the conflict graph will depend both on the underlying network as well as the causal estimand under investigation.
See Figure~\ref{fig:conflict-graph-examples} for specific examples of conflict graphs.
Below, we present the conflict graphs for the causal estimands of interest presented in Section~\ref{sec:causal-estimands}.

\begin{itemize}
	\item \textbf{Global Average Treatment Effect}: 
	The conflict graph for the global average treatment effect is $\cH = G^2$, where there is an edge between $i$ and $j$ if they are within a distance of at most $2$ in $G$.
	To see this, suppose that $i$ and $j$ are within a distance of at most $2$ in $G$ so that they share a common neighbor $r \in \widetilde{N}(i) \cap \widetilde{N}(j)$.
	If unit $i$ receives the ``all treated'' exposure then $z_r = 1$ so that it is impossible for unit $j$ to receive the ``all control'' exposure, which would require $z_r = 0$.
	For this reason, units $i$ and $j$ are in conflict.
	
	\item \textbf{Direct Effect}:
	The conflict graph for the direct effect is $\cH = G^1$ where there is an edge between $i$ and $j$ if they are within a distance at most $1$ in $G$, i.e. $G^1$ is the original graph $G$ with additional self-edges.
	To see this, suppose that $i$ and $j$ are within a distance of at most $1$ so that $j \in \widetilde{N}(i)$.
	If unit $i$ receives the ``direct treatment'' exposure then $z_i = 1$ and $z_j = 0$, which means that unit $j$ cannot also receive the ``direct treatment'' exposure simulateously.
	In contrast to the GATE, units $i$ and $j$ which are distance $2$ in $G$ are not in conflict for the direct effect, as both contrastive exposures require that the common neighbors $r \in \widetilde{N}(i) \cap \widetilde{N}(j)$ receive $z_r = 0$, so that there is no conflict.
	
	\item \textbf{Spill-Over Effect}: 
	The conflict graph for the spill-over effect, as defined in Section~\ref{sec:causal-estimands} depends on the overlapping structure of the seed subjects. 
	Two distinct units $i \neq j$ are adjacent in the conflict graph if either $i \in M_j$, $j \in M_i$, or $M_i \cap M_j \neq \emptyset$.
	If $M_i$ contains the entire neighborhood $N(i)$ then $\cH = G^2$ and $\lamH = \lambda(G^2$).
	On the other hand, suppose that seed subjects form pairs i.e. $\abs{M_i} = 1$ with the property that if $M_i = \setb{j}$ then $M_j = \setb{i}$.
	In this case, $\cH$ is a perfect matching between these pairs so that $\lamH = 2$.
\end{itemize}

Our analysis relies on insights from spectral graph theory and so we quickly review a few relevant concepts here and refer readers to \citet{Spielman2019Book}.
The adjacency matrix $\adjH$ of the conflict graph is a symmetric $n$-by-$n$ matrix where $\adjH(i,j) = 1$ if $i$ and $j$ are adjacent in $\adjH$ and $0$ otherwise.
Because $\adjH$ is symmetric, all of its eigenvalues are real and its eigenvectors form an orthonormal basis.
Because each unit is in conflict with itself, $\cH$ has self-loops for every vertex and $\adjH(i,i) = 1$ for every $i \in [n]$.
We remark that it is more common in the literature to study graphs without self-loops.
Adding self-loops to every vertex of a graph is equivalent to adding the identity matrix to the adjacency, which simply shifts the eigenvalues by $+1$.
Thus, some familiar statements about eigenvalues of an adjacency matrix may feature an additional $+1$ due to the addition of self loops.

The largest eigenvalue in absolute value will be positive and we denote this by $\lamH$.
Because all diagonal entries of $\adjH$ are $1$, we have that $\lamH \geq 1$.
The largest eigenvalue $\lamH$ lies between the average and the maximum degree, i.e. $d_{\textrm{avg}}(\cH) \leq \lamH \leq d_{\textrm{max}}(\cH)$, where the degree of a vertex in the conflict graph $\cH$ counts the self-loop.
If all vertices in $\cH$ had the same degree $d$, then $ d_{\textrm{avg}}(\cH) = d_{\textrm{max}}(\cH) = d$, so that $\lamH = d$ as well.
However, if the underlying graph exhibits degree heterogeneity (i.e. $d_{\textrm{avg}}(G) \ll d_{\textrm{max}}(G)$) then the conflict graph will typically also experience such degree heterogeneity (i.e. $d_{\textrm{avg}}(\cH) \ll d_{\textrm{max}}(\cH)$).
In this case, it may happen that $d_{\textrm{avg}}(\cH)$, $\lamH$, and $d_{\textrm{max}}(\cH)$ are all of different asymptotic orders.
As an extreme example, if $\cH$ is a star graph then $d_{\textrm{avg}}(\cH) = \bigTheta{1}$ and $d_{\textrm{max}}(\cH) = \bigTheta{n}$ while $\lamH = \bigTheta{\sqrt{n}}$.
As the level of degree heterogeneity increase, the difference between the maximum degree and the maximum eigenvalue becomes more apparent.

\ifnum \value{spacesave}=1 {
} \else {
	The Conflict Graph Design will require $\lamH$ as input.
	One of the simplest methods for computing the largest eigenvalue is the power method \citep{Golub96} and its various modifications.
	In practice, $\lamH$ may be numerically computed using standard solvers e.g. LAPACK \citep{LAPACK}.
} \fi

Within the literature on causal inference under interference, several authors have proposed alternative graphs which are also meant to describe relevant aspects of the interference.
The interference dependence graph considered by \citet{Saevje2021Average} for analyzing estimates of the Expected Average Treatment Effect contains the units as vertices where a directed edge from $i$ to $j$ means that the potential outcome function $y_j$ depends on treatment $z_i$.
\citet{puelz2022graph} introduce the null-exposure graph for constructing powerful permutation-type hypothesis tests.
The null-exposure graph is bipartite with units $i \in[n]$ and assignments $\vec{z} \in \setb{0,1}^n$ as the two vertex sets, where an edge between $i$ and $\vec{z}$ means that $i$ receives one of the relevant exposures $e_1$, $e_0$ under assignment $\vec{z}$.
The dependency graph is a general type of construction widely used to establish consistency and CLTs of effect estimators \citep[see e.g.][]{Aronow2017Estimating, Ogburn2024Causal, Harshaw2023Design}.
Its vertices are units and the edge set encodes independence of individual effect estimators under the experimental design.
Note that the dependency graph depends on the experimental design, whereas the conflict graph depends only on fundamental unobservability between units' exposures and does not involve any aspect of randomization.
The differences between all of these graph reflects the different purposes they serve.
For example, the  salient feature of the null-exposure graph is its maximal bicliques, which provide exactly the information relevant for constructing permutation tests \citep{puelz2022graph}.
On the other hand, the bicliques in the null exposure graph are not directly relevant for constructing an experimental design to obtain the $\lamH / n$ rates we describe below.

\subsection{Importance Orderings} \label{sec:importance-ordering}

A key aspect of the Conflict Graph Design is a method for resolving conflicts among neighboring units in the conflict graph which seek to be assigned incompatible exposures.
In order to resolve these conflicts, the Conflict Graph Design uses what we refer to as an importance ordering.
In this section, we provide a definition of importance orderings as well as two constructions.

An \emph{ordering} of the experimental subjects $i \in [n]$ is a bijection $\pi : [n] \to [n]$ that maps units to orderings, i.e. $\pi(i)$ is the position of unit $i$ in the ordering and $\pi^{-1}(p)$ is the unit in position $p$ of the ordering.
We take the convention that the earlier units in the ordering are the most \emph{important}, i.e. $\pi^{-1}(1)$ is the most important unit and $\pi^{-1}(n)$ is the least important unit.
Given an ordering and a conflict graph, we define $\mathcal{N}^\pi_b(i)$ to be the more important neighbors, i.e.
\ifnum \value{spacesave}=1 {
	$
	\mathcal{N}^\pi_b(i) = \setb{ j \in \mathcal{N}(i) : \pi(j) < \pi(i) } 
	$.
} \else {
	\[
	\mathcal{N}^\pi_b(i) = \setb[\big]{ j \in \mathcal{N}(i) : \pi(j) < \pi(i) } \enspace.
	\]
	\vspace{-12pt}
} \fi	
\begin{definition}[Importance Ordering] \label{def:importance-ordering}
	An ordering $\pi$ is an \emph{importance ordering} with respect to conflict graph $\cH$ if $\abs{ \mathcal{N}_b^\pi(i) } \leq \lamH-1$ for all units $i \in [n]$.
\end{definition}
In an importance ordering, each unit has at most $\lamH-1$ more important neighbors.
We remark that it is not a priori clear that an ordering with this property exists for all graphs.
In particular, recall that under degree heterogeneity of the conflict graph, the largest eigenvalue $\lamH$ will typically be asymptotically dominated by the maximum degree $d_{\max}(\mathcal{H})$.
In these cases, an importance ordering ensures that for each of the large degree nodes, only a fraction of their neighbors will be more important.
In this case, a random ordering would (with high probability) not form an importance ordering, as each high degree node would have a constant probability of having half of its neighbors appear before it in the ordering.
Thus, the condition of an importance ordering is more restrictive than it may appear at first glance.
We provide two constructions of importance orderings, both of which are derived using insights from spectral graph theory.

We first present an importance ordering based on the leading eigenvector of the adjacency matrix.
Suppose that the conflict graph $\cH$ is connected.
By the Perron-Frobenius theorem, the leading eigenvector $\ev \in \Reals^n$ of the adjacency matrix $\adjH$ may be chosen to have positive entries.
The \emph{eigenvector ordering} $\eigpi$ is obtained by sorting the entries of $\ev$ from highest to lowest, breaking ties arbitrarily.
In other words, $\argmax_{i \in [n]} \ev(i)$ is placed in the first position in the ordering and $\argmin_{i \in [n]} \ev(i)$ is placed in the last position.
If the conflict graph $\cH$ is not connected, then the eigenvector ordering is obtained by arbitrarily concatenating the eigenvector ordering of all its connected components.

\expcommand{\eigenordering}{%
	The eigenvector ordering $\eigpi$ is an importance ordering.
}
\begin{proposition} \label{prop:eigenvector-ordering-is-important}
	\eigenordering
\end{proposition}

The Conflict Graph Design requires an importance ordering be specified.
For the purposes of our analyses, any valid importance ordering may be used.
Some experimenters may prefer to use the eigenvector ordering as it may be readily obtained via a power iteration algorithm.
In Appendix~\suppref{sec:sequential-degree-ordering-appendix}, we present a second construction of an importance ordering based on sequentially removing minimum degree vertices in the conflict graph.
This construction may be preferable for larger networks because there are no issues with numerical accuracy and it may be obtained in $\bigO{(n+m) \log(n)}$ time, where $m$ is the number of edges in $\cH$.

% connection to Ugander & Yin (2023) spectral weights, found in Section 4.2
The eigenvector ordering bears resemblance to the spectral weighting technique used in the Randomized Graph Cluster Randomization (RGCR) designs of \citet{Ugander2023Randomized}.
While both approaches aim to increase the individual probabilities that each unit receives either exposure, the importance ordering has the benefit of controlling the pair-wise exposure probabilities for pairs of non-conflicting units (Lemma~\ref{lemma:prob-of-desired-exposure-events}).
Additionally, the RGCR designs are developed only for the Global Average Treatment Effect and will not generally attain $\lamH / n$ rates of consistency.

The notion of importance ordering has interesting connections to spectral graph theory.
\citet{Wilf1967Eigenvalues} showed that the chromatic number of a graph $G$ is at most $\lambda(G) + 1$ and his proof relies on the sequential degree-based importance ordering.
One may view the eigenvector ordering described in this section as an alternative proof of Wilf's theorem, which may be of independent interest.

\subsection{Conflict Graph Design} \label{sec:our-design}

\begin{figure}[t]
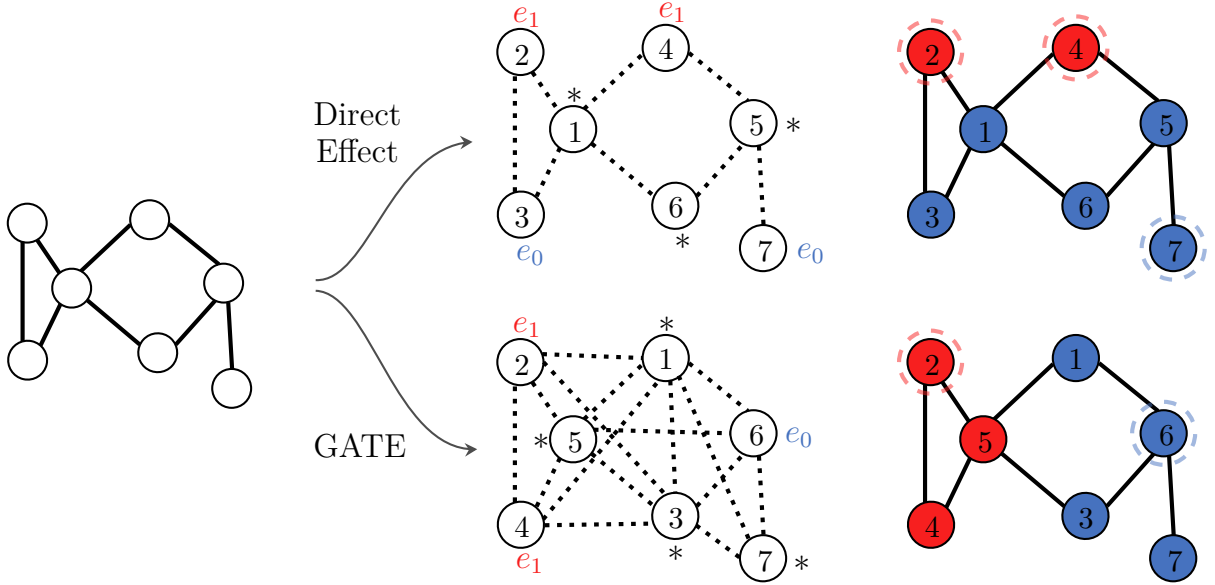

	\centering
	\resizebox{\textwidth}{!}{
		\tikzset{every picture/.style={line width=0.75pt}} %set default line width to 0.75pt

\begin{tikzpicture}[x=0.75pt, y=0.75pt, yscale=-0.8,xscale=0.8]
% , show background rectangle

\begin{scope}[xshift=-110, xscale=0.7, yscale=0.7]
    \input{\figpath/algorithm/graph}
\end{scope}

\begin{scope}[xshift=120, yshift=-70, yscale=0.5, xscale=0.5]
    \input{\figpath/algorithm/dte-step1}
\end{scope}

\begin{scope}[xshift=305, yshift=-70, yscale=0.5, xscale=0.5]
    \input{\figpath/algorithm/dte-step2}
\end{scope}

\begin{scope}[xshift=120, yshift=70, yscale=0.5, xscale=0.5]
    \input{\figpath/algorithm/gate-step1}
\end{scope}

\begin{scope}[xshift=305, yshift=70, yscale=0.5, xscale=0.5]
    \input{\figpath/algorithm/gate-step2}
\end{scope}

\begin{scope}[xshift=-80, yshift=-70, xscale=0.7, yscale=0.8]
    \draw [black!70]   (257,215.8) .. controls (302.54,217.38) and (321.03,117.04) .. (389.7,111.93) ;
    \draw [shift={(391.8,111.8)}, rotate = 177.4] [fill=black!70  ][line width=0.08]  [draw opacity=0] (10.72,-5.15) -- (0,0) -- (10.72,5.15) -- (7.12,0) -- cycle    ;
    %Curve Lines [id:da17867640169249244] 
    \draw [black!70]   (257,223.8) .. controls (304.14,225.38) and (324.98,343.8) .. (393.7,343.06) ;
    \draw [shift={(395.8,343)}, rotate = 177.4] [fill=black!70  ][line width=0.08]  [draw opacity=0] (10.72,-5.15) -- (0,0) -- (10.72,5.15) -- (7.12,0) -- cycle    ;    
\end{scope}

\draw (70,-30) node [anchor=north west][inner sep=0.75pt]   [align=center] {\large Direct\\ \large Effect};

\draw (70,170) node [anchor=north west][inner sep=0.75pt]   [align=center] {\large GATE};

\end{tikzpicture}
	}
	\caption{
		Illustration of the Conflict Graph Design for Direct Effect and GATE.
		The left panel contains the original graph.
		The middle panel shows the conflict graph, an importance ordering, and the first step of the algorithm: sampling a realization of the desired exposures $U_i$.
		The right panel shows the resulting intervention on the original graph, where a dashed circle around unit $i$ indicates that the event $E_{(i, k)}$ occurred.}
	\label{fig:algorithm}
\end{figure}

The Conflict Graph Design is presented in Algorithm~\ref{alg:conflict-graph-design}.
The design requires the conflict graph $\cH$, its largest eigenvalue $\lamH$, and an importance ordering $\pi$ as input.
A random intervention $Z \in \mathcal{Z} = \setb{0,1}^n$ is produced using a two-step procedure.
Figure~\ref{fig:algorithm} contains a visual depiction of the algorithm for the direct effect and GATE.

\expcommand{\cgdalgobody}{%
	\KwIn{Importance ordering $\pi$, maximum eigenvalue $\lamH$, and conflict graph $\mathcal{H}$.}
	\KwOut{Random intervention $Z \in \mathcal{Z} = \setb{0,1}^n$}
	Set sampling parameter $r = 2$\;
	Sample desired exposure variables $U_1, \dots, U_n$ independently and identically as
	$$
	U_i \gets
	\left\{
	\begin{array}{lr}
		e_1 & \text{with probability } \frac{1}{r \cdot 2 \lamH}\\
		e_0 & \text{with probability } \frac{1}{r \cdot 2 \lamH}\\
		* & \text{with probability } 1 - \frac{1}{ r \cdot \lamH }
	\end{array}
	\right.
	$$
	Initialize the intervention vector $Z \gets \vec{0}$. \;
	\For{$i =1 \dots n$}{
		\If{$U_i = e_k \in \setb{e_1, e_0}$ and $U_j = *$ for all $j \in \mathcal{N}^\pi_b(i)$\label{algline:exposure-condition}}{
			Update intervention vector: set $Z_j = z_{i}^{(k)}(j)$ for all $j \in \widetilde{N}(i)$
			\label{algline:set-intervention} \;
		}
	}
}
\begin{restatable}{algorithm}{cgdalgo}
    \DontPrintSemicolon
    \caption{\ourdesign{}}\label{alg:conflict-graph-design}
    \cgdalgobody
\end{restatable}

In the first step, each unit $i \in [n]$ receives a \emph{desired exposure variable} $U_i$, which are independent and identical random variables taking values in $ \setb{e_1, e_0, *}$.
As the name suggests, the desired exposure variables $U_1 \dots U_n$ represent the exposures that we would like to assign to each unit, where $e_1$ and $e_0$ correspond to the two contrastive exposures under consideration and ``$*$'' is a ``null option''.
The probability of each of these values is selected according to the largest eigenvalue $\lamH$ and a sampling parameter $r \geq 1$.
These probabilities are well-defined because $\lamH \geq 1$.
While the sampling parameter $r \geq 1$ can be adjusted, we recommend setting $r=2$ as this yields the near-optimal constants in our analysis (see Theorem~\ref{thm:variance-analysis-finite-sample} and following discussion).
It will often be the case that neighboring units $i$ and $j$ in the conflict graph receive desired exposure variables $U_i$ and $U_j$ which no intervention can realize simultaneously.

In the second step, these conflicts among incompatible desired exposure variables are resolved.
The conflict resolution scheme is simple: each unit defers to the "desires" of their more important neighbors.
If a unit $i \in [n]$ receives a non-null desired exposure variable $U_i \in \setb{e_1, e_0}$, then we look to see the desired exposure variables of his more important neighbors.
If each more important neighbor receives the null desired exposure variable (i.e. $U_j = *$ for all $j \in \mathcal{N}_b^\pi(i)$) then the algorithm updates the intervention vector $Z$ so that unit $i$ is guaranteed to receive the desired exposure $U_i$.
On the other hand, if at least one of the more important neighbors also receives a non-null desired exposure variable $U_j \in \setb{e_1, e_0}$, then no update to the intervention vector is made on behalf of unit $i$.

One could interpret the Conflict Graph Design as aiming to emulate an infeasible Bernoulli design where each unit receives exposures in an i.i.d. fashion.
Of course, such a design is infeasible due to the nature of interference and so conflicts will necessarily arise.
The primary goal of the Conflict Graph Design is to mitigate the effect of these conflicts on the variance of an effect estimator, presented more formally in the next section.
We remark that when the graph $G$ has no edges (which corresponds to no interference) and the sampling parameter is set as $r=1$, then the Conflict Graph Design is precisely the Bernoulli design where treatments are assigned independently with probability $1/2$.

Sampling an intervention from the Conflict Graph Design is computationally efficient, requiring only $\bigO{n + m}$ operations where $m$ is the number of edges in the conflict graph $\cH$.
The pre-processing step of constructing the conflict graph $\cH$, computing its largest eigenvalue $\lamH$, and constructing an importance ordering $\pi$ are the more computationally intensive steps, though they can also be carried out efficiently using existing methods.
Moreover, these pre-processing computations only need to be performed once.
The linear time sampling is helpful, for example, when testing sharp null hypothesis via re-sampling.

\subsection{Effect Estimator} \label{sec:effect-estimator}

In order to estimate the aggregated treatment effect, we use a modified Horvitz--Thompson estimator.
Before presenting our estimator, it will be useful to review the standard Horvitz--Thompson estimator for the arbitrary neighborhood interference model \citep{Aronow2017Estimating}.
The  \emph{standard Horvitz--Thompson estimator} is defined as 
\begin{equation}\label{eq:standard_ht}
\eate_{\textrm{standard}}
= \frac{1}{n} \sum_{i=1}^n Y_i \cdot \paren[\Bigg]{ 
	\frac{\indicator[\big]{d_i(Z) = d_i(\zonei)}}{\Pr[\big]{d_i(Z) = d_i(\zonei)}} 
	-  
	\frac{\indicator[\big]{d_i(Z) = d_i(\zzeroi)}}{\Pr[\big]{d_i(Z) = d_i(\zzeroi)}} 
}
\enspace.
\end{equation}
Here, the observed outcome $Y_i$ is used in the estimator under the event that $i$ receives one of the two contrastive exposures.
We will use a modified Horvitz--Thompson estimator which is constructed specifically for the Conflict Graph Design.
For each unit $i \in [n]$ and exposure $k \in \setb{0,1}$, define the \emph{desired exposure event} $E_{(i,k)}$ as
\[
E_{(i,k)} = \setb[\Big]{ U_i = e_k \text{ and } U_j = * \text{ for all } j \in \mathcal{N}_b^\pi(i)  }
\enspace.
\]
The Conflict Graph Design ensures that under the desired exposure event $E_{(i,k)}$, unit $i$ receives the contrastive exposure $e_k$, i.e. $d_i(Z) = d_i(\contrastz{i}{k})$ and so $Y_i = y_i(e_k)$.

\expcommand{\desiredexposures}{%
	Under the event $E_{(i,k)}$, we have that $d_i(Z) = d_i(\contrastz{i}{k})$.
}
\begin{lemma} \label{lem:desired_exposures}
	\desiredexposures
\end{lemma}

Let us explain the intuition behind Lemma~\ref{lem:desired_exposures}.
Suppose $E_{(i,k)}$ occurs.
Then, in Line~\ref{algline:set-intervention}, the algorithm will assign individual assignments to unit $i$ and its neighbors so that it receives the desired exposure.
We claim that the algorithm will not ``over-write'' these individual assignments later in the execution.
To see this, observe that only neighbors of $i$ in the conflict graph could potentially cause such a change.
However, event $E_{(i,k)}$ ensures that the more important neighbors will not affect the neighborhood of $i$, since they receive value $*$. 
As for the less important neighbors, they ``yield'' to the desires of their more important neighbor $i$ and will also not modify the neighborhood.
This proof highlights the role of the null option ``$*$'' in this design.
Intuitively, when an important node is assigned $*$, then it does not have any desired exposure, thus allowing less important nodes to receive their desired exposures.
In that way, all nodes have a reasonable chance (quantified in Lemma~\ref{lemma:prob-of-desired-exposure-events} below) of receiving their desired exposures in the final treatment vector. 

By the preceding discussion, we know that all units $i$ for which $E_{(i,k)}$ occurs will receive their desired exposure and thus their observed outcome is relevant for estimating the effect. Thus, a reasonable approach would be to base our estimator on the observed outcomes of these units. 
Indeed, the \emph{modified Horvitz--Thompson estimator} $\eate$ is defined using the desired exposure events:
\[
\eate = \frac{1}{n} \sum_{i=1}^n Y_i \cdot \paren[\Bigg]{ 
	\frac{\indicator[\big]{E_{(i,1)}}}{\Pr[\big]{E_{(i,1)}} } 
	-  
	\frac{\indicator[\big]{E_{(i,0)}}}{\Pr[\big]{E_{(i,0)}} } 
}
\enspace.
\]
The modified Horvitz--Thompson estimator requires not only the observed outcomes $Y_1 \dots Y_n$, but also the values of the desired exposure variables $U_1 \dots U_n$ and the importance ordering $\pi$, both of which are required to evaluate the indicator $\indicator{E_{(i,k)}}$.
While standard Horvitz--Thompson estimators require computationally expensive Monte Carlo sampling to estimate exposure probabilities under complex designs, the probabilities of the desired exposure events under the Conflict Graph Design may be computed exactly from the importance ordering:

\expcommand{\probexposure}{%
	For unit $i \in[n]$ and exposure $k \in \setb{0,1}$, the probability of $E_{(i,k)}$ is
	\[
	\Pr[\big]{ E_{(i,k)} } = \frac{1}{r \cdot 2 \lamH} \cdot \paren[\Big]{ 1 - \frac{1}{r \cdot \lamH} }^{\abs{ \mathcal{N}_b^\pi(i)}}
	\]
	For a pair of distinct units $i \neq j \in [n]$ and exposures $k, \ell \in \setb{0,1}$, the probability of the intersection of desired exposure events is $\Pr{ E_{(i,k)} \cap E_{(j,\ell)}  } = 0$ if $i$ and $j$ are adjacent in $\cH$ and otherwise
	\[
	\Pr[\big]{ E_{(i,k)} \cap E_{(j,\ell)} } 
	= \paren[\Big]{ \frac{1}{r \cdot 2 \lamH} }^2 \cdot 
	\paren[\Big]{ 1 - \frac{1}{r \cdot \lamH} }^{\abs{ \mathcal{N}_b^\pi(i) \cup \mathcal{N}_b^\pi(j)}}  
	\enspace.
	\]
}
\begin{lemma} \label{lemma:prob-of-desired-exposure-events}
	\probexposure
\end{lemma}

\begin{corollary}
	There exist positive constants $c_1 \leq c_2$ such that for every unit $i \in [n]$ and relevant exposure $k \in \setb{0,1}$, the desired exposure probability is bounded:
	$c_1 \cdot \frac{1}{\lamH} \leq \Pr{ E_{(i,k)} } \leq c_2 \cdot \frac{1}{\lamH}$.
\end{corollary}

The following proposition shows that the estimator is unbiased under the proposed design.
The key idea in the proof is to use Lemma~\ref{lemma:prob-of-desired-exposure-events} to show that each of the individual desired exposure events has positive probability.

\expcommand{\unbiasedlemma}{%
	Under arbitrary neighborhood interference (Assumption~\mainref{assumption:ani-model}), the
	\ifnum \value{spacesave}=1 {
	} \else {
		modified Horvitz--Thompson
	} \fi
	estimator is unbiased under the Conflict Graph Design: $\E{\eate} = \ate$.
}
\begin{proposition} \label{prop:unbiased-estimator}
	\unbiasedlemma
\end{proposition}

Unlike the standard estimator, the modified Horvitz--Thompson estimator only uses the outcome $Y_i$ under the desired exposure events $E_{i,1}$ and $E_{i,0}$.
As we shall see in the analysis presented in the next section, these desired exposure events have the benefit of removing unwanted correlations between exposures from the estimator.
However, this can have the following awkward consequence: if a unit receives the contrastive exposure $e_k$ by accident{\textemdash}that is, outside of the desired exposure event $E_{i,k}${\textemdash}then it will not be included in the estimator.
In Figure~\ref{fig:conflict-graph-examples}, this occurs for subjects 3 and 6 in the direct effect and for subjects 4 and 7 under the GATE.
While ignoring some of the observed outcomes in this manner may seem counterproductive, we show in Section~\mainref{sec:alternative-estimators} that it is actually necessary to achieve the proposed rates of estimation under the Conflict Graph Design.
	
	\section{Analysis of Point Estimation} \label{sec:estimation-analysis}

We now present our analysis of the effect estimator under the Conflict Graph Design.
We begin by presenting a finite sample analysis in Section~\ref{sec:finite-sample-analysis} and then interpret the results within an asymptotic framework in Section~\ref{sec:asymptotic-analysis}.
Finally, alternative estimators and their failure modes are discussed in Section~\ref{sec:alternative-estimators}.

\subsection{Finite Sample Analysis} \label{sec:finite-sample-analysis}

% a sentence or two of introduction
The following theorem provides an analysis of the finite sample variance, which serves as the foundation for the main contributions of the paper.
Recall that Assumption~\mainref{assumption:ani-model} is the arbitrary neighborhood interference model.

\expcommand{\varianceanalysisfinitesample}{%
	Suppose Assumption~\mainref{assumption:ani-model} holds.
	Under the Conflict Graph Design, the finite sample variance of the modified Horvitz--Thompson estimator is bounded as
	\[
	\Var{\eate} \leq \frac{ \cleanconst \cdot \lamH }{n}  \cdot \paren[\Big]{ \frac{1}{n} \sum_{i=1}^n y_i(e_1)^2 + \frac{1}{n} \sum_{i=1}^n y_i(e_0)^2 }
	\enspace.
	\]
}
\begin{theorem} \label{thm:variance-analysis-finite-sample}
	\varianceanalysisfinitesample
\end{theorem}

% interpret the aspects of the theorem
The upper bound on the variance consists of two parts.
The first part contains the dependence on the global connectivity of the conflict graph (i.e. as measured by $\lamH$), which itself depends on the interplay between the underlying network $G$ and the causal estimand under investigation.
The second part depends on the magnitude of the relevant potential outcomes, as measured by the squared second moment.
Such a separation between these quantities allows us to obtain a tighter asymptotic analysis, though we defer these interpretations to the next section.

% the sampling parameter discussion
In Algorithm~\mainref{alg:conflict-graph-design}, we suggest using a sampling parameter $r=2$ as it approximately optimizes the constant $\cleanconst$.
Our analysis shows that any value $r \in [1.8, 2.8]$ achieves the same result, and specifically that $r \approx 2.19$ attains the optimized constant of $\finalconst$.
Different choices of $r \geq 1$ will affect the constant appearing in the bound, but not the rates.

% why is this theorem important within the literature ?
As discussed in Section~\mainref{sec:main-application}, the Conflict Graph Design and effect estimator achieve improved precision across a variety of well-studied causal estimands arising in network experiments.
A key insight facilitating this improvement has been to forgo the standard dependency graph analysis for a new style of \emph{operator norm analysis}.
This insight manifests itself in our work in two ways.
From the perspective of analysis, the operator norm approach is tighter and thus{\textemdash}when appropriately carried out{\textemdash}results in the derivation of better rates.
Even more so, focusing on the operator norm analysis suggests new algorithmic approaches for constructing experimental designs.
The result is improved precision through both the construction of new experimental designs and improved analyses.

We believe that the operator norm analysis will be of independent interest in the design and analysis of randomized experiments more broadly.
For this reason, we sketch a proof of Theorem~\mainref{thm:variance-analysis-finite-sample} in the remainder of this section.
To focus the exposition on the most central aspects of the proof, we prove a bound with larger constants.

% we are going to be giving a proof sketch
The first step of analyzing the variance is to study the associated covariance terms.
A key benefit of the proposed design and estimator is that we can obtain a fairly tight analysis of all of the relevant covariance terms between the inverse probability weighted indicators, $\indicator{E_{(i,k)}} / \Pr{ E_{(i,k)} }$.
The following lemma provides a rough characterization which is sufficient for the purposes of the proof sketch.
We use $d_{\cH}(i,j)$ to denote the minimum path distance between units $i$ and $j$ in the conflict graph $\cH$.

\expcommand{\covarianceterms}{%
	For each unit $i \in [n]$ and contrastive exposure $k \in\setb{0,1}$, the variance of each probability weighted indicator is bounded as
	\[
	\Var[\Big]{ \frac{ \indicator{ E_{(i,k)} } }{ \Pr{ E_{(i,k)} } } } \leq 5 \lamH \enspace.
	\]
	For each pair of distinct units $i \neq j \in [n]$ and contrastive exposure $k, \ell \in \setb{0,1}$, the covariance between probability weighted indicators may be characterized according to their distance in $\cH$:
	\begin{align*}
		\Cov[\Big]{ \frac{ \indicator{ E_{(i,k)} } }{ \Pr{ E_{(i,k)} } }, \frac{\indicator{ E_{(j,\ell)} } }{ \Pr{ E_{(j,\ell)} } } } 
		&= \left\{
		\begin{array}{lr}
			-1 & \text{ if } d_{\cH}(i,j) = 1\\
			0 & \text{ if } d_{\cH}(i,j) \geq 3
		\end{array}
		\right.\\
		\abs[\Bigg]{\Cov[\Big]{ \frac{ \indicator{ E_{(i,k)} } }{ \Pr{ E_{(i,k)} } }, \frac{\indicator{ E_{(j,\ell)} } }{ \Pr{ E_{(j,\ell)} } } } }
		&\leq 6 \cdot \frac{ \abs[\big]{ \mathcal{N}(i) \cap \mathcal{N}(j) } }{\lamH}
		\quad \text{for } d_{\cH}(i,j) = 2 \enspace.
	\end{align*}
}
\begin{lemma} \label{lemma:individual-covariance-terms}
	\covarianceterms
\end{lemma}

The proof of Lemma~\mainref{lemma:individual-covariance-terms} requires a direct analysis of covariance terms using the closed form for the desired exposure event probabilities (Lemma~\mainref{lemma:prob-of-desired-exposure-events}).
The proof crucially uses the defining feature of importance orderings (i.e. at most $\lamH$ more important neighbors) when carrying out several Taylor series-style arguments.
With Lemma~\mainref{lemma:individual-covariance-terms} in hand, we are ready to provide a proof sketch of Theorem~\mainref{thm:variance-analysis-finite-sample}.

\begin{proof}[Proof Sketch of Theorem~\mainref{thm:variance-analysis-finite-sample}]
	For simplicity, we will show the result with a larger constant: 
	\ifnum \value{spacesave}=1 {
		$
		\Var{\eate} \leq(24 \lamH / n) \cdot \braces{ \frac{1}{n} \sum_{i=1}^n y_i(e_1)^2 + \frac{1}{n} \sum_{i=1}^n y_i(e_0)^2 }
		$.
	} \else {
		\[
		\Var{\eate} \leq \frac{24 \cdot \lamH }{n} \cdot \paren[\Big]{ \frac{1}{n} \sum_{i=1}^n y_i(e_1)^2 + \frac{1}{n} \sum_{i=1}^n y_i(e_0)^2 }
		\enspace.
		\]
	} \fi
	We begin by writing the estimator as the difference of two estimators $\eate = \eate^{(1)} - \eate^{(0)}$ where each estimator is the average of individual outcome estimates, i.e. $\eate^{(k)} = (1/n) \sum_{i=1}^n \eate_i^{(k)}$ where $\eate_i = Y_i \cdot \indicator{E_{(i,k)}} / \Pr{ E_{(i,k)} }$.
	Under arbitrary neighborhood interference (Assumption~\mainref{assumption:ani-model}), the individual outcome estimator is $\eate^{(k)}_i = y_i(e_k) \cdot \indicator{E_{(i,k)}} / \Pr{ E_{(i,k)} }$ so that the only random aspect is the event indicator.
	The variance of the estimator decomposes as
	\[
	n \cdot \Var{\eate^{(k)}} = \frac{1}{n} \sum_{i=1}^n \sum_{j=1}^n y_i(e_k) y_j(e_k) \Cov[\Big]{ \frac{ \indicator{ E_{(i,k)} } }{ \Pr{ E_{(i,k)} } }, \frac{\indicator{ E_{(j,k)} } }{ \Pr{ E_{(j,k)} } } } 
	\enspace.
	\]
	We can write this quadratic form in matrix notation as
	\[
	n \cdot \Var{ \eate^{(k)} } = 
	\begin{bmatrix}
		\frac{1}{\sqrt{n}} \vec{y}(e_k) 
	\end{bmatrix}^\tran 
	\mat{C}_{(k)}
	\begin{bmatrix}
		\frac{1}{\sqrt{n}} \vec{y}(e_k) 
	\end{bmatrix}
	\text{  where  }
	\mat{C}_{(k)}(i,j) = \Cov[\Big]{ \frac{ \indicator{ E_{(i,k)} } }{ \Pr{ E_{(i,k)} } }, \frac{\indicator{ E_{(j,k)} } }{ \Pr{ E_{(j,k)} } } } 
	\enspace.
	\]
	We seek a bound of the form $n \cdot \Var{\eate^{(k)}} \leq C \cdot \paren{ (1/n) \sum_{i=1}^n y_i(e_k)^2 }$.
	The smallest $C$ for which this bound holds is precisely the $\ell_2 \to \ell_2$ \emph{operator norm} of $\mat{C}_{(k)}$, denoted $\norm{ \mat{C}_{(k)} }$.
	Because $\mat{C}_{(k)}$ is symmetric positive semidefinite, the operator norm is equal to its largest eigenvalue.
	Using the characterization of the covariance terms in Lemma~\mainref{lemma:individual-covariance-terms}, we can decompose the covariance matrix as $\mat{C}_{(k)} = \mat{C}_{(k), 0} + \mat{C}_{(k), 1} + \mat{C}_{(k), 2}$, where $\mat{C}_{(k), r}$ contains the covariance between units which are distance $r$ away in the conflict graph $\cH$.
	The operator norm is subadditive so that 
	\ifnum \value{spacesave}=1 {
		$
		\norm{ \mat{C}_{(k)} } \leq \norm{ \mat{C}_{(k),0} } + \norm{ \mat{C}_{(k),1} } + \norm{ \mat{C}_{(k), 2} }
		$.
	} \else {
		\[
		\norm[\big]{ \mat{C}_{(k)} } \leq \norm[\big]{ \mat{C}_{(k),0} } + \norm[\big]{ \mat{C}_{(k),1} } + \norm[\big]{ \mat{C}_{(k), 2} }
		\enspace.
		\]
	} \fi
	Thus, it suffices to bound each operator norm individually.
	
	Our remaining analysis follows from the characterization of terms in Lemma~\mainref{lemma:individual-covariance-terms}.
	The first matrix $\mat{C}_{(k),0}$ is a diagonal matrix whose entries are bounded as $\Var{ \indicator{ E_{(i,k)} } /  \Pr{ E_{(i,k)} } } \leq 5 \lamH$ so that $\norm{ \mat{C}_{(k),0} } \leq 5 \lamH$.
	The second matrix is equal to the negative adjacency matrix of $\mathcal{H}$, i.e. $\mat{C}_{(k),1} = -\adjH$ so that $\norm{ \mat{C}_{(k),1} } = \lamH$.
	The third matrix $\mat{C}^{(k),2}$ has entries whose absolute value are bounded as $(6 / \lamH) \abs{ \mathcal{N}(i) \cap \mathcal{N}(j) }$.
	This intersection term counts the number of paths of length $2$ between $i$ and $j$, which is precisely the entries of the squared adjacency matrix $\adjH^2$ which has largest eigenvalue $\lamH^2$.
	Thus, the operator norm of the third matrix is bounded as $\norm{ \mat{C}_{(k),2} } \leq (6 / \lamH) \norm{ \adjH^2 } = 6 \lamH$.
	
	This analysis yields that the overall operator norm is bounded as $\norm{ \mat{C}_{(k)} } \leq 12 \lamH$.
	The desired result follows by using $\Var{\eate} = \Var{\eate^{(1)} - \eate^{(0)}} \leq 2 \Var{\eate^{(1)}} + 2 \Var{\eate^{(0)}}$, where the last inequality follows from the Cauchy-Schwartz and AM-GM inequalities.
\end{proof}
It is worth highlighting the benefits of the above approach based on understanding the operator norm, as opposed to the more standard analysis of the variance using dependency graphs.
Previous approaches focus on identifying all pairs of units that are not independent and upper bounding their covariance by the maximum variance of a unit using the Cauchy-Schwartz inequality.
This approach yields a bound on the overall variance that scales with some power of the maximum degree of the graph, $d_{\max}$.
In contrast, we obtain a more fine-grained characterization of the covariance between units depending on their distance and relative topology in the graph. 
This in turn allows us to obtain tighter control of the operator norm without any $d_{\max}$ dependencies.
Section~\suppref{sec:dependency-graph} of the supplement contains a formal comparison with the dependency graph method.

The supplement contains additional technical results.
In Section~\suppref{sec:supp-number-of-exposed-units}, we show that the number of subjects who receive the desired exposure $e_k$, i.e.  $T_k = \sum_{i=1}^n \indicator{E_{(i,k)}}$, concentrates around its mean, which is of order $n / \lamH$.
\ifnum \value{spacesave}=1
{In Section~\suppref{sec:alternative-estimators}, we also show that the standard Horvitz--Thompson estimator which uses the exposure events, as opposed to the desired exposure events used by the modified estimator, can achieve exponentially worse rates for certain classes of graphs.}
\else
{}
\fi

\subsection{Asymptotic Analysis} \label{sec:asymptotic-analysis}

In this section, we interpret the finite sample bound on the variance within an asymptotic framework. As we have mentioned in Section~\mainref{sec:asymptotic_framework_intro}, we consider triangular array asymptotic, where we have a sequence of graphs $G^{(n)}$ and potential outcome functions $y^{(n)}$, where the index $n$ is the number of nodes. For simplicity of presentation in the sequel, we will sometimes suppress the dependence on $n$, whenever it is clear from the context. 
To proceed, we begin by positing a second moment assumption on the outcomes.

\begin{assumption}[Bounded Second Moments] \label{assumption:bounded-second-moment}
	The second moments of the potential outcomes are bounded: $\frac{1}{n} \sum_{i=1}^n y_i(e_k)^2 = \bigO{1}$ for each $k \in \setb{0,1}$.
\end{assumption}

The second moment assumption is mild and should be viewed as analogous to the familiar bounded second moment assumption in an i.i.d. sampling-based framework.
Assumption~\mainref{assumption:bounded-second-moment} is required for $\sqrt{n}$-consistent estimation of treatment effects in a design-based framework without interference.
This bounded second moment assumption allows for some potential outcomes to grow larger as the sample size increases.
For example, Assumption~\mainref{assumption:bounded-second-moment} permits that outcomes scale proportionally with their degrees, provided that the degree distribution of the network has bounded second moment.
Several prior works place a substantially stronger uniform bound (i.e. $p=\infty$ moment bound) on all potential outcomes which prevents any such growth in the magnitude of outcomes \citep[e.g.,][]{Ugander2023Randomized, Harshaw2023Design}.
We remark further that we do not posit any convergence of the second moments, only their asymptotic boundedness.

\expcommand{\varianceasymptotic}{%
	Suppose Assumptions~\mainref{assumption:ani-model} and \mainref{assumption:bounded-second-moment} hold.
	Under the Conflict Graph Design, the modified Horvitz--Thompson estimator converges in mean square at a $\sqrt{ \lamH / n }$-rate:
	\[
	\limsup_{n \to \infty} \sqrt{ \frac{n}{\lamH} } \cdot \E[\Big]{ \paren[\big]{ \ate - \eate }^2 }^{1/2} < \infty \enspace.  
	\]
}
\begin{theorem} \label{thm:variance-analysis-asymptotic}
	\varianceasymptotic
\end{theorem}

Theorem~\mainref{thm:variance-analysis-asymptotic} shows that the modified Horvitz--Thompson estimator achieves a $\sqrt{ \lamH / n }$-rate of convergence under the Conflict Graph Design.
The rate depends on the Conflict Graph $\cH$, which itself depends on the underlying graph $G$ and the causal estimand under investigation.
As discussed in Section~\mainref{sec:main-application}, these rates demonstrate the ways in which more ``local'' effects (e.g. direct effects) are easier to estimate than more ``global'' effects (e.g. global average treatment effect).
Likewise, as the underlying network $G$ becomes more connected, so too does the conflict graph $\cH$.
Importantly, these rates do not depend on the maximum degree of $G$, which is one of the strengths of the proposed design and analysis over prior works.

It may be the case that for some sufficiently well-connected sequence of graphs (e.g. the complete graph), consistent estimation of any causal effect will be impossible.
By turning to the conflict graph, we can obtain sufficient conditions for consistent effect estimation which depend both on the underlying network and the estimand.

\begin{assumption}[Sublinear Connectivity] \label{assumption:lam-bound}
	The conflict graph satisfies $\lamH = \littleO{n}$.
\end{assumption}

\begin{corollary} \label{corollary:effect-estimator-consistency}
	Suppose Assumptions~\mainref{assumption:ani-model}-\mainref{assumption:lam-bound}.
	Under the Conflict Graph Design, the modified Horvitz--Thompson estimator is consistent in mean square: $\lim_{n \to \infty} \E{ \paren{\ate - \eate}^2 } = 0$.
\end{corollary}

Corollary~\mainref{corollary:effect-estimator-consistency} demonstrates that positing that the eigenvalue of the dependency graph $\lamH$ is asymptotically dominated by $n$ is a sufficient condition for consistent effect estimation.
As discussed in Section~\mainref{sec:main-application}, this condition can generally be interpreted for specific estimands.
To the best of our knowledge, these are the weakest conditions for consistent estimation of commonly studied causal effects in network experiments, (e.g. global treatment effects, direct effects).

\subsection{Alternative Estimators} \label{sec:alternative-estimators}

As we previously discussed in Section~\mainref{sec:effect-estimator}, the modified Horvitz-Thompson estimator $\eate$ was chosen over the standard one in order to avoid unwanted correlations in the exposures of different nodes, which affect the variance of the estimator.
However, it is natural to wonder whether this reflects a limitation in our analysis of the variance or whether the standard estimator actually incurs worse precision guarantees.

In the following theorem, we show that the variance of the standard Horvitz--Thompson estimator can be substantially larger than its modified counterpart under the Conflict Graph Design. 
For simplicity, we focus on the direct effect, although similar illustrations can be constructed for a variety of estimands.
Specifically, we construct an asymptotic sequence of graphs and potential outcome functions satisfying Assumptions~\mainref{assumption:ani-model}-\mainref{assumption:lam-bound} such that the modified Horvitz--Thompson estimator converges in mean square at a rate of $\bigO{n^{-1/4}}$ under the Conflict Graph Design while the standard Horvitz--Thompson estimator converges in mean square at an \emph{exponentially slower} rate of $\bigOmega{ \paren{\log(n)}^{1/2} }$.

%%%%%%%%%%%%%%%%%%%%%%%%%%%%%
% WARNING: MANUALLY COPY THIS INTO APPENDIX!!
%%%%%%%%%%%%%%%%%%%%%%%%%%%%%
\expcommand{\truehtlowerbound}{%
	Consider the direct treatment effect $\ate$, as defined in Section~\mainref{sec:causal-estimands}. 
	There is a sequence $G^{(n)}$ of graphs with a triangular sequence $y_i^{(n)}$ of potential outcome functions satisfying Assumptions~\mainref{assumption:ani-model}-\mainref{assumption:lam-bound}, such that the following holds under the Conflict Graph Design:
	\[
	\limsup_{n\to \infty} n^{1/4}\cdot \E[\Big]{ \paren[\big]{ \ate - \eate }^2 }^{1/2} < \infty
	\quad \text{while} \quad
	0 < \liminf_{n\to \infty} \sqrt{\log n}\cdot \E[\Big]{ \paren[\big]{ \ate - \eate_{\textrm{standard}} }^2 }^{1/2}
	\enspace. 
	\]
}
\begin{theorem} \label{thm:true_Horvitz_Thompson_counterexample}
	\truehtlowerbound
\end{theorem}

Theorem~\ref{thm:true_Horvitz_Thompson_counterexample} implies that further assumptions on the graph sequence are needed, in order to establish that $\eate_{\textrm{standard}}$ achieves the same consistency guarantees as $\eate$. 
The main reason why $\eate_{\textrm{standard}}$ has a slow rate for the sequence of graphs in Theorem~\ref{thm:true_Horvitz_Thompson_counterexample} is that there is a growing gap between the maximum degree and the maximum eigenvalue of the graph.
The following Theorem shows that if this is not the case, then indeed $\eate_{\textrm{standard}}$ enjoys the same consistency guarantees as $\eate$, at least in the case of the direct effect.  

%%%%%%%%%%%%%%%%%%%%%%%%%%%%%
% WARNING: MANUALLY COPY THIS INTO APPENDIX!!
%%%%%%%%%%%%%%%%%%%%%%%%%%%%%
\expcommand{\standardhtdregular}{%
	Consider the direct treatment effect $\ate$, as defined in Section~\mainref{sec:causal-estimands}. 
	Suppose Assumptions~\mainref{assumption:ani-model}-\mainref{assumption:lam-bound} hold and, in addition, $d_{\textrm{max}}(\cH) = \bigTheta{\lamH}$. 
	Then, under the Conflict Graph Design the standard Horvitz--Thompson estimator converges in mean square at $\sqrt{\lamH / n}$ rate:
	\ifnum \value{spacesave}=1 {
		$\limsup_{n\to \infty} \sqrt{n / \lamH} \cdot \E{ \paren[\big]{ \ate - \eate_{\textrm{standard}} }^2 }^{1/2} < \infty$.
	} \else {
		\[
		\limsup_{n\to \infty} \sqrt{\frac{n}{\lamH}} \cdot \E[\Big]{ \paren[\big]{ \ate - \eate_{\textrm{standard}} }^2 }^{1/2} < \infty \enspace. 
		\]
	} \fi
}
\begin{theorem} \label{thm:true_HT_d_regular}
	\standardhtdregular
\end{theorem}

Observe that under the assumption that $d_{\textrm{max}}(\cH) = \bigTheta{\lamH}$, then a $\sqrt{\lamH / n}$ rate is equivalent to a $\sqrt{d_{\max}(\cH) / n}$ rate.
In particular, this implies that if the graph sequence is $d$-regular with $d = \littleO{n}$, then consistent estimation is possible using the standard Horvitz--Thompson estimator.
On the technical side, the proof of Theorem~\ref{thm:true_HT_d_regular} is significantly more involved than Theorem~\mainref{thm:variance-analysis-finite-sample}.
\ifnum \value{spacesave}=1
{We refer to Section~\suppref{sec:supp-true-ht-d-regular-proof} of the supplement for more details.}
\else
{
	As discussed above, this is expected, since $\eate_{\textrm{standard}}$ depends on exposure events, which are significantly more correlated than the desired exposure events using the modified estimator.
	Under the Conflict Graph Design, there exist additional correlations between nodes that are at distance $3$ and $4$ in $G$.
	In order to control these different kinds of correlations, we again break down the covariance matrix into parts, each one encoding the dependence of nodes at a given distance in $G$.
	To analyze the correlations that appear at every distance, we devise a novel coupling method based on interpolating the original graph into one with simpler correlations and using path counting arguments for paths of length $3$ or $4$ in order to bound the maximum eigenvalue.  
}
\fi

\section{Inferential Methods} \label{sec:inference}

In this section, we propose confidence intervals for the aggregated treatment effect based on the modified Horvitz-Thompson estimator under the Conflict Graph Design and analyze their asymptotic coverage.
Section~\mainref{sec:variance-estimation} describes a conservative variance estimator which serves as the basis for the proposed intervals. 
In Section~\mainref{sec:intervals}, we propose and analyze confidence intervals obtained from Chebyshev's inequality as well as standard Wald-type intervals.

For asymptotic analyses, we need to introduce two additional assumptions: bounded 4th moments and non-super-efficiency.
Both of these assumptions are asymptotic in nature.

\begin{assumption}[Bounded Fourth Moments] \label{assumption:bounded-fourth-moments}
	The fourth moments of the potential outcomes are bounded: $\frac{1}{n} \sum_{i=1}^n y_i(e_k)^4 = \bigO{1}$ for each $k \in \setb{0,1}$.
\end{assumption}

The bounded fourth moment assumption prevents outliers among the potential outcomes.
Such an assumption is required in order to consistently estimate the variance of the effect estimator.
Although it is stronger than the second moment assumptions given in Assumption~\mainref{assumption:bounded-second-moment}, it is weaker than requiring an absolute bound on all outcomes.
In this way, Assumption~\mainref{assumption:bounded-fourth-moments} still allows for growth of some of the outcomes.

The second assumption is \emph{non-super-efficiency}, which places lower bounds on the variance of the effect estimator.

\begin{assumption}[Non-Superefficient] \label{assumption:not-superefficient}
	The variance of the modified Horvitz--Thompson estimator is bounded below as $\Var{\eate} \geq \bigOmega{\lamH / n}$.
\end{assumption}

Theorem~\mainref{thm:variance-analysis-asymptotic} shows that under the bounded second moment assumption, the variance of the modified Horvitz--Thompson estimator is bounded as $\Var{\eate} \leq \bigO{\lamH / n}$.
Assumption~\mainref{assumption:not-superefficient} posits that the variance is bounded below by the same rate.
While a non-superefficiency assumption is typically superfluous in an i.i.d. sampling based framework, some form of this an assumption is required for valid inference in design-based inference settings \citep[see e.g., ][]{Aronow2017Estimating, Leung2022Causal, Saevje2023Misspecified, Harshaw2022Design}.
For example, Assumption~\mainref{assumption:not-superefficient} rules out the case where all outcomes are equal to zero so that the estimator is zero with probability 1.
Generally speaking, Assumption~\mainref{assumption:not-superefficient} is ruling out potential outcomes which collude with the experimental design to achieve exceptionally fast rates of convergence, which are considered to be knife edge cases not relevant to practice.

%%\chriscomment{Add a slightly longer paragraph that explains that both outcomes hold with exponentially high probability under certain conditions.}
To better contextualize these additional assumptions, suppose that potential outcomes are drawn i.i.d. from a super-population.
Under standard assumptions on the super-population (e.g. existence of fourth moments), then simple applications of the law of large numbers and second moment method imply that Assumptions~\mainref{assumption:bounded-fourth-moments} and ~\mainref{assumption:not-superefficient} both hold with probability tending to $1$ as $n \to \infty$.
We provide the details in Section~\suppref{sec:iid-outcomes} of the Supplementary Materials.

\subsection{Variance Estimation} \label{sec:variance-estimation}

Standard methods for constructing confidence intervals from a point estimator require an estimate of its variance.
Unfortunately, neither unbiased and nor consistent variance estimation is possible within a design based framework.
The reason is that the variance $\Var{\eate}$ contains many products of fundamentally unobservable terms such as $y_i(e_1) \cdot y_i(e_0)$.
Under arbitrary neighborhood interference, there are additional unobservable products of outcomes, including those between neighboring units in the conflict graph.
Given that consistent variance estimation is impossible, experimenters often opt for conservative estimates of the variance, i.e. those that are upwardly biased.
Using these conservative variance estimates results in inferential procedures which are similarly conservative.

The standard approach to constructing a conservative variance estimator is to first construct an estimable upper bound on the variance; that is, one that does not contain products of unobservable terms.
%One of the earliest estimable variance bounds is due to \citep{Neyman1923} in the context of the difference-in-means estimator under a completely randomized design with no interference.
While a variety of estimable variance upper bounds have been proposed in the no-interference setting \citep[see e.g.,][]{Neyman1923, Kempthorne1955Randomization, Wilk1955Randomzation, Robins1988Confidence, Abadie2008Estimation, Imai2008Variance, Aronow2014Sharp}, there are considerably fewer methods for obtaining estimable upper bounds under interference.
Aronow and Samii (2013, 2017) % it's annoying I did this manually, but I'm not sure how else to make it work
provide a general recipe for obtaining variance bounds using the AM-GM inequality, but this has been shown to be overly conservative, especially in interference settings.
\citet{Harshaw2021Optimized} propose a systematic approach for obtaining variance bounds under exposure mapping models of interference.
However, their approach is computationally expensive and typically does not yield a closed form solution, which poses challenges for ensuring consistent estimation.

We propose a variance bound based on the operator norm of the covariance matrix of inverse probability weighted indicators.
Let $\varM$ be a $2n$-by-$2n$ matrix whose entries are indexed by $(i,k),(j,\ell)$ with $i,j \in [n]$ and $k, \ell \in \setb{0,1}$ and are given by
\ifnum \value{spacesave}=1 {
	$
	\varM \paren{ (i,k), (j,\ell) } 
	=
	(-1)^{\indicator{k \neq \ell}} \Cov{ \frac{\indicator{ E_{(i,k)} }}{ \Pr{ E_{(i,k)} } } ,  \frac{\indicator{ E_{(j,\ell)} }}{ \Pr{ E_{(j,\ell)} } }}
	$,
} \else {
	\[
	\varM \paren[\Big]{ (i,k), (j,\ell) } 
	=
	(-1)^{\indicator{k \neq \ell}} \Cov[\Big]{ \frac{\indicator{ E_{(i,k)} }}{ \Pr{ E_{(i,k)} } } ,  \frac{\indicator{ E_{(j,\ell)} }}{ \Pr{ E_{(j,\ell)} } }}
	\]
} \fi
and let $\lamV$ be its largest eigenvalue.
The variance bound is given by
\[
n \cdot \vb \triangleq \lamV \cdot \braces[\Bigg]{ \frac{1}{n} \sum_{i=1}^n y_i(e_1)^2 + \frac{1}{n} \sum_{i=1}^n y_i(e_0)^2 }
\enspace.
\]

The variance bound $\vb$ contains only squared individual outcomes, which are each observed with positive probability.
The principle behind our construction of this variance bound is the same principle behind our analysis:
the effect estimator is linear in the observed outcomes so that the variance is a quadratic form in the potential outcomes.
Thus, the operator norm of this quadratic form yields an upper bound.
Moreover, this upper bound does not feature problematic unobservable terms which cannot be estimated.
While Theorem~\mainref{thm:variance-analysis-finite-sample} derives the upper bound $\lamV \leq \cleanconst \cdot \lamH$, we can use the largest eigenvalue directly for the purposes of variance estimation.
Using the exact eigenvalue has the additional benefit that the variance bound is sharp, i.e. equality holds for some potential outcomes.

\expcommand{\conservativevbistight}{%
	The variance bound is conservative: $\Var{\eate} \leq \vb$.
	Moreover, equality holds for potential outcomes which align with the leading eigenvector of $\varM$.
}
\begin{proposition} \label{prop:conservative-vb-is-tight}
	\conservativevbistight
\end{proposition}

In order to estimate the variance bound, we propose a Horvitz--Thompson style estimator.
The variance estimator takes a similar form as the effect estimator:
\begin{align*}
	n \cdot \evb 
	=
	\lamV \cdot \paren[\Bigg]{ \frac{1}{n} \sum_{i=1}^n Y_i^2 \braces[\Bigg]{ \frac{ \indicator{ E_{(i,1)} } }{ \Pr{ E_{(i,1)} }}  + \frac{ \indicator{ E_{(i,0)} } }{ \Pr{ E_{(i,0)} }} } }
	\enspace.
\end{align*}
In addition to the observed outcomes $Y_1 \dots Y_n$ and intervention $Z$, the variance estimator $\evb$ requires knowledge of the probability of desired exposure events $\Pr{ E_{i,k} }$.
As shown in Lemma~\mainref{lemma:prob-of-desired-exposure-events}, these probabilities have simple closed form expressions which may be easily computed.
This is different from most experimental designs under network interference, which require estimating design probabilities through Monte Carlo simulation.

The proposition below shows that the variance estimator is an unbiased estimate of the variance bound and that the expected variance estimator has the same $\bigO{\lamH / n}$ rates as the true variance, which ensures that the width of confidence intervals will inherit corresponding rates.
We remark that the Aronow-Samii variance bound would be much larger, typically scaling as $\bigOmega{ d_{\max}(\cH) / n }$.
This would effectively lose the benefit of the improved rates of the Conflict Graph Design.

\expcommand{\unbiasedvarest}{%
	Under Assumption~\mainref{assumption:ani-model}, the variance estimator is unbiased for the variance bound: $\E{ \evb } = \vb$.
	Under Assumptions~\mainref{assumption:ani-model}-\mainref{assumption:bounded-second-moment}, the expected variance bound has the rate: $\E{ \evb } = \bigO{ \lamH / n }$.
}
\begin{proposition} \label{prop:unbiased-var-est}
	\unbiasedvarest
\end{proposition}

We now seek to bound the error of the variance estimator to the variance bound.
The rate guaranteed by Theorem~\mainref{thm:variance-analysis-asymptotic} together with the non-superefficiency assumption (Assumption~\mainref{assumption:not-superefficient}) ensure that the variance goes down at a rate of $\bigTheta{ \lamH / n }$.
Thus, the most relevant error of the variance estimator is that after a $n / \lamH$ normalization.

\expcommand{\varestconsistency}{%
	Under Assumptions~\mainref{assumption:ani-model} and \mainref{assumption:bounded-fourth-moments}, 
	\ifnum \value{spacesave}=1 {
		the error of the normalized variance estimator is
		$\abs{\frac{n}{\lamH} \cdot \evb - \frac{n}{\lamH} \cdot \vb}
		= \bigOp{ \sqrt{\frac{\lamH}{n} } }$.
	} \else {
		the additive error between the normalized variance estimator and the variance bound scales as
		\[
		\abs[\Big]{\frac{n}{\lamH} \cdot \evb - \frac{n}{\lamH} \cdot \vb}
		= \bigOp[\Bigg]{ \sqrt{\frac{\lamH}{n} \ } }
		\enspace.
		\]
	} \fi
}
\begin{proposition} \label{prop:var-est-consistency}
	\varestconsistency
\end{proposition}

Just as the point estimator converges as a rate of $\bigO{\lamH / n}$, so too does the normalized variance estimator. 
Thus, the $\lamH = \littleO{n}$ condition on the graph is sufficient for both consistent point estimation and consistent variance estimation.
Proposition~\mainref{prop:var-est-consistency}, non-superefficiency, and the continuous mapping theorem imply that the variance estimator is stable, i.e. $\vb / \evb \xrightarrow{p} 1$.

\subsection{Confidence Intervals} \label{sec:intervals}

In this section, we propose two types of confidence intervals for the aggregated treatment effect and analyze their asymptotic coverage.
For a given level $\alpha \in (0,1]$, we define the intervals
\ifnum \value{spacesave}=1 {
	\[
	C_{\textrm{Cheb}}(\alpha) = \eate \pm \frac{1}{\sqrt{\alpha}} \cdot \sqrt{\evb}
	\quadand
	C_{\textrm{Wald}}(\alpha) = \eate \pm \Phi^{-1}(1 - \alpha/2) \cdot \sqrt{\evb}
	\enspace,
	\]
} \else {
	\begin{align*}
		C_{\textrm{Cheb}}(\alpha) &= \eate \pm \frac{1}{\sqrt{\alpha}} \cdot \sqrt{\evb} \\
		C_{\textrm{Wald}}(\alpha) &= \eate \pm \Phi^{-1}(1 - \alpha/2) \cdot \sqrt{\evb}
		\enspace,
	\end{align*}
} \fi
where $\Phi^{-1} : (0,1) \to \Reals$ is the quantile function of a standard normal.
The first interval is based on a Chebyshev bound while the second interval is a standard Wald-type interval based on a normal approximation.
By Proposition~\ref{prop:unbiased-var-est}, the expected width of both intervals is $\bigO{\sqrt{\lamH / n}}$; however, the dependence on $\alpha$ will be different.
The Chebyshev intervals will be wider than the Wald type intervals and this difference becomes more pronounced for smaller $\alpha$.
For example, at level $\alpha = 0.05$ we have that $ \frac{1}{\sqrt{\alpha}} \approx 4.47$ and $\Phi^{-1}(1 - \alpha/2) \approx 1.96$ so that the Chebyshev interval is roughly $2.28$ times larger than the Wald type interval.
However, we shall see below that the narrower Wald-type intervals require stronger conditions on the underlying graph in order to achieve asymptotic coverage.

We begin by demonstrating asymptotic coverage of the Chebyshev-based intervals.

\expcommand{\intervalschebyshev}{%
	Under Assumptions~\mainref{assumption:ani-model}-\mainref{assumption:not-superefficient}, the Chebyshev intervals asymptotically cover at the nominal rates: for each $\alpha \in (0,1]$, $\liminf_{n \to \infty} \Pr{ \ate \in C_{\textrm{Cheb}}(\alpha) } \geq 1 - \alpha$.
}
\begin{theorem} \label{theorem:intervals-cover-chebyshev}
	\intervalschebyshev
\end{theorem}

The Chebyshev intervals require relatively weak assumptions on the underlying network.
This is because the interval itself is based on a finite sample tail bound which holds for any random variable and so the only substantive requirement is that the variance estimator converges appropriately.
As we saw in the previous section, this does not require stronger conditions on the underlying network than what is required for consistent point estimation.

Wald-type intervals are based on a normal approximation and so they are only appropriate when the standardized estimator is asymptotically normal.
Theorem~\mainref{thm:clt} below provides sufficient conditions for a CLT of the standardized effect estimator under the Conflict Graph Design, which are stronger than the conditions for consistent point estimation.
The proof uses Stein's method via the dependency graph technique \citep{Ross2011Fundamentals}, which has been commonly employed in network causal inference \citep[see e.g.,][]{Chin2019Central,Ogburn2024Causal,Aronow2017Estimating,Harshaw2023Design}.

\expcommand{\clttheorem}{%
	Suppose that for the maximum degree $d_{\textrm{max}}$ we have $d_{\textup{max}}(\cH) = o(n^{1/9})$.
	Then, under Assumptions~\mainref{assumption:ani-model}-\mainref{assumption:not-superefficient}, the standardized estimator is asymptotically normal under the Conflict Graph Design: $(\eate - \ate) / \sqrt{ \Var{\eate} } \xrightarrow{d} \mathcal{N}(0,1)$.
}
\begin{theorem} \label{thm:clt}
	\clttheorem
\end{theorem}

\expcommand{\intervalswald}{%
	Under Assumptions~\mainref{assumption:ani-model}-\mainref{assumption:not-superefficient} and $d_{\textup{max}}(\cH) = o(n^{1/9})$, the Wald intervals asymptotically cover at the nominal rates: for each $\alpha \in (0,1]$, $\liminf_{n \to \infty} \Pr{ \ate \in C_{\textrm{Wald}}(\alpha) } \geq 1 - \alpha$.
}
\begin{corollary} \label{corollary:intervals-cover-wald}
	\intervalswald
\end{corollary}

The condition that the maximum degree of the conflict graph is asymptotically bounded as $d_{\max}(\cH) = \littleO{n^{1/9}}$ required by Theorem~\mainref{thm:clt} and Corollary~\mainref{corollary:intervals-cover-wald} for a CLT and asymptotic coverage of Wald-type intervals is substantially stronger than what is required for consistency of point estimation and variance estimation.
This is due, in some part, to fundamental limitations of the proof technique underlying Theorem~\mainref{thm:clt}, i.e. Stein's method via a dependency graph.
However, we believe that stronger conditions on the underlying network are necessary for a standard CLT to hold.
The following proposition demonstrates that even when the conditions for consistency are met, the standardized estimator may not converge in distribution.

\expcommand{\cltcounter}{%
	Suppose the causal estimand is the direct effect. 
	There exists a sequence of graphs with $\lamH = \Theta\lp(\sqrt{n}\rp)$ and a sequence of outcome functions $y^{(n)}$ satisfying Assumptions~\mainref{assumption:ani-model}-\mainref{assumption:not-superefficient}, such that the standardized estimator $(\eate - \ate) / \sqrt{\Var{\eate}}$ does not converge in distribution under the Conflict Graph Design.
}
\begin{proposition} \label{prop:clt-counter-example}
	\cltcounter
\end{proposition}

The sequence of graphs alluded to in Proposition~\mainref{prop:clt-counter-example} is based on a simple construction, where for even and odd $n$ we choose graphs from two different sequences. 
For even $n$, we select a star graph and for odd $n$ we select a graph that is formed by starting from a clique on $\sqrt{n}$ nodes and attaching to each node a separate $\sqrt{n}$-node clique.
Along the two subsequences, the standardized estimator converges to two different limiting distributions, thus the overall sequence does not converge.
\ifnum \value{spacesave}=1
{}
\else
{
The issue here is that even when the largest eigenvalue is modest (i.e. $\lamH = \bigO{\sqrt{n}}$), the maximum degree could vary between $\bigTheta{\sqrt{n}}$ and $\bigTheta{n}$.
These variations in the maximum degree can actually translate into variations in the limiting distributions, which is at the heart of this illustration.
}
\fi
This counter example demonstrates that constructing Wald type intervals through standard techniques requires either stronger assumptions than point estimation or a different approach all together.

Whether experimenters ought to use Chebyshev intervals or Wald-type intervals is an intricate question.
On the one hand, the Chebyshev interval is appropriate whenever the graph is sufficiently well behaved so that precise estimation of the aggregate effect is possible.
On the other hand, while Wald-type intervals seem only to be applicable for a restricted class of graphs, they are narrower and thus more desirable.
At the present moment, we recommend that experimenters run numerical simulations using realistically generated outcomes to determine whether the Wald type intervals may cover at the nominal rates for their specific experiment.

\ifnum \value{spacesave}=1
{Due to space constraints, we defer numerical simulations which evaluate the proposed methods to Section~\suppref{sec:simulations} of the supplement.}
\else
{}
\fi
	
	\section{Numerical Simulations} \label{sec:simulations}

In this section, we present numerical simulations to evaluate the proposed methods.
We focus our simulations on the global average treatment effect and the direct treatment effect, given that they are the most well-studied in the literature.
In our simulations, we vary the sample size from $n=500$ to $n=3,000$.

We construct networks from a preferential attachment model \citep{Barabasi1999Emergence, Easley2010Networks}, which has been used to model the World Wide Web, genetic networks, and social networks to name a few.
A network on $n$ vertices is sampled from the preferential attachment model given parameters $r$ and $m$ by the following iterative procedure: the graph is initialized to have $m$ nodes with no edges.
At each iteration $t=m+1, \dots n$, the $t$th node is added to the network and attached to $m$ nodes, which are selected randomly and proportional to $\textrm{deg}_t(i)^r$.
In this way, new nodes are more likely to connect with nodes which already have many connections, a preference that increases with the parameter $r$.
In our simulations, we set $m = 4$ and we selected the exponent parameters $r = 1.0$ for Global Average Treatment Effect and $r = 1.5$ for the Direct Treatment Effect.
The motivation for selecting these values of $r$ was to create problem instances which displayed the following properties: (i) $\lamH \ll n$ so that consistency is possible and (ii) $\lamH \ll \dmax(\cH)$ so that the problem instance is challenging.

For both estimands of interest, we construct potential outcomes according to the arbitrary neighborhood interference model, i.e. Assumption~\mainref{assumption:ani-model}.
Potential outcomes are constructed by randomly sampling the ``control exposure'' outcomes as $y_i(e_0)$ and the ``treatment exposure'' outcome $y_i(e_1)$ as
\ifnum \value{spacesave}=1 {
	$y_i(e_0) = \alpha_{i,1}$ and $y_i(e_1) = \alpha_{i,2} \cdot \sqrt{ \textrm{deg}_G(i) }$,
} \else {
	\[
	y_i(e_0) = \alpha_{i,1} 
	\quadand 
	y_i(e_1) = \alpha_{i,2} \cdot \sqrt{ \textrm{deg}_G(i) }
	\enspace,
	\]
} \fi
where the coefficients are sampled independently across units as $\alpha_{i,1} \sim \mathcal{N}(1,1)$ and $\alpha_{i,2} \sim \mathcal{N}(2,1)$.
Constructing outcomes in this way creates outcome heterogeneity, providing a challenging problem instance where nodes with high degree also have large individual treatment effects.
This construction of the outcomes also ensures the second moments will be bounded, i.e. Assumption~\mainref{assumption:bounded-second-moment} holds.
Although the network and outcomes are constructed randomly, this random construction occurs only once and the reported statistics (e.g. variance, coverage, etc) are conditioned on this single random construction of networks and outcomes.

Our simulations feature both standard Horvitz--Thompson estimators and the modified Horvitz--Thompson estimator introduced in Section~\mainref{sec:effect-estimator}.
As discussed in Section~\mainref{sec:effect-estimator}, the probabilities of desired exposure used in the modified Horvitz--Thompson estimator can be computed exactly.
However, the probabilities of actual exposure used in the standard Horvitz--Thompson estimator require Monte Carlo simulation across all designs, for which we use $k = 10,000$ draws from the designs.
These $k = 10,000$ random draws are also used to report the statistical properties of the designs e.g. variance, interval widths, coverage, etc.
Due to space constraints, we focus on the variance of the experimental designs here in the main body and report on the inferential procedures and practical recommendations in the Supplementary Material.

\begin{figure}[t]
	\centering
	\begin{subfigure}[t]{0.49\textwidth}
		\centering
		\includegraphics[width=\textwidth]{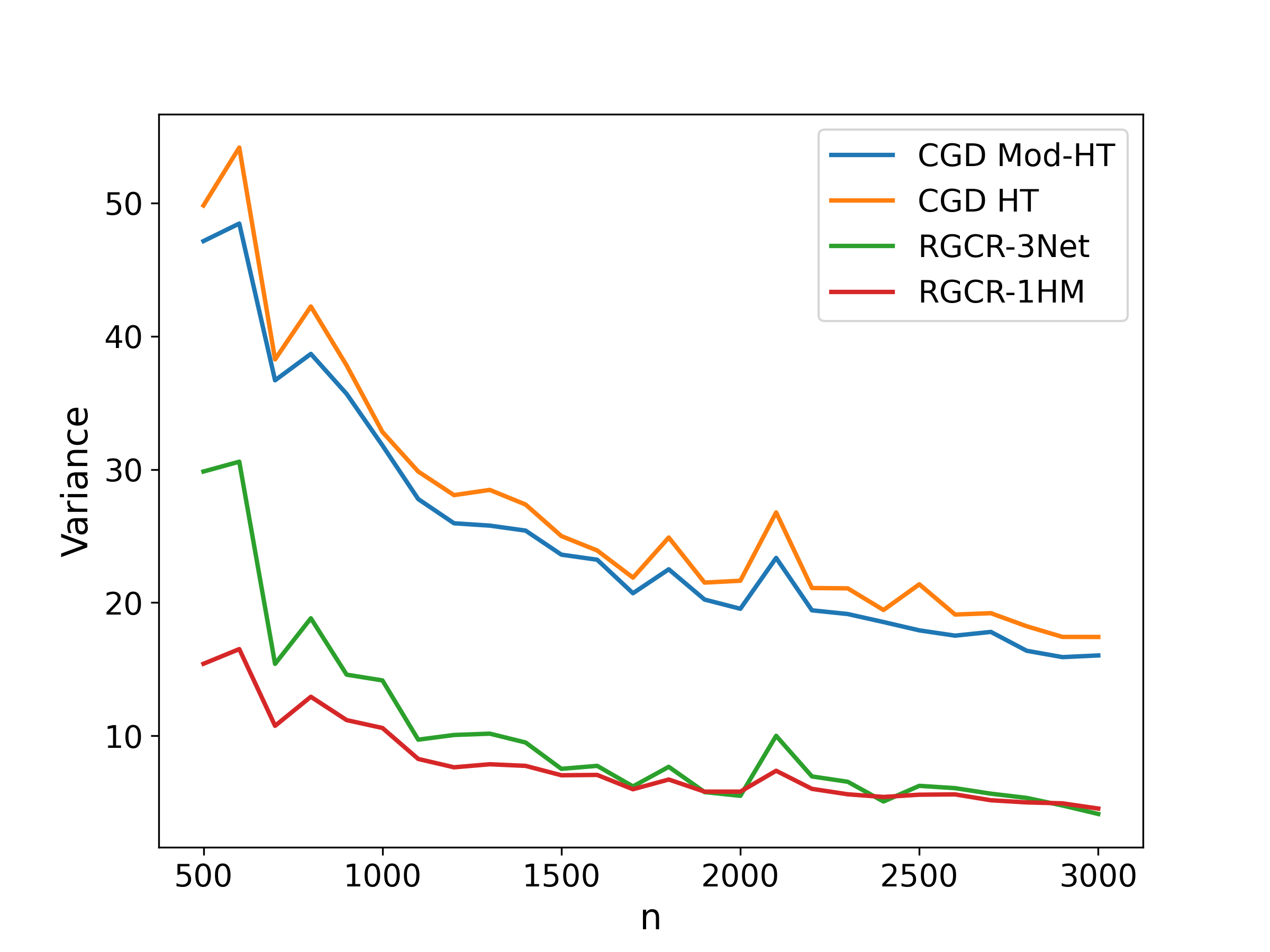}
		\caption{Global Average Treatment Effect}
		\label{fig:gate-variances}
	\end{subfigure}%
	~ 
	\begin{subfigure}[t]{0.49\textwidth}
		\centering
		\includegraphics[width=\textwidth]{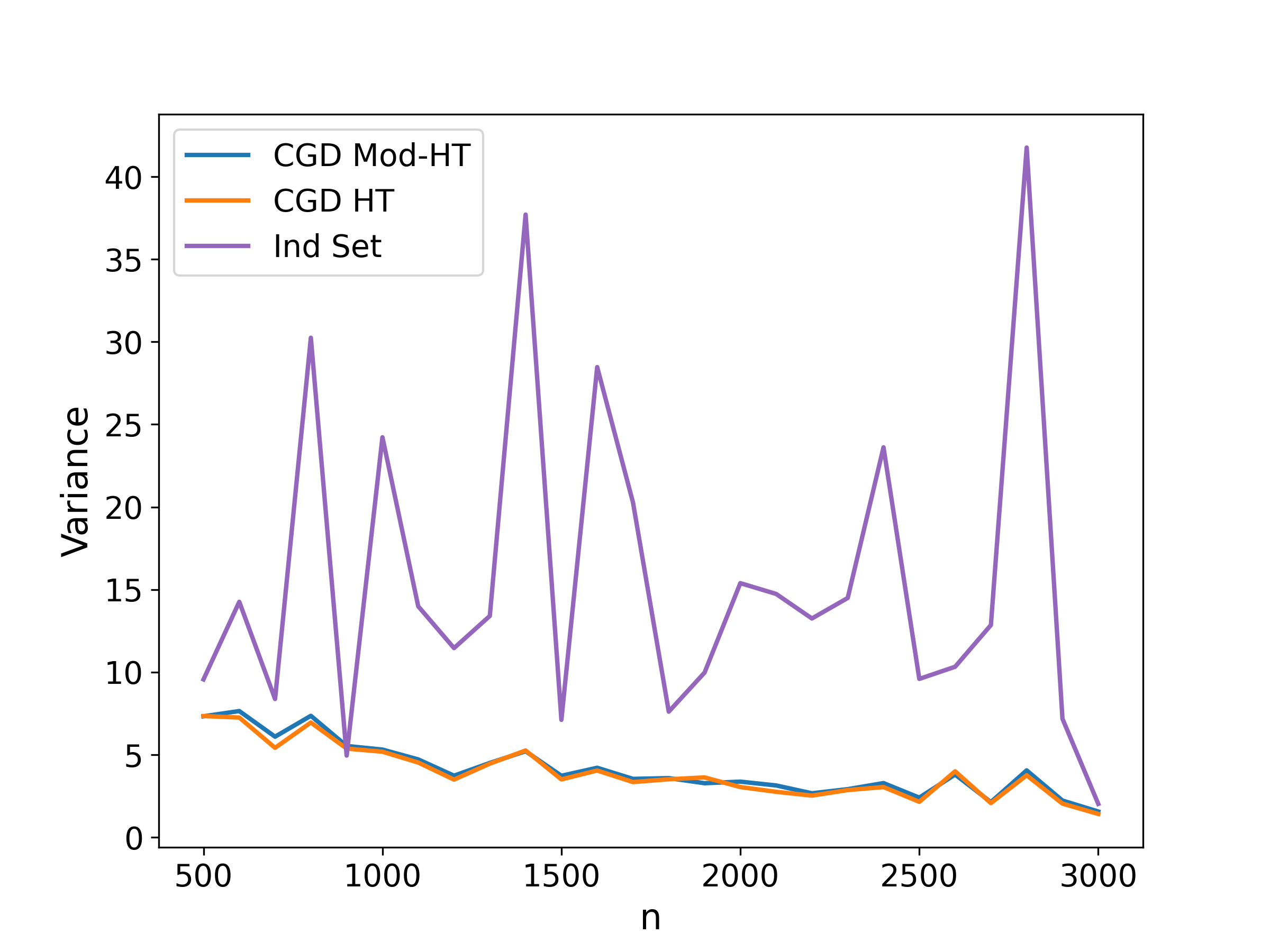}
		\caption{Direct Treatment Effect}
		\label{fig:dte-variances}
	\end{subfigure}
	\caption{
		Comparison of the variance of various experimental designs.
		Figure~\ref{fig:gate-variances} compares RGCR and Conflict Graph Design for estimating the Global Average Treatment Effect.
		Figure~\ref{fig:dte-variances} shows the Independent Set Design and Conflict Graph Design for estimating  the Direct Treatment Effect.
	}
	\label{fig:variance-comparison}
\end{figure}

\paragraph{Global Average Treatment Effect:}
In addition to the Conflict Graph Design for the global average treatment effect, we implemented 1-Hop and 3-Net Randomized Graph Cluster Randomization (RGCR) designs with spectral weighting \citep{Ugander2023Randomized}, which are current state-of-the-art experimental designs for this effect.
Figure~\ref{fig:gate-variances} shows a plot of the variance for each design from which we see several findings.
The modified Horvitz--Thompson estimator achieves a slightly lower variance under the Conflict Graph Design than its standard counterpart, though both are comparable.
More interesting is that the RGCR designs achieve variances which are roughly 4 times lower than that of the Conflict Graph Design.
This improvement of RGCR over CGD is not a general phenomenon: in Section~\suppref{sec:supp-3net-counterexample} of the supplement, we construct families of graphs where RGCR incurs an exponentially worse rate than the CGD.
Nevertheless, we feel that this empirical observation may be insightful for future development of experimental designs.

\paragraph{Direct Treatment Effect:}
In addition to the Conflict Graph Design for the direct treatment effect, we implemented the Independent Set design of \citet{Karwa2018Systematic}.
Among the few designs targeted for estimating the direct effect, we chose this one as it had been most extensively evaluated through numerical simulations.
Figure~\ref{fig:dte-variances} shows a plot of the variance for each design, which demonstrates two findings.
First, the modified and standard Horvitz--Thompson estimators are comparable under the Conflict Graph Design.
We also find that the Conflict Graph Design achieves a variance which is roughly between 3-7 times lower than that of the Independent Set Design.
This is because the Independent Set Design will assign treatment exposure to high degree nodes with low probability, thereby incurring high variance as these are precisely the nodes which have large outcomes.
We found that this difference becomes more pronounced for larger values of the exponent $r$.
	
	\section{Conclusion} \label{sec:conclusion}

In this work, we have presented the Conflict Graph Design and a modified Horvitz—Thompson estimator which are constructed to estimate a single experimenter-specified causal effect in a network experiment to high precision.
We have shown that the methods attain $\Var{\eate} \leq \bigO{\lamH / n}$ rates, which depend not only on the underlying network but also the causal effect.
This general approach to experiment design yields improved rates for the most well-studied causal effects as well as new results for causal effects which have received less attention in the literature on experiment design.
We have also proposed confidence intervals which feature similar rates and provided sufficient conditions for their asymptotic validity.
Finally, we believe that the operator norm analysis carried out in this work will prove to be a useful technique for analyzing future methods under interference.

The most intriguing open question raised by this work is whether the $\lamH / n$ rate is optimal or whether it can be improved.
Broadly speaking, resolving questions of optimality would require the development of lower bounds on rates of estimation in network experiments.
This is an especially challenging task given the lack of well-developed efficiency theory for design-based inference, which stands in stark contrast to the mature efficiency theory developed within a super-population framework \citep[see e.g.,][]{Kennedy2016Semiparametric}. 
Another open question is the development of tighter Wald-type intervals in the $\lamH = \littleO{n}$ consistency regime, where Proposition~\mainref{prop:clt-counter-example} shows that the standardized estimator will generally not have a limiting distribution.
We believe this requires the modification or development of specific theoretical tools beyond the usual application of Stein’s method in causal inference under interference.

The Conflict Graph Design and the modified Horvitz--Thompson estimator have certain limitations which may also form the basis for future work.
For example, these methods do not use covariate information, which is often available to the experimenter.
The AIPW estimator, which uses a fitted regression model to improve precision, is the most common way to use covariate information to improve precision.
Understanding when such adjusted estimators can improve precision in network experiments is an open question \citep[see e.g.][]{Gao2025Causal}.
However, we suspect that such improvements will be in efficiency (i.e. constant terms) rather than the rates, which are primarily driven by the experimental design.
A second limitation is that the proposed methods are designed to estimate a single causal effect to high precision.
It will typically be ill advised to run the Conflict Graph Design for a primary estimand when multiple secondary estimands are also of interest, due to possible positivity violations.
It is an interesting open problem is to construct a network experimental design to increase the precision of multiple causal effect estimators simultaneously.
It seems likely to us that this would involve a trade-off: the more estimands that are considered, the worse the rate of estimation for each of them must be.
	
	% references
	\printbibliography
	
	\newpage
		
	\appendix
	
	\addcontentsline{toc}{section}{Appendix} % Add the appendix text to the document TOC
	\part{Appendix} % Start the appendix part
	\parttoc % Insert the appendix TOC
	\newpage
	
	\section{Additional Results and Discussion} \label{sec:supp-additional-results}

\subsection{Interference and Mediation} \label{sec:supp-mediation}

In this section, we draw some comparisons between two subfields of causal inference: network interference and mediation analysis.
Broadly speaking, these two fields are distinct from each other.
At the same time, they share many of the same terminology such as ``direct effects'' and ``fundamental unobservability''.
This is unfortunate because the underlying problems and solutions referenced by these shared terms differ substantially between the two fields.
The goal of this section is to clarify these differences.

In the context of mediation analysis, the well-studied natural direct and indirect effects are defined as
\[
\textrm{NDE} = \E{ Y(1, M(0)) - Y(0, M(0)) }
\quadand
\textrm{NIE} = \E{ Y(1, M(1)) - Y(1, M(0))}
\enspace.
\]
For a comprehensive treatment of mediation, we refer the reader to the excellent textbook of \citet{VanderWeele2015Explanation}.
In the context of mediation analysis, the term ``fundamental unobservability'' refers to the fact that neither of the cross-world counterfactuals $Y(1, M(0))$ and $Y(0, M(1))$ can be observed in the data if $M(1) \neq M(0)$.
This issue persists even in a randomized experiment because it is impossible to realize treatment $Z = 1$ and mediator $M = M(0)$ simultaneously.
This means that without further assumptions, the natural direct and indirect effects are not identifiable, i.e. no unbiased estimator exists.
The typical solution is to posit exclusion restrictions or cross-world independence assumptions so that unbiased estimation of the causal effect is possible.

Following \citet{Aronow2017Estimating}, the majority of the subsequent literature on interference has focused on estimating contrastive effects, defined as:
\[
\ate = \frac{1}{n} \sum_{i=1}^n y_i(e_1) - y_i(e_0) 
\enspace.
\]
In the context of our paper, the term ``fundamental unobservability'' refers to the fact that there are pairs of potential outcomes $y_i(e_k)$ and $y_j(e_\ell)$ which cannot be \emph{observed simultaneously}.
For example, the two potential outcomes $y_i(e_1)$ and $y_i(e_0)$ for a single unit are not simultaneously observable because unit $i$ cannot receive both exposures $e_1$ and $e_0$ simultaneously.
Under network interference, potential outcomes $y_i(e_1)$ and $y_j(e_0)$ between neighboring units $i$ and $j$ will not be simultaneously observable if there is no intervention $\vec{z}$ which results in unit $i$ receiving exposure $e_1$ and unit $j$ receiving exposure $e_0$.
In this work, we introduce the conflict graph for the purpose of encoding this ``fundamental observability'', i.e. the conflict graph has edges between pairs of units with simultaneously unobservable potential outcomes.

In the context of network experiments, this problem of fundamental unobservability is not one of identification: indeed, randomization alone is enough to ensure that the contrastive effect can be unbiasedly estimated.
Instead, the fundamental unobservability introduces a problem of \emph{precision}.
If two potential outcomes between two units cannot be simultaneously observed, then this means that their corresponding terms in an estimator will necessarily be highly correlated.
The solution to this problem is not to place additional assumptions (e.g. exclusion restrictions, cross-world independence, etc), but rather to carefully construct the randomization of treatment so that the effect estimator is as precise as possible.
In this work, we have shown that under the Conflict Graph Design, it is possible to construct an unbiased effect estimator whose variance is of order $\lamH / n$.

In summary, the term ``fundamental unobservability'' references different problems{\textemdash}with different solutions{\textemdash} in the contexts of mediation and interference.
In mediation analysis, cross-world outcomes are not directly observable without further assumptions, which represents a problem of identification.
In network interference, pairs of potential outcomes are not simultaneously observable, which only poses a problem for precision.
While the two fields are largely distinct, connections between them serve as a fruitful area for research.
Initial works in this direction include \citet{Vanderweele2013Mediation} and \citet{Imai2020Identification}.

\subsection{Extension to Contrastive Marginal Effects} \label{sec:extention-to-marginal-effects}

In this section, we demonstrate how the Conflict Graph Design may be used to estimate the more general class of contrastive marginal casual effects.
All the results in this section may be viewed as strict generalizations of the main results presented in the paper for the class of contrastive effects.

\subsubsection{Marginal Contrastive Effects}

We begin by defining the class of marginal contrastive effects.
Suppose that for each unit $i \in [n]$, the experimenter specifies two contrasting distributions $\mathcal{D}_{i,0}$ and $\mathcal{D}_{i,1}$  over intervention vectors $\vec{z} \in \setb{0,1}^n$.
We define the corresponding \emph{contrastive marginal effect} as
\[
\ate = \frac{1}{n} \sum_{i=1}^n 
	\Esub[\big]{ \vec{z} \sim \mathcal{D}_{i,1} }{ y_i(\vec{z}) } 
	- \Esub[\big]{ \vec{z} \sim \mathcal{D}_{i,0} }{ y_i(\vec{z}) }
	\enspace.
\]
In other words, the contrastive marginal effect is the difference in the expected potential outcomes under the contrasting distributions $\mathcal{D}_{i,1}$ and $\mathcal{D}_{i,0}$, averaged over all units in the experiment.
We view the contrasting distributions as being defined separately from the experimental design.
For example, the class of contrastive effects studied in the main paper correspond to the case where each of the contrasting distributions places all its mass on one treatment assignment.

\begin{figure}[t]
	\centering
	
	% First subfigure
	\begin{subfigure}{0.24\textwidth}
		\centering
		
		\begin{tikzpicture}[every node/.style={circle, draw}]
			
			\definecolor{treated}{rgb}{1, 0.2, 0.2}
			\definecolor{control}{rgb}{0.3, 0.6, 1}
			
			\node[fill=control] (1) at (0,0) {};
			\node[fill=control] (2) at (1,1) {};
			\node[fill=control] (3) at (1,-1) {};
			\node[fill=treated] (4) at (-1,1) {};
			\node[fill=treated] (5) at (-1, -1) {};
			
			\draw (1) -- (2);
			\draw (1) -- (3);
			\draw (1) -- (4);
			\draw (1) -- (5);
		\end{tikzpicture}
		
		\caption{Exposure $e_1$}
		\label{fig:subim1}
	\end{subfigure}%
	\hfill
	% Second subfigure
	\begin{subfigure}{0.24\textwidth}
		\centering
		\begin{tikzpicture}[every node/.style={circle, draw}]
			\definecolor{treated}{rgb}{1, 0.2, 0.2}
			\definecolor{control}{rgb}{0.3, 0.6, 1}
			
			\node[fill=control] (1) at (0,0) {};
			\node[fill=treated] (2) at (1,1) {};
			\node[fill=control] (3) at (1,-1) {};
			\node[fill=treated] (4) at (-1,1) {};
			\node[fill=control] (5) at (-1, -1) {};
			
			\draw (1) -- (2);
			\draw (1) -- (3);
			\draw (1) -- (4);
			\draw (1) -- (5);
		\end{tikzpicture}
		\caption{Exposure $e_2$}
		\label{fig:subim2}
	\end{subfigure}%
	\hfill
	% Third subfigure
	\begin{subfigure}{0.24\textwidth}
		\centering
		\begin{tikzpicture}[every node/.style={circle, draw}]
			\definecolor{treated}{rgb}{1, 0.2, 0.2}
			\definecolor{control}{rgb}{0.3, 0.6, 1}
			
			\node[fill=control] (1) at (0,0) {};
			\node[fill=control] (2) at (1,1) {};
			\node[fill=treated] (3) at (1,-1) {};
			\node[fill=treated] (4) at (-1,1) {};
			\node[fill=control] (5) at (-1, -1) {};
			
			\draw (1) -- (2);
			\draw (1) -- (3);
			\draw (1) -- (4);
			\draw (1) -- (5);
		\end{tikzpicture}
		\caption{Exposure $e_3$}
		\label{fig:subim3}
	\end{subfigure}
	\hfill
	% Fourth subfigure
	\begin{subfigure}{0.24\textwidth}
		\centering
		\begin{tikzpicture}[every node/.style={circle, draw}]
			\definecolor{treated}{rgb}{1, 0.2, 0.2}
			\definecolor{control}{rgb}{0.3, 0.6, 1}
			
			\node[fill=control] (1) at (0,0) {};
			\node[fill=treated] (2) at (1,1) {};
			\node[fill=control] (3) at (1,-1) {};
			\node[fill=control] (4) at (-1,1) {};
			\node[fill=treated] (5) at (-1, -1) {};
			
			\draw (1) -- (2);
			\draw (1) -- (3);
			\draw (1) -- (4);
			\draw (1) -- (5);
		\end{tikzpicture}
		\caption{Exposure $e_4$}
		\label{fig:subim4}
	\end{subfigure}
	
	\caption{Four distinct exposures which all correspond to ``50\% of neighbors are treated''}
	\label{fig:distinct-exposures}
\end{figure}

Certain causal questions may lend themselves more naturally to the class of contrastive marginal effects.
For example, consider the question of ``what is the effect of having 50\% treated neighbors''?
The central difficulty of writing this as a contrastive effect of the form $\frac{1}{n} \sum_{i=1}^n y_i(e_1) - y_i(e_0)$ is that under the arbitrary neighborhood interference model, ``50\% treated neighbors
'' does not correspond to a unique exposure $e_1$.
Indeed, there will typically be many exposures that may satisfy this condition (see Figure~\ref{fig:distinct-exposures}) and it is not clear which one we ought to use as $e_1$.
One approach forward is to re-interpret the causal question as implicitly marginalizing over all the interventions $z$ in which ``50\% of neighbors are treated'' is true.
In other words, we may consider the marginal contrastive effect obtained by choosing the contrastive distributions as follows: $D_{i,1}$ is uniform on the set of assignments $\setb{ \vec{z} : \frac{1}{N(i) }\sum_{j \in N(i) } \vec{z}_j = 1/2 }$ and $D_{i,0}$ places all mass on assignment $\vec{z} = \vec{0}$.
This example shows that this broader class of estimands is more expressive.

This definition of marginalized contrastive treatment effect is independent of the arbitrary neighborhood interference model.
However, under the arbitrary neighborhood interference model, the class of marginal contrastive effects may equivalently be defined with respect to distributions on exposures, i.e.
\[
\ate = \frac{1}{n} \sum_{i=1}^n 
\Esub[\big]{ e_1 \sim \mathcal{D}_{i,1} }{ y_i(e_1) } 
- \Esub[\big]{ e_0 \sim \mathcal{D}_{i,0} }{ y_i(e_0) }
\enspace.
\]
Without loss of generality, we will continue to work with the definition which uses contrastive distributions over treatment assignments.

\subsubsection{Generalized Conflict Graph}

The conflict graph introduced in the main paper depended on both the underlying network and the contrastive effect.
In this section, we generalize that definition to be appropriate for the class of marginalized contrastive effect.
We begin by defining what it means for two units to be in conflict:

\begin{definition}[Conflicting Units]
	Two experimental units $i, j \in [n]$ are in \emph{conflict} if there exists treatment assignments $\vec{z}_i$ in the support of $\mathcal{D}_{i,k}$ and $\vec{z}_j$ in the support of $\mathcal{D}_{j,\ell}$ such that there is no assignment vector $\vec{z}$ such that $d_i(\vec{z}) = d_i(\vec{z}_i)$ and $d_j(\vec{z}) = d_i(\vec{z}_j)$.
\end{definition}

Following the main paper, we define the \emph{conflict graph} $\mathcal{H}$ to be a graph with units $i \in [n]$ as vertices with an edge between units $i$ and $j$ if they are in conflict.
As before, the conflict graph depends both on the underlying network and the causal effect.

\subsubsection{Generalized Conflict Graph Design and Estimator}

The generalized Conflict Graph Design is given formally as Algorithm~\ref{alg:cgd-marginal}.
The procedure is similar to the original conflict graph design: the ``desired exposures'' are drawn i.i.d. and conflicts are resolved via an importance ordering.
The key difference is that if a unit $i \in [n]$ is selected to receive $e_k$ (and all of its more important neighbors receive `*') then the experimenter will draw a treatment assignment $\vec{z} \sim \mathcal{D}_{i,k}$, and this will be used to determine treatment given to unit $i$ and its neighborhood.

\begin{algorithm}
    \DontPrintSemicolon
    \caption{Conflict Graph Design for Contrastive Marginal Effects}\label{alg:cgd-marginal}
	\KwIn{Importance ordering $\pi$, maximum eigenvalue $\lamH$, conflict graph $\mathcal{H}$ and distributions $\mathcal{D}_{i,0}, \mathcal{D}_{i,1}$ for all $i \in [n]$.}
	\KwOut{Random intervention $Z \in \mathcal{Z} = \setb{0,1}^n$}
	Set sampling parameter $r = 2$\;
	Sample desired exposure variables $U_1, \dots, U_n$ independently and identically as
	$$
	U_i \gets
	\left\{
	\begin{array}{lr}
		e_1 & \text{with probability } \frac{1}{r \cdot 2 \lamH}\\
		e_0 & \text{with probability } \frac{1}{r \cdot 2 \lamH}\\
		* & \text{with probability } 1 - \frac{1}{ r \cdot \lamH }
	\end{array}
	\right.
	$$\;
    Sample variables $Z^{(i,0)} \sim \mathcal{D}_{i,0}$ and $Z^{(i,1)} \sim \mathcal{D}_{i,1}$ for all $i \in [n]$ independently as 
    $$
    Z^{(i,0)} \sim \mathcal{D}_{i,0}\quad, \quad Z^{(i,1)} \sim \mathcal{D}_{i,1}
    $$\;
	Initialize the intervention vector $Z \gets \vec{0}$. \;
	\For{$i =1 \dots n$}{
		\If{$U_i = e_k \in \setb{e_1, e_0}$ and $U_j = *$ for all $j \in \mathcal{N}^\pi_b(i)$\label{algline:exposure-condition}}{
			Update intervention vector: set $Z_j = Z^{(i,k)}(j)$ for all $j \in \widetilde{N}(i)$
			\label{algline:set-intervention} \;
		}
	}
\end{algorithm}

The modified Horvitz--Thompson estimator may be derived in a similar way.
We define the desired exposure events as
\begin{align*}
    E_{(i,1)} & = \setb{ U_i = e_1, U_j = * \text{ ,for all } j \in \mathcal{N}^\pi_b(i)} \\
    E_{(i,0)} & = \setb{ U_i = e_0, U_j = * \text{ ,for all } j \in \mathcal{N}^\pi_b(i)}
    \enspace.
\end{align*}
The estimator is defined as
\begin{equation}\label{eq:eate-marginal}
    \eate = \frac{1}{n} \sum_{i=1}^n Y_i \cdot \paren[\Big]{ \frac{\indicator{E_{(i,1)}}}{\Pr{E_{(i,1)}}}  -  \frac{\indicator{E_{(i,0)}}}{\Pr{E_{(i,0)}}} }
\end{equation}

\subsubsection{Analysis of Bias and Variance}

In the setting of marginalized contrastive effects, we will find it helpful to further divide the desired exposure events $E_{(i,0)}, E_{(i,1)}$ according to which particular exposure will be observed.
Thus, for each unit $i$ and exposure $z \in \{0,1\}^{|\tilde{N}(i)|}$, we define the events
\begin{align*}
	E_{(i,1,z)} & = \setb{ U_i = e_1, U_j = * \text{ for all } j \in \mathcal{N}^\pi_b(i), Z^{(i,1)} = z } \\
	E_{(i,0,z)} & = \setb{ U_i = e_0, U_j = * \text{ for all } j \in \mathcal{N}^\pi_b(i), Z^{(i,0)} = z }
\end{align*}
We begin by first establishing a convenient expression for the probabilities of events $E_{(i,1,z)}, E_{(i,0,z)}$. 
The proof is very similar to that of Lemma~\mainref{lemma:prob-of-desired-exposure-events}.
\begin{lemma}\label{lem:probabilities-marginal}
    Under the Conflict Graph Design, for every node $i$, every contrast $k \in \{0,1\}$ and every exposure $z \in \{0,1\}^{|\tilde{N}(i)|}$, we have
    \begin{gather*}
        \Pr{E_{(i,k,z)}} = \Pr{E_{(i,k)}} \cdot \Prsub{\mathcal{D}_{(i,k)}}{z} 
        = \frac{1}{2r\lambda(\cH)} \paren[\Big]{1 - \frac{1}{r\lambda(\cH)}}^{|\mathcal{N}_b^\pi(i)|} \cdot \Prsub{\mathcal{D}_{(i,k)}}{z}
        \enspace.
    \end{gather*}
\end{lemma}
\begin{proof}
    It has already been established in Lemma~\mainref{lemma:prob-of-desired-exposure-events} for the usual Conflict Graph Design that 
    \[
    \Pr{E_{(i,k)}} = \frac{1}{2r\lambda(\cH)} \paren[\Big]{1 - \frac{1}{r\lambda(\cH)}}^{|\mathcal{N}_b^\pi(i)|} \enspace.
    \]
    This of course continues to hold with the design of Algorithm~\ref{alg:cgd-marginal}, since the event is exactly the same.
    In order for $E_{(i,k,z)}$ to occur, we need $E_{(i,k)}$ to occur and also $Z^{(i,k)} = z$. Since $E_{(i,k)}$ depends on $U$ variables, that are sampled independently of the $Z$ variables, we have
    \[
    \Pr{E_{(i,k,z)}} = \Pr{E_{(i,k)}} \cdot \Pr{Z^{(i,k)} = z} = \Pr{E_{(i,k)}} \cdot \Prsub{\mathcal{D}_{(i,k)}}{z} \enspace.
    \qedhere
    \]
\end{proof}

In the next Lemma, we establish that the estimator $\eate$ is unbiased, similarly to how we did for the modified Horvitz-Thompson estimator in Proposition~\mainref{prop:unbiased-estimator}.

\begin{proposition}\label{prop:eate-marginal-unbiased}
Under arbitrary neighborhood interference (Assumption~\mainref{assumption:ani-model}), the modified Horvitz--Thompson estimator $\eate$ is unbiased under the Conflict Graph Design: $\E{\eate} = \ate$.
\end{proposition}

\begin{proof}
    We have
    \begin{align}
    \E{\eate} & = \frac{1}{n} \sum_{i=1}^n \E[\Bigg]{\frac{\indicator{E_{(i,1)}}}{\Pr{E_{(i,1)}}} \cdot Y_i  - \frac{\indicator{E_{(i,0)}}}{\Pr{E_{(i,0)}}} \cdot Y_i}\nonumber\\
    &= \frac{1}{n} \sum_{i=1}^n \frac{\E[\big]{\indicator{E_{(i,1)}}\cdot Y_i}}{\Pr{E_{(i,1)}}}   - \frac{\E[\big]{\indicator{E_{(i,0)}}\cdot Y_i}}{\Pr{E_{(i,0)}}} \label{eq:eate-simplified}
    \end{align}
    Let us know compute $\E[\big]{\indicator{E_{(i,k)}}\cdot Y_i}$. We have
    \begin{align*}
        \E[\big]{\indicator{E_{(i,k)}}\cdot Y_i} &= \E[\Big]{\sum_{z \in \{0,1\}^{|\tilde{N}(i)|}} \indicator{E_{(i,k,z)}}\cdot Y_i}\\
        &= \sum_{z \in \{0,1\}^{|\tilde{N}(i)|}} \E[\big]{\indicator{E_{(i,k,z)}}}\cdot y_i(z)\\
        &= \sum_{z \in \{0,1\}^{|\tilde{N}(i)|}} \Pr{E_{(i,k)}}\Prsub{\mathcal{D}_{(i,k)}}{z}\cdot y_i(z)\enspace,
    \end{align*}
    where in the last step we used Lemma~\ref{lem:probabilities-marginal}. 
    Plugging this inside \eqref{eq:eate-simplified} yields the result.

    % In particular, as we have seen in the proof of REF, we know that event $E_{(i,1)}$ occurs with probability 
    % \begin{align*}
    %     \E{\eate} & = \frac{1}{n} \sum_{i=1}^n \E[\Bigg]{\sum_{z \in \{0,1\}^{|\tilde{N}(i)|}} \Prsub{\mathcal{D}_{(i,1)}}{z}\cdot\frac{\indicator{E_{(i,1,z)}}}{\Pr{E_{(i,1,z)}}} \cdot Y_i  - \sum_{z \in \{0,1\}^{|\tilde{N}(i)|}} \Prsub{\mathcal{D}_{(i,0)}}{z}\cdot\frac{\indicator{E_{(i,0,z)}}}{\Pr{E_{(i,0,z)}}} \cdot Y_i} \\
    %     &= \frac{1}{n} \sum_{i=1}^n \sum_{z \in \{0,1\}^{|\tilde{N}(i)|}} \Prsub{\mathcal{D}_{(i,1)}}{z}\cdot \E[\Bigg]{\frac{\indicator{E_{(i,1,z)}}}{\Pr{E_{(i,1,z)}}} \cdot Y_i}  - \sum_{z \in \{0,1\}^{|\tilde{N}(i)|}} \Prsub{\mathcal{D}_{(i,0)}}{z}\cdot \E[\Bigg]{\frac{\indicator{E_{(i,0,z)}}}{\Pr{E_{(i,0,z)}}} \cdot Y_i} \\
    %     &= \frac{1}{n} \sum_{i=1}^n \sum_{z \in \{0,1\}^{|\tilde{N}(i)|}} \Prsub{\mathcal{D}_{(i,1)}}{z}\cdot y_i(z)  - \sum_{z \in \{0,1\}^{|\tilde{N}(i)|}} \Prsub{\mathcal{D}_{(i,0)}}{z}\cdot y_i(z) \\
    %     &= \frac{1}{n} \sum_{i=1}^n \Esub{ z \sim \mathcal{D}_{i,1} }{ y_i(z) } - \Esub{ z \sim \mathcal{D}_{i,0} }{ y_i(z) } \\
    %     &= \ate
    %     \end{align*}
\end{proof}

We can also establish an upper bound on the variance of the modified Horvitz--Thompson estimator under the generalized Conflict Graph Design, similarly to Theorem~\mainref{thm:variance-analysis-finite-sample}.
The theorem below is nearly identical to Theorem~\mainref{thm:variance-analysis-finite-sample}, except that the variance is bounded by the expected second moment with respect to the contrastive distributions $\mathcal{D}_{i,1}$ and $\mathcal{D}_{i,0}$.
In this way, the magnitude of the potential outcomes is measured with respect to the contrastive distributions.

\begin{theorem}\label{thm:variance-bound-marginal-effects}
Under arbitrary neighborhood interference (Assumption~\mainref{assumption:ani-model}), there exists a constant $C > 0$ such that the variance of the modified Horvitz--Thompson estimator $\eate$ under the Conflict Graph Design is bounded as
\[
    \Var{\eate} 
    \leq C \cdot \frac{\lamH}{n} \cdot 
    \braces[\Bigg]{
    \frac{1}{n} \sum_{i=1}^n 
    \Esub{\vec{z} \sim \mathcal{D}_{i,1}}{ y_i(\vec{z} )^2 }  
    + \Esub{\vec{z} \sim \mathcal{D}_{i,0}}{ y_i(\vec{z} )^2 }
	}
    \enspace.
\]
\end{theorem}
\begin{proof}
    As we did in the proof of unbiasedness in Lemma~\ref{prop:eate-marginal-unbiased}, it will be convenient to write the estimator in the following form.
    \[
    \eate = \frac{1}{n} \sum_{i=1}^n \sum_{z \in \{0,1\}^{|\tilde{N}(i)|}} \Prsub{\mathcal{D}_{(i,1)}}{z}\cdot\frac{\indicator{E_{(i,1,z)}}}{\Pr{E_{(i,1,z)}}} \cdot y_i(z)  - \sum_{z \in \{0,1\}^{|\tilde{N}(i)|}} \Prsub{\mathcal{D}_{(i,0)}}{z}\cdot\frac{\indicator{E_{(i,0,z)}}}{\Pr{E_{(i,0,z)}}} \cdot y_i(z)
    \]
    We start by decomposing our estimator and estimand as
    \begin{gather*}
    \eate = \\
     \underbrace{\frac{1}{n} \sum_{i=1}^n \sum_{z \in \{0,1\}^{|\tilde{N}(i)|}} \Prsub{\mathcal{D}_{(i,1)}}{z}\cdot\frac{\indicator{E_{(i,1,z)}}}{\Pr{E_{(i,1,z)}}} y_i(z)}_{\eate_1} - \underbrace{\frac{1}{n} \sum_{i=1}^n \sum_{z \in \{0,1\}^{|\tilde{N}(i)|}} \Prsub{\mathcal{D}_{(i,0)}}{z}\cdot\frac{\indicator{E_{(i,0,z)}}}{\Pr{E_{(i,0,z)}}} y_i(z)}_{\eate_0}\\
    \ate = \underbrace{\frac{1}{n} \sum_{i=1}^n \Esub{ z \sim \mathcal{D}_{i,1} }{ y_i(z) }}_{\ate_1} - \underbrace{\frac{1}{n} \sum_{i=1}^n \Esub{ z \sim \mathcal{D}_{i,0} }{ y_i(z) }}_{\ate_0}
    \end{gather*}
    Clearly, the proof of Proposition~\ref{prop:eate-marginal-unbiased} shows that $\E{\eate_1} = \ate_1$ and $\E{\eate_0} = \ate_0$. Thus, we have that 
    \begin{align*}
        \Var{\eate} & = \Var{\eate_1 - \eate_0} \\
        & = \Var{\eate_1} + \Var{\eate_0} - 2\Cov{\eate_1, \eate_0} \\
        & \le 2\Var{\eate_1} + 2\Var{\eate_0}\enspace.
    \end{align*}
    In the last step, we used the Cauchy-Schwarz inequality to bound the covariance term and then the AM-GM inequality.
    Thus, it suffices to bound the variance of $\eate_1$ and $\eate_0$ separately. We will bound the variance of $\eate_1$, and a similar bound for $\eate_0$ follows analogously.

    We now expand the variance of $\eate_1$ as
    \begin{align}
        \Var{\eate_1} & = \Var[\Bigg]{\frac{1}{n} \sum_{i=1}^n \sum_{z \in \{0,1\}^{|\tilde{N}(i)|}} \Prsub{\mathcal{D}_{(i,1)}}{z}\cdot\frac{\indicator{E_{(i,1,z)}}}{\Pr{E_{(i,1,z)}}} y_i(z)} \nonumber\\
        & = \frac{1}{n^2} \sum_{i=1}^n \sum_{j=1}^n \sum_{z \in \{0,1\}^{|\tilde{N}(i)|}} \sum_{z' \in \{0,1\}^{|\tilde{N}(j)|}} C_{(i,j,z,z')} \cdot y_i(z) y_j(z')\enspace,
    \end{align}
        where we defined the covariance terms for every $i,j \in [n]$, $z \in  \{0,1\}^{|\tilde{N}(i)|}$ and $z' \in \{0,1\}^{|\tilde{N}(j)|}$ as
        \[
        C_{(i,j,z,z')} = \Prsub{\mathcal{D}_{i,1}}{z}\cdot \Prsub{\mathcal{D}_{j,1}}{z}\cdot \Cov[\Bigg]{\frac{\indicator{E_{(i,1,z)}}}{\Pr{E_{(i,1,z)}}}, \frac{\indicator{E_{(j,1,z')}}}{\Pr{E_{(j,1,z')}}}}
        \enspace.
        \]
        Let us now compute these covariance terms.
        We denote by $d(i,j)$ the distance between $i,j$ in the Conflict Graph $\cH$.
        Throughout the remainder of the proof, we use $C > 0$ to denote a (possibly changing) constant which depends only on the choice of $r$ in the Conflict Graph Design.
        
        \textbf{Single Unit}:
        First, if $i=j$ and $z=z'$, we have
        \begin{align*}
            C_{(i,i,z,z)} & = \Prsub{\mathcal{D}_{i,1}}{z}^2 \cdot \Var[\Big]{\frac{\indicator{E_{(i,1,z)}}}{\Pr{E_{(i,1,z)}}}} \\
            & = \Prsub{\mathcal{D}_{i,1}}{z}^2 \cdot \paren[\Big]{\frac{1}{\Pr{E_{(i,1,z)}}} - 1} \\
            & = \frac{\Prsub{\mathcal{D}_{i,1}}{z}}{\Pr{E_{(i,1)}}} - \Prsub{\mathcal{D}_{i,1}}{z}^2
        \end{align*}
        Now suppose $i = j$ but $z \neq z'$. Clearly, the events $E_{(i,1,z)} , E_{(i,1,z')}$ are disjoint, so the covariance term is
        \[
        C_{(i,i,z,z')} = - \Prsub{\mathcal{D}_{i,1}}{z}^2
        \enspace.
        \]
        Thus, for the terms $i = j$ in variance decomposition, we have
        \begin{align}
        	&\frac{1}{n^2} \sum_{i=1}^n \sum_{z \in \{0,1\}^{|\tilde{N}(i)|}} \sum_{z' \in \{0,1\}^{|\tilde{N}(j)|}} C_{(i,j,z,z')} \cdot y_i(z) y_i(z') \\
            &\frac{1}{n^2} \sum_{i=1}^n \paren[\Bigg]{\sum_{z \in \{0,1\}^{|\tilde{N}(i)|}} \frac{\Prsub{\mathcal{D}_{i,1}}{z}}{\Pr{E_{(i,1)}}} y_i(z)^2 - \paren[\Big]{\sum_{z \in \{0,1\}^{|\tilde{N}(i)|}} \Prsub{\mathcal{D}_{i,1}}{z} y_i(z) }^2}\nonumber\\
            &\quad\leq \frac{1}{n^2} \sum_{i=1}^n \sum_{z \in \{0,1\}^{|\tilde{N}(i)|}} \frac{\Prsub{\mathcal{D}_{i,1}}{z}}{\Pr{E_{(i,1)}}} y_i(z)^2 \nonumber \\
            &\quad \leq \frac{C \cdot \lamH}{n^2} \sum_{i=1}^n \sum_{z \in \{0,1\}^{|\tilde{N}(i)|}} \Prsub{\mathcal{D}_{i,1}}{z} y_i(z)^2
            &\text{(Lemma~\mainref{lemma:individual-covariance-terms})}
             \nonumber\\
            &\quad = \frac{C \cdot \lamH}{n} \cdot \frac{1}{n} \sum_{i=1}^n \Esub[\big]{\vec{z} \sim \mathcal{D}_{i,1}}{ y_i(\vec{z})^2 } \label{eq:variance-ii}
            \enspace.
        \end{align}

		\textbf{Neighboring Units}:
        Let us now focus on distinct pairs of units $i,j \in [n]$ which are neighbors in the conflict graph $\cH$, i.e. $d(i,j)=1$.
        For these units, the desired exposure events $E_{(i,1,z)}, E_{(j,1,z')}$ are disjoint.
        Thus, the covariance terms for any $z,z'$ are given as
        \[
        C_{i,j,z,z'} = - \Prsub{\mathcal{D}_{i,1}}{z} \Prsub{\mathcal{D}_{j,1}}{z'}
        \enspace.
        \]
        Thus, the corresponding terms in the variance can be written as 
        \begin{align}
        &\frac{1}{n^2} \sum_{i \neq j: d(i,j) = 1} \sum_{z \in \{0,1\}^{|\tilde{N}(i)|}} \sum_{z' \in \{0,1\}^{|\tilde{N}(j)|}} C_{(i,j,z,z')} \cdot y_i(z) y_j(z') \\
        &\quad = \frac{1}{n^2} \sum_{i \neq j: d(i,j) = 1} \sum_{z \in \{0,1\}^{|\tilde{N}(i)|}} \sum_{z' \in \{0,1\}^{|\tilde{N}(j)|}} \paren[\big]{-\Prsub{\mathcal{D}_{i,1}}{z} \Prsub{\mathcal{D}_{j,1}}{z'} y_i(z)y_j(z')}\\
        &\quad = - \frac{1}{n^2} \sum_{i \neq j: d(i,j) = 1} \paren[\Big]{\sum_{z \in \{0,1\}^{|\tilde{N}(i)|}} \Prsub{\mathcal{D}_{i,1}}{z} y_i(z)} \paren[\Big]{\sum_{z' \in \{0,1\}^{|\tilde{N}(i)|}} \Prsub{\mathcal{D}_{i,1}}{z'} y_j(z')} \\
        &\quad= - \frac{1}{n} \sum_{i=1}^n \sum_{j=1}^n A_{\cH}(i,j) \cdot
         \paren[\Big]{ \frac{1}{\sqrt{n}}  \Esub{\vec{z} \sim \mathcal{D}_{i,1}}{ y_i(\vec{z}) } }
         \paren[\Big]{ \frac{1}{\sqrt{n}}  \Esub{\vec{z} \sim \mathcal{D}_{j,1}}{ y_j(\vec{z}) } } \\
         &\quad \leq  \frac{\lamH}{n} \cdot \frac{1}{n} \sum_{i=1}^n \Esub{\vec{z} \sim \mathcal{D}_{i,1}}{ y_i(\vec{z}) }^2 \\
         	\intertext{by definition of operator norm}\\
         &\quad \leq \frac{\lamH}{n} \cdot \frac{1}{n} \sum_{i=1}^n \Esub{\vec{z} \sim \mathcal{D}_{i,1}}{ y_i(\vec{z})^2 }\\
         	\intertext{by Jensen's inequality.}
        \end{align}

		\textbf{Non-neighboring Units}:
        Finally, let us focus on pairs of units $i,j$ that are at least distance $2$ in the Conflict Graph $\cH$. 
        For any $z,z'$, the correlation between events $E_{(i,1,z)}, E_{(j,1,z')}$ is bounded similarly to that of $E_{(i,1)},E_{j,1}$, as the following calculation shows:
        \begin{align*}
            C_{i,j,z,z'} &= \Prsub{\mathcal{D}_{i,1}}{z}\cdot \Prsub{\mathcal{D}_{j,1}}{z}\cdot\Cov[Bigg]{\frac{\indicator{E_{(i,1,z)}}}{\Pr{E_{(i,1,z)}}}, \frac{\indicator{E_{(j,1,z')}}}{\Pr{E_{(j,1,z')}}}} \\
            &= \Prsub{\mathcal{D}_{i,1}}{z}\cdot \Prsub{\mathcal{D}_{j,1}}{z}\cdot \paren[\Big]{\frac{\Pr{E_{(i,1,z)} \cap E_{(j,1,z')}}}{\Pr{E_{(i,1,z)}}\Pr{E_{(j,1,z')}}} - 1}\\
            &= \Prsub{\mathcal{D}_{i,1}}{z}\cdot \Prsub{\mathcal{D}_{j,1}}{z}\cdot \paren[\Big]{\frac{\Prsub{\mathcal{D}_{i,1}}{z}\cdot \Prsub{\mathcal{D}_{j,1}}{z} \cdot \Pr{E_{(i,1)} \cap E_{(j,1)}}}{\Prsub{\mathcal{D}_{i,1}}{z}\cdot \Prsub{\mathcal{D}_{j,1}}{z} \cdot \Pr{E_{(i,1)}} \cdot \Pr{E_{(j,1)}}} - 1}\\
            &= \Prsub{\mathcal{D}_{i,1}}{z}\cdot \Prsub{\mathcal{D}_{j,1}}{z}\cdot \paren[\Big]{\frac{ \Pr{E_{(i,1)} \cap E_{(j,1)}}}{ \Pr{E_{(i,1)}} \cdot \Pr{E_{(j,1)}}} - 1}\\
            &\leq C \Prsub{\mathcal{D}_{i,1}}{z}\cdot \Prsub{\mathcal{D}_{j,1}}{z}\cdot \frac{|\mathcal{N}_b^\pi(i)\cap \mathcal{N}_b^\pi(j)|}{\lamH}\enspace,
        \end{align*}
        where in the final inequality we used Lemma~\mainref{lemma:individual-covariance-terms} for some constant $C$. 
        Thus, the remaining terms in the variance decomposition may be bounded as
        \begin{align*}
            &\frac{1}{n^2}\sum_{i \neq j: d(i,j) \geq 2} \sum_{z \in \{0,1\}^{|\tilde{N}(i)|}} \sum_{z' \in \{0,1\}^{|\tilde{N}(j)|}}  y_i(z)y_j(z') C_{i,j,z,z'}\\
            &\quad \leq \frac{C}{n^2} \sum_{i \neq j: d(i,j) \geq 2} \sum_{z \in \{0,1\}^{|\tilde{N}(i)|}} \sum_{z' \in \{0,1\}^{|\tilde{N}(j)|}} \Prsub{\mathcal{D}_{i,1}}{z} \Prsub{\mathcal{D}_{j,1}}{z'} |y_i(z)y_j(z')| \frac{|\mathcal{N}(i)\cap \mathcal{N}(j)|}{\lamH}\\
            &\quad = \frac{C}{n^2} \sum_{i \neq j: d(i,j) \geq 2} \frac{|\mathcal{N}(i)\cap \mathcal{N}(j)|}{\lamH} \paren[\Big]{\sum_{z \in \{0,1\}^{|\tilde{N}(i)|}} \Prsub{\mathcal{D}_{i,1}}{z} |y_i(z)|} \paren[\Big]{\sum_{z' \in \{0,1\}^{|\tilde{N}(i)|}} \Prsub{\mathcal{D}_{i,1}}{z'} |y_j(z')|} \\
            &\quad = \frac{C}{n \cdot \lamH}  \sum_{i \neq j: d(i,j) \geq 2} \abs[\big]{ \mathcal{N}(i) \cap \mathcal{N}(j) }
            \paren[\Big]{ \frac{1}{\sqrt{n}} \Esub[\big]{\vec{z} \sim \mathcal{D}_{i,1}}{ \abs{y_i(z)} } }
            \paren[\Big]{ \frac{1}{\sqrt{n}} \Esub[\big]{\vec{z} \sim \mathcal{D}_{j,1}}{ \abs{y_j(z)} } } \\
            &\quad \leq \frac{C}{n \cdot \lamH} \sum_{i=1}^n \sum_{j=1}^n A_{\cH}^2(i,j)
            \paren[\Big]{ \frac{1}{\sqrt{n}} \Esub[\big]{\vec{z} \sim \mathcal{D}_{i,1}}{ \abs{y_i(z)} } }
            \paren[\Big]{ \frac{1}{\sqrt{n}} \Esub[\big]{\vec{z} \sim \mathcal{D}_{j,1}}{ \abs{y_j(z)} } } \\
            &\quad \leq \frac{C}{n \cdot \lamH} \cdot \lamH^2 \cdot \frac{1}{n} \sum_{i=1}^n \Esub[\big]{\vec{z} \sim \mathcal{D}_{i,1}}{ \abs{y_i(\vec{z})} }^2 \\
            \intertext{by definition of the operator norm}\\
            &\quad \leq \frac{C \cdot \lamH}{n} \cdot \frac{1}{n} \sum_{i=1}^n \Esub[\big]{\vec{z} \sim \mathcal{D}_{i,1}}{y_i(\vec{z})^2 }\\
            \intertext{by Jensen's inequality.}
        \end{align*}
        
        Thus, by combining the analysis for single units, neighboring units, and non-neighboring units, we have that the variance is bounded as desired, i.e. there exists a constant $C$ such that
        \[
        \Var{\eate_1}
        \leq C \frac{\lamH}{n} \cdot \frac{1}{n} \sum_{i=1}^n \Esub{\vec{z} \sim \mathcal{D}_{i,1}}{ y_i(\vec{z})^2 }
        \enspace.
        \]
        Similar arguments can be made for $\Var{\eate_2}$ and thus the proof is complete. 
    \end{proof}

Theorem~\ref{thm:variance-bound-marginal-effects} is a finite-sample bound that holds for any network.
If we want to interpret this to asymptotically mean that the variance is of order at most $\lamH / n$, then we need to assume that the second moment of the outcomes with respect to the contrastive distributions is asymptotically bounded, which is a generalization of Assumption~\mainref{assumption:bounded-second-moment}.

\subsection{Concentration of the Number of Exposed Units} \label{sec:supp-number-of-exposed-units}

In this section, we analyze the concentration properties of the number of units which are assigned to their desired exposure.
Formally, for each exposure $k \in \{0,1\}$, we define the random variable
\[
    T_k = \sum_{i=1}^n \indicator{E_{(i,k)} } \enspace.
\]
$T_k$ is exactly the number of units assigned to their desired exposure $e_k$ under the Conflict Graph Design.
These are also the number of units that will be used by the modified Horvitz-Thompson estimator.
In this section, we will prove the following theorem, which shows that $T_k$ concentrates around its expectation $\E{T_k}$, which is of order $n / \lamH$.

\begin{proposition}\label{thm:concentration-exposed-units}
	There exists constants $c$, $C$, and $K$ such that the following holds:
    Under the Conflict Graph Design, for either exposure $k \in \setb{0,1}$ and for any $\alpha \in (0,1)$, with probability at least $1-\alpha$ it holds that
    \[
        c \cdot \frac{n}{\lamH} - \sqrt{\frac{1}{\alpha} }\cdot \sqrt{K \cdot \frac{n}{\lamH}}\leq T_k \leq C \cdot \frac{n}{\lamH} + \sqrt{\frac{1}{\alpha} }\cdot \sqrt{K \cdot \frac{n}{\lamH}}\enspace.
    \]
\end{proposition}

Proposition~\ref{thm:concentration-exposed-units} shows that in the consistency regime where $\lamH = \littleO{n}$, we have that $T$ concentrates around its mean with probability tending to 1.
In the remainder of this section, we discuss how to prove this result.
Without loss of generality, we focus on $k=1$ and drop the corresponding subscript when it is clear from context.

To prove Proposition~\ref{thm:concentration-exposed-units}, we will analyze the mean and variance of $T$ and then apply Chebyshev's inequality.
To this end, it will be helpful to recognize that $T$ is equivalent to the modified Horvitz--Thompson estimator for a specific choice of potential outcomes.
Consider the potential outcomes $y_i(e_1) = \Pr{E_{(i,1)}}$ and $y_i(e_0) = 0 $.
For this choice of potential outcomes, the Horvitz--Thompson estimator is related to the number of treated units as $T = n \cdot \eate$, i.e.
\[
\eate 
= \frac{1}{n} \sum_{i=1}^n \braces[\Bigg]{y_i(e_1) \cdot \frac{\indicator{E_{(i,1)}}}{\Pr{\indicator{E_{(i,1)}}}} - y_i(e_0) \cdot \frac{\indicator{E_{(i,0)}}}{\Pr{\indicator{E_{(i,0)}}}} }
= \frac{1}{n} \sum_{i=1}^n \indicator{E_{(i,1)}}
= \frac{1}{n} \cdot T
\enspace.
\]
One can use this relationship $T = n \cdot \eate$ and existing analysis for $\eate$ in the main body to prove high probability bounds on $T$.

We start by computing the expectation of $T$.
This is straighforward using the lower bounds for the probabilities of exposure as well as linearity of expectation.

\begin{lemma}\label{lem:expectation_exposed}
    Under the Conflict Graph Design, there exists constants $c$ and $C$ such that
    \[
       c \cdot \frac{n}{\lamH} \leq \E{T}  \leq C \cdot \frac{n}{\lamH} \enspace.
    \]
\end{lemma}
\begin{proof}
    The proof follows directly from the linearity of expectation and the upper and lower bounds on the probabilities of exposure from Lemma~\mainref{lemma:individual-covariance-terms}.
\end{proof}

Next, we upper bound the variance of $T$. 
Because $T$ is essentially an instance of the Horvitz--Thompson estimator, we can directly use the result of  Theorem~\mainref{thm:variance-analysis-finite-sample} to obtain an upper bound on the variance:

\begin{lemma}\label{lem:variance_exposed}
	There exists a constant $K$ such that
    under the Conflict Graph Design
    \[
        \Var{T} \leq K \cdot \frac{n}{\lamH} \enspace.
    \]
\end{lemma}

Thus, combining Lemmas~\ref{lem:expectation_exposed} and~\ref{lem:variance_exposed}, we can show that $T_k$ concentrates around its expectation using Chebyshev's inequality.
We give the proof of the main result below.

\begin{proof}[Proof of Theorem~\ref{thm:concentration-exposed-units}]
    Using Chebyshev's inequality and Lemma~\ref{lem:variance_exposed}, we have that for any $\alpha \in (0,1)$
    \begin{align*}
        \Pr[\Bigg]{\abs{T_k - \E{T_k}} \geq \sqrt{\frac{K(r) n}{\alpha\lamH} }} &\leq \frac{\Var{T_k}}{\frac{K(r)n}{\alpha\lamH}} 
        \leq \alpha \enspace.
    \end{align*}
    The result then follows by combining this with the bounds on $\E{T_k}$ from Lemma~\ref{lem:expectation_exposed}.
\end{proof}

\subsection{Comparison to Dependency Graph Analysis}\label{sec:dependency-graph}

In this section, we elaborate on the dependency graph method for analyzing the rates of consistency in network experiments.
Variants of this method have been widely used (either explicitly or implicitly) throughout the literature on interference \citep[see e.g.][]{Aronow2017Estimating, Ogburn2024Causal, Harshaw2023Design, Ugander2023Randomized}.
We show that the dependency graph analysis is loose when applied to the Conflict Graph Design, incurring an additional $d_{ \max{} }(\cH^2)$ factor.

We begin by introducing a commonly used notion of dependency graph.

\begin{definition}\label{def:dependency-graph}
	Given a set of random variables $X_1 \cdots X_n$, a graph $\mathcal{D} = (V,E)$ with vertices $V = \setb{1 \dots n}$ is a \emph{dependency graph} if the absence of an edge between $i$ and $j$ implies that random variables $X_i$ and $X_j$ are independent.
\end{definition}

\newcommand{\maxdepdeg}{d_{\max{}}(\mathcal{D})}

We denote the degree of a random variable $X_i$ in the dependency graph as $d_{\mathcal{D}}(i)$.
The maximum degree $\maxdepdeg = \max_{i \in [n]} d_{\mathcal{D}}(i)$ of the dependency graph is a measure of the independence of the set of random variables $X_1 \dots X_n$.
Definition~\ref{def:dependency-graph} describes the simplest form of a dependency graph using only pair-wise independence, which suffices for the purpose of analyzing the variance.
More nuanced forms of dependency graphs involving higher order independence are used in central limit theorems (see e.g. the definition of dependency graph used in the proof of the CLT in Section~\suppref{sec:supp-clt-proof}).

We seek to bound the variance of the modified Horvitz--Thompson estimator.
Recall that this estimator can be decomposed into the average of individual estimators, i.e.
\[
\eate = \frac{1}{n} \sum_{i=1}^n Y_i \cdot \paren[\Bigg]{ 
	\frac{\indicator[\big]{E_{(i,1)}}}{\Pr[\big]{E_{(i,1)}} } 
	-  
	\frac{\indicator[\big]{E_{(i,0)}}}{\Pr[\big]{E_{(i,0)}} } 
}
= \frac{1}{n} \sum_{i=1}^n \eate_i
\enspace,
\]
where the desired exposure events are defined as 
\[
E_{(i,k)} = \setb[\Big]{ U_i = e_k \text{ and } U_j = * \text{ for all } j \in \mathcal{N}_b^\pi(i)  }
\enspace.
\]
We will consider a dependency graph $\mathcal{D}$ on the individual estimators $\eate_1 \dots \eate_n$.
Note that the dependency graph $\mathcal{D}$ encodes the independence of the individual estimators, whereas the conflict graph $\cH$ encodes only fundamental unobservability of pairs of potential outcomes.
This two graphs are distinct and, in fact, will typically not be equal in network experiments.
We return to this point more later in the section.

We will also define $\gamma$ to be the inverse of the smallest relevant exposure probability, i.e.
\[
\gamma = \max_{i \in [n], k \in \setb{0,1}} \frac{1}{\Pr{ E_{(i,k)} }}
\enspace.
\]
This quantity is relevant for bounding the variance of an individual effect estimator, as shown in the following lemma.
We remark that while we focus on the Conflict Graph Design, the same principle may be used for an arbitrary design.

\begin{lemma} \label{lemma:individual-var-bound-dep-graph}
	Suppose the arbitrary neighborhood interference model holds (Assumption~\mainref{assumption:ani-model}).
	Under the Conflict Graph Design, the variance of each individual effect estimator may be bounded as
	\[
	\Var{\eate_i} \leq \gamma \cdot \paren[\Big]{ y_i(e_1)^2 + y_i(e_0)^2 }
	\enspace.
	\]
\end{lemma}
\begin{proof}
	Recall that the individual estimator $\eate_i$ is unbiased for the individual effect $\ate_i$.
	Under the arbitrary neighborhood interference assumption,
	the variance may be calculated as
	\begin{align*}
		\Var{\eate_i} 
		&= \E[\big]{ \paren[\big]{\eate_i - \ate_i}^2 }  \\
		&= \E[\Big]{ \paren[\Big]{ 
				y_i(e_1) \cdot \braces[\Big]{ \frac{\indicator{E_{(i,1)}}}{\Pr{E_{(i,1)}}} - 1 } 
				-
				y_i(e_0) \cdot \braces[\Big]{ \frac{\indicator{E_{(i,0)}}}{\Pr{E_{(i,0)}}} - 1 } 
			}^2 } \\
		&=
		y_i(e_1)^2 \cdot \E[\Big]{ \braces[\Big]{ \frac{\indicator{E_{(i,1)}}}{\Pr{E_{(i,1)}}} - 1 }^2  }
		+
		y_i(e_0)^2 \cdot \E[\Big]{ \braces[\Big]{ \frac{\indicator{E_{(i,0)}}}{\Pr{E_{(i,0)}}} - 1 }^2  }\\
		&\quad \quad + 2 y_i(e_1) y_i(e_0)
		\intertext{Calculating these expectations, we have that}
		&=
			y_i(e_1)^2 \cdot \paren[\Big]{ \frac{1}{\Pr{E_{(i,1)}}} - 1 }
			+ y_i(e_0)^2 \cdot \paren[\Big]{ \frac{1}{\Pr{E_{(i,0)}}} - 1 }
			+ 2y_i(e_1) y_i(e_0)
			\intertext{Finally, an application of the AM-GM inequality to the third term yields}
		&\leq y_i(e_1)^2 \cdot \frac{1}{\Pr{E_{(i,1)}}} + y_i(e_0)^2 \cdot \frac{1}{\Pr{E_{(i,0)}}} \\
		&\leq \gamma \cdot \paren[\big]{ y_i(e_1)^2 + y_i(e_0)^2 }
		\enspace,
	\end{align*}
	where the final inequality used the definition of $\gamma$.
\end{proof}

The following proposition is the general form of the dependency graph analysis.
Although we focus on the Conflict Graph Design, the idea can be more generally applied to an arbitrary design \citep{Aronow2017Estimating, Harshaw2022Design}.
The key idea is that Cauchy-Schwarz and AM-GM inequalities are applied to covariance terms which are not guaranteed to be zero by the dependency graph, and Lemma~\ref{lemma:individual-var-bound-dep-graph} to bound the individual variances.

\begin{proposition}[Dependency Graph Analysis]  \label{thm:dependency-graph-general}
	Suppose the arbitrary neighborhood interference model holds (Assumption~\mainref{assumption:ani-model}).
	Under the Conflict Graph Design, the variance of the modified Horvitz--Thompson estimator is bounded as
	\[
	\Var{\eate} \leq
	\frac{\maxdepdeg}{n} \cdot 
	\gamma \cdot 
	\braces[\Big]{ \frac{1}{n} \sum_{i=1}^n y_i(e_1)^2 + y_i(e_0)^2 }
	\enspace.
	\]
\end{proposition}
\begin{proof}
	We begin by decomposing the variance of the sum into the sum of the relevant covariance terms.
	We use the notation $i \sim_{\mathcal{D}} j$ to denote that units $i$ and $j$ are adjacent in the dependency graph $\mathcal{D}$.
	\begin{align*}
		\Var{\eate} 
		&= \Var[\big]{ \frac{1}{n} \sum_{i=1}^n \eate_i } \\
		&= \frac{1}{n^2} \sum_{i=1}^n \sum_{j=1}^n \Cov{ \eate_i, \eate_j } \\
		&= \frac{1}{n^2} \sum_{i=1}^n \sum_{j=1}^n \indicator{i \sim_{\mathcal{D}} j} \cdot \Cov{ \eate_i, \eate_j } 	
			\enspace,
			\intertext{where the last line follows by definition of the edges in the dependency graph.
			In particular, we have that $\indicator{i \sim_{\mathcal{D}} j} \cdot \Cov{ \eate_i, \eate_j } 	= \Cov{ \eate_i, \eate_j }$ because the absence of an edge means that the variables are independent and thus the covariance is zero.
			Using the Cauchy-Schwarz and AM-GM inequalities, we have that
		}
		&\leq \frac{1}{n^2} \sum_{i=1}^n \sum_{j=1}^n \indicator{i \sim_{\mathcal{D} }j} \cdot \sqrt{ \Var{\eate_i} \cdot \Var{\eate_j} }
			&\text{(Cauchy-Schwarz)} \\
		&\leq \frac{1}{n^2} \sum_{i=1}^n \sum_{j=1}^n \indicator{i \sim_{\mathcal{D} }j} \cdot \frac{1}{2} \paren[\Big]{ \Var{\eate_i} + \Var{\eate_j} }
		&\text{(AM-GM)} \\
		&= \frac{1}{n^2} \sum_{i=1}^n d_{\mathcal{D}}(i) \cdot \Var{\eate_i} 
			\enspace,
		\intertext{
			where the last equality follows by collecting terms and using the definition of the degrees in the dependency graph.
			Now, we bound each of the degrees by the maximum degree and apply the result of Lemma~\ref{lemma:individual-var-bound-dep-graph} to bound the individual variances to obtain
		}
		&\leq \frac{\maxdepdeg}{n} \cdot \frac{1}{n} \sum_{i=1}^n \Var{\eate_i}
			&\text{(def of $\maxdepdeg$)} \\
		&\leq \frac{\maxdepdeg}{n} \cdot \gamma \cdot \frac{1}{n} \sum_{i=1}^n y_i(e_1)^2 + y_i(e_0)^2
		\enspace,
			&\text{(Lemma~\ref{lemma:individual-var-bound-dep-graph})}
	\end{align*}
	which completes the proof.
\end{proof}

Proposition~\ref{thm:dependency-graph-general} contains the general dependency graph analysis.
In order to compare this to the operator norm analysis, we need to determine the values of the inverse probabilities $\gamma$ and the maximum degree of the dependency graph $\maxdepdeg$.
This is done in the following corollary.

\newcommand{\cgdmaxdep}{d_{\max{}}(\cH^2)}

\begin{corollary}[CGD Dependency Graph Analysis] \label{corollary:cgd-dep-analysis}
	Suppose the arbitrary neighborhood interference model holds (Assumption~\mainref{assumption:ani-model}).
	There exists a constant $C > 0$ such that the variance of the modified Horvitz--Thompson estimator under the Conflict Graph Design is bounded as
	\[
	\Var{\eate} \leq
	C \cdot \frac{\lamH}{n} \cdot 
	\cgdmaxdep \cdot 
	\braces[\Big]{ \frac{1}{n} \sum_{i=1}^n y_i(e_1)^2 + y_i(e_0)^2 }
	\enspace,
	\]
	where $\cH^2$ is the graph obtained by connecting all nodes which are at most distance two away in the conflict graph $\cH$ and $\cgdmaxdep$ is its maximum degree.
\end{corollary}
\begin{proof}
	First, we use Lemma~\mainref{lemma:individual-covariance-terms} to bound the inverse probabilities by $\gamma \leq C \cdot \lamH$.
	Next, we recognize that two individual effect estimators are independent under the Conflict Graph Design if and only if they are greater than two hops away in the conflict graph $\cH$.
	Thus, $\mathcal{D} = \cH^2$.
\end{proof}

Corollary~\ref{corollary:cgd-dep-analysis} contains the complete dependency graph analysis applied to the Conflict Graph Design.
The key difference between the dependency graph analysis and the operator norm analysis is that the former incurs an additional $\cgdmaxdep$ factor.
This factor $\cgdmaxdep$ will typically be orders of magnitude larger than $\lamH$.

For example, consider the estimation of the Global Average Treatment Effect (GATE) where the underlying graph is denoted as $G$.
Then, the additional factor $\cgdmaxdep$ is at most $d_{\max{}}(G)^4$, and this bound will be tight for most regular graphs.
This is exactly the extra term appearing in the analysis of \citet{Ugander2023Randomized}, because they employ the dependency graph method.
In general, the dependency graph method will incur polynomial dependencies on the maximum degree in the graph $G$.
This is because the dependency graph cannot distinguish small correlations from large ones.
In contrast, using the operator norm analysis, we obtain refined bounds on the correlations depending on the distance between nodes in the graph, which yields an overall better analysis of the rates.

\paragraph{Illustrative Example}
Consider the direct treatment effect $\tau = (1/n) \sum_{i=1}^n y_i(e_1) - y_i(e_0)$, where $e_1$ is the exposure that you are treated and your neighbors receive control and $e_0$ is the exposure that you and all your neighbors receive control.
Suppose that the network $G$ is the star graph and that treatment is assigned according the Conflict Graph Design.

\begin{figure}[t]
	\centering
	
	% First subfigure
	\begin{subfigure}{0.3\textwidth}
		\centering
		
		\begin{tikzpicture}[every node/.style={circle, draw}]
			
			\node (0) at (0,0) {};
			\node (1) at (0,1) {};
			\node (2) at (0.866, 0.5) {};
			\node (3) at (0.866,-0.5) {};
			\node (4) at (0, -1) {};
			\node (5) at (-0.866, -0.5) {};
			\node (6) at (-0.866, 0.5) {};
			
			\draw (0) -- (1);
			\draw (0) -- (2);
			\draw (0) -- (3);
			\draw (0) -- (4);
			\draw (0) -- (5);
			\draw (0) -- (6);
		\end{tikzpicture}
		
		\caption{Original Network $G$}
		\label{fig:illustration-original-network}
	\end{subfigure}%
	\hfill
	% Second subfigure
	\begin{subfigure}{0.3\textwidth}
		\centering
		\begin{tikzpicture}[every node/.style={circle, draw}]
			\node (0) at (0,0) {};
			\node (1) at (0,1) {};
			\node (2) at (0.866, 0.5) {};
			\node (3) at (0.866,-0.5) {};
			\node (4) at (0, -1) {};
			\node (5) at (-0.866, -0.5) {};
			\node (6) at (-0.866, 0.5) {};
			
			\draw (0) -- (1);
			\draw (0) -- (2);
			\draw (0) -- (3);
			\draw (0) -- (4);
			\draw (0) -- (5);
			\draw (0) -- (6);
		\end{tikzpicture}
		\caption{Conflict Graph $\cH$}
		\label{fig:illustration-conflict-graph}
	\end{subfigure}%
	\hfill
	% Third subfigure
	\begin{subfigure}{0.3\textwidth}
		\centering
		\begin{tikzpicture}[every node/.style={circle, draw}]
			\node (0) at (0,0) {};
			\node (1) at (0,1) {};
			\node (2) at (0.866, 0.5) {};
			\node (3) at (0.866,-0.5) {};
			\node (4) at (0, -1) {};
			\node (5) at (-0.866, -0.5) {};
			\node (6) at (-0.866, 0.5) {};
			
			\draw (0) -- (1);
			\draw (0) -- (2);
			\draw (0) -- (3);
			\draw (0) -- (4);
			\draw (0) -- (5);
			\draw (0) -- (6);
			\draw (1) -- (2);
			\draw (1) -- (3);
			\draw (1) -- (4);
			\draw (1) -- (5);
			\draw (1) -- (6);
			\draw (2) -- (3);
			\draw (2) -- (4);
			\draw (2) -- (5);
			\draw (2) -- (6);
			\draw (3) -- (4);
			\draw (3) -- (5);
			\draw (3) -- (6);
			\draw (4) -- (5);
			\draw (4) -- (6);
			\draw (5) -- (6);
		\end{tikzpicture}
		\caption{Dependency Graph $\mathcal{D}$}
		\label{fig:illustration-dependency-graph}
	\end{subfigure}
	
	\caption{
		Figure~\ref{fig:illustration-original-network} depicts a star graph as the original underlying network $G$.
		Figure~\ref{fig:illustration-conflict-graph} depicts the conflict graph $\cH$ for the direct effect, which in this case is the original graph $G$.
		Figure~\ref{fig:illustration-for-dep-graph-analysis} depicts the dependency graph $\cD$ which is the complete graph, as the treatment assignment of each node is correlated under the CGD.
	}
	\label{fig:illustration-for-dep-graph-analysis}
\end{figure}

Figure~\ref{fig:illustration-for-dep-graph-analysis} shows the underlying graph $G$, the conflict graph $\cH$, and the dependency graph $\cD$ for this example.
The conflict graph $\cH$ is simply the original graph $G$ because each leaf is in conflict with the central node, but pairs of leaf nodes are not themselves in conflict, i.e. any pair of estimand-relevant exposures can be simultaneously observed for leaf nodes.
On the other hand, the dependency graph $\cD$ is fully connected: the central node's exposure is of course correlated with each leaf's exposure, but in fact each pair of leaves has correlated exposures through the central node, their common neighbor.
For this reason, the dependency graph analysis yields a much more pessimistic analysis of the variance:
\[
\underbrace{\Var{\eate} \leq \frac{d_{\max}(\cD)}{n} \cdot \gamma}_{\textrm{Dependency Graph Analysis}} = \underbrace{ \bigO{\sqrt{n}}}_{\textrm{Star Graph}}
\quadtext{while}
\underbrace{\Var{\eate} \leq \frac{\lamH}{n}}_{\textrm{Operator Norm Analysis}} = \underbrace{\bigO[\Big]{\frac{1}{\sqrt{n}}}}_{\textrm{Star Graph}}
\]
In each of the steps above, the first inequality is due to the proof technique and the equality is simply a calculation for this illustrative example.
Note that the dependency graph analysis cannot show that the estimator is consistent, whereas the operator norm analysis establishes consistency at a $n^{-1/4}$ rate.

\subsection{Comparison with \cite{Ugander2023Randomized}} \label{sec:supp-3net-counterexample}

In the simulations presented in Section~\ref{sec:supp-simulations}, we observed that the Randomized Graph Cluster Randomization designs (RGCR) of \cite{Ugander2023Randomized} achieved lower variance than the Conflict Graph Design. 
This prompts the question of whether the particular selection of graphs for these simulations favors the RGCR designs, or whether these designs also achieve $\lamH/n$ rates in general.
We resolve this question in this section, by showing that the RGCR design does not achieve the $\lamH/n$ rate in general.
In particular, we construct a family of graphs $\{G_n\}_{n=1}^\infty$ where the 1-hop-max RGCR design with Horvitz-Thompson estimator performs \emph{exponentially worse} than the Conflict Graph Design with the modified Horvitz-Thompson estimator. 

Recall that the GATE is defined as

\[
\tau = \frac{1}{n} \sum_{i=1}^n \paren[\Big]{y_i(\vec{1}) - y_i(\vec{0})}
\]

We first state the main result that we will prove in this section. We use $\Esub{\operatorname{CGD}}{\cdot}, \Esub{\operatorname{1-hm}}{\cdot}$ to denote expectation with respect to the Conflict Graph Design and the 1-hop-max design, respectively. 
We will denote by $\eate_{\operatorname{1-hm}}$ the estimator used in \cite{Ugander2023Randomized} together with 1-hop-max design.
This is given by the standard Horvitz-Thompson estimator for the GATE and takes the form
\[
\eate_{\operatorname{1-hm}} = \frac{1}{n} \sum_{i=1}^n
Y_i \cdot  \paren[\Big]{\frac{\indicator{E_{(i,1)}}}{\Pr{E_{(i,1)}}} - \frac{\indicator{E_{(i,0)}}}{\Pr{E_{(i,0)}}}}
\enspace,
\]
where the events $E_{(i,1)}$ and $E_{(i,0)}$ are simply the events that a node receives the all treatment or the all control exposure, i.e.
\begin{equation}
\begin{aligned}
E_{(i,1)} = \{z \in \{0,1\}^n : z_i = 1, z_j = 1 \forall j \in N(i)\}\\
E_{(i,0)} = \{z \in \{0,1\}^n : z_i = 0, z_j = 0 \forall j \in N(i)\}
\end{aligned}
\label{eq:events_def}
\end{equation}
We note that the same notation has been used in Section~\mainref{sec:effect-estimator} to define the desired exposure events used in the modified Horvitz-Thompson estimator used with the Conflict Graph Design.
We emphasize that for the purposes of this section, the notation $E_{(i,1)}, E_{(i,0)}$ will always be defined as in \eqref{eq:events_def}.

We also note that $\eate_{\operatorname{CGD}}$ will denote the modified Horvitz-Thompson estimator that is used together with the Conflict Graph Design, as defined in Section~\mainref{sec:effect-estimator}.
As in Theorem~\mainref{thm:variance-analysis-asymptotic}, we will assume that the second moments of the potential outcomes are bounded.

\begin{theorem}\label{thm:1hopmax-counterexample}
    Let $\ate$ be the Global Average Treatment Effect. There is a sequence $G_n$ of graphs with maximum degree $d_n = \bigTheta{n^{1/6}}$, with a triangular sequence $y_i^{(n)}$ of potential outcome functions satisfying Assumptions~\mainref{assumption:ani-model} and~\mainref{assumption:bounded-second-moment}, such that the following hold
    \begin{gather}
        \Esub{\operatorname{CGD}}{\paren[\big]{\tau - \eate_{\operatorname{CGD}}}^2}^{1/2} = \bigO{n^{-1/6}} \label{eq:rate-cgd} \\
         \text{ while } \nonumber\\
         \Esub{\operatorname{1-hm}}{\paren[\big]{\tau - \eate_{\operatorname{1-hm}}}^2}^{1/2} = \bigOmega[\Big]{\frac{1}{\log^2 n}} \enspace. \label{eq:rate-1hm}
    \end{gather}
\end{theorem}

This implies that the rate achieved by the 1-hop-max design is exponentially worse than that achieved by the Conflict Graph Design in this family of graphs. 
This shows that the 1-hop-max design does not achieve comparable performance to the Conflict Graph Design for all graphs. In particular, the simulations in Section~\ref{sec:supp-simulations} involve graphs that do not reflect the worst-case performance of the 1-hop-max design.

\begin{figure}[H]
	\centering
	\includegraphics[width=0.6\textwidth]{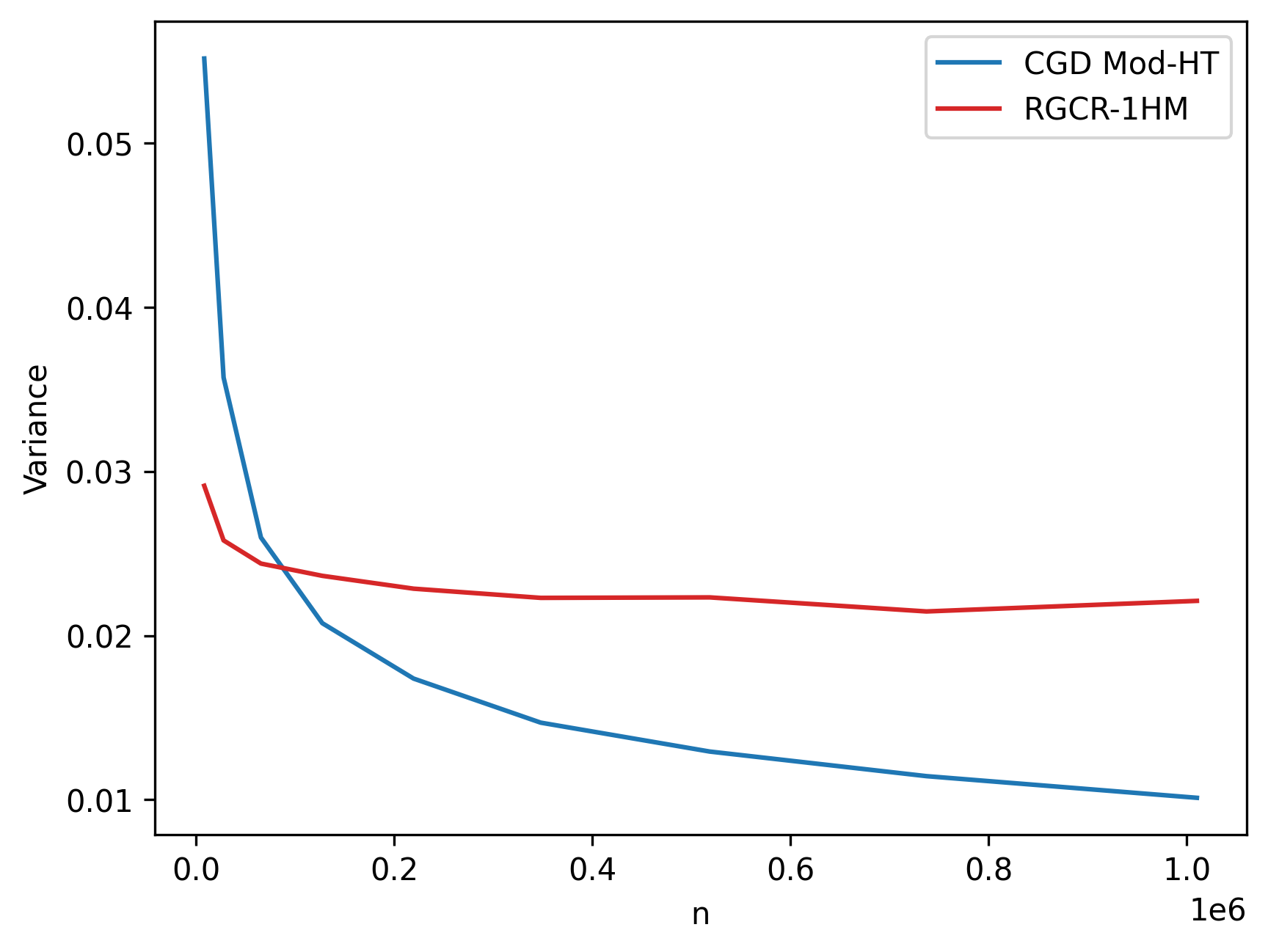}
	\caption{Comparison of the variance achieved by the 1-hop-max and the conflict graph designs, as a functions of the number of nodes $n$.}
	\label{fig:1hm-vs-cgd-bat-graph}
\end{figure}

For completeness, we have also experimentally verified this difference in performance between these two designs. In Figure~\ref{fig:1hm-vs-cgd-bat-graph}, we compare the variance of estimating the GATE using the 1-hop-max design and the Conflict Graph Design on graphs from the family $G_n$ that is defined in the proof of Theorem~\ref{thm:1hopmax-counterexample}.
We observe that the Conflict Graph Design converges at a polynomial rate, while the convergence of the 1-hop-max design is barely noticeable even for graphs with millions of nodes. This is consistent with the predictions of Theorem~\ref{thm:1hopmax-counterexample}.

In the rest of the section, we will establish a variety of useful lemmas and will present the proof of Theorem~\ref{thm:1hopmax-counterexample} in the end.
We start by describing the 1-hop-max design from \cite{Ugander2023Randomized}.
We will use the notation $B_l(i)$  from \cite{Ugander2023Randomized} to denote the nodes that are at distance at most $l$ from $i$, including $i$ itself. Note that $B_1(i) = N(i) \cup \{i\} = \tilde{N}(i)$.

For a given graph $G$, the 1-hop-max design starts by computing the weights for each node in the graph, which are the entries of the leading eigenvector $w \in \Reals^n$ of the adjacency matrix of $G^2$. 
The algorithm then works in two phases. In the first phase, it samples a random variable $X_i \sim \beta(w_i,1)$ independently for each node $i$. 
In the second phase, it assigns each node $i$ a cluster label $c_i$, which is the identity of the node in $B_1(i)$ with the largest $X_j$ value (breaking ties arbitrarily).
After all nodes have been assigned a cluster label, the algorithm assigns treatment or control to each cluster independently with probability $1/2$. We provide the pseudocode of the 1-hop-max design in Algorithm~\ref{alg:1hm}.

\begin{algorithm}
    \DontPrintSemicolon
    \caption{\textsc{Weighted 1-hop-max Clustering}}\label{alg:1hm}
    \KwIn{Graph $G = (V,E)$, weights $\vec{w} \in \Reals^n$ (leading eigenvector of adjacency matrix of $G^2$)}
    \KwOut{Random intervention $Z \in \mathcal{Z} = \setb{0,1}^n$}
    \For{$i \in V$}{
        Sample $X_i \sim \beta(w_i,1)$ independently
    }
    \For{$i \in V$}{
        $c_i \gets \argmax_{j \in B_1(i)} X_j$ \tcp*{break ties arbitrarily}
    }
    \For{each distinct cluster label $k$ in $\{c_i : i \in V\}$}{
        Sample $Y_k \sim \operatorname{Bernoulli}(1/2)$ independently\\
        \For{each $j$ with $c_j = k$}{
            $Z_j \gets Y_k$
        }
    }
\end{algorithm}

% It then orders the nodes in decreasing order of the values of the variables $X_i$. Afterwards, it forms a set $S$ by going through the nodes one by one in this ordering starting from the first: if the current node $i$ in the ordering is unmarked, it is added in $S$ and all nodes at distance at most $2$ from $i$ that are not already in $S$ are marked. 
% After the set $S$ has been formed, the algorithm proceeds by assigning all remaining nodes not in $S$ to their closest node in $S$ in graph distance in $G$, forming clusters. 
% We call the nodes in $S$ the \emph{centers} of the clusters.

The key difficulty when analyzing the performance of the 1-hop-max design comes from the fact that the clusters are random and do not necessarily contain the entire neighborhood of a node. It is even possible that a node $j$ is assigned cluster $c_j = i$, but the node $i$ is assigned to a different cluster. 
To make these calculations easier, it would help if there is some kind of uniformity in the graph. 
We now describe the class of graphs we will be considering.

     We define the following family $G_n $ of graphs, indexed by the number of nodes $n$. For each $n$, there is an associated parameter $d_n$ that is used to define the graph.
     The graph is depicted in Figure~\ref{fig:gate-batgraph} and we provide a formal description below:
    
    The vertex set $V$ is partitioned as 
    \[
    V = \{l,r\} \cup U \cup \paren[\big]{\cup_{i \in U}(L_i \cup R_i)} \cup C \enspace,
    \]
    where $|U| = d_n, |L_i| = |R_i| = d_n/2, |C| = d_n^3$. 
    We denote $L = \cup_{i \in U}L_i$ and $R = \cup_{i \in U} R_i$. From the above it follows that $|L| = |R| = d_n^2/2$.
    The graph has the following edges.
    \begin{enumerate}
        \item There is an edge between $l$ (respectively $r$) and every node in $L$ (respectively $R$) .
        \item Every node $i \in U$ is connected to all nodes in $L_i \cup R_i$. 
        \item Every node $i \in L \cup R$ is connected to $d$ distinct nodes in $C$. Call $C_i$ the neighbors of $i$ in $C$.
    \end{enumerate}
Thus, the graph $G_n$ has total number of vertices $n = 2 + d_n + d_n^2 + d_n^3 = d_n^3 + \littleO{d_n^3}$, thus we have $d_n = n^{1/3} + \littleO{n^{1/3}}$.

Intuitively, the structure of $G_n$ is similar to a three-level tree, but where the root is also connected to the second level.
The reason for this is that the nodes in $L \cup R$ will end up in the cluster of $l$ or $r$ with sufficiently high probability. In turn, this will correlate the exposure events for the nodes in $U$.

\begin{figure}
	\centering
    \resizebox{\textwidth}{!}{
        \input{\figpath/appendix/gate-batgraph.tex}
    }
	\caption{The graph $G^{(n)}$ used in the proof of Theorem~\ref{thm:1hopmax-counterexample}. It contains a subset of nodes $U$, that each have $d$ neighbors, where half of the neighbors are also connected to a hub node $l$ and the other half to a hub node $r$. Each of the neighbors of nodes in $U$ is also connected to $d$ distinct nodes in $C$. The proof will proceed by showing that the exposures of nodes in $U$ are highly correlated under the 1-hop-max design, leading to high variance for the Horvitz-Thompson estimator.}
	\label{fig:gate-batgraph}
\end{figure}

We start by listing some important properties about the spectrum of this graph. In particular, we are interested in the maximum eigenvalue of the adjacency matrix, as well as in the relative magnitudes of the entries in the leading eigenvector, since these are used in the 1-hop-max design. For simplicity,
we will omit the dependence on $n$ in the notation of the graph and the degree in the remainder of this section.

\begin{lemma}\label{lem:family}
    Let $G$ be the graph in the family defined earlier with $n$ nodes, where $d= n^{1/3}(1 + o(1))$. Then, the following properties hold:
    \begin{enumerate}
        \item\label{it:maxeig} The maximum eigenvalue of the adjacency matrix of $G^{2}$ is $\lambda(G^2) = d^2/2 + d/2 + O(1)$.
        \item\label{it:eigvec} If $w \in \Reals^n$ is the leading eigenvector of the adjacency matrix of $G^{2}$ with unit norm, then we have for all $i$,
        \[
        w_i = \frac{1}{\sqrt{n}}\cdot \left\{
    \begin{array}{ll}
        \Theta(1) &\text{ if } i \in L \cup R \cup \{l,r\} \\
        \Theta(1/d) &\text{ if } i \in U\\
        \Theta(1/d^2) &\text{ if } i \in C
        \end{array}
    \right.
        \]
    \end{enumerate}
\end{lemma}
\begin{proof}
    Let us denote by $\lambda$ the maximum eigenvalue of the adjacency matrix of $G^2$ for simplicity.
    First of all, note that due to symmetry, all nodes in $L \cup R \cup \{l,r\}$ have the same entry in the leading eigenvector, which we denote by $x_L$. Similarly, all nodes in $U$ have the same entry $x_U$ and all nodes in $C$ have the same entry $x_C$.
    Also, $l,r$ have the same entry $x_l$. 
    In $G^2$, nodes $l,r$ are connected to all nodes in $U$, as well as $d^3/2$ nodes in $C$ each. A node in $U$ is connected to $l,r$, all nodes in $L_i \cup R_i$ and $d^2$ nodes in $C$. A node in $L_i$ (respectively $R_i$) is connected to $l$ (respectively $r$), one node in $U$, all other nodes in $L$ (respectively $R$), $d/2$ nodes in $R$ and $d$ nodes in $C$. Finally, a node in $C$ is connected to either $l$ or $r$, one node in $U$, one node in $L$ or $R$ and $d-1$ other ``sibling'' nodes in $C$.
    We can therefore write the conditions for the leading eigenvector as the following system of equations:
    \begin{equation}
    \begin{aligned}
        \lambda x_l &= \frac{d^2}{2} x_L + dx_U + \frac{d^3}{2} x_C \\
        \lambda x_L &= \paren[\big]{\frac{d^2}{2} + \frac{d}{2} -1}x_L + x_l + x_U + d x_C \\
        \lambda x_U &= 2x_l + d x_L + d^2 x_C\\
        \lambda x_C &= x_L + x_U + x_l + (d-1)x_C
    \end{aligned}\label{eq:eig-system}
    \end{equation}
    By solving this system, we arrive at the following polynomial equation for $\lambda$:
    \begin{gather*}
    4\lambda^{4}
    -(2d^{2} + 6d - 8)\lambda^{3}
    -(6d^{2} + 22d - 4)\lambda^{2}
    +(d^{5} + 3d^{4} - 6d^{3} - 6d^{2} - 20d)\lambda\\
    +(4d^{5} - 16d^{3} + 12d^{2} - 8d) = 0
    \end{gather*}
    We can determine the asymptotic behavior of the roots of this polynomial by using the method of dominant balance (see e.g. Chapter 7 in \cite{bender2013advanced}). In particular, suppose $\lambda = \alpha d^\gamma + \littleO{d^{\gamma}}$ for some $\alpha, \gamma > 0$. Plugging this into the polynomial and keeping only the dominant terms, we get
    \[
    4\alpha^{4} d^{4\gamma} - 2\alpha^{3} d^{2 + 3\gamma} - 6\alpha^{2} d^{2 + 2\gamma} + \alpha d^{5 + \gamma} + 4 d^{5} = 0
    \]
    To have a balance, we need at least two terms to have the same exponent, otherwise one term will dominate the others for large $d$. Note that if $\gamma > 2$, the first term $d^{4\gamma}$ is strictly dominating all the other terms, thus we must have $\gamma \leq 2$. Let's try $\gamma = 2$, since we care about the largest root. 
    Then, the dominant terms are the first and second terms $4\alpha^{4} d^{8}$ and $- 2\alpha^{3} d^{8}$. Balancing these two terms gives $\alpha = 1/2$. Thus, we have found one root with asymptotic behavior $\lambda = d^{2}/2 + \littleO{d^{2}}$. To compute the second order term, we can assert that $\lambda = d^{2}/2 + \beta d^{\delta} + \littleO{d^{\delta}}$ for some $\beta, \delta$ with $\delta < 2$. Plugging this back into the polynomial and keeping only the dominant terms, we get
    \[
    2\beta d^{6 + \delta} - \frac{3}{2} \beta d^{6 + \delta} - \frac{3}{4} d^7 - \frac{3}{2} d^6 + \frac{1}{2} d^7 + 4d^5  = 0
    \]
    The only way to have a balance of the leading order terms is to have $\delta = 1$. Balancing the terms of order $d^7$ gives
    \[
    2 \beta - \frac{3}{2} \beta - \frac{3}{4} + \frac{1}{2} = 0 \enspace, 
    \]
    which implies $\beta = 1/2$. Thus, we have established that $\lambda = d^{2}/2 + d/2 + \littleO{d}$.

    Finally, to establish the scaling of the entries of the leading eigenvector, we can plug in the value of $\lambda$ back into the system of equations \eqref{eq:eig-system} and solve for the ratios between $x_l, x_L, x_U, x_C$. Specifically, we can start from the fourth equation in \eqref{eq:eig-system} to express $x_C$ in terms of $x_L, x_U, x_l$. By the asymptotics on $\lambda$ that we just established, we get 
    \[
    x_C = \frac{2}{d^2} (1 + o(1))\paren[\big]{x_L + x_U + x_l} \enspace.
    \]
    Now we plug this value to the third equation, which gives 
    \[
    x_U = \frac{8}{d^2} (1 + o(1)) x_l + \frac{2}{d} (1 + o(1)) x_L \enspace.
    \]
    Thus, the $x_U$ term is orders of magnitude smaller than both $x_l,x_L$. Finally, plugging these estimates in the first equation yields
    \[
    \frac{d^2}{2} (1 + o(1)) x_l = \frac{d^2}{2} x_L + \frac{8}{d} (1 + o(1)) x_l + 2 (1 + o(1)) x_L +2d (1 + o(1)) (x_l + x_L)
    \]
    which implies that $x_L = (1 + o(1)) x_l$. Thus, we get 
    \[
    x_U = \frac{2}{d}(1 + o(1))x_l \quad \text{ and } \quad x_C = \frac{4}{d^2}(1 + o(1)) x_l \enspace.
    \]
    Normalizing for unit norm gives the desired result.
\end{proof}

We are now ready to analyze the performance of the 1-hop-max design on this family of graphs.
To analyze the variance of the estimator $\eate_{\operatorname{1-hm}}$, 
we will first require some technical lemmas about Beta and exponential distributions.
The first is about the distribution of the minimum of a subset of exponential random variables.

\begin{lemma}\label{lem:exp-min}
    Let $T_1,\ldots, T_n$ be independent random variables where $T_i \sim \operatorname{Exp}(w_i)$, where $\operatorname{Exp}(\lambda)$ is the exponential distribution with parameter $\lambda$ and density function $x\mapsto \lambda e^{-\lambda x}$ for $x > 0$. Then, 
    \[
    \Pr{T_1 < \min_{i > 1}T_i} = \frac{w_1}{\sum_{i=1}^n w_i}
    \]
\end{lemma}
\begin{proof} 
    We use the well-known fact that a random variable $T$ follows $\operatorname{Exp}(w)$ distribution if and only if  $X = e^{-T}$ follows $\beta(w,1)$ distribution. Thus, under the transformation $X_i = e^{-T_i}$, the probability can be rewritten as 
    \[
    \Pr{X_1 > \max_{i > 1}X_i}
    \]
    and we can use Theorem 4.10 from \cite{Ugander2023Randomized} to conclude the proof.
\end{proof}

The second technical lemma is a generalization about the rank statistics of independent random variables that follow the exponential distribution.

\begin{lemma}\label{lem:order}
    Let $T_1,\ldots, T_n$ be independent random variables where $T_i \sim \operatorname{Exp}(w_i)$, where $\operatorname{Exp}(\lambda)$ is the exponential distribution with parameter $\lambda$ and density function $x\mapsto \lambda e^{-\lambda x}$ for $x > 0$. 
    Furthermore, suppose that $w_1 \leq \rho \cdot \min_{i > 1}w_i$, for some $\rho > 0$. 
    Let 
    \[
    T_{(1)},\ldots, T_{(n)}
    \]
    be the variables $T_1,\ldots, T_n$ ordered from smallest to largest, with $T_{(1)}$ being the smallest and $T_{(n)}$ the largest. 
    Then, for any $r \in [n]$,
    \[
    \Pr{T_1 \leq T_{(r)}} \leq \rho \cdot \frac{r}{n - r + 1} \enspace.
    \]
\end{lemma}
\begin{proof}
    We use the memoryless property of the exponential distribution, which asserts that for a random variable $Y$ that follows an exponential distribution and any $s, t > 0$, we have
    \[
    \Pr{Y > s+t | Y > s} = \Pr{Y > t}\enspace.
    \]
    Imagine we sample each $T_i$ by using an exponential ``clock'', where the clock fires once at a random time $T_i$ and $\Pr{T_i \leq t} = 1 - e^{-w_i t}$ is given as in the exponential distribution.
    Then, we can imagive we have $n$ independent clocks, one associated to each random variable $T_i$. 
    We will say $T_i$ fires if its associated clock fires.
    The time is partitioned in rounds. The first round is the time until the earliest clock fires, the second round is the time between the earliest and second earliest clocks etc. Let $A_t$ be the event that the clock assigned to $T_1$ fires in one of the first $t$ rounds, which equivalently says that $1$ appears in one of the first $t$ positions in the ordering. Then, the statement of the lemma is about upper bounding $\Pr{A_r}$. 
    By the law of total probability, we can write
    \[
    \Pr{A_r} = \Pr{A_1} + \Pr{A_2 | A_1^c} + \ldots \Pr{A_r|A_{r-1}^c} \enspace.
    \]
    Let us bound $\Pr{A_j | A_{j-1}^c}$ for each $j$. 
    This is the event that $T_1$ fires in the $j$-th round. To analyze that, let us first condition on the fact that the remaining clocks after the $j-1$-th round form the subset $L_j$ and $1 \in L_j$. Conditioned on that event, by the memoryless property of the exponential distribution, we know that the law of the variables in $L_j$ after the start of the $j$-th round is still the same law of an exponential distribution. Thus, by Lemma~\ref{lem:exp-min}, the probability that $1$ is the clock that fires in the $j$-th round, conditioned on the subset $L_j$, is 
    \[
    \frac{w_1}{\sum_{i \in L_j} w_i} \leq \rho\cdot \frac{1}{|L_j|} = \rho \cdot \frac{1}{n - j + 1}
    \]
    Since this holds for all possible conditionings of the subset $L_j$, we conclude that
    \[
    \Pr{A_j|A_{j-1}^c} \leq \rho \cdot \frac{1}{n - j + 1}
    \]
    Therefore,
    \[
    \Pr{A_r} \leq \rho \cdot \sum_{j=1}^r \frac{1}{n - j + 1} \leq \frac{\rho \cdot r}{n-r+1}
    \]
\end{proof}

The third lemma involves Beta distributions and will be useful when we calculate joint probabilities of exposure.

\begin{lemma}\label{lem:beta-joint}
    Suppose we have an index set $V = \{i,j\}\cup A \cup B \cup C$, where $A,B,C$ are pairwise disjoint. For every $i\in V$, we sample independently $X_i \sim \beta(w_i,1)$, where $w_i > 0$ is some positive weight. For a subset $R \subseteq V$, denote $w_R = \sum_{k \in R} w_k$. Then
    \[
    \Pr{X_i > \max_{k \in A \cup B} X_k \textrm{ and } X_j > \max_{k \in B \cup C} X_k} = \frac{w_iw_j}{w_V}\paren[\Big]{\frac{1}{w_i + w_A + w_B} + \frac{1}{w_j + w_B + w_C}}
    \]
\end{lemma}
\begin{proof}
    If $R$ is an arbitrary subset, a well-known property of Beta distributions (see Theorem 4.10 in \cite{Ugander2023Randomized}) is that $\max_{k \in R}X_k \sim \beta(w_R, 1)$. 
    Thus, we can rewrite the target probability as
    \begin{multline*}
    \Pr{X_i > \max(Y_A,Y_B) \textrm{ and } X_j > \max(Y_B, Y_C)} \\
    = \int_0^1 \int_0^1 w_i w_j x_i^{w_i - 1} x_j^{w_j -1} \Pr{Y_A < x_i} \Pr{Y_C < x_j} \Pr{Y_B < \min(x_i,x_j)} dx_i dx_j
    \end{multline*}
    where $Y_A \sim \beta(w_A,1), Y_B \sim \beta(w_B,1),Y_C \sim \beta(w_C,1)$. Using the convenient expression for the distribution function of a Beta distribution, we have
    \begin{multline*}
    \int_0^1 \int_0^1 w_i w_j x_i^{w_i - 1} x_j^{w_j -1} \Pr{Y_A < x_i} \Pr{Y_C < x_j} \Pr{Y_B < \min(x_i,x_j)} dx_i dx_j \\
    = 
    \int_0^1 \int_0^1 w_i w_j x_i^{w_i - 1} x_j^{w_j -1} x_i^{w_A} x_j^{w_C} \min(x_i,x_j)^{w_B} dx_i dx_j
    \end{multline*}
    Splitting the region $[0,1]^2$ into subregions $x_i < x_j$ and $x_i \geq x_j$, we can calculate
    \begin{align*}
    &\int_0^1 \int_0^{x_j} w_i w_j x_i^{w_i - 1} x_j^{w_j -1} x_i^{w_A} x_j^{w_C} x_i^{w_B} dx_i dx_j \\
    &\quad = 
    \frac{w_iw_j}{w_i+w_A + w_B} \int_0^1 x_j^{w_j + w_C + w_i + w_A + w_B - 1} dx_j\\
    &\quad = \frac{w_iw_j}{(w_i+w_A + w_B)w_V}
    \end{align*}
    Similarly
    \begin{align*}
        \int_0^1 \int_0^{x_i} w_i w_j x_i^{w_i - 1} x_j^{w_j -1} x_i^{w_A} x_j^{w_C} x_j^{w_B} dx_i dx_j = \frac{w_iw_j}{(w_j + w_B+w_C)w_V}
    \end{align*}
    Overall, this yields
    \[
    \Pr{X_i > \max_{k \in A \cup B} X_k \textrm{ and } X_j > \max_{k \in B \cup C} X_k} = \frac{w_iw_j}{w_V}\paren[\Big]{\frac{1}{w_i + w_A + w_B} + \frac{1}{w_j + w_B + w_C}}
    \qedhere
    \]
\end{proof}

We now establish a few crucial properties of the 1-hop-max design.
In particular, the outcomes in the proof of Theorem~\ref{thm:1hopmax-counterexample} will be chosen so that they are supported on nodes in $U$. Thus, we need to establish an upper bound on the probability that a given node in $U$ gets the all treatment exposure.
First, we recall a lower bound on the probability of exposure from \cite{Ugander2023Randomized}.

\begin{lemma}[Theorem 4.11 in \cite{Ugander2023Randomized}]\label{lem:ugander}
    Let $G$ be a graph, on which we run the 1-hop-max design. Then, for each node $i$ in $G$, we have
    \[
    \Pr{E_{(i,1)}} \geq \frac{1}{1 + \lambda(G^2)} \enspace,
    \]
    where $\lambda(G^2)$ is the maximum eigenvalue of the adjacency matrix of $G^2$.
\end{lemma}

% In the proof of Lemma~\ref{lem:ugander}, an important event is the one where $X_i$ is the largest value in the $2$-hop neighborhood of $i$, which implies that $i$ and all of its neighbors will be assigned to the same cluster. Thus, let us formally define these events for each $i$:

% \[
% F_i = \{X_i > \max_{j \in N(i)}X_j\}
% \]

We would like to establish an upper bound on the probability of exposure that is of the same order $\bigO{1/d^2}$ for any $i \in U$. Intuitively, the reason is that if $X_i$ does not dominate the values of its neighbors, then all neighbors will likely be assigned to different clusters, due to the local tree-like structure of the graph around $i$. This means that for $i$ to get the all treatment exposure, all these clusters must be assigned treatment, which happens with exponentially small probability. The details are given below. 

\begin{lemma}\label{lem:exposure-upper}
Let $G$ be the graph with $n$ nodes described in the beginning of the section. Suppose we run the 1-hop-max design on this graph. Then, for any $i \in U$,
\[
\Pr{E_{(i,1)}} = \bigO[\Big]{ \braces[\Big]{ \frac{\log d}{d} }^2} \enspace.
\]
\end{lemma}

\begin{proof}
     By construction, we know that the degree of a node $i \in U$ in $G$ is $d$.  
    For an integer $r \in [d]$, define the events
    \begin{gather}\label{eq:F-event}
    F_{r,L} := \{ \text{There are at least } r \text{ nodes } j \in L_i \text{ such that } X_j > X_l\}\\
    F_{r,R} := \{ \text{There are at least } r \text{ nodes } j \in R_i \text{ such that } X_j > X_r\}
    \end{gather}
    In other words, if $F_{r,L}$ (respectively $F_{r,R}$)occurs, then there are at least $r$ neighbors of $i$ in $L_i$ (respectively $R_i$) with larger values of $X$ than $l$ (respectively $r$). 
    We will first show that at least one of $F_{r,L}, F_{r,R}$ holds with high probability.
    First of all, it will be convenient to use the fact that if a variable $X$ follows $\beta(w,1)$ distribution, then $T = - \log X$ follows $\textrm{Exp}(w)$ distribution. Let us define the variables $T_i = -\log X_i$ for all $i$. 
    Then, $F_{r,L}^c$ is the event that $l$ appears among the top $r$ values of $\{X_j, j \in L_i\}\cup \{X_l\}$, or equivalently, among the $r$ smaller values of $\{T_j, j \in L_i\}\cup \{T_l\}$.
    By property~\ref{it:eigvec} of Lemma~\ref{lem:family}, we know that $w_l= \bigO{\min_{j \in L_i}w_j}$.
    Thus, by Lemma~\ref{lem:order}, we have that 
    \[
    \Pr{F_{r,L}^c}  = \bigO[\Big]{\frac{r}{d-r+1}}
    \]
    The same logic yields
    \[
    \Pr{F_{r,R}^c}  = \bigO[\Big]{\frac{r}{d-r+1}}
    \]
    Notice that the events $F_{r,L}$ and $F_{r,R}$ depend on disjoint subsets of variables, thus are independent.
    This implies that 
    \[
    \Pr{F_{r,L}^c \cap F_{r,R}^c} = \Pr{F_{r,L}^c} \cdot \Pr{F_{r,R}^c} = \bigO[\Big]{\paren[\Big]{\frac{r}{d-r+1}}^2}\enspace.
    \]
    We now define the following events
    \begin{gather*}
    G_{r,L} = \{ \text{There are at least } r \text{ nodes } j \in L_i \text{ such that } X_j > X_i\}\\
    G_{r,R} = \{ \text{There are at least } r \text{ nodes } j \in R_i \text{ such that } X_j > X_i\}
    \end{gather*}
    By property~\ref{it:eigvec} we know that $w_i = \bigO{\min_{j \in L_i} w_j / d}$. Thus, applying Lemma~\ref{lem:order}, we get
    \begin{gather*}
    \max\paren[\big]{\Pr{G_{r,L}^c} , \Pr{G_{r,R}^c}} = \bigO[\Big]{\frac{r}{d(d-r+1)}}
    \end{gather*}
    Thus, using the union bound we have that 
    \begin{equation}\label{eq:bound-few-important-neighbors}
    \Pr{(F_{r,L}^c \cap F_{r,R}^c )\cup G_{r,L}^c \cup G_{r,R}^c} =  \bigO[\Big]{\frac{r}{d(d-r+1)}} + \bigO[\Big]{\paren[\Big]{\frac{r}{d-r+1}}^2}
    \end{equation}
    Now, consider the complement event, i.e. 
    \[
   A := \paren[\big]{(F_{r,L}^c \cap F_{r,R}^c )\cup G_{r,L}^c \cup G_{r,R}^c}^c = (F_{r,L} \cup F_{r,R}) \cap G_{r,L} \cap G_{r,R}
    \]

    If $A$ holds, this means that either in the left or in the right set of neighbors of $i$, there are at least $r$ of them that have bigger values than both $X_i$ and $X_l$ or $X_r$. 
    % Let us assume without loss of generality that this happens for the left side, since the argument for the right side is identical.
    % We show that if $r$ is sufficiently large, then the probability of exposure of $i$ is very small. 
    Let $S_{i,L} \subseteq L_i$ be the set of left neighbors with larger value, i.e.
    \[
    S_{i,L} := \{j \in L_i: X_j > \max\paren{X_i,X_l}\} \enspace.
    \]
    We define the set $S_{i,R}$ analogously for the neighbors on the right side.
    % By assumption, we have that $|S_{i,L}| \geq r$.
    Notice that for every node $j \in S_{i,L}$ we have $c_j \neq i,l$, since $X_j > \max(X_i,X_l)$. 
    Also, by construction, the only common neighbors of $j,k \in L_i$ are $i,l$. Thus, we conclude that every node in $S_{i,L}$ will be assigned to a separate cluster. 
    Now, let us condition on the set $S_{i,L}$. Since each cluster is treated independently with probability $1/2$, the probability that all nodes in $S_{i,L}$ receive treatment is $(1/2)^{|S_{i,L}|}$. 
    Therefore, we just proved that
    \[
    \Pr{E_{(i,1)}|S_{i,L}} \leq \paren[\Big]{\frac{1}{2}}^{|S_{i,L}|} \enspace.
    \]
    Now, we suppose that $A$ holds. By the preceding discussion, this implies $|S_{i,L}| \geq r$ or $|S_{i,R}| \geq r$.
    Without loss of generality, let's assume that $|S_{i,L}| \geq r$. Since the above bound on the probability holds for any given $S_{i,L}$ that we condition on, we conclude that 
    \begin{equation*}
    \Pr{E_{(i,1)} \cap F_{r,L}  \cap G_{r,L} \cap G_{r,R}} \leq \paren[\Big]{\frac{1}{2}}^{r} \enspace.
    \end{equation*}
    Exactly the same logic applies for the right side $R$, so using a union bound we get
    \begin{equation}\label{eq:bound-many-important-neighbors}
        \Pr{E_{(i,1)} \cap (F_{r,L} \cup F_{r,R}) \cap G_{r,L} \cap G_{r,R}} \leq \paren[\Big]{\frac{1}{2}}^{r-1} \enspace.
    \end{equation}
    Thus, combining inequalites \eqref{eq:bound-many-important-neighbors} and \eqref{eq:bound-few-important-neighbors} we get for every $r$
    \[
    \Pr{E_{(i,1)}} = \bigO[\Bigg]{\paren[\Bigg]{\frac{1}{2}}^{r-1} +  \frac{r}{d(d-r+1)} + \paren[\Big]{\frac{r}{d-r+1}}^2}
    \]
    Choosing $r = 3 \log d$ in the above yields the desired result.
\end{proof}

We next show a lower bound for the joint probability of two nodes in $U$ to receive the all treatment exposure. 
This is where we crucially need to use the extra nodes $l,r$. Without them, this probability would be of order $O(1/d^4)$. However, with nodes $l,r$ connected to all nodes in $L_i,R_i$ respectively, for all $i \in U$, it turns out the answer is $\Theta(1/d^2)$.

\begin{lemma}\label{lem:exposure-lower}
    Let $i,j \in U$, where $U$ is as in Lemma~\ref{lem:family}. Let $E_{(i,1)}$ and $E_{(j,1)}$ be the events that nodes $i$ and $j$ receive the all treatment exposure under the 1-hop-max design on $G_n$. Then, we have
    \[
    \Pr{E_{(i,1)} \cap E_{(j,1)}} = \bigOmega[\Big]{\frac{1}{d^2}}\enspace.
    \]
\end{lemma}
\begin{proof}
    For an arbitrary subset of nodes $R$, we abuse notation and denote by $N(R)$ the union of neighbors of nodes in $R$.
    Define the events
    \begin{gather*}
    H_L = \{X_l > \max(X_i,X_j,\max_{k \in L_i \cup L_j \cup N(L_i) \cup N(L_j)}X_k)\}\\
    H_R = \{X_r > \max(X_i,X_j,\max_{k \in R_i \cup R_j \cup N(R_i) \cup N(R_j)}X_k)\}
    \end{gather*}
    We claim that if $H_L$ (respectively $H_R$) holds, then all nodes in $L_i \cup L_j$ (respectively $R_i \cup R_j$) will be assigned to the cluster $l$ (respectively $r$). To see this, note that for $k \in L_i$, the possible clusters it could be assigned to are all their neighbors and itself. By the definition of the event $H_L$, $X_l$ has the largest value among all neighbors of $k$ and $k$ itself, so $c_k = l$. Similarly we argue that if $k\in L_j$ and $H_L$ holds then $c_k = l$. Also, analogously we can argue for $H_R$. 
    Thus, if $H_L,H_R$ both hold, then all nodes in $L_i \cup L_j$ will be in the cluster of $l$ and all nodes in $R_i \cup R_j$ in the cluster of $r$. 
    Thus, conditioned on that event, the probability that all nodes in $L_i \cup L_j \cup R_i \cup R_j$ receive treatment is at least $1/4$. The nodes $i,j$ might be in different clusters, so the probability that they are both also assigned treatment is again at least $1/4$. 
    Overall, this shows that 
    \[
    \Pr{E_{(i,1)} \cap E_{(j,1)}} \geq \frac{1}{16} \Pr{H_L \cap H_R}
    \]
    Thus, the claim reduces to showing that $\Pr{H_L \cap H_R} = \bigOmega{1/d^2}$.
    Notice that the event $H_L \cap H_R$ can be written exactly in the form of Lemma~\ref{lem:beta-joint}, $i = l, j = r$ and 
    \begin{gather*}
        A = \paren{L_i \cup L_j \cup N(L_i) \cup N(L_j)} \setminus \{i,j,l,r\}\\
        C = \paren{R_i \cup R_j \cup N(R_i) \cup N(R_j)} \setminus \{i,j,l,r\}\\
        B = \{i,j\}
    \end{gather*}
    Since the 2-hop neighborhoods of $i,j$ are tree-like, it is clear that $A,B,C$ are pairwise disjoint. 
    Thus, applying Lemma~\ref{lem:beta-joint} yields
    \begin{gather*}
    \Pr{H_L \cap H_R} = \frac{w_lw_r}{w_l + w_r + w_i + w_j + \sum_{k \in N(i) \cup N(j) \cup (N(N(i)) \cup N(N(j)) )\setminus \{i,j,l,r\}} w_k }\\
    \cdot \left(\frac{1}{w_l + w_i + w_j + \sum_{k \in \cup L_i \cup L_j \cup (N(L_i) \cup N(L_j) )\setminus \{i,j,l,r\}} w_k} \right.\\
    + \left. \frac{1}{w_r + w_i + w_j + \sum_{k \in \cup R_i \cup R_j \cup (N(R_i) \cup N(R_j) )\setminus \{i,j,l,r\}} w_k}\right)
    \end{gather*}
    We now use property~\ref{it:eigvec} of Lemma~\ref{lem:family}. This says that for $k,r \in N(i) \cup N(j) \cup \{l,r\} $ we have $w_k / w_r = \bigTheta{1}$ and for $k \in N(i)\cup N(j) \cup \{l,r\}$ and $r \in N(N(i)) \cup N(N(j)) \setminus \{l,r\}$ $w_k /w_r = \bigTheta{d^2}$, wince $r$ belongs to the bottom layer of the graph $C$.
    Finally, by construction we have that $|N(N(i))|, |N(N(j))| = \bigO{d^2}$. Putting all of this together, we conclude that 
    \[
    \Pr{H_L \cap H_R} 
    = \frac{\bigTheta{d^2}}{\bigTheta{d^2}}\paren[\Big]{\frac{1}{\bigTheta{d^2}} + \frac{1}{\bigTheta{d^2}}}
    = \bigTheta[\Big]{\frac{1}{d^2}}
    \enspace.
    \qedhere
    \]
\end{proof}

We now have all the necessary tools for the proof of our main result of the section.

% \begin{theorem}\label{thm:3net-counterexample}
%     Let $G_n$ be the graph constructed in Lemma~\ref{lem:family} and consider the 1-hop-max design on $G_n$ with the standard Horvitz-Thompson estimator $\eatetrue$. Then, for the sequence of potential outcomes defined above, we have
%     \[
%     \Var{\eatetrue} = \bigOmega[\Big]{\frac{d^3}{n \log^4 d} } \enspace.
%     \]
% \end{theorem}
\begin{proof}[Proof of Theorem~\ref{thm:1hopmax-counterexample}]
We will use the sequence of graphs $G_n$ constructed in the beginning of the section, where $d_n = (1+o(1))n^{1/3}$.
Let us first discuss the performance of the Conflict Graph Design on this sequence of graphs. By Lemma~\ref{lem:family} we have that $\lambda(G_n^2)= O(d_n^2)$.
Thus, if the potential outcomes satisfy Assumption~\mainref{assumption:bounded-second-moment}, then by Theorem~\mainref{thm:variance-analysis-finite-sample} , the mean-squred error of the modified Horvitz-Thompson estimator $\eate_{\operatorname{CGC}}$ is bounded as
\[
\Esub{\operatorname{CGD}}{\paren[\big]{\tau - \eate_{\operatorname{CGD}}}^2} = \bigO[\Big]{\frac{\lambda(G_n^2)}{n}} = \bigO[\Big]{\frac{d^2}{n}}\enspace,
\]
which implies \eqref{eq:rate-cgd}.
We now turn to analyzing the performance of the standard Horvitz-Thompson estimator $\eate_{\operatorname{1-hm}}$ under the 1-hop-max design on the sequance of graphs $G_n$. 

We begin by specifying the sequence of potential outcomes that we will be considering. We will focus on nodes in $U$, so only the outcomes for these nodes will be non-zero. Also, we will focus on the all treatment exposure only, so the all control exposure outcomes will be zero for all nodes.
Specifically, for each node $i \in U$, we set $y_i(\vec{1}) = \sqrt{n/d}$.
In other words,

\[
y_i^{(n)}(\vec{0}) = 0 \quad, \quad 
y_i^{(n)}(\vec{1}) = \left\{ 
\begin{array}{ll}
    \sqrt{\frac{n}{d}} &\text{ , if } i \in U\\
    0 & \text{ , otherwise}
\end{array}
\right.
\]
Clearly Assumption~\mainref{assumption:bounded-second-moment} is satisfied for this choice of outcomes, since
\begin{align*}
\frac{1}{n} \sum_{i=1}^n \paren[\Big]{y_i(\vec{1})^2 + y_i(\vec{0})^2} 
&= \sum_{i \in U} \frac{n}{d}
= 1 \enspace.
\end{align*}
As stated in \cite{Ugander2023Randomized}, the estimator $\eate_{\operatorname{1-hm}}$ is unbiased for the GATE $\ate$. Thus, to understand the mean-squared error, it suffices to analyze the variance of $\eate_{\operatorname{1-hm}}$.
We start by writing the variance of the estimator as
    \begin{align*}
    \Var{\eate_{\operatorname{1-hm}}} &= \Var[\Big]{\frac{1}{n} \sum_{i=1}^n \frac{\indicator{E_{(i,1)}}}{\Pr{E_{(i,1)}}} Y_i }\\
    &= \frac{1}{n^2} \sum_{i=1}^n \sum_{j=1}^n \Cov[\Big]{\frac{\indicator{E_{(i,1)}}}{\Pr{E_{(i,1)}}} Y_i, \frac{\indicator{E_{j,1}}}{\Pr{E_{j,1}}} Y_j }\\
    &= \frac{1}{n^2} \sum_{i \in U} \sum_{j\in U} \paren[\Bigg]{\frac{\Pr{E_{(i,1)}\cap E_{j,1}}}{\Pr{E_{(i,1)}\cdot \Pr{E_{j,1}}}} - 1 }\cdot y_i(\vec{1})y_j(\vec{1})\\
    &= \frac{1}{dn} \sum_{i \in U} \sum_{j\in U} \paren[\Bigg]{\frac{\Pr{E_{(i,1)}\cap E_{j,1}}}{\Pr{E_{(i,1)}\cdot \Pr{E_{j,1}}}} - 1 }\enspace.
    \end{align*}
    Now, using the bounds on the probabilities of exposure from Lemmas~\ref{lem:exposure-lower} and ~\ref{lem:exposure-upper}, we have that for each $i,j \in U$ with $i \neq j$,
    \[
    \frac{\Pr{E_{(i,1)}\cap E_{j,1}}}{\Pr{E_{(i,1)}\cdot \Pr{E_{j,1}}}} - 1 = \bigOmega[\Big]{\frac{\frac{1}{d^2}}{\paren[\Big]{\frac{\log^2 d}{d^2}}^2}} = \bigOmega[\Big]{\frac{d^2}{\log^4 d}} \enspace.
    \]
    If $i =j$, then similarly we get from Lemma~\ref{lem:exposure-upper}
    \[
    \frac{1}{\Pr{E_{(i,1)}}} - 1 = \bigOmega[\Big]{\frac{d^2}{\log^2 d}} = \bigOmega[\Big]{\frac{d^2}{\log^4 d}} 
    \]
    Thus, we conclude that 
    \[
    \Var{\eatetrue} = \frac{1}{dn} \cdot d^2 \cdot \bigOmega[\Big]{\frac{d^2}{\log^4 d}} = \bigOmega[\Big]{\frac{d^3}{n \log^4 d}} \enspace.
    \]
    Using the fact that $d_n = (1+o(1))n^{1/3}$ and taking the limit $n\to \infty$ in the above implies \eqref{eq:rate-1hm}.
\end{proof}

	\section{Additional Simulation Results} \label{sec:supp-simulations}

In this section, we present additional discussion of the simulation results.
We focus here on reporting aspects of the inferential procedures.
For all simulations, we report intervals for the $95\%$ confidence level.

See Section~\suppref{sec:simulations} for the detailed discussion of the networks that are used in the simulation.
In these extended simulations, we consider two types of outcomes:
\begin{itemize}
	\item \textbf{Large Outliers}: $y_i(e_0) = \alpha_{i,1}$ and $y_i(e_1) = \alpha_{i,2} \cdot \sqrt{ \textrm{deg}_G(i) }$
	\item \textbf{Medium Outliers}: $y_i(e_0) = \alpha_{i,1}$ and $y_i(e_1) = \alpha_{i,2} \cdot \paren{ \textrm{deg}_G(i) }^{1/4}$
\end{itemize}
where the coefficients are sampled independently across units as $\alpha_{i,1} \sim \mathcal{N}(1,1)$ and $\alpha_{i,2} \sim \mathcal{N}(2,1)$.
In each of these settings, the potential outcome under control exposure $e_0$ is kept roughly constant while the potential outcome under treatment exposure $e_1$ is proportional to some power of the unit's degree in the network.
By the construction of the preferential attachment network, one can verify that the Large Outliers construction ensures boundedness of the second moment (Assumption~\mainref{assumption:bounded-second-moment})
but the fourth moment (Assumption~\mainref{assumption:bounded-fourth-moments}) will generally not be bounded.
On the other hand, the Medium Outliers construction ensures boundedness of 4th moments but not necessarily boundedness of higher moments.
In the sense that moment conditions reflect outlier behavior, these two outcome settings reflect different outlier behavior.

Recall that boundedness of the second moment was required for $\sqrt{n / \lamH}$-consistency (Assumption~\mainref{assumption:bounded-second-moment}) and that boundedness of the fourth moment (Assumption~\mainref{assumption:bounded-fourth-moments}) was a sufficient condition both for consistent variance estimation and the Central Limit Theorem, which played a key role in the construction of valid confidence intervals.
Thus, we may anticipate that while the Large Outliers setting allowed for consistent point estimation, the inferential procedures will only work for Medium Outliers.

Recall that the Central Limit Theorem (Theorem~\mainref{thm:clt}) also required a strong condition on the maximum degree of the conflict graph, i.e. $\dmax(\cH) = \littleO{n^{1/9}}$.
We feel that this reflects a limitation of the proof technique rather than a necessary condition for a CLT.
In the numerical simulations, the assumption $\dmax(\cH) = \littleO{n^{1/9}}$ is far from holding and so these simulations may be viewed as a ``stress test'' of the inferential procedures when this assumption is not met.
On the other hand, the fourth moment condition seems both necessary and sufficient for obtaining precise variance estimators at the given rates (Proposition~\mainref{prop:var-est-consistency}).
For this reason, we vary the moments of the outcomes.

Throughout our simulations, we examine the following confidence intervals:
\begin{itemize}
	\item \textbf{Wald (exact $\Varsym$)}: A Wald confidence interval where the exact variance is used. This is an oracle confidence interval because the exact variance is not known to the experimenter. This reflects the behavior of the ``true'' standardized estimator.
	\item  \textbf{Wald (exact $\vb$)}: A Wald confidence interval where the exact variance bound is used. This is an oracle confidence interval because the exact variance bound is not known to the experimenter. This reflects the behavior of the  ``variance bound'' standardized estimator.
	\item \textbf{Wald ($\evb$)}: A Wald confidence interval using the conservative variance estimator, proposed in Section~\mainref{sec:intervals}.
	\item \textbf{Chebyshev (exact $\Varsym$)}: A Chebyshev confidence interval where the exact variance is used. This is an oracle confidence interval because the exact variance is not known to the experimenter.
	\item  \textbf{Chebyshev (exact $\vb$)}: A Chebyshev confidence interval where the exact variance bound is used. This is an oracle confidence interval because the exact variance bound is not known to the experimenter.
	\item \textbf{Chebyshev ($\evb$)}: A Chebyshev confidence interval using the conservative variance estimator, proposed in Section~\mainref{sec:intervals}.
\end{itemize}

\subsection{Direct Treatment Effect}

Let us first consider the direct treatment effect.
We begin by examining the inferential methods in the presence of Large Outliers.

% FIGURE 1:
% for the  Large Outliers, create the subplots
% (a) widths (b) coverage
\begin{figure}[H]
	\centering
	\begin{subfigure}{0.49\textwidth}
		\centering
		\includegraphics[width=\textwidth]{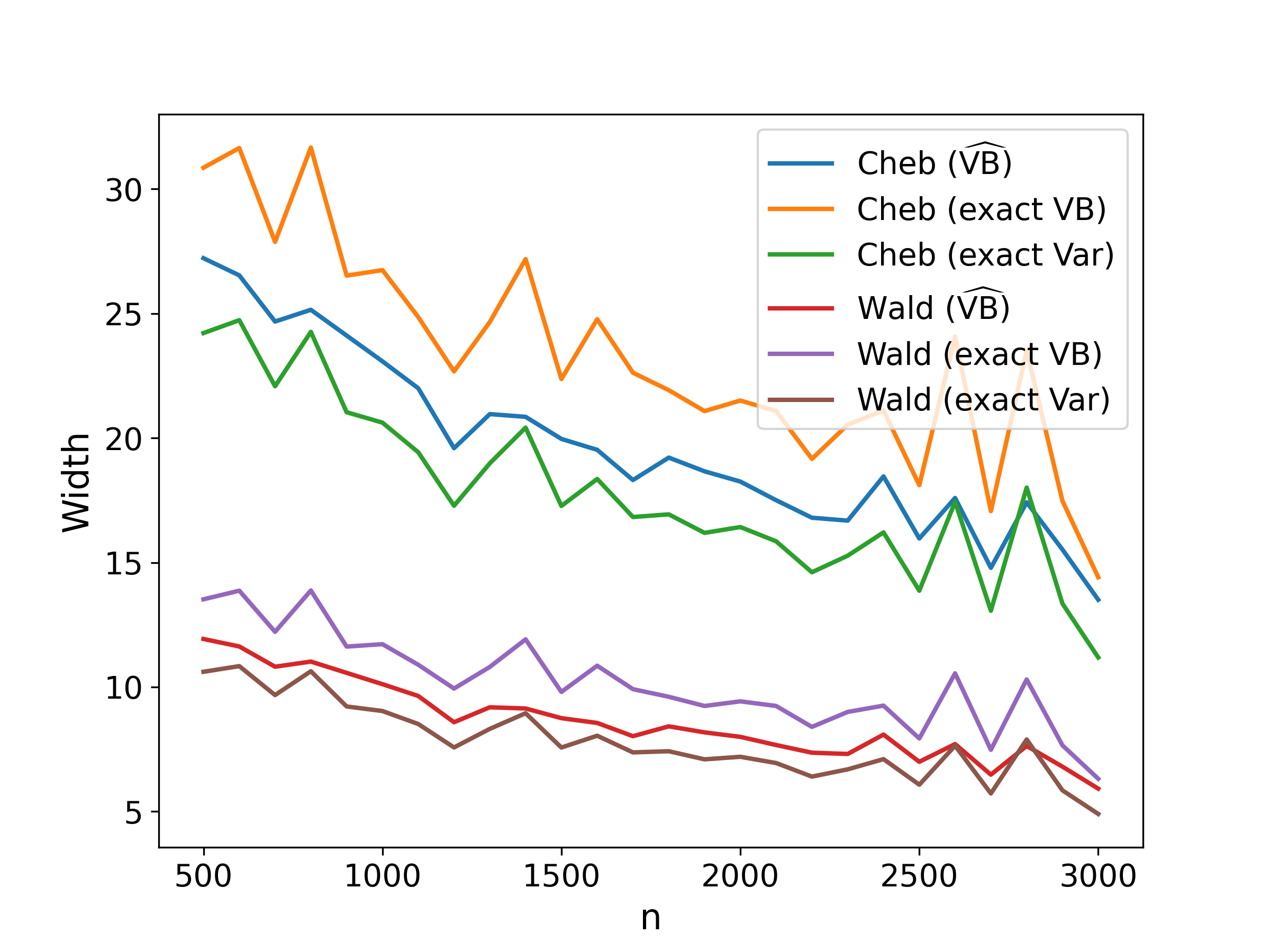}
		\caption{Expected CI Widths}
		\label{fig:dte-large-widths}
	\end{subfigure}%
	~ 
	\begin{subfigure}{0.49\textwidth}
		\centering
		\includegraphics[width=\textwidth]{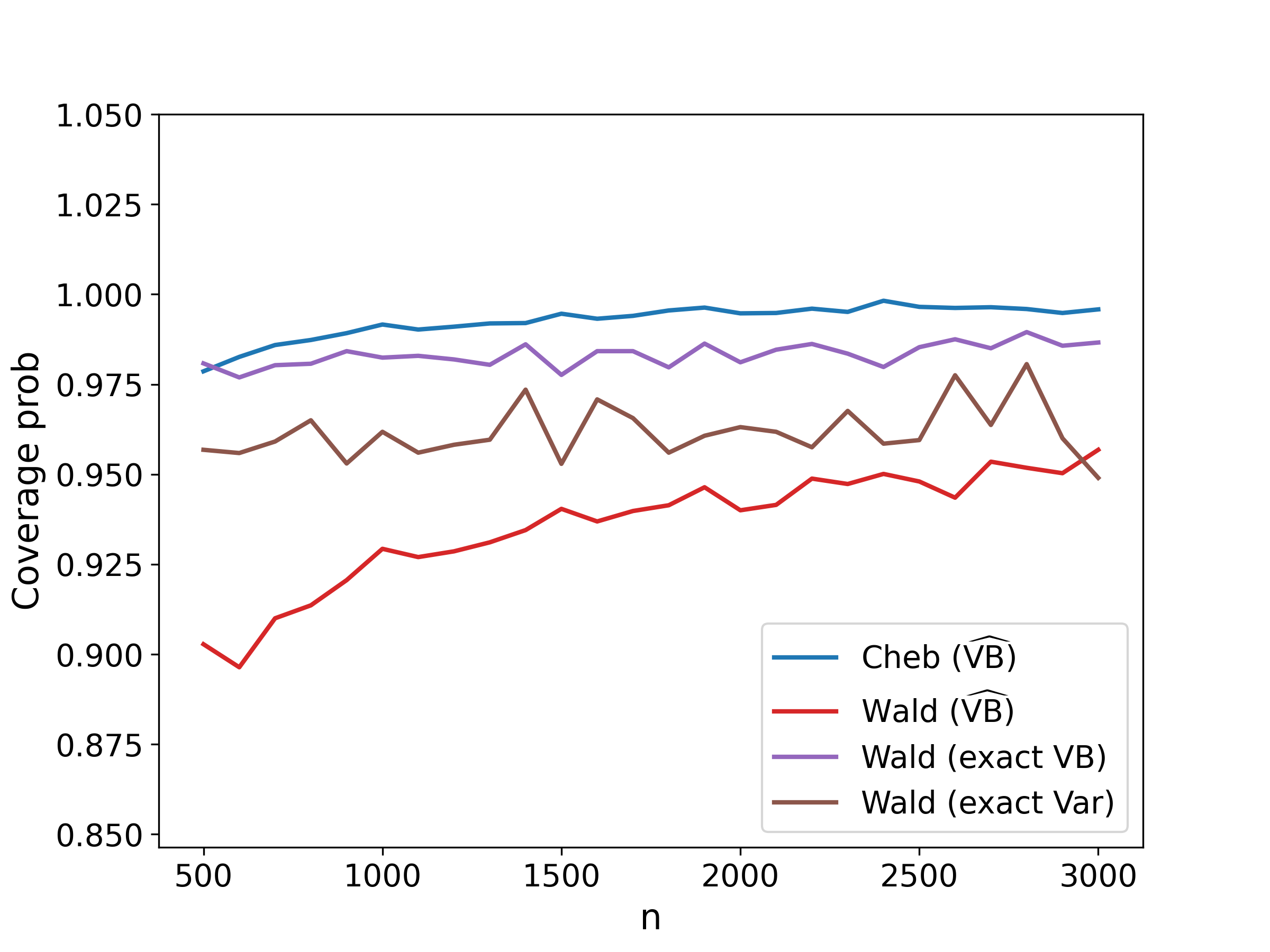}
		\caption{Coverage}
		\label{fig:dte-large-covers}
	\end{subfigure}
	\caption{Confidence Interval Statistics for Direct Treatment Effect in the Large Outliers setting.}
	\label{fig:dte-large-cis}
\end{figure}

Figure~\ref{fig:dte-large-cis} contains a plot of the expected interval widths and the coverage as $n$ grows.
In terms of interval widths, we see what is to be expected: namely, that the Chebyshev-types widths are larger than the Wald-type widths and within these classes, intervals based on the variance bound are larger than intervals based on the exact variance.
We see that the Chebyshev intervals cover above the nominal rate, which is to be expected.
The oracle Wald intervals cover slightly above the nominal rate while the actual Wald intervals cover significantly below the nominal rate.

% FIGURE 2: 
% for the Large Outliers, creaet te subplots
%  (a) histogram (b) var( ratio )
\begin{figure}[H]
	\centering
	\begin{subfigure}{0.49\textwidth}
		\centering
		\includegraphics[width=\textwidth]{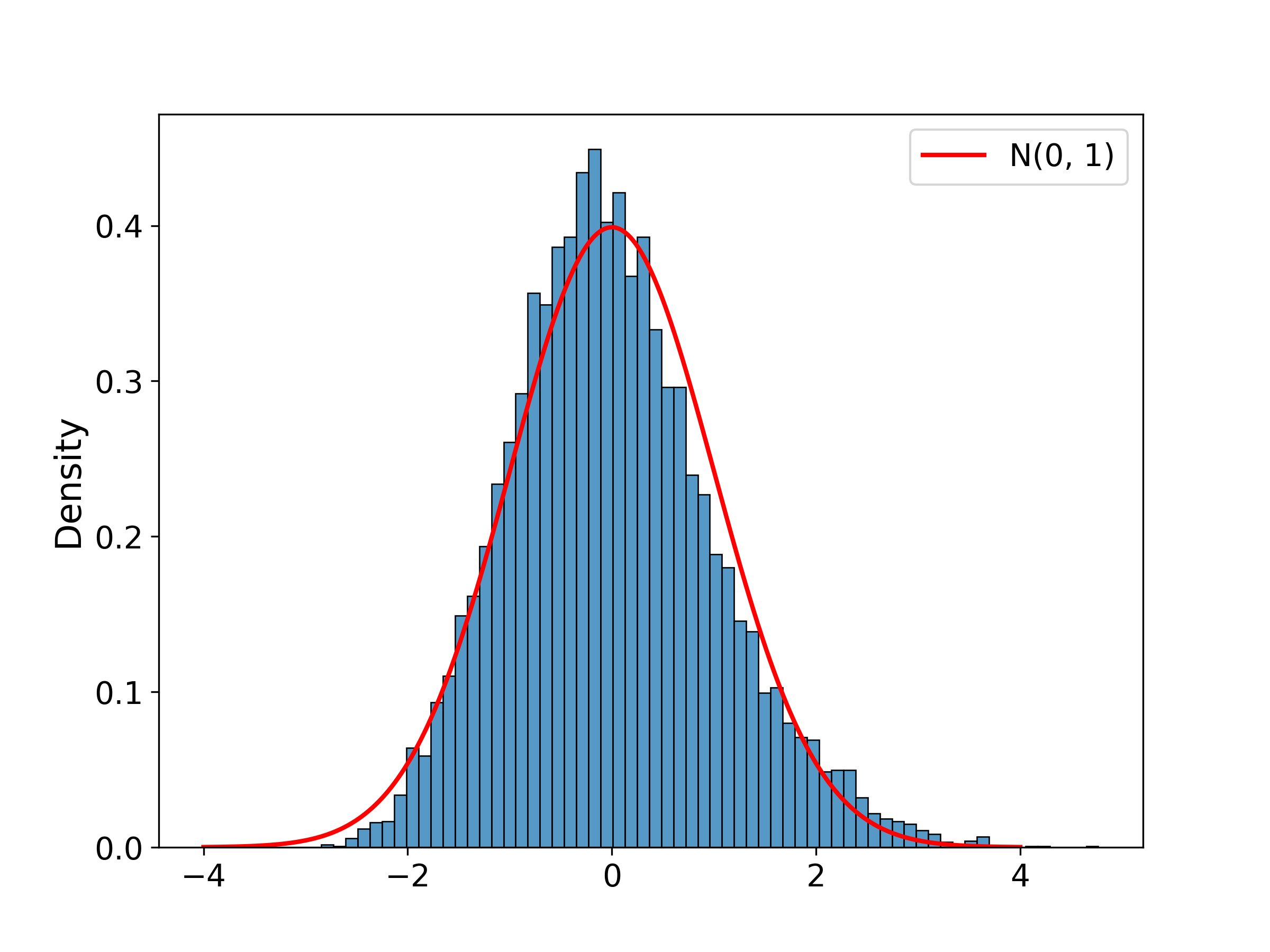}
		\caption{Standardized Estimator Histogram}
		\label{fig:dte-large-histogram}
	\end{subfigure}%
	~ 
	\begin{subfigure}{0.49\textwidth}
		\centering
		\includegraphics[width=\textwidth]{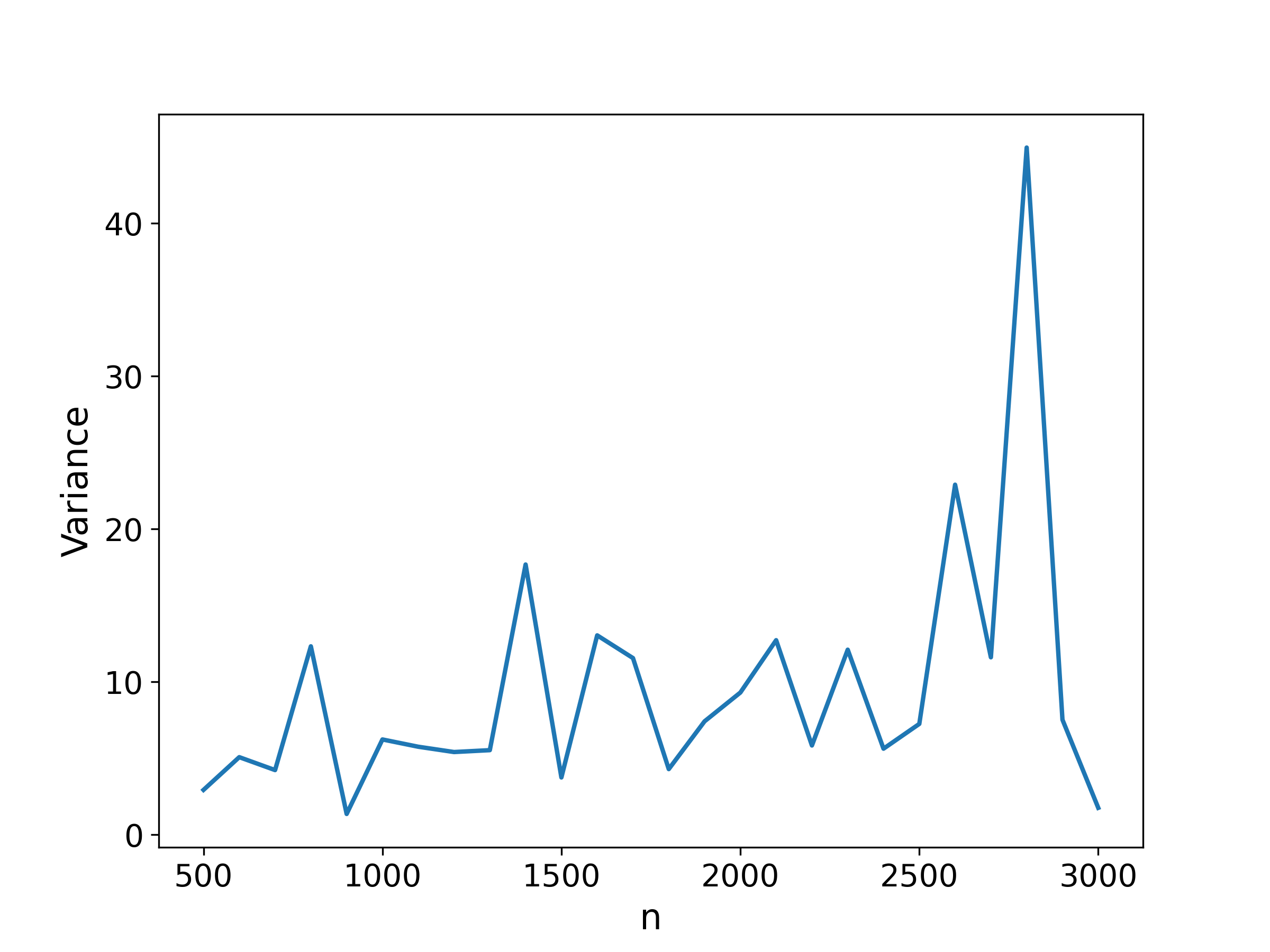}
		\caption{Variance of the ratio $\evb / \vb$}
		\label{fig:dte-large-varratio}
	\end{subfigure}
	\caption{Additional interval statistics for Direct Treatment Effect in the Large Outliers setting.}
	\label{fig:dte-large-consistency}
\end{figure}

Our goal now is to explain the behavior of the Wald-type intervals.
We begin by looking at the histogram of the standardized estimator $\paren{\eate - \ate} / {\sqrt{\Var{\eate}}}$ at $n=3,000$ samples, contained in Figure~\ref{fig:dte-large-histogram}.
Although the histograms are approximately a standard normal, they are different in two ways:
(1) the standardized effect estimator is slightly more concentrated around 0 for moderate values of the error (e.g. less than 2 in absolute value) and (2) far away from 0, the right tails are thicker than a standard normal.

We attribute these irregularities to the high degree nodes, which also have large outcomes.
We know that such behavior can theoretically occur even when the estimator is consistent.
See Proposition~\mainref{prop:clt-counter-example} and its proof in Section~\suppref{sec:supp-clt-counterexample} for more rigorous discussion.
Roughly speaking, the high degree nodes contribute significantly to the standardized error when assigned to treatment exposure, but this is a low probability event.
Informally speaking, this results in the standardized error distribution being narrower than $\mathcal{N}(0,1)$ around 0, but having little peaks far away from 0, though these are typically vanishing in large samples.
This phenomenon explains the histogram we see in Figure~\ref{fig:dte-large-covers} and also why the oracle Wald intervals over-cover ever so slightly.

This does not explain, however, why the actual Wald intervals severely undercover.
To understand this better, we turn our attention to the variance estimator, $\evb$.
The usual Slutsky-style arguments for asymptotic coverage require that $\evb / \vb$ converges in probability to 1.
While it is (strictly speaking) not a necessary condition, examining the variance of the ratio $\evb / \vb$ will help us understand whether this concentration is happening.
As we see in Figure~\ref{fig:dte-large-varratio}, this variance is not decreasing with the sample size.
We believe that this is the explanation for the severe under-coverage of the Wald intervals.
In particular, we believe that the variance is hard to estimate because of the presence of Large Outliers, i.e. unbounded fourth moments.
This simulation provides numerical evidence to affirm that the bounded fourth moments really are necessary for the statistical validity of the proposed inferential methods.

% Figure 3:
% for the Medium outliers, create the subplots
% : (a) widths (b) coverage
\begin{figure}[H]
	\centering
	\begin{subfigure}{0.49\textwidth}
		\centering
		\includegraphics[width=\textwidth]{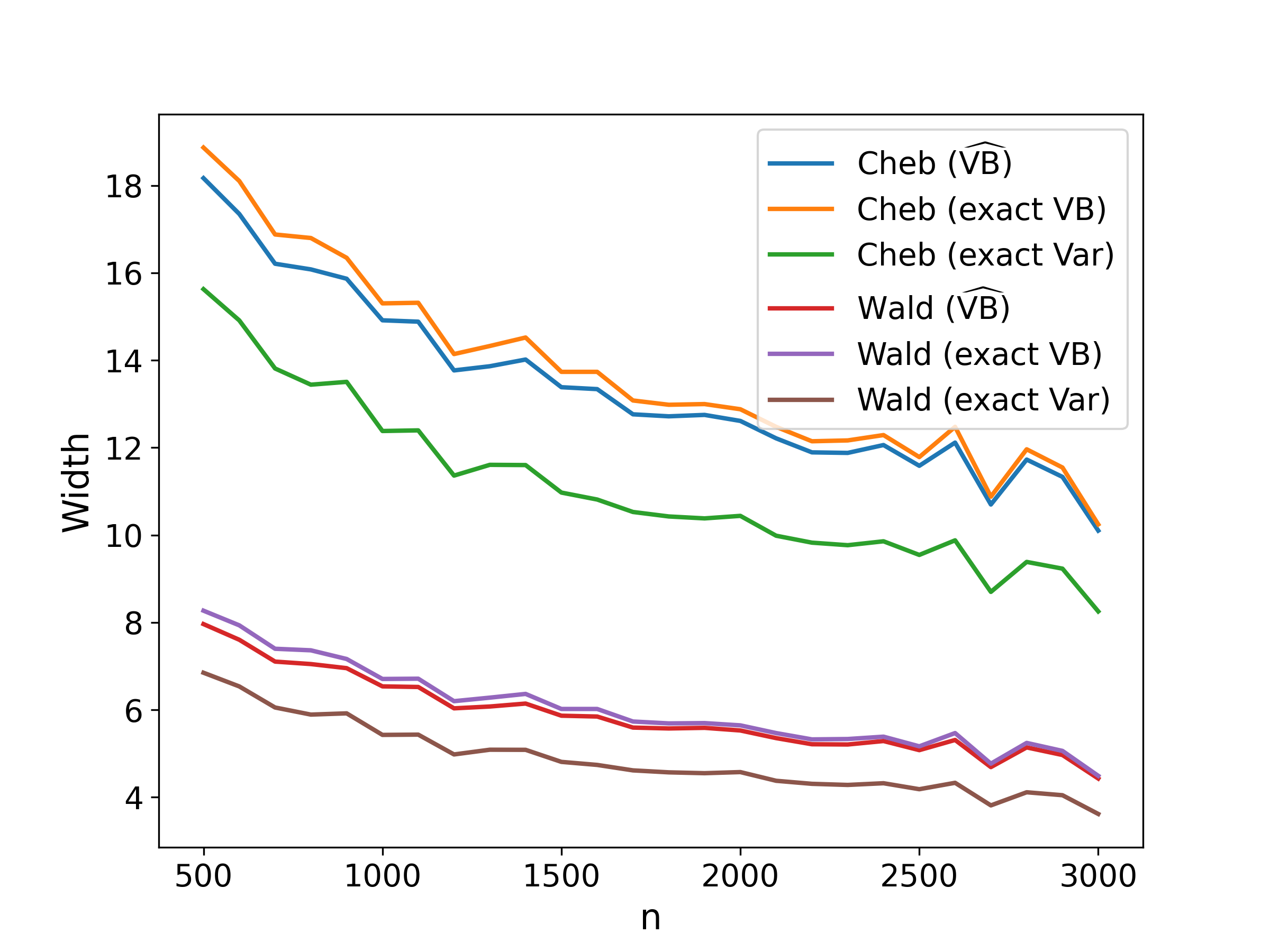}
		\caption{Expected CI Widths}
		\label{fig:dte-med-widths}
	\end{subfigure}%
	~ 
	\begin{subfigure}{0.49\textwidth}
		\centering
		\includegraphics[width=\textwidth]{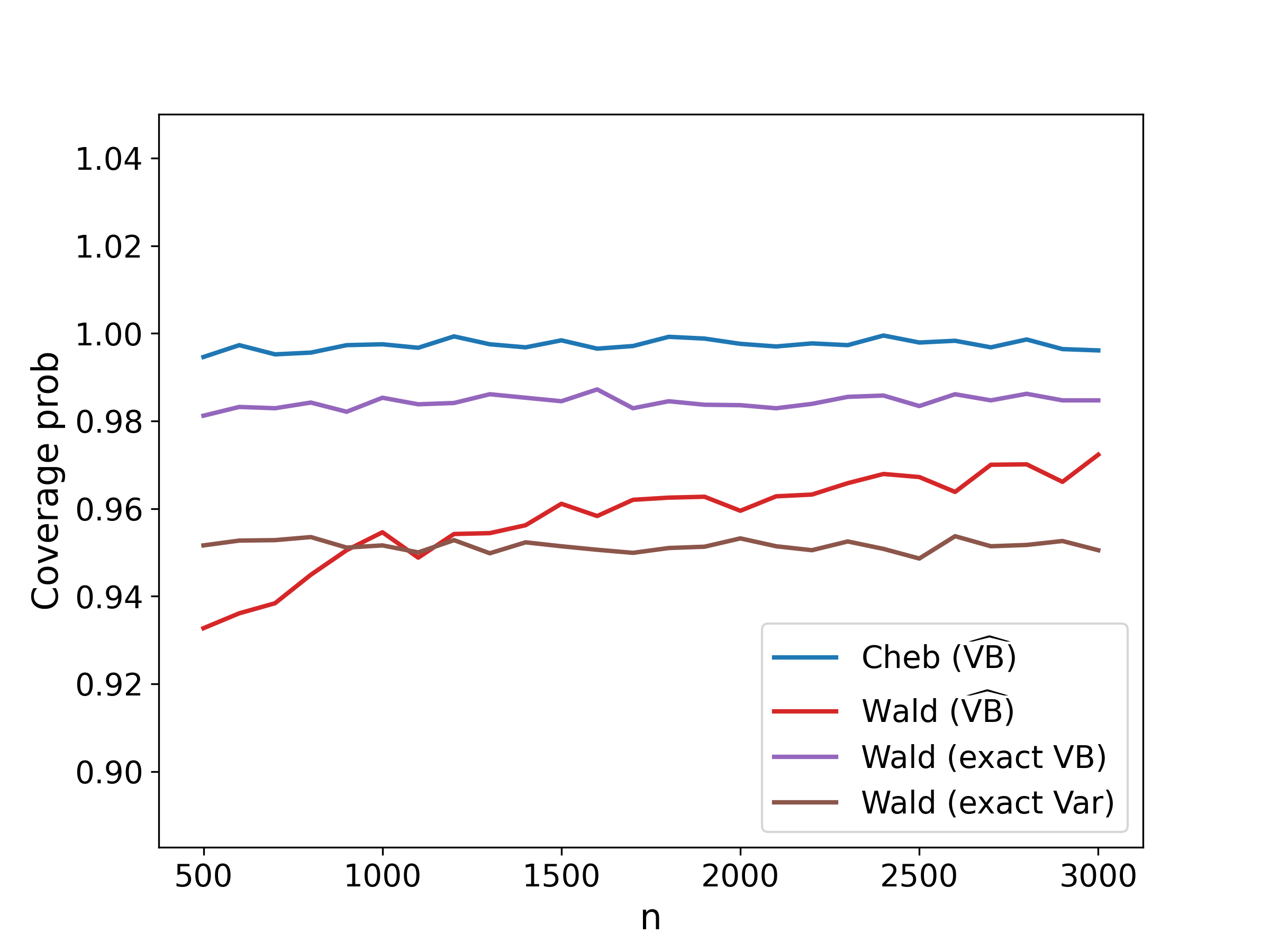}
		\caption{Coverage}
		\label{fig:dte-med-covers}
	\end{subfigure}
	\caption{Additional interval statistics for Direct Treatment Effect in the Medium Outliers setting.}
	\label{fig:dte-med-cis}
\end{figure}

Thus, we are motivated to investigate the statistical validity of the proposed inferential methods when the fourth moments are bounded.
See Figure~\ref{fig:dte-med-cis} for the expected width and coverage of the intervals for the Medium Outlier setting as $n$ grows.
We see that the widths behave as expected: Chebyshev intervals are wider than Wald intervals and within these classes the intervals using variance bound is larger than those using the true variance.
We see that the coverage is now behaving as expected as well.
At around $n=1,000$, the Wald intervals cover at the nominal level. 
As before, the Chebyshev intervals over-cover at all included sample sizes.

% Figure 4: 
% for the Medium outliers, create the following subplots
% (a) histogram  (b) variance of ratio()
\begin{figure}[H]
	\centering
	\begin{subfigure}{0.49\textwidth}
		\centering
		\includegraphics[width=\textwidth]{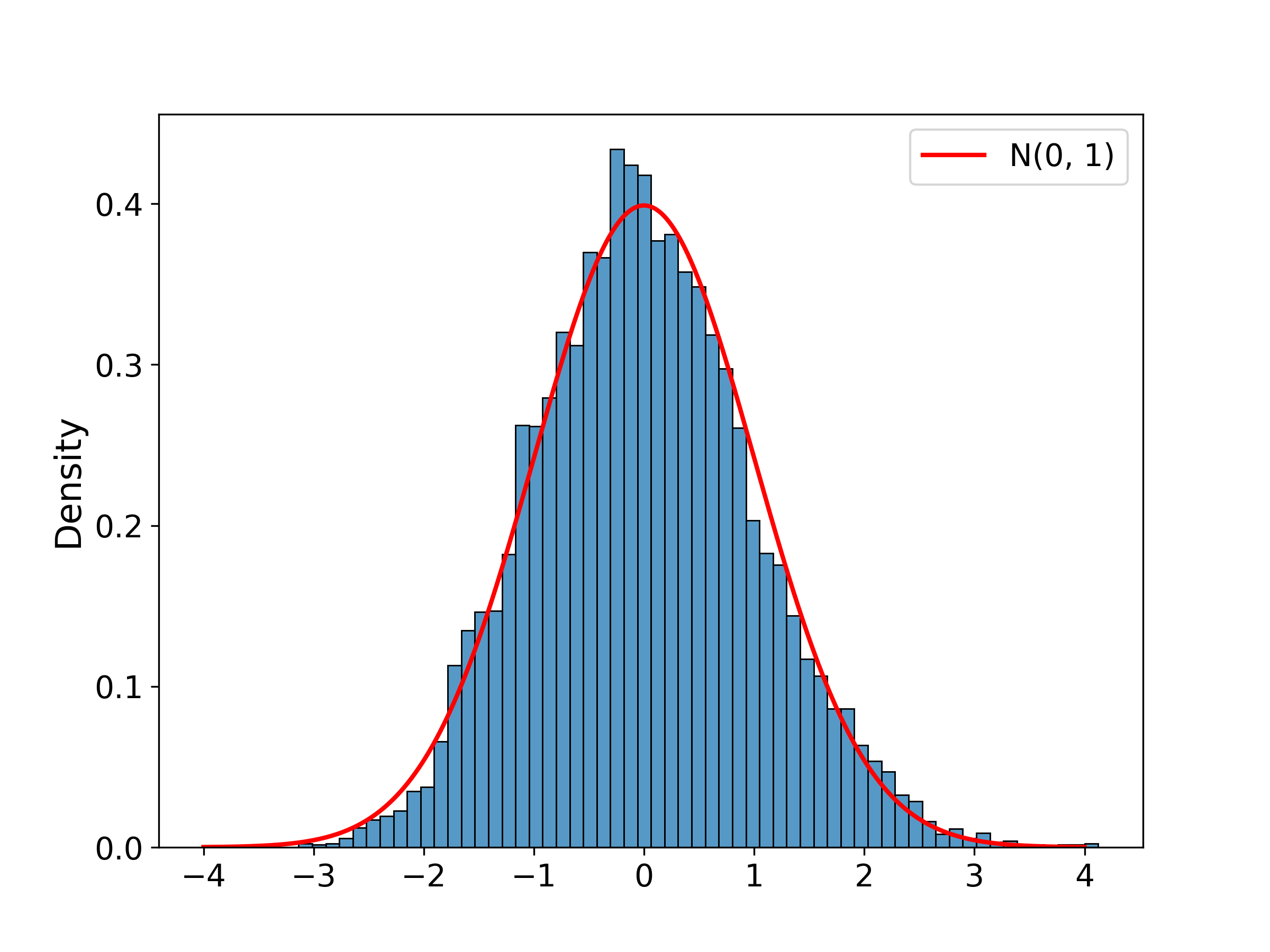}
		\caption{Standardized Estimator Histogram}
		\label{fig:dte-med-histogram}
	\end{subfigure}%
	~ 
	\begin{subfigure}{0.49\textwidth}
		\centering
		\includegraphics[width=\textwidth]{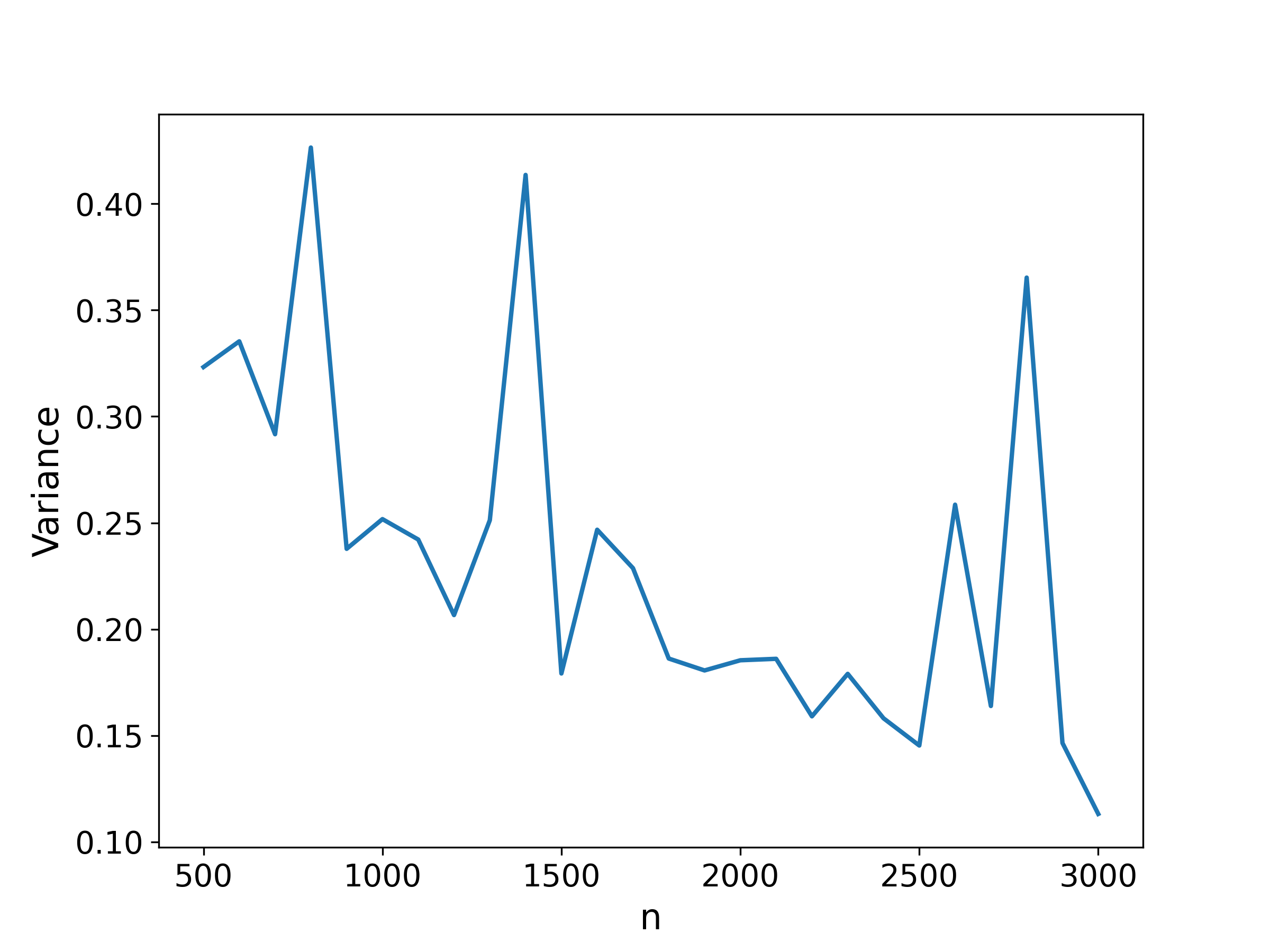}
		\caption{Variance of the ratio $\evb / \vb$}
		\label{fig:dte-med-varratio}
	\end{subfigure}
	\caption{Consistency Statistics for Direct Treatment Effect in the Medium Outliers setting.}
	\label{fig:dte-med-consistency}
\end{figure}

Let us confirm that both ingredients of the inferential procedure are working.
In Figure~\ref{fig:dte-med-histogram} we see that the histogram of the standardized error is approximately normal.
There is but the slightest hint of a bump in the  right tail, reflecting the presence of a few high degree nodes.
Moreover, we see in Figure~\ref{fig:dte-med-varratio} that the variance of the ratio $\evb / \vb$ is roughly decreasing so that convergence of the variance estimator appears to be happening.
There are some bumps in this deceasing plot, but we attribute this to the random generation of the network.

% FIgure 5:
% for the Medium Outliers: 
% plot the variance comparison
\begin{figure}[H]
	\centering
	\includegraphics[width=0.6\textwidth]{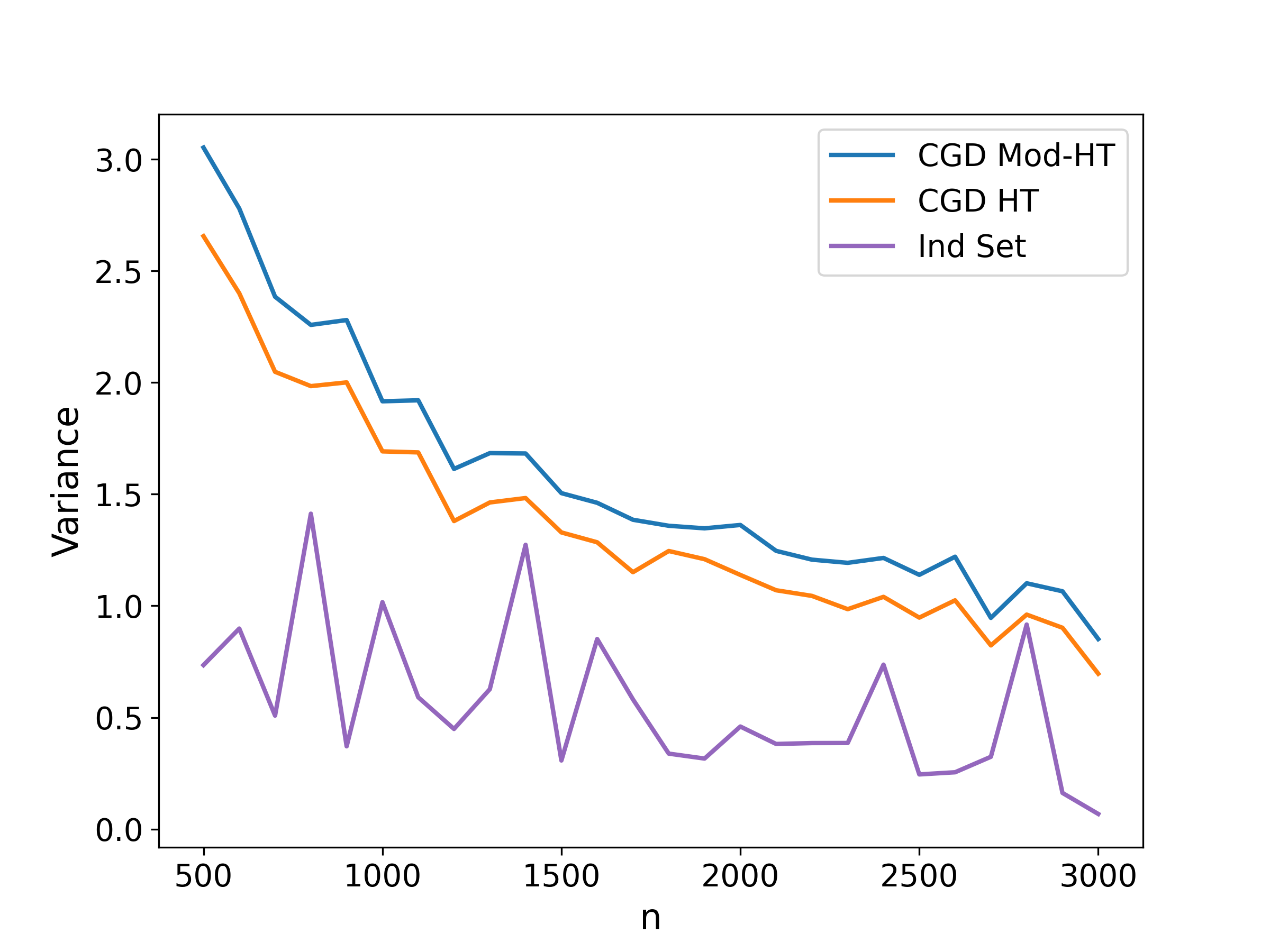}
	\caption{Comparison of the variance of estimating DTE between the various experimental designs in the Medium Outliers setting.}
	\label{fig:dte-med-variances}
\end{figure}

But how does the Conflict Graph Design behave relative to the benchmark Independent Set Design?
Figure~\ref{fig:dte-med-variances} shows that in the setting of Medium Outliers, the Independent Set Design achieves a lower variance than the Conflict Graph Design with modified or standard Horvitz--Thompson estimators.
This improvement is by a factor of about 2-3.
For this simulation set-up, the Conflict Graph Design achieves improved variance in the presence of outliers but the benchmark design achieves lower variance when fewer outliers are present.

\subsection{Global Average Treatment Effect}

We now consider the Global Average Treatment Effect.
Roughly speaking, the findings above remain more or less the same: the Wald-type intervals under-cover due to statistical fluctuations in the variance estimators arising from large outliers.
As the outliers are decreased, the coverage are trending towards the nominal levels.
The main difference here is that even in the medium outlier setting, the sample size does not appear large enough for the coverage to approach the nominal levels.

% Figure 6: 
% for the Large Outliers, create the following subplots:
% (a) width (b) coverage 
% (c) histogram (d) var of ratio
\begin{figure}[H]
	\centering
	\begin{subfigure}{0.49\textwidth}
		\centering
		\includegraphics[width=\textwidth]{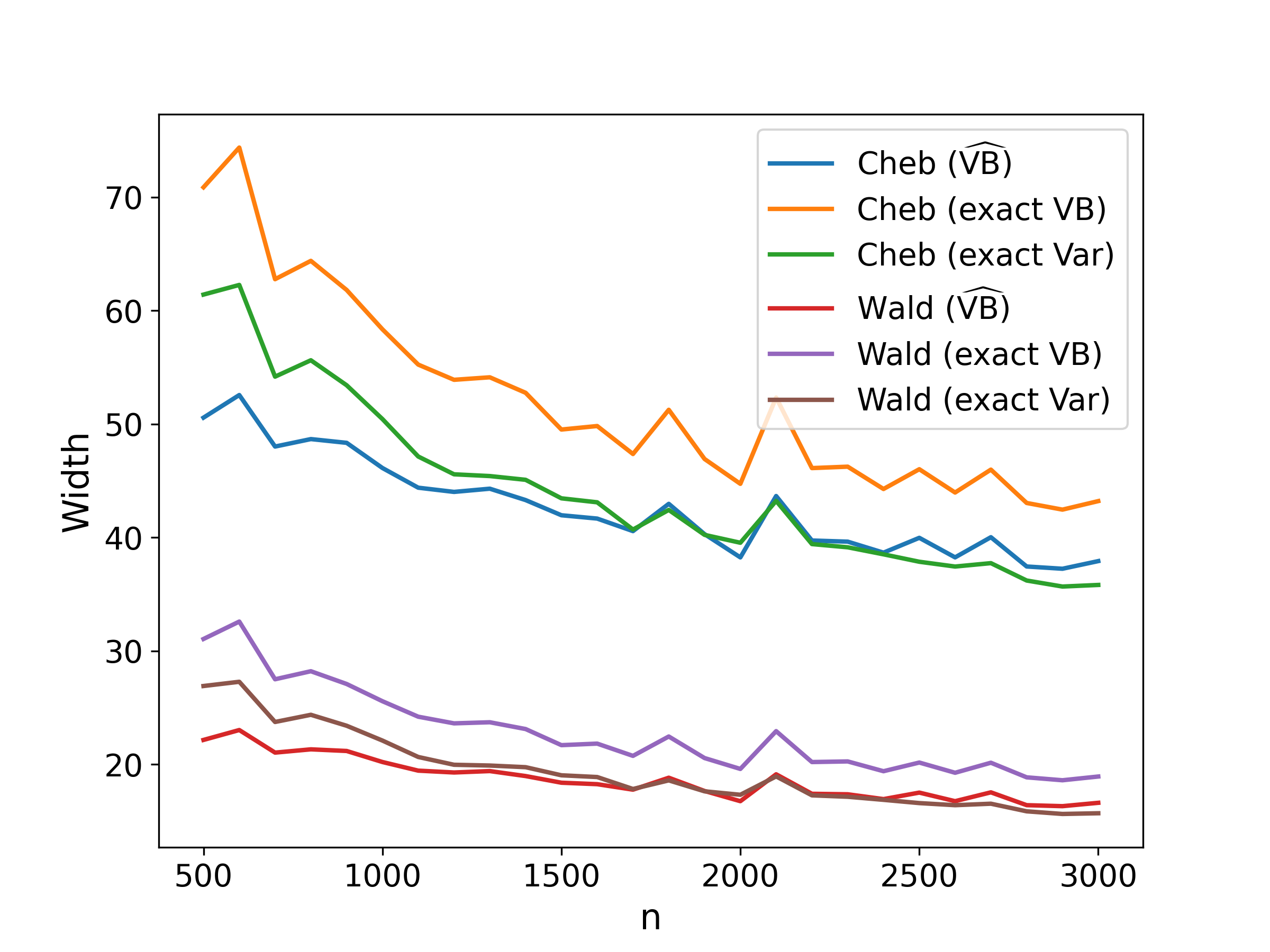}
		\caption{Expected CI Widths}
		\label{fig:gate-large-widths}
	\end{subfigure}%
	~ 
	\begin{subfigure}{0.49\textwidth}
		\centering
		\includegraphics[width=\textwidth]{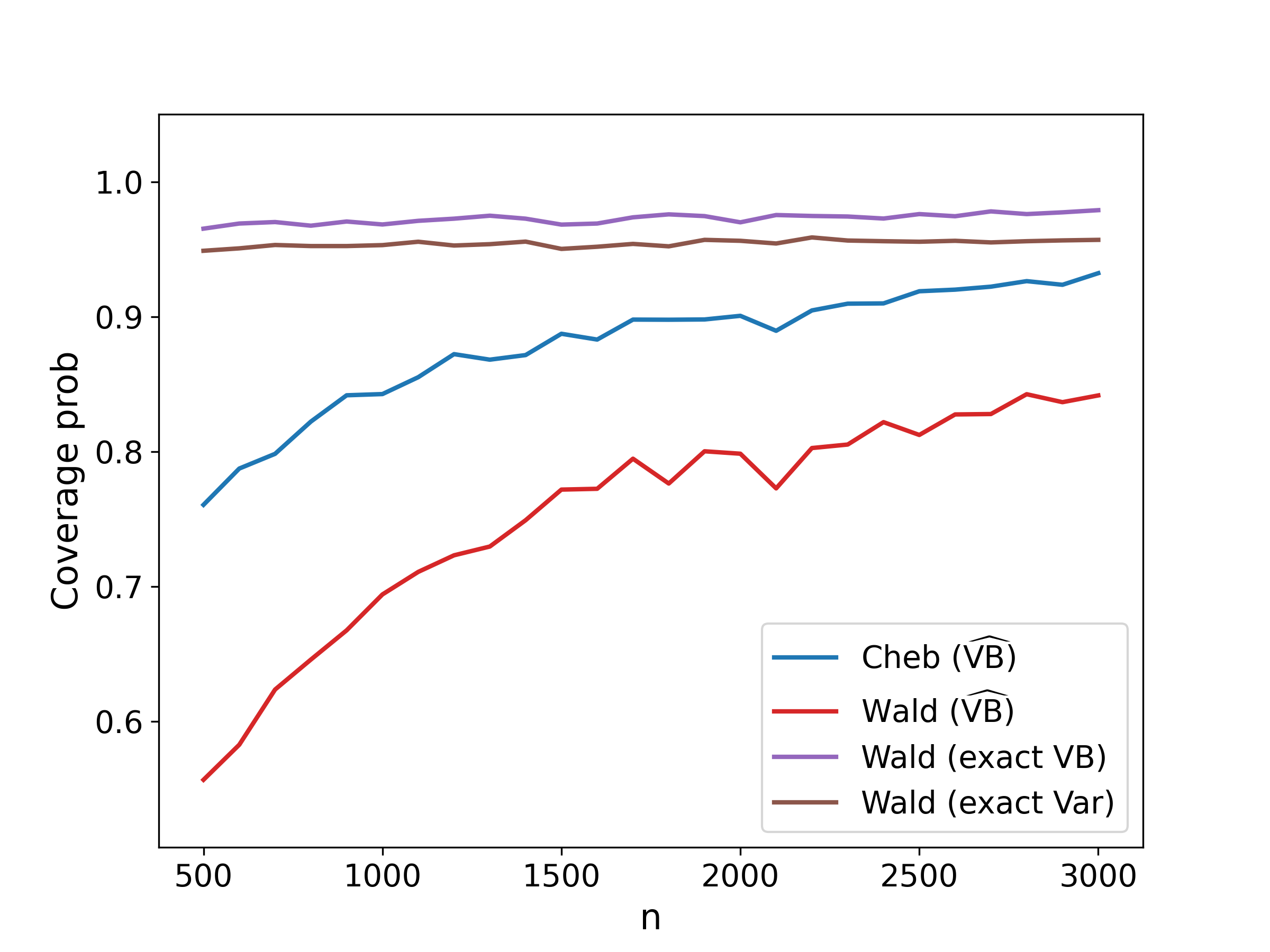}
		\caption{Coverage}
		\label{fig:gate-large-covers}
	\end{subfigure}
	\begin{subfigure}{0.49\textwidth}
		\centering
		\includegraphics[width=\textwidth]{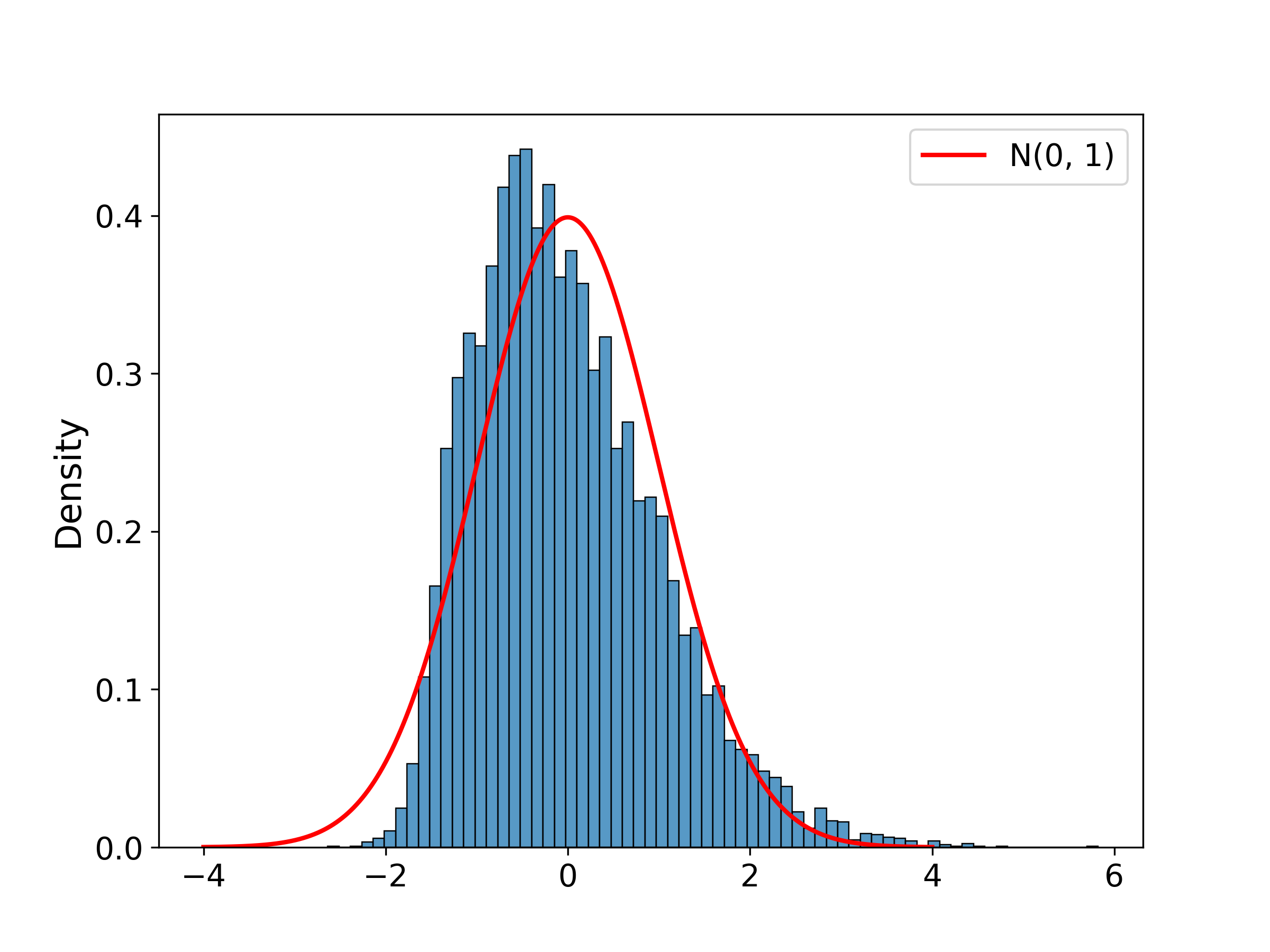}
		\caption{Standardized Estimator Histogram}
		\label{fig:gate-large-histogram}
	\end{subfigure}%
	~ 
	\begin{subfigure}{0.49\textwidth}
		\centering
		\includegraphics[width=\textwidth]{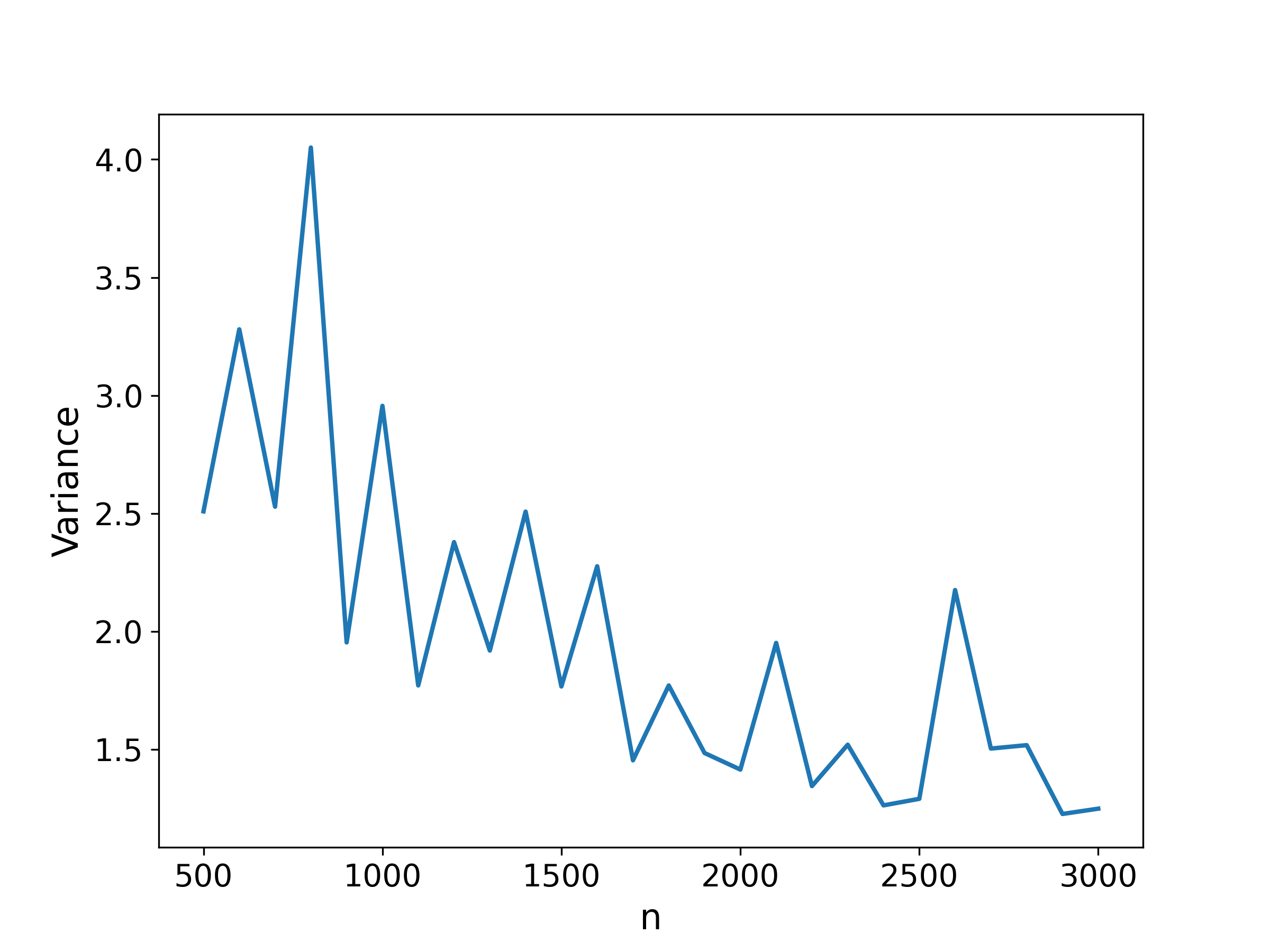}
		\caption{Variance of the ratio $\evb / \vb$}
		\label{fig:gate-large-varratio}
	\end{subfigure}
	\caption{Confidence Interval Statistics for Global Average Treatment Effect in the Large Outliers setting.}
	\label{fig:gate-large-cis}
\end{figure}

Figure~\ref{fig:gate-large-cis} shows the statistical summaries of the inferential procedures under the Large Outliers setting.
While the oracle interval widths look as expected (e.g. Chebyshev is larger than Wald), the widths of the actual intervals display a certain peculiarity: the actual intervals have a width that is closer to the actual variance than the variance bound.
Moreover, the actual intervals severely under-cover while the oracle Wald interval (exact variance) covers at the nominal level.
Both of these peculiarities can be explain in the same way as before; namely, the variance estimator has too much statistical fluctuation.
This is corroborated in two ways.
First, Figure~\ref{fig:gate-large-histogram} that shows the standardized estimator is approximately normal, albeit with a right skew due to high degree nodes with large outcomes.
Even so, the tail approximation obtained from a normal works sufficiently well in the sense that the oracle Wald interval covers at the nominal rate.
Second, Figure~\ref{fig:gate-large-varratio} demonstrates that $\evb / \vb$ has variance which is not decreasing sufficiently fast.
We conjecture that this is again due to the Large Outliers, i.e. the fourth moments are not bounded.

% Figure 7:
% for the Medium Outliers, create the following subplots:
% (a) width (b) coverage 
% (c) histogram (d) var of ratio
\begin{figure}[H]
	\centering
	\begin{subfigure}{0.49\textwidth}
		\centering
		\includegraphics[width=\textwidth]{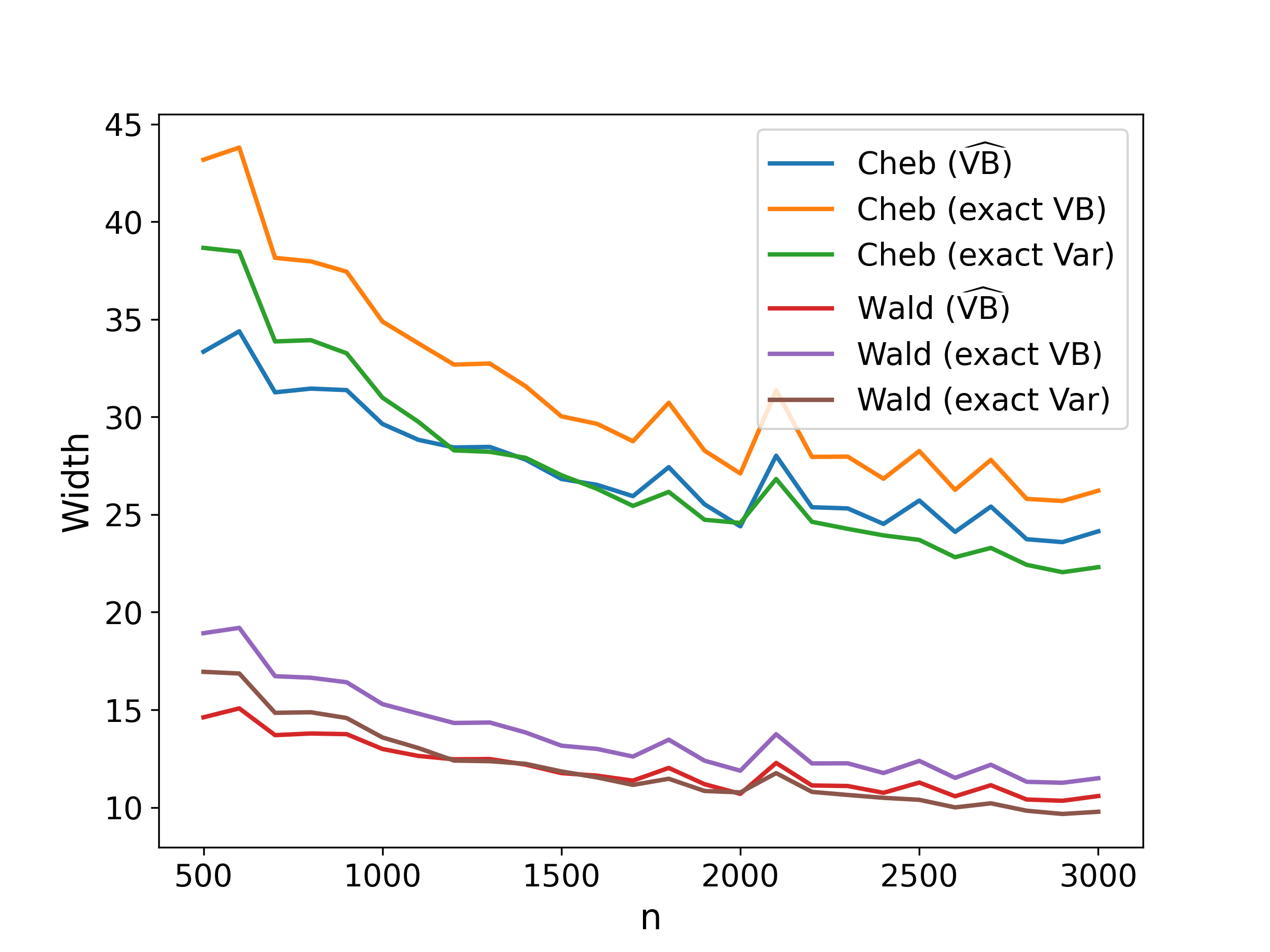}
		\caption{Expected CI Widths}
		\label{fig:gate-med-widths}
	\end{subfigure}%
	~ 
	\begin{subfigure}{0.49\textwidth}
		\centering
		\includegraphics[width=\textwidth]{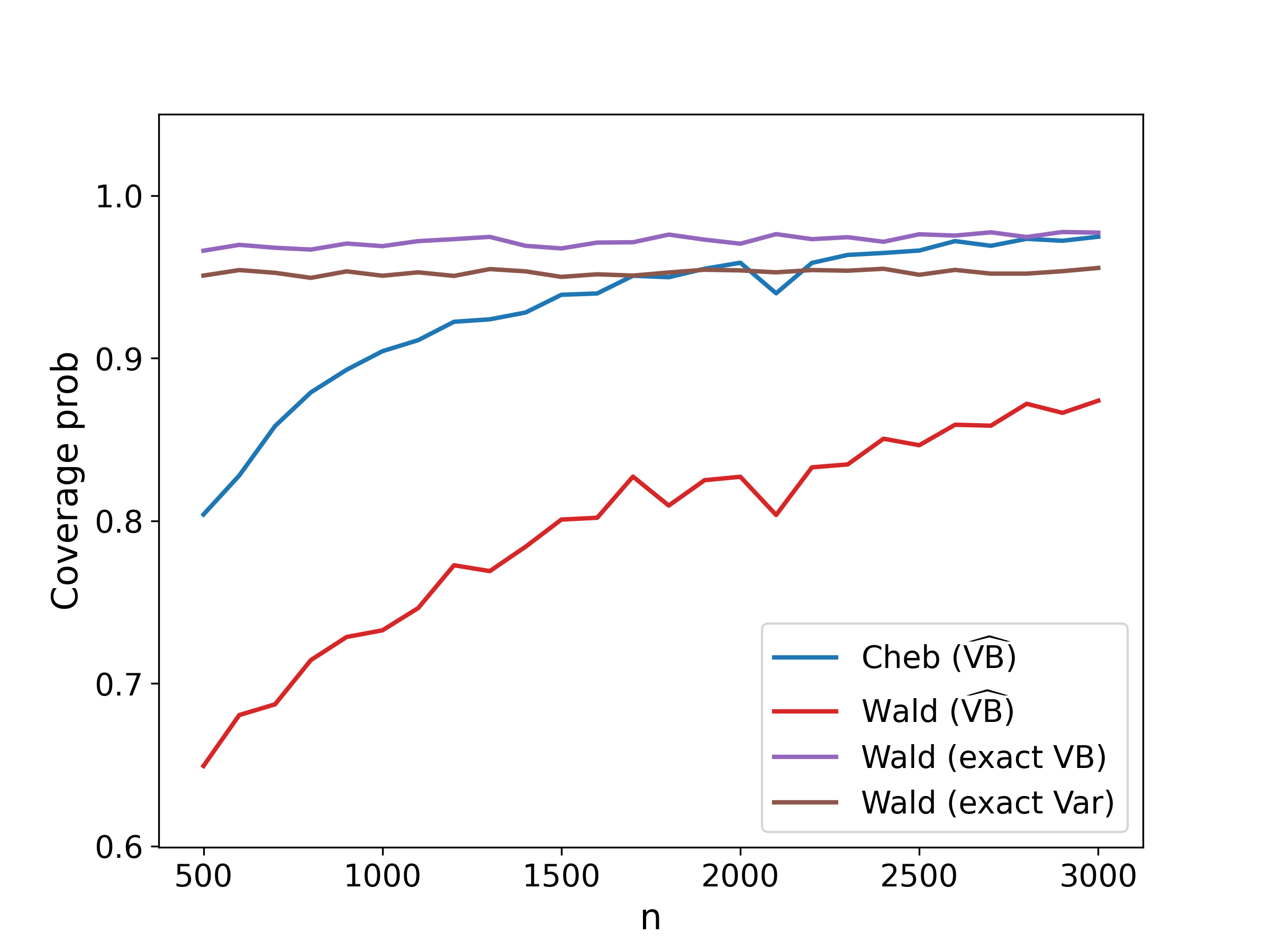}
		\caption{Coverage}
		\label{fig:gate-med-covers}
	\end{subfigure}
	\begin{subfigure}{0.49\textwidth}
		\centering
		\includegraphics[width=\textwidth]{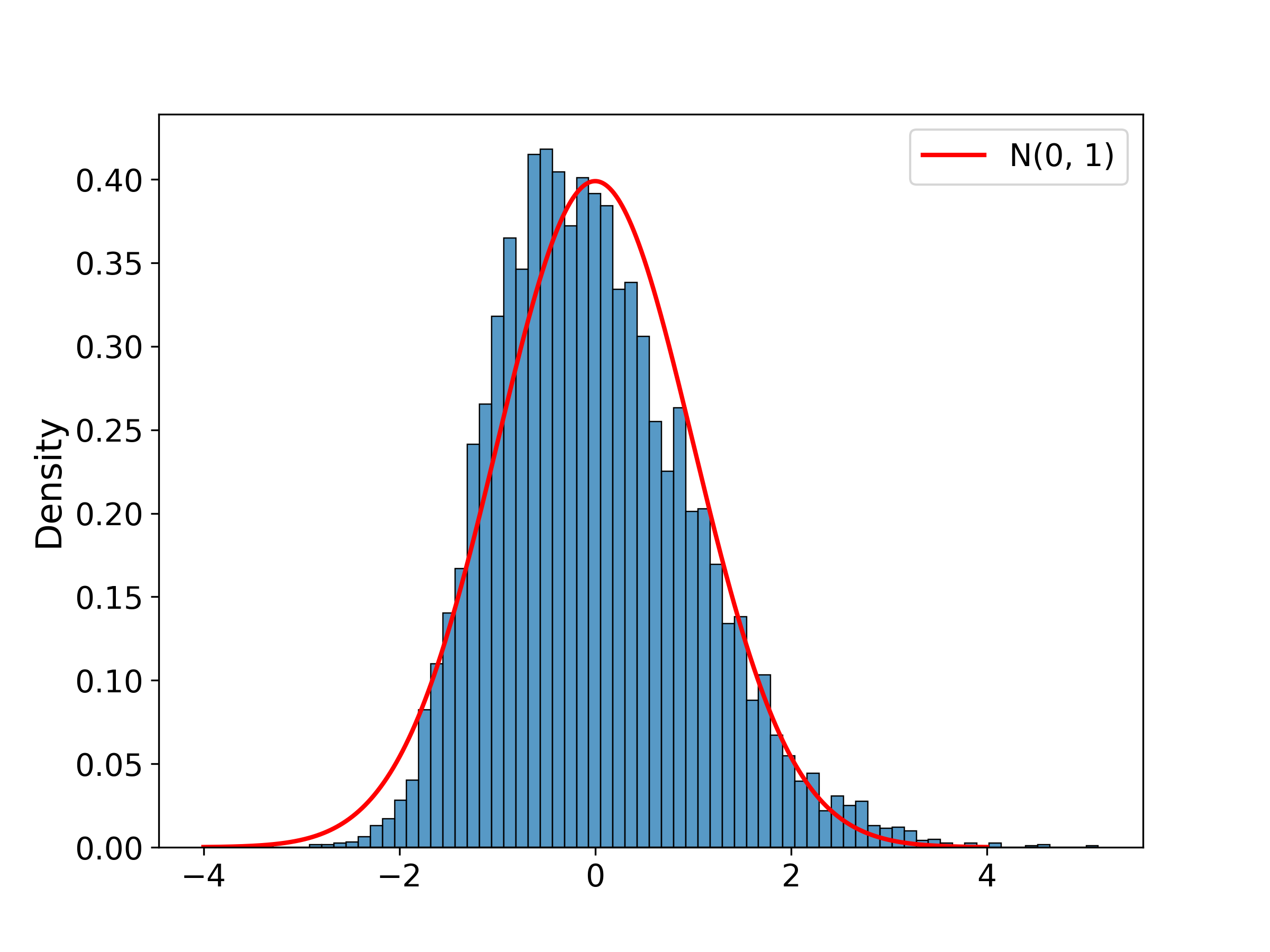}
		\caption{Standardized Estimator Histogram}
		\label{fig:gate-med-histogram}
	\end{subfigure}%
	~ 
	\begin{subfigure}{0.49\textwidth}
		\centering
		\includegraphics[width=\textwidth]{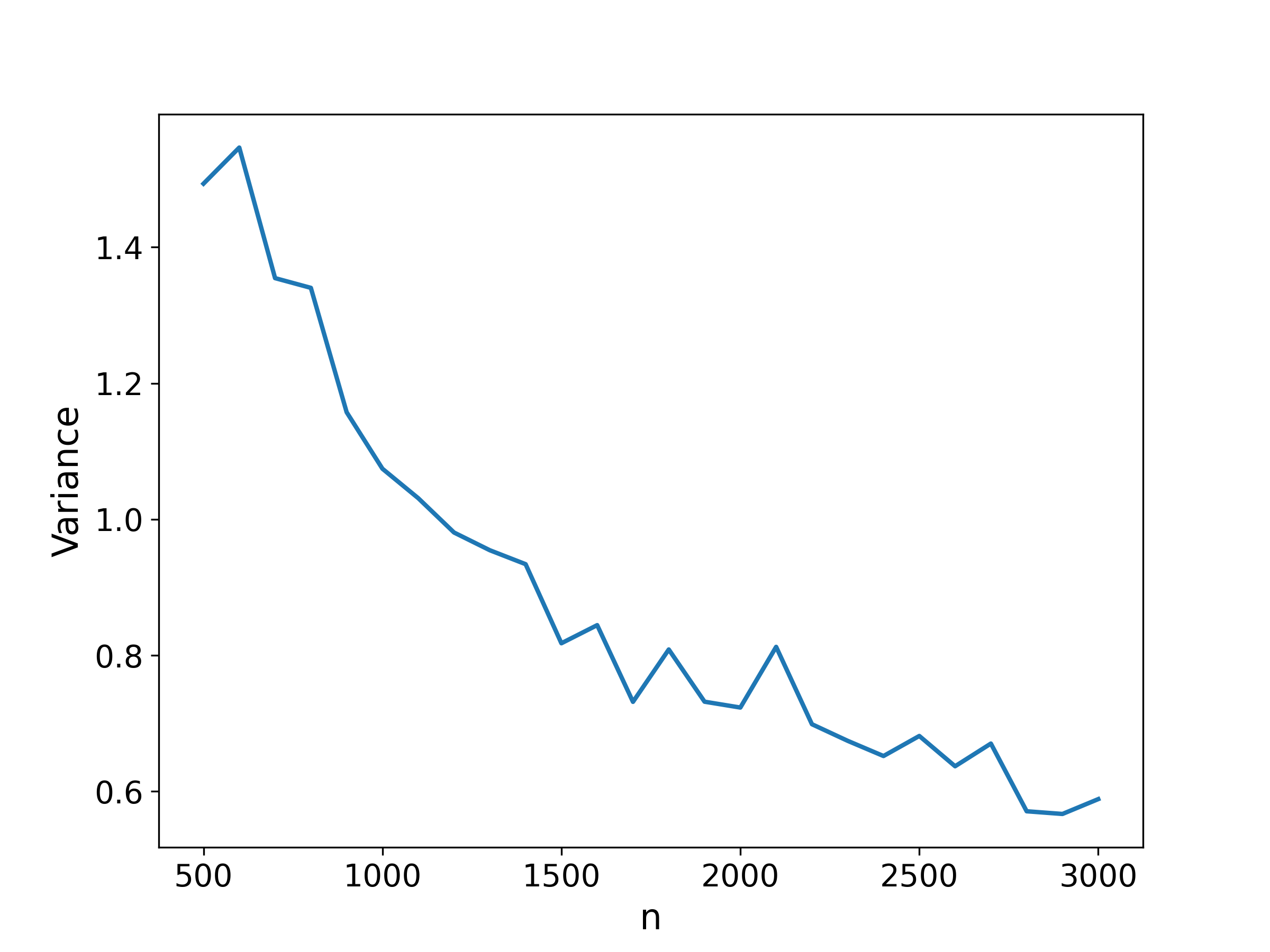}
		\caption{Variance of the ratio $\evb / \vb$}
		\label{fig:gate-med-varratio}
	\end{subfigure}
	\caption{Additional interval statistics for Global Average Treatment Effect in the Medium Outliers setting.}
	\label{fig:gate-med-cis}
\end{figure}

To test this explanation, we investigate the performance of the inferential procedures in the setting of Medium Outliers.
Figure~\ref{fig:gate-med-cis} shows the relevant statistical summaries.
The figures look roughly the same as in the case of Large Outliers, except that the coverage is slightly better, though still under-covering.
This under-coverage of the Wald intervals was initially puzzling to us because our simulation results indicate that both (1) the normal approximation is relatively good (2) the variance estimator has statistical fluctuation which is decreasing with the sample size.
The remaining explanation is that the statistical fluctuation, while decreasing, is not small enough for $\evb$ to be a sufficiently good approximation of $\vb$ for the purposes of coverage of the interval.

To test this explanation, we turn to our theoretical predictions.
In the proof of coverage for the Wald-style intervals (Corollary~\mainref{corollary:intervals-cover-wald}), we show that the following holds in the asymptotic limit:
\[
\Pr{ \ate \in C_{\textrm{Wald}}(\alpha) } 
\approx 1 - 2 \cdot \paren[\Bigg]{1-\Phi_\sigma\paren[\Big]{\Phi_1^{-1}(1 - \alpha/2)\cdot\sqrt{1-\epsilon}}}
\quadwhere
\abs[\Big]{ \frac{\evb}{\vb} -1 } \approx \epsilon 
\]
and $\Phi_\sigma^{-1}$ is the normal quantile function with $\sigma^2 = \Var{\eate} / \vb$ to reflect the increase due to the variance bound.
Proposition~\mainref{prop:var-est-consistency} shows that under the fourth moment condition on the outcomes, the quantity $\epsilon$ will be on the order $\bigOp{\sqrt{\lamH / n}}$.
Thus, we derived the predicted coverage rate:
\[
\predcover \triangleq 1 - 2 \cdot \paren[\Bigg]{1-\Phi_\sigma\paren[\Big]{\Phi_1^{-1}(1 - \alpha/2)\cdot\sqrt{1-c \cdot \frac{\lamH}{n}}}}
\enspace,
\]
where $c$ is a constant that we will choose.
Our goal was then to test whether this predicted coverage rate matched the true coverage rate of our method, for some value of $c$.

% Figure 8: 
% for the  Medium Outliers, create the following subplots:
% (a) predicted coverage vs all coverage for DTE (b) predicted coverage vs all coverage for GATE 
\begin{figure}[H]
	\centering
	\begin{subfigure}{0.49\textwidth}
		\centering
		\includegraphics[width=\textwidth]{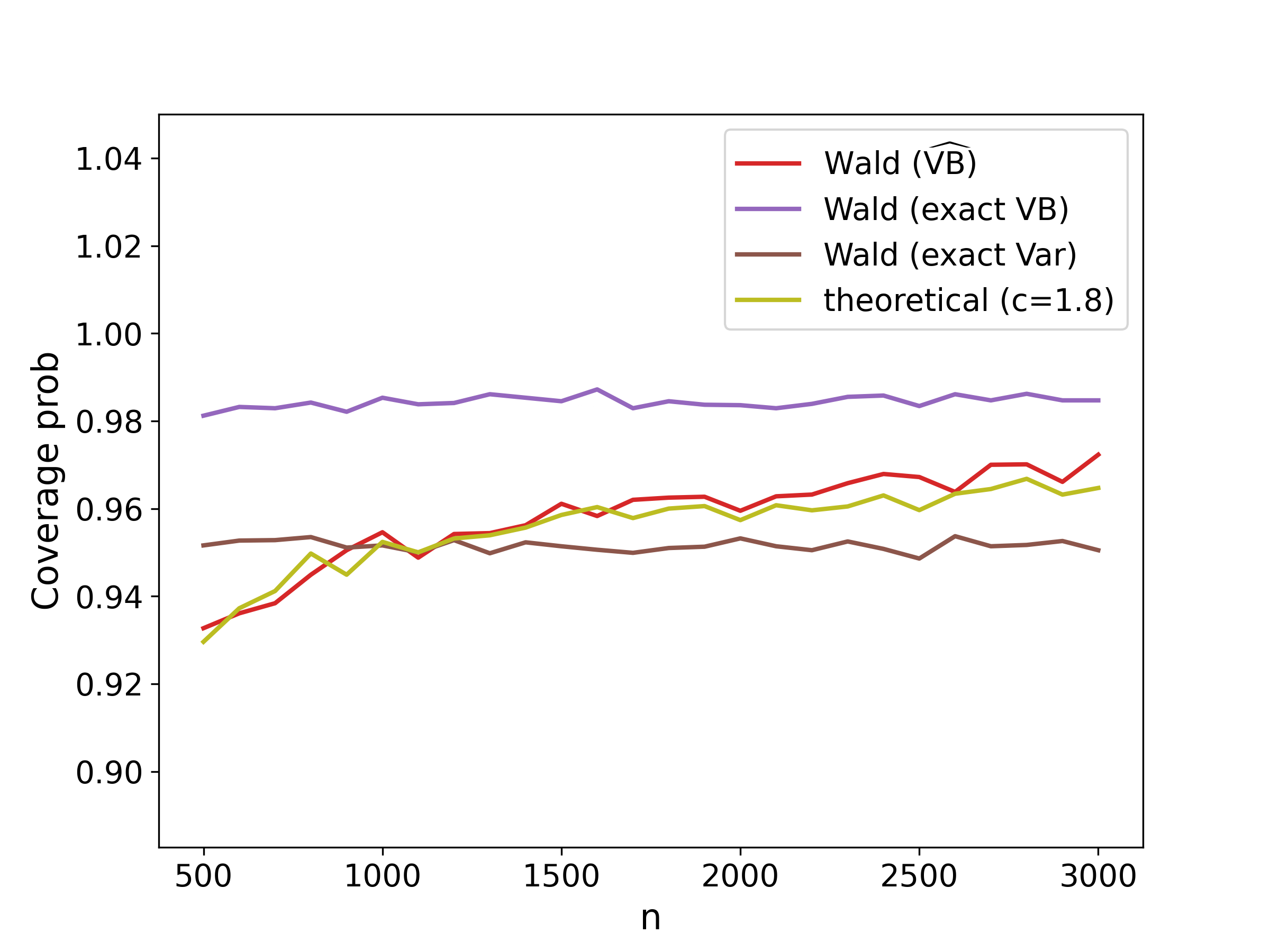}
		\caption{Direct Treatment Effect}
		\label{fig:dte-med-theorcover}
	\end{subfigure}%
	~ 
	\begin{subfigure}{0.49\textwidth}
		\centering
		\includegraphics[width=\textwidth]{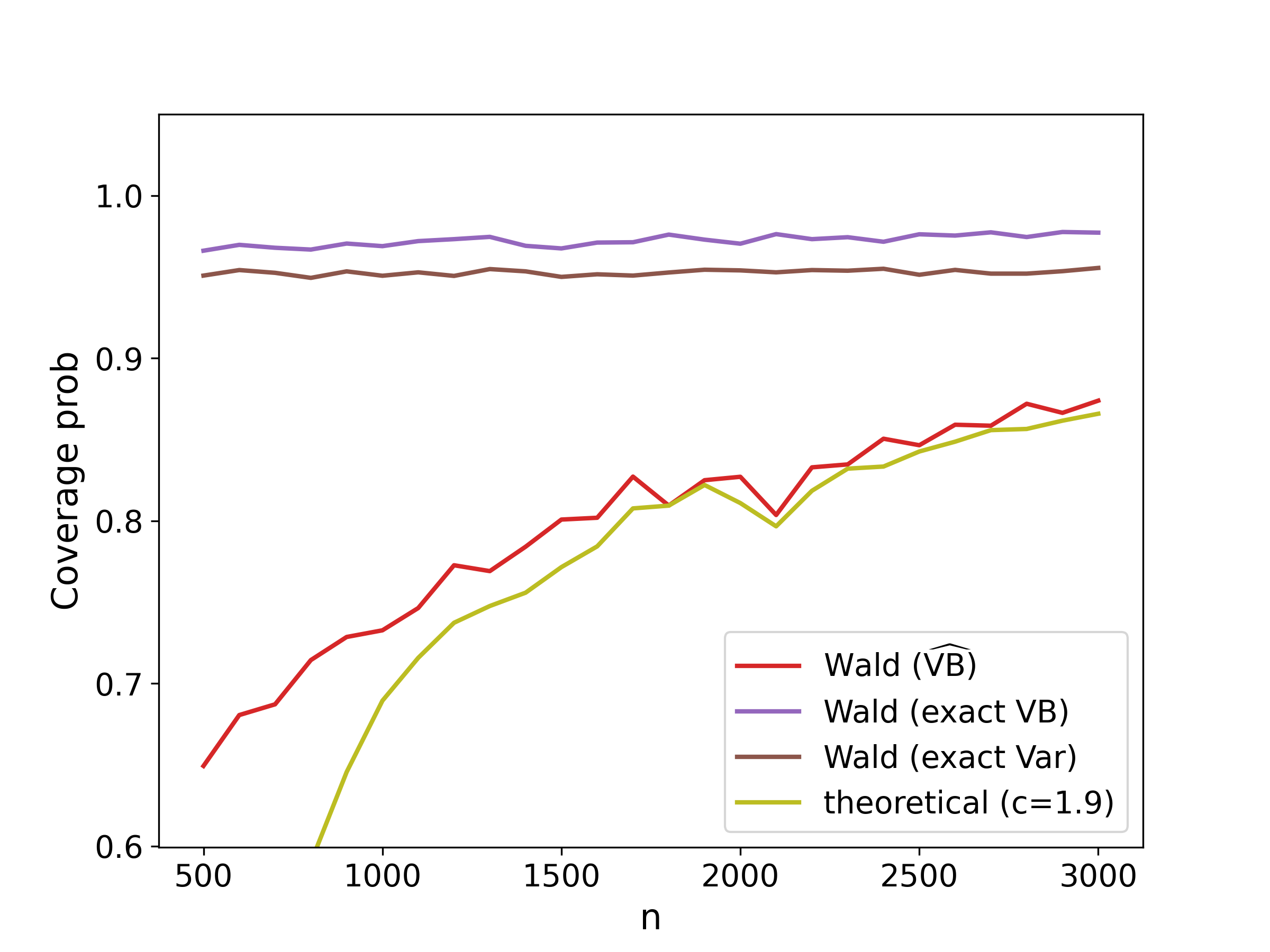}
		\caption{Global Average Treatment Effect}
		\label{fig:gate-med-theorcover}
	\end{subfigure}
	\caption{Comparison of theoretically predicted coverage with true coverage in the Medium Outliers setting.}
	\label{fig:med-theorcover}
\end{figure}

Figure~\ref{fig:med-theorcover} plots the predicted coverage against the true coverage of the Wald intervals for the setting of Medium Outliers in both effects.
We see in both figures that the predicted coverage probability matches the true coverage probability of the Wald intervals quite well, for the chosen constants.
Figure~\ref{fig:dte-med-theorcover} confirms that the coverage for the Direct Treatment Effect is at the nominal levels.
At the same time, Figure~\ref{fig:gate-med-theorcover} confirms that the coverage for the Global Average Treatment Effect is not yet at the nominal levels.

So what explains this difference in coverage between the two effects considered in the simulation?
In short, it is due to the difference in the rate of convergence of the variance estimators.
This difference is driven by the difference in the rates $\sqrt{\lamH / n}$ between the two settings.
Recall that the Conflict Graph for the direct effect is $\cH = G$ while the Conflict Graph for the Global Average Treatment Effect is the two-hop graph, i.e. $\cH = G^2$.

% Figure 9:
% plot sqrt(lam / n) for the DTE and GATE
% if the plot isn't clear, then plot on a log - linear scale (i.e. y axis is log, x axis is linear)
% xlabel: n
% ylabel sqrt(lam / n), but in latex
\begin{figure}[H]
	\centering
	\includegraphics[width=0.6\textwidth]{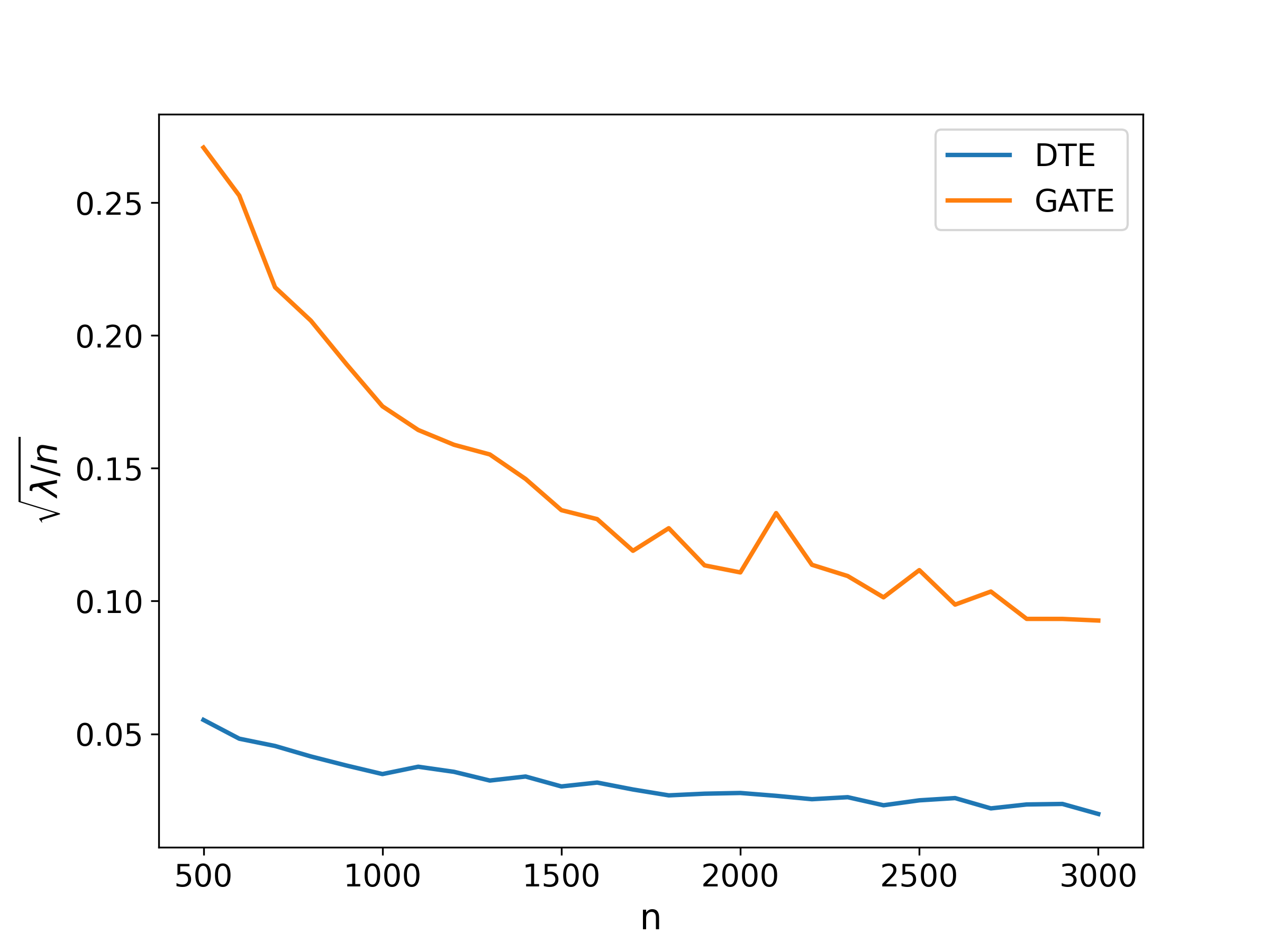}
	\caption{Comparison of $\lamH / n$ rates between the graphs used for DTE and GATE estimation.}
	\label{fig:lamn-rate-comparison}
\end{figure}

Figure~\ref{fig:lamn-rate-comparison} demonstrates this difference in the rates.
Because $\lamH / n$ converges to zero much faster for the direct treatment effect, the asymptotics kick in faster than for the Global Average Treatment Effect.
This example highlights that more global effects are indeed more difficult to estimate than more local effects.
This should be taken into account when conducting power analyses before running an experiment.

% Figure 10:
% create the following subplots:
% (a) variance comparison for Medium Outliers, GATE
\begin{figure}[H]
	\centering
	\includegraphics[width=0.6\textwidth]{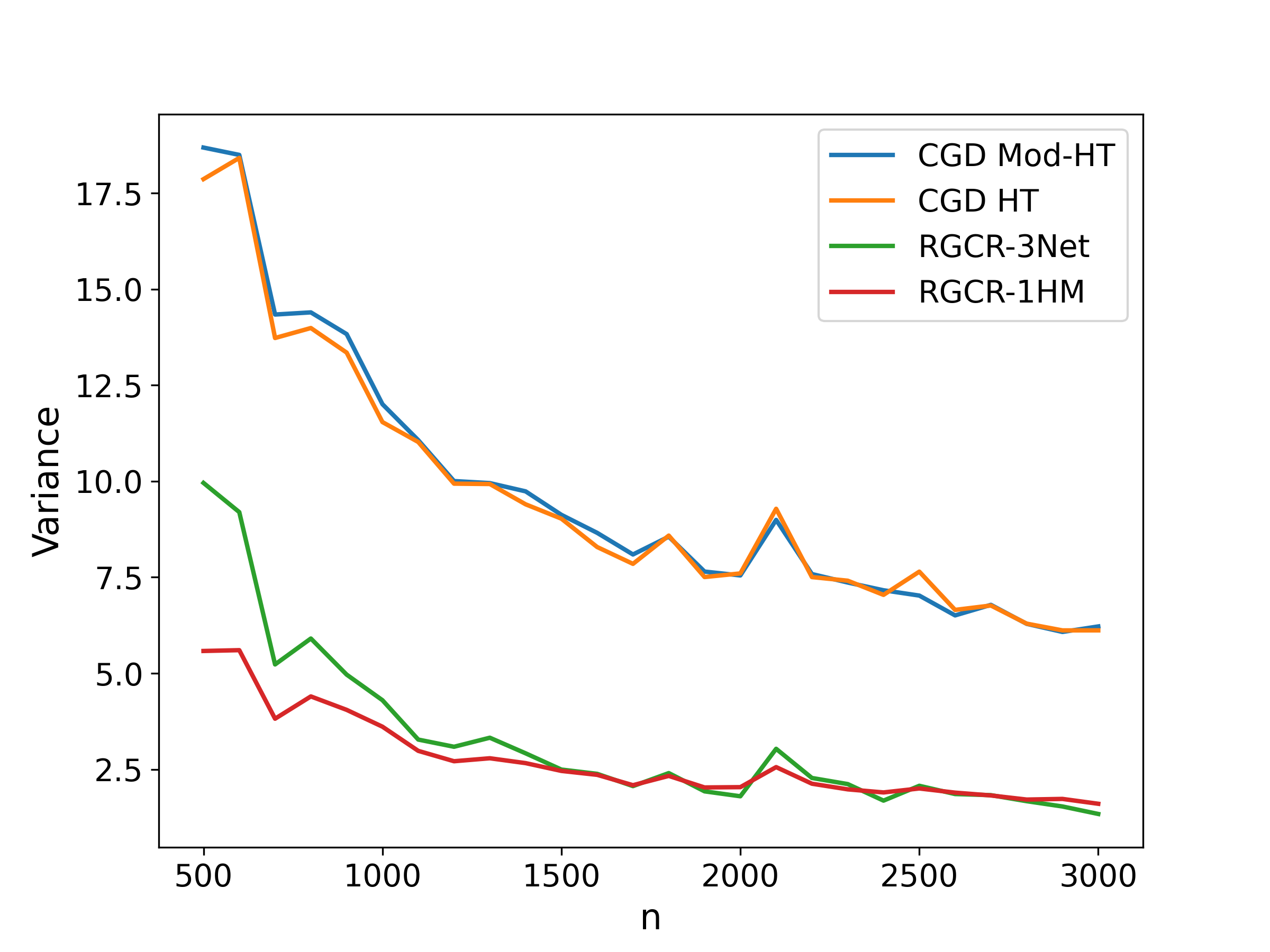}
	\caption{Comparison of the variance of estimating GATE between the various experimental designs in the Medium Outliers setting.}
	\label{fig:gate-med-variances}
\end{figure}

Figure~\ref{fig:gate-med-variances} shows that in the presence of fewer outliers, the RGCR designs continue to achieve smaller variance than the Conflict Graph Design.

\subsection{Practical Considerations}

The simulation exercise reveals several practical considerations.
First, that although the Conflict Graph Design achieves improved rates of convergence, it is always worthwhile to benchmark against alternative designs.
Second, that the performance of a design against \emph{worst-case} potential outcomes satisfying bounded second moment may not reflect the performance of a design against \emph{typical} potential outcomes.
As always, pre-experiment simulation studies are recommended.

From a methodological perspective, we see that the biggest benefit may come from devising new inferential methods that work in the presence of larger outliers.
This is especially important when the outliers may also have high degrees in the graph.
It remains an open question to offer a more systematic treatment of inference in these high-degree regimes.
Until then, coverage is typically guaranteed by Chebyshev interval and this may be a safer bet for many practitioners. 
Moreover, the theoretical approximations for coverage probabilities $\predcover$ derived above may be useful for power calculations until more rigorous methods are made available.
	
	\section{Conflict Graph Design: Preliminary Lemmas}

In this Section, we collect the proofs of some preliminary lemmas about the Conflict Graph Design and the modified Horvitz-Thompson estimator. 
Let's start by introducing some notation that will be helpful for the remainder of this Appendix. For a given conflict graph $\mathcal{H}$ and importance ordering $\pi$, we say that unit $i$ is a \emph{parent} of unit $j$, or $j$ is a \emph{child} of $i$, if and only if $i \in \mathcal{N}_b^{\pi}(j)$. 

\subsection{Desired Exposures (Lemma~\mainref{lem:desired_exposures})}
We start by proving Lemma~\mainref{lem:desired_exposures}, which says that if event $E_{(i,k)}$ holds for a unit $i$, then it will receive the desired exposure in the final treatment assignment. 

\begin{reflemma}{\mainref{lem:desired_exposures}}
	\desiredexposures
\end{reflemma}

\begin{proof}

Suppose $E_{(i,k)}$ occurs. Then, the condition in Line~\mainref{algline:exposure-condition} will be satisfied for a unit $i$. 
Then, Line~\mainref{algline:set-intervention} will be executed, which will cause the algorithm to assign individual assignments to unit $i$ and its neighbors so that it receives the desired exposure. In other words, at the end of the iteration corresponding to $i$, we have $\forall j \in \widetilde{N}(i): Z_j = \contrastz{i}{k}(j)$. 

We claim that the algorithm will not ``over-write'' these individual assignments later in the execution.
To see this, first observe that only neighbors of $i$ in the conflict graph could potentially cause such a change.
Indeed, if a unit $r$ is not a neighbor of $i$ in the conflict graph, then if Line~\mainref{algline:set-intervention} was executed for $r$, it would set $\forall j \in \widetilde{N}(r): Z_j = \contrastz{r}{k}(j)$. Since $r$ and $j$ are not neighbors in the conflict graph, by definition this means that no entries of the neighborhood $\widetilde{N}(i)$ of $i$ would be overwritten by a different value.

Let us now focus on a node $r \in \mathcal{N}(i)$. If $r$ is more important than $i$, meaning $r \in \mathcal{N}_b^\pi(i)$, then
 event $E_{(i,k)}$ ensures that Line~\mainref{algline:set-intervention} will not be executed for $r$, since $U_r = *$. Thus, it will not affect the neighborhood of $i$. 
If $r$ is less important than $i$ in $\pi$, meaning $i \in \mathcal{N}_b^\pi(r)$, then 
again event $E_{(i,k)}$ ensures that Line~\mainref{algline:set-intervention} will not be executed for $r$, since $U_i \neq *$. Thus, it will also not affect the neighborhood of $i$. 
Intuitively, less important units ``yield'' to the desires of their more important neighbors. 

Overall, this shows that in the final treatment vector $Z$ the assignment in $\widetilde{N}(i)$ will not be modified after $i$. Thus, we still have $\forall j \in \widetilde{N}(i): Z_j = \contrastz{i}{k}(j)$ and the proof is complete. 
\end{proof}

\subsection{Expressions for Desired Exposure Probabilities (Lemma~\mainref{lemma:prob-of-desired-exposure-events})}

In this Section, we provide simple expressions for the desired exposure probabilities $\Pr{E_{(i,k)}}$ used in the effect estimator.
These closed form expressions will help us bounding the variance of the effect estimator.
Because the desired exposure probabilities have these simple closed forms, the modified Horvitz--Thompson estimator does not require any computationally expensive procedure to compute these probabilities, e.g. Monte Carlo simulations. 

\begin{reflemma}{\mainref{lemma:prob-of-desired-exposure-events}}
	\probexposure
\end{reflemma}

\begin{proof}
    We start by computing the probability of event $E_{(i,k)}$. By definition, and since the $U_i$ variables are independent, we have
     \begin{align*}
     \Pr{E_{(i,k)}} &= \Pr[\Big]{U_i = e_k, U_j = * \forall j \in \mathcal{N}_b^\pi (i)} \\
     &= \Pr[\Big]{U_i = e_k} \prod_{j \in \mathcal{N}_b^\pi(i)} \Pr[\Big]{U_j = *}\\
     &= \frac{1}{2 r \lamH} \paren[\Big]{1 - \frac{1}{r\lamH}}^{|\mathcal{N}_b^\pi (i)|}\,,
     \end{align*}
    which proves the first claim. 
    
    For a pair of units $i,j$ that are connected by an edge in $\mathcal{H}$, if $E_{(i,k)}, E_{(j,l)}$ both hold, then $U_i \neq *, U_j \neq *$. As we already observed in the proof of Lemma~\mainref{lem:desired_exposures}, either $i \in \mathcal{N}_b(j)$ or $j \in \mathcal{N}_b(i)$ occur.
    Suppose $i \in \mathcal{N}_b(j)$ without loss of generality. Then, $E_{(j,l)}$ dictates that $U_i = *$ since $i$ is a parent of $j$, which is a contradiction. Thus, $E_{(i,k)} \cap E_{(j,l)} = \emptyset$. 
    
    Finally, suppose $i$ and $j$ are not adjacent in $\mathcal{H}$. 
    Then, we can write
    \begin{align*}
        \Pr[\Big]{E_{(i,k)} \cap E_{(j,l)}} &= \Pr[\Big]{U_i = e_k , U_j = e_l, U_r = * \forall r \in \mathcal{N}_b^\pi(i), U_s = * \forall s \in \mathcal{N}_b^\pi(j)} \\
        &= \Pr[\Big]{U_i = e_k , U_j = e_l, U_r = * \forall r \in \mathcal{N}_b^\pi(i) \cup \mathcal{N}_b^\pi(j)}\\
        &= \Pr[\Big]{U_i = e_k} \Pr[\Big]{U_j = e_l} \prod_{r \in \mathcal{N}_b^\pi(i) \cup \mathcal{N}_b^\pi(j)} \Pr[\Big]{U_r = *}\\
        &= \paren[\Big]{ \frac{1}{2 r \lamH} }^2 \cdot 
        \paren[\Big]{ 1 - \frac{1}{r \lamH} }^{\abs{ \mathcal{N}_b^\pi(i) \cup \mathcal{N}_b^\pi(j)}}  
        \enspace.
    \end{align*}
    
    \end{proof}

\subsection{Effect Estimator is Unbiased (Proposition~\mainref{prop:unbiased-estimator})}

Next, using Lemma~\mainref{lem:desired_exposures}, we prove that our estimator is unbiased.

\begin{refproposition}{\mainref{prop:unbiased-estimator}}
	\unbiasedlemma
\end{refproposition}

\begin{proof}
    By Lemma~\mainref{lem:desired_exposures}, we know that under the event $E_{(i,k)}$, then $d_i(Z) = d_i(\contrastz{i}{k})$.
    By Assumption~\mainref{assumption:ani-model}, this implies that $Y_i = y_i(Z) = y_i(\contrastz{i}{k})$.
    Thus, we always have the identity $Y_i \indicator{E_{(i,k)}} = y_i(\contrastz{i}{k})\indicator{E_{(i,k)}}$ for $k \in \{0,1\}$.
    Furthermore, by Lemma~\mainref{lemma:prob-of-desired-exposure-events} we have $\Pr{E_{(i,k)}} > 0$, for all $i \in [n]$ and $k \in \{0,1\}$.
    The above observations allow us to write
    \begin{align*}
        \E{\hat{\tau}} &= \frac{1}{n} \sum_{i= 1}^n \E[\Bigg]{Y_i \lp( \frac{\indicator{E_{(i,1)}}}{\Pr{E_{(i,1)}}} - \frac{\indicator{E_{(i,0)}}}{\Pr{E_{(i,0)}}} \rp)}\\
        &= \frac{1}{n} \sum_{i= 1}^n \lp( y_i(\contrastz{i}{1})  \frac{\E[\big]{\indicator{E_{(i,1)}}}}{\Pr{E_{(i,1)}}} - y_i(\contrastz{i}{0}) \frac{\E[\big]{\indicator{E_{(i,0)}}}}{\Pr{E_{(i,0)}}} \rp) \\
        &= \tau
    \end{align*}
\end{proof}

\subsection{Eigenvector Importance Ordering (Proposition~\mainref{prop:eigenvector-ordering-is-important})}

In this section, we prove that the eigenvector ordering $\eigpi$ defined in Section~\mainref{sec:importance-ordering} is an importance ordering.
For continuity, the proof is stated in terms of conflict graphs; however, the proof works for arbitrary graphs more generally.

\begin{refproposition}{\mainref{prop:eigenvector-ordering-is-important}}
	\eigenordering
\end{refproposition}

\begin{proof}
	
	Assume that the conflict graph $\cH$ is connected.
	We will see at the end of the proof how the general graphs can be handled in terms of connected components.
	Because $\cH$ is connected, the Perron Frobenius theorem states the eigenvector $\ev$ corresponding to the largest eigenvalue can be chosen to have positive entries, i.e. $\ev(i) > 0$ for all $i \in [n]$.
	
	Recall that $\eigpi$ is an importance ordering if 
	\[
	\abs{\mathcal{N}_b^{\eigpi}(i)} \leq \lamH
	\quadwhere
	\mathcal{N}_b^{\eigpi}(i) = \setb[\Big]{ j \in \mathcal{N}(i) : \eigpi(j) < \eigpi(j) }
	\enspace.
	\]
	Let us also define the neighbors which come after unit $i$ in the ordering as
	\[
	\mathcal{N}_a^{\eigpi}(i) = \setb[\Big]{ j \in \mathcal{N}(i) : \eigpi(j) > \eigpi(j) }
	\enspace,
	\]
	which are the children of unit $i \in [n]$.
	Let us now suppress the dependence on $\eigpi$ for notational convenience.
	Using the definition of the eigenvector $\ev(i)$ and the fact that each vertex has a self-loop, we have that
	\begin{equation}
	\lamH \ev(i) 
	= \ev(i) + \sum_{j \in \mathcal{N}(i)} \ev(j) 
	= \ev(i) + \sum_{j \in \mathcal{N}_b(i)} \ev(j) + \sum_{j \in \mathcal{N}_a(i)} \ev(j) 
	\geq \ev(i) + \abs[\big]{ \mathcal{N}_b(i) } \ev(i) \label{eq:before_neighbors}
	\end{equation}
	The last inequality follows since by definition $\ev(j) \geq \ev(i)$ for all $j \in \mathcal{N}_b(i)$.
	From the assumption that $\cH$ is connected, we have that $\ev(i) > 0$ for all $i$ so that we can divide both sides of \eqref{eq:before_neighbors} with $\ev(i)$.
	Rearranging terms in this way yields that 
	\[
	\abs[\big]{ \mathcal{N}_b(i) } \leq \lamH - 1  \enspace.
	\]
	This establishes that if $\cH$ is connected, then the eigenvector ordering $\eigpi$ is an importance ordering.

	Now, suppose $\cH$ is not connected.
	Let $\cH_1,\ldots, \cH_k$ be the connected components, with corresponding eigenvector orderings $\eigpi^{(1)}, \ldots, \eigpi^{(k)}$.
	Each $\cH_i$ contains $s_i$ nodes.
	In that case, the eigenvector ordering $\eigpi$ for the entire graph $\cH$ is defined as the ``concatenation'' of these orderings, i.e.
	\[
	\eigpi (i) = \sum_{j=1}^{r_{i-1}}s_j + \eigpi^{(r_i)}(i), \text{ where } i \in \cH_{r_i}
	\]
	We have already shown that each $\eigpi^{(i)}$ is an importance ordering for the corresponding $\cH_i$.
	Since there are no edges between components, the defining property of importance ordering is not affected for each component.
	In other words, for each node $i$, $\mathcal{N}_b(i)$ remains the same in $\eigpi$ and $\eigpi^{(r_i)}$, where $r_i$ is the component that contains $i$.
	This concludes the claim in the general case.
\end{proof}

\subsection{Sequential Degree Importance Ordering}\label{sec:sequential-degree-ordering-appendix}

We present a second construction of an importance ordering based on sequentially removing minimum degree vertices in the conflict graph.
We refer to this construction as the \emph{sequential degree ordering} $\degpi$ and the formal definition appears in Algorithm~\ref{alg:degree-ordering}.
The ordering is constructed in reverse order over the iterations of the algorithm.
In the first iteration, a vertex of minimum degree in $\cH$ is placed last in the ordering.
We then remove this minimum degree vertex and its edges from $\cH$.
This process repeats so that the vertex of minimum degree is continually added to the lowest available position in the ordering.
Even though this was not stated in terms of importance orderings, the key ideas for the proof can be found in \citep{Wilf1967Eigenvalues}. We give the details for completeness.
\begin{algorithm}
	\DontPrintSemicolon
	\caption{\textsc{Sequential-Degree-Ordering}}\label{alg:degree-ordering}
	\KwIn{Conflict graph $\cH$.}
	\KwOut{Sequential degree ordering $\degpi$}
	Initialize graph $\cH^{(1)} = \cH$\;
	Initialize empty ordering $\degpi(i) = \emptyset$ for all $i \in [n]$\;
	\For{$i =1 \dots n$}{
		Let $j$ be a minimum degree node in $\cH^{(i)}$, breaking ties arbitrarily. \;
		Update ordering $\degpi(j) = n-i+1$\;
		Construct $\cH^{(i+1)}$ by removing node $j$ and its edges from $\cH^{(i)}$.\;
	}
\end{algorithm}

\begin{proposition}[\citet{Wilf1967Eigenvalues}] \label{prop:sequential-degree-ordering-is-important}
	The sequential degree ordering $\degpi$ is an importance ordering.
\end{proposition}
\begin{proof}
    For each $i \in [n]$, we know that it is chosen in order $\degpi(i)$. This means that at the time $i$ is selected by the algorithm, there are $n - \degpi(i)$ nodes that have already been selected as less important. Let $A_i$ denote this set of units, and $B_i = [n] \setminus A_i \setminus \{i\}$. To prove that $\degpi$ is an importance ordering, we need to argue that $i$ has at most $\lamH-1$ neighbors in $B_i$. This exactly corresponds to the degree of $i$ in the induced subgraph $\cH^{n -\degpi(i)+1}$. By definition of Algorithm~\ref{alg:degree-ordering}, $i$ is the node of minimum degree in the induced subgraph $\cH^{n -\degpi(i)+1}$. We know that the minimum degree in $\cH^{n -\degpi(i)+1}$ is at most $\lamH^{(i)}-1$, the maximum eigenvalue of the adjacency matrix of the graph $\cH^{n -\degpi(i)+1}$(we have to subtract $1$ because of self-loops). Since this adjacency matrix is a principal submatrix of $\mat{A}_\cH$, by restricting only to rows and columns that lie in $B_i \cup \{i\}$, by the Cauchy Interlacing Theorem we have that $\lamH^{(i)} \leq \lamH$. We conclude that the degree of $i$ in $\cH^{n -\degpi(i)+1}$ is at most $\lamH-1$, which completes the proof. 
\end{proof}

The notion of importance ordering has some interesting connections to techniques from Spectral Graph Theory. In particular, as we observed, the fact that 
$\degpi$ is an importance ordering essentially follows from the proof of Wilf's Theorem \citep{Wilf1967Eigenvalues}, which states that the chromatic number of a graph $G$ is at most $\lambda(G)+1$, where as usual $\lambda(G)$ is the maximum eigenvalue of the adjacency matrix. Indeed, an importance ordering can be used to construct a coloring of the graph. In \citet{Wilf1967Eigenvalues}, a coloring is constructed by iterating over all nodes in the order dictated by $\degpi$, starting from the most important one. For each node $i$, there are at most $\lambda(G)$ neighbors that are more important, which have already been colored. Thus, there is always at least one color that $i$ can be assigned to, without violating the validity of the coloring.
In this way, the proof that the eigenvector ordering $\eigpi$ is an importance ordering provides an alternative proof of Wilf's theorem, which may be of independent interest.

	\section{Analysis of Effect Estimator}

In this section, we will prove Theorem~\mainref{thm:variance-analysis-finite-sample} which bounds the finite sample variance of the effect estimator under the Conflict Graph Design.
We restate the theorem below for convenience. 

\begin{reftheorem}{\mainref{thm:variance-analysis-finite-sample}}
	\varianceanalysisfinitesample
\end{reftheorem}

\noindent 
We also restate the Conflict Graph Design here for reference. 

% re-state for arXiv, otherwise just write again for journal
\ifnum \value{spacesave}=1 {
	\begin{algorithm}
		\DontPrintSemicolon
		\caption{\ourdesign{}}
		\cgdalgobody
	\end{algorithm}
} \else {
	\cgdalgo*
} \fi

We first present some important Lemmas that will be useful for the proof of Theorem~\mainref{thm:variance-analysis-finite-sample}. Finally, in Section~\suppref{sec:finite-sample-proof} we present the proof of the statement. 

\subsection{Expression for Covariance Terms}

It will be convenient for the analysis of $\Var{\eate}$ to obtain closed form expressions for the terms that appear in the variance. 
This is done in the following Lemma.

\begin{lemma}\label{lem:Cij_expression}
	Suppose we run the Conflict Graph Design with some importance ordering $\pi$ and constant $r > 0$ with $r \lamH > 1$. 
	For each unit $i \in [n]$ and contrastive exposure $k \in\setb{0,1}$, the variance of each probability weighted indicator can be expressed as
	\[
	\Var[\Big]{ \frac{ \indicator{ E_{(i,k)} } }{ \Pr{ E_{(i,k)} } } } = \frac{2r \lamH}{\paren[\Big]{1 - \frac{1}{r \lamH}}^{|\mathcal{N}_b^\pi (i)|}} -1 \enspace.
	\]
	Furthermore, for any unit $i$ we have 
	\[
	\Cov[\Big]{ \frac{ \indicator{ E_{(i,0)} } }{ \Pr{ E_{(i,0)} } }  , \frac{ \indicator{ E_{(i,1)} } }{ \Pr{ E_{(i,1)} } }  } = -1\enspace.
	\] 
	For each pair of distinct units $i \neq j \in [n]$ and contrastive exposure $k, \ell \in \setb{0,1}$, the covariance between probability weighted indicators may be written as:
	\begin{align*}
		\Cov[\Big]{ \frac{ \indicator{ E_{(i,k)} } }{ \Pr{ E_{(i,k)} } }, \frac{\indicator{ E_{(j,\ell)} } }{ \Pr{ E_{(j,\ell)} } } } 
		&= \left\{
		\begin{array}{lr}
			-1 & \text{ if } d_{\cH}(i,j) = 1\\
			\paren[\Big]{1 - \frac{1}{r \lamH}}^{- |\mathcal{N}_b^\pi(i) \cap \mathcal{N}_b^\pi(j)|} - 1 & \text{ if } d_{\cH}(i,j) \geq 2
		\end{array}
		\right.
	\end{align*}
\end{lemma}
\begin{proof}
The proof follows immediately from the expressions of the desired exposure probabilities provided in Lemma~\mainref{lemma:prob-of-desired-exposure-events}. In particular, for every unit $i \in [n]$, 
\[
	\Var[\Big]{ \frac{ \indicator{ E_{(i,k)} } }{ \Pr{ E_{(i,k)} } } }  
	= \frac{1}{\Pr{ E_{(i,k)} }} - 1
	= \frac{2r \lamH}{\paren[\Big]{1 - \frac{1}{r \lamH}}^{|\mathcal{N}_b^\pi (i)|}}-1
	\enspace.
\]
Since $E_{(i,0)} \cap E_{(i,1)} = \emptyset$, we have that
\[
C_{(i,0),(i,1)} 
= \Cov[\Big]{ \frac{ \indicator{ E_{(i,0)} } }{ \Pr{ E_{(i,0)} } }  , \frac{ \indicator{ E_{(i,1)} } }{ \Pr{ E_{(i,1)} } }  }
= - \E[\Big]{\frac{ \indicator{ E_{(i,0)} } }{ \Pr{ E_{(i,0)} } } }\E[\Big]{\frac{ \indicator{ E_{(i,1)} } }{ \Pr{ E_{(i,1)} } } } 
= -1
\enspace.
\]
Finally, let $i \neq j$ be distinct units. If $d_{\mathcal{H}}(i,j) = 1$, then $i$ and $j$ are adjacent in $\mathcal{H}$, which means by Lemma~\mainref{lemma:prob-of-desired-exposure-events} that
\[
\Cov[\Big]{ \frac{ \indicator{ E_{(i,k)} } }{ \Pr{ E_{(i,k)} } }  , \frac{ \indicator{ E_{(j,\ell)} } }{ \Pr{ E_{(j,\ell)} } }  }
 = - \E[\Big]{\frac{ \indicator{ E_{(i,k)} } }{ \Pr{ E_{(i,k)} } } }\E[\Big]{\frac{ \indicator{ E_{(j,\ell)} } }{ \Pr{ E_{(j,\ell)} } } } 
 = -1
 \enspace.
\]
If $d_{\mathcal{H}}(i,j) \geq 2$, then again by Lemma~\mainref{lemma:prob-of-desired-exposure-events}
\begin{align*}
 \Cov[\Big]{ \frac{ \indicator{ E_{(i,k)} } }{ \Pr{ E_{(i,k)} } }  , \frac{ \indicator{ E_{(j,\ell)} } }{ \Pr{ E_{(j,\ell)} } }  } 
 &= \frac{\Pr{E_{(i,k)} \cap E_{(j,\ell)}}}{\Pr{E_{(i,k)}} \Pr{E_{(j,\ell)}}} - 1 \\
&= \frac{\paren[\Big]{ \frac{1}{2 r \lamH} }^2 \cdot 
\paren[\Big]{ 1 - \frac{1}{r \lamH} }^{\abs{ \mathcal{N}_b^\pi(i) \cup \mathcal{N}_b^\pi(j)}}}{\paren[\Big]{ \frac{1}{2 r \lamH} }^2 \cdot 
\paren[\Big]{ 1 - \frac{1}{r \lamH} }^{\abs{ \mathcal{N}_b^\pi(i)} + \abs{\mathcal{N}_b^\pi(j)}}} - 1  \\
&= \paren[\Big]{1 - \frac{1}{r \lamH}}^{- |\mathcal{N}_b^\pi(i) \cap \mathcal{N}_b^\pi(j)|} - 1
\enspace. \qedhere
\end{align*}

\end{proof}

Lemma~\suppref{lem:Cij_expression} shows that the covariance between units depends on their distance in the conflict graph $\cH$. This will prove useful when bounding the variance, as we will be able to group correlations between nodes according to their distance. 

\subsection{Bounding the Covariance Terms (Lemma~\mainref{lemma:individual-covariance-terms})}

An important step when bounding $\Var{\eate}$ will be to obtain precise bounds on the covariance terms that appear. An important feature of the Conflict Graph Design is that we can provide such bounds, which will depend on the relative topology of $i$ and $j$ in the conflict graph. 
Indeed, Lemma~\mainref{lemma:individual-covariance-terms} provides a characterization of the covariance terms. 
Here, we prove a stronger version of Lemma~\mainref{lemma:individual-covariance-terms} with better constants that depend on the parameter $r$ as well. The proof follows by using the expression for the covariance terms provided by Lemma~\suppref{lem:Cij_expression}, together with several linearization arguments.

\begin{reflemma}{\mainref{lemma:individual-covariance-terms}*}
	Suppose we run the Conflict Graph Design with some importance ordering $\pi$ and constant $r > 1$. 
	For each unit $i \in [n]$ and contrastive exposure $k \in\setb{0,1}$, the variance of each probability weighted indicator is bounded as
	\[
	\Var[\Big]{ \frac{ \indicator{ E_{(i,k)} } }{ \Pr{ E_{(i,k)} } } } \leq \constvar \lamH \enspace.
	\]
	For each pair of distinct units $i \neq j \in [n]$ and contrastive exposure $k, \ell \in \setb{0,1}$, the covariance between probability weighted indicators may be characterized according to their distance in $\cH$:
	\begin{align*}
		\Cov[\Big]{ \frac{ \indicator{ E_{(i,k)} } }{ \Pr{ E_{(i,k)} } }, \frac{\indicator{ E_{(j,\ell)} } }{ \Pr{ E_{(j,\ell)} } } } 
		&= \left\{
		\begin{array}{lr}
			-1 & \text{ if } d_{\cH}(i,j) = 1\\
			0 & \text{ if } d_{\cH}(i,j) \geq 3
		\end{array}
		\right.\\
		\abs[\Bigg]{\Cov[\Big]{ \frac{ \indicator{ E_{(i,k)} } }{ \Pr{ E_{(i,k)} } }, \frac{\indicator{ E_{(j,\ell)} } }{ \Pr{ E_{(j,\ell)} } } } }
		&\leq \constcovar \cdot \frac{ \abs[\big]{ \mathcal{N}(i) \cap \mathcal{N}(j) } }{\lamH}
		\quad \text{for } d_{\cH}(i,j) = 2 \enspace.
	\end{align*}
\end{reflemma}

\begin{proof}
	For event $E_{(i,k)}$, by the expression from Lemma~\mainref{lemma:prob-of-desired-exposure-events} we have that
	\[
		\Var[\Big]{ \frac{ \indicator{ E_{(i,k)} } }{ \Pr{ E_{(i,k)} } } }  
		= \frac{1}{\Pr{ E_{(i,k)} }} - 1
		\leq \frac{2r \lamH}{\paren[\Big]{1 - \frac{1}{r \lamH}}^{|\mathcal{N}_b^\pi (i)|}}
		\leq \frac{2r \lamH}{\paren[\Big]{1 - \frac{1}{r \lamH}}^{\lamH}}
		\enspace.
	\]
	The last inequality follows since $\pi$ is an importance ordering and thus $|\mathcal{N}_b^\pi (i)|\leq \lamH - 1 < \lamH$. 
	We distinguish cases according to whether the conflict graph has a non-self edge or not. 
	If it doesn't, then the matrix $\adjH$ is the identity and $\lamH = 1$. Then, the variance bound takes the form
	\[
	\frac{2r}{1 - \frac{1}{r}} \lamH \leq \constvar \lamH
	\enspace.
	\]
	If $\cH$ contains at least one non-self edge, then clearly $\lamH \geq 2$. 
	In that case, we show that the following function is decreasing with $\lambda$. 
	\[
	f(\lambda) = \paren[\Big]{1 - \frac{1}{r \lambda}}^{-\lambda+1}\enspace.
	\]
	Since $\lamH \geq 2$, the first part of the Lemma will follow immediately. 
	To prove monotonicity, we examine the first derivative.
	\[
	f'(\lambda) = \paren[\Big]{- \ln \paren[\big]{1 - 1/(r\lambda)} - \frac{1}{r\lambda - 1}} f(\lambda) < 0 \enspace.
	\]
	The above follows from the inequality $- \ln (1-1/x) \leq 1/(x-1)$ for all $x > 1$. 

	Moving to the second part of the Lemma, for $i,j$ with $d_\mathcal{H}(i,j) =1$ the claim has already been established in Lemma~\suppref{lem:Cij_expression}. For $i,j$ with  $d_\mathcal{H}(i,j) \geq 2$, the claim is non-trivial only if $\cH$ contains at least one non-self edge. Thus, in the sequel we assume that is the case, which implies in particular that $\lamH \geq 2$. We would like to prove that for a suitable constant $C = C(r) >0$ 

	\begin{align*}
		\Cov[\Big]{ \frac{ \indicator{ E_{(i,k)} } }{ \Pr{ E_{(i,k)} } }, \frac{\indicator{ E_{(j,\ell)} } }{ \Pr{ E_{(j,\ell)} } } } =
		\paren[\Big]{1 - \frac{1}{r\lamH}}^{-|\mathcal{N}_b^\pi(i) \cap \mathcal{N}_b^\pi(j)|} - 1 \leq C(r) \frac{|\mathcal{N}_b^\pi(i) \cap \mathcal{N}_b^\pi(j)|}{\lamH}\,,
	\end{align*}
	where the first equality comes from Lemma~\suppref{lem:Cij_expression}. 
	Consider the function 
	\[
	g(x) = \paren[\Big]{1 - \frac{1}{r\lamH}}^{-x} - 1 - C(r) \frac{x}{\lamH} \enspace.
	\]
	We will show that for all $0 < x < \lamH$ we have $g(x) \leq 0$ for a suitable constant $C(r)$. Then the result would follow, since 
	$|\mathcal{N}_b^\pi(i) \cap \mathcal{N}_b^\pi(j)| \leq \lamH-1 < \lamH$ (since $\pi$ is an importance ordering). 
	Again by taking derivative with respect to $x$, we get
	\[
	g'(x) = -  \ln\paren[\Big]{1 - \frac{1}{r\lamH}} \paren[\Big]{1 - \frac{1}{r\lamH}}^{-x} - \frac{C(r)}{\lamH} \enspace.
	\]
	Since $x \leq \lamH$, in order to have $g(x) \leq 0$ for all $x$, it suffices to pick a constant $C(r)$ such that
	\[
	- \lamH \ln\paren[\Big]{1 - \frac{1}{r\lamH}} \paren[\Big]{1 - \frac{1}{r\lamH}}^{-\lamH} \leq C(r) \enspace.
	\]
	Indeed, the left hand side is bounded by a constant for all $\lamH \geq 2$, and it attains it's maximum value when $\lamH = 2$. Thus, it suffices to pick
	\[
	C(r) = \constcovar\enspace,
	\]
	and the proof is complete by noticing that $|\mathcal{N}_b^\pi(i) \cap \mathcal{N}_b^\pi(j)| \leq |\mathcal{N}(i) \cap \mathcal{N}(j)|$. 
\end{proof}

\subsection{Linear Algebra Preliminaries}

Our approach to bounding the variance relies on the operator norm approach.
As such, we will need to bound the largest eigenvalue of a block matrix in terms of the eigenvalues of the blocks.
We now state and prove two lemmas that allows us to do so.
Both will be useful once we break down the variance in different terms.

\begin{lemma}\label{lemma:operator_norm_diagonal}
	Let $\mat{D}^{(1)}, \mat{D}^{(2)}, \mat{D}^{(3)} \in \R^{n \times n}$ be diagonal matrices.
	Then, the operator norm of the block matrix
	\[
	\begin{pmatrix}
		\mat{D}^{(1)} &\mat{D}^{(3)}\\
		\mat{D}^{(3)} &\mat{D}^{(2)}
	\end{pmatrix}
	\]
	is at most $\max_{i \in [n]} ( \max(|\mat{D}^{(1)}_{ii}|, |\mat{D}^{(2)}_{ii}|)) + \max_{i \in [n]} |\mat{D}^{(3)}_{ii}|$. 
	\end{lemma}
	\begin{proof}
	We can write the block matrix as a sum
	\[
		\begin{pmatrix}
			\mat{D}^{(1)} &\mat{D}^{(3)}\\
			\mat{D}^{(3)} &\mat{D}^{(2)}
		\end{pmatrix} = 
		\begin{pmatrix}
			\mat{D}^{(1)} &0\\
			0 &\mat{D}^{(2)}
		\end{pmatrix} +
		\begin{pmatrix}
			0&\mat{D}^{(3)}\\
			\mat{D}^{(3)} &0
		\end{pmatrix}
	\]
	Clearly, the operator norm of the first matrix is $\max_{i \in [n]} ( \max(|\mat{D}^{(1)}_{ii}|, |\mat{D}^{(2)}_{ii}|))$ and of the second matrix $\max_{i \in [n]} |\mat{D}^{(3)}_{ii}|$. The result follows from the subadditivity of operator norm. 
	\end{proof}

	\begin{lemma}\label{lem:block_operator_norm}
		Suppose $\mat{S} \in \R^{n \times n}$ is a symmetric matrix and $\mat{S}' \in \R^{2n \times 2n}$ is the following block matrix
		\[
		\mat{S}' = \begin{pmatrix}
			\mat{S} &-\mat{S}\\
			-\mat{S} &\mat{S}
		\end{pmatrix}
		\]
		Then
		\[
		\norm{\mat{S}'} = 2 \norm{\mat{S}}
		\]
	\end{lemma}
	\begin{proof}
	Since $\mat{S}$ is symmetric, let $\lambda_1,\ldots, \lambda_n$ be the eigenvalues of $\mat{S}$ with corresponding eigenvectors $\vx_1,\ldots, \vx_n \in \Reals^n$. We can easily find the eigendecomposition of $\mat{S}'$ based on the one of $\mat{S}$. Indeed, for each $\vx_i$, consider the vector $\vy_i = (\vx_i,\vx_i) \in \Reals^{2n}$. We can check that $\vy_i$ is an eigenvector of $\mat{S}'$ with eigenvalue $0$. 
	Also, let's consider the vector $\vy'_i = (\vx_i, -\vx_i)$. Then, we can check that $\vy'_i$ is an eigenvector of $\mat{S}'$ with eigenvalue $2\lambda_i$. Also, all these vectors are pairwise orthogonal. Thus, the spectrum of $\mat{S}'$ is the spectrum of $\mat{S}$ multiplied by $2$ and with an added eigenvalue $0$ of multiplicity $n$. The result follows. 
	\end{proof}

\subsection{Finite Sample Variance Analysis (Theorem~\mainref{thm:variance-analysis-finite-sample})}\label{sec:finite-sample-proof}

We are now ready to prove Theorem~\mainref{thm:variance-analysis-finite-sample}, which constitutes the main result of the paper.

\begin{proof}[Proof of Theorem~\mainref{thm:variance-analysis-finite-sample}]

To begin our analysis, let us first decompose the modified Horvitz--Thompson estimator.
Observe that the estimator may be written as the average of individual treatment effect estimators, 
\[
\eate = \frac{1}{n} \sum_{i=1}^n \eate_i
\quadwhere
\eate_i = y_i(e_1) \cdot \frac{ \indicator{ E_{(i,1)} } }{ \Pr{ E_{(i,1)} } } 
- y_i(e_0) \cdot \frac{ \indicator{ E_{(i,0)} } }{ \Pr{ E_{(i,0)} } }
\enspace,
\]
where we have used the arbitrary neighborhood exposure assumption (Assumption~\mainref{assumption:ani-model}), as observed in the proof of Proposition~\mainref{prop:unbiased-estimator}, to write the estimator in terms of the (deterministic) potential outcomes $y_i(e_1)$ and $y_i(e_0)$.
The only random parts of the estimator are therefore the inverse probability weighted indicators.
Thus, the variance of the estimator may be decomposed as the sum of individual covariance terms of the individual inverse probability weighted indicators:
\begin{equation}\label{eq:var_expression}
n \cdot \Var{\eate} 
= \frac{1}{n} \sum_{i=1}^n \sum_{j=1}^n \sum_{\substack{k \in \setb{0,1} \\ \ell \in \setb{0,1}}} (-1)^{k - \ell} y_i(e_k) y_j(e_\ell) \Cov[\Bigg]{ \frac{ \indicator{ E_{(i,k)} } }{ \Pr{ E_{(i,k)} } }  , \frac{ \indicator{ E_{(j,\ell)} } }{ \Pr{ E_{(j,\ell)} } }  }
\enspace.
\end{equation}
In order to analyze the variance, we must characterize these covariance terms.
Throughout this section, we shall use the shorthand that for a pair of units $i,j \in [n]$ and contrastive exposures $k,\ell \in \setb{0,1}$, the covariance terms are denoted as  

$$
C_{(i,k),(j,\ell)} \triangleq \Cov[\Big]{ \frac{ \indicator{ E_{(i,k)} } }{ \Pr{ E_{(i,k)} } }  , \frac{ \indicator{ E_{(j,\ell)} } }{ \Pr{ E_{(j,\ell)} } }  }
\enspace.
$$

A crucial observation here is that expression \eqref{eq:var_expression} can be seen as a quadratic form of the potential outcomes. 
In particular, let us define $\vec{\alpha} \in \R^{2n}$ to be the vector that contains all $2n$ relevant potential outcomes that determine the effect $\tau$, as follows:
\begin{equation}
    \vec{\alpha}_{i} = \begin{cases}
        \frac{1}{\sqrt{n}}y_i(e_0) & \text{if } 0 \leq i \leq n, \\
        \frac{1}{\sqrt{n}}y_{i-n}(e_1)   & \text{if } n+1 \leq i \leq 2n.
      \end{cases}
\end{equation}
We define matrices $\varM^{(kl)} \in \R^{n \times n}$ for $k,l\in \{0,1\}$ as 
\begin{equation}\label{eq:Rkl}
    \varM_{ij}^{(kl)} = (-1)^{k+l} C_{(i,k),(j,l)}
    \enspace.
\end{equation}
Then, \eqref{eq:var_expression} can also be written in the following matrix form:
\begin{equation}\label{eq:var_matrix}
\Var{\tau} = \frac{1}{n} \vec{\alpha}^\top \underbrace{\begin{pmatrix}
    \varM^{(00)} &\varM^{(01)}\\
    \varM^{(10)} &\varM^{(11)}
\end{pmatrix}}_{\varM} \vec{\alpha}
\enspace.
\end{equation}
 
\noindent
We now notice that 
\[
\norm{\vec{\alpha}}_2^2 = \frac{1}{n} \sum_{i=1}^n y_i(e_1)^2 + \frac{1}{n} \sum_{i=1}^n y_i(e_0)^2
\enspace.
\]
Thus, we have that 
\begin{equation}\label{eq:variance-as-eigenvalue}
\Var{\eate} 
= \frac{1}{n} \vec{\alpha}^\top \varM \vec{\alpha} 
\leq \frac{1}{n} \cdot \sup_{\vec{x} \in \Reals^n} \frac{\vec{x}^\top \varM \vec{x}}{\norm{\vec{x}}_2^2}\cdot \norm{\vec{\alpha}}_2^2 
= \frac{1}{n} \cdot \norm{\varM} \cdot \paren[\Bigg]{\frac{1}{n} \sum_{i=1}^n y_i(e_1)^2 + \frac{1}{n} \sum_{i=1}^n y_i(e_0)^2}\enspace.
\end{equation}
In the above, $\norm{\varM}$ is the operator norm of $\varM$.
Because $\varM$ is symmetric and positive semidefinite, the operator norm is equal to the largest eigenvalue of $\varM$, denoted by $\lambda_{\textrm{max}}(\varM)$. 
Thus, proving Theorem~\mainref{thm:variance-analysis-finite-sample} is equivalent to bounding $\lambda_{\textrm{max}}(\varM)$ by a multiple of $\lamH$. 
We focus on this task in the remainder of the proof. 

The strategy we will employ will be to split $\varM$ into a sum of matrices. Each matrix has entries only for pairs $i,j$ of nodes at a particular distance $0$, $1$ or $2$ away in the graph. This is justified by Lemma~\suppref{lem:Cij_expression}, which shows that the values of the entries depend on the distance of the nodes in the conflict graph $\mathcal{H}$. Notice that if $d_\cH(i,j)\geq 3$, then the events $E_{(i,k)}, E_{(j,\ell)}$ are independent, so $C_{(i,k),(j,\ell)} = 0$.
This greatly simplifies the task of bounding the variance.
Hence, we split each matrix $\varM^{(kl)}$ into the sum of three matrices, which contain the entries of $\varM$ for pairs of units at distance $0,1$ or $2$ away. According to the previous observations, we can write

\begin{equation}\label{eq:var_block_final}
    \varM =  \underbrace{\begin{pmatrix}
        \mat{T}^{(0)} &-I\\
        -I &T^0
    \end{pmatrix}}_{\mat{R}^{(0)}} + \underbrace{\begin{pmatrix}
        \mat{T}^{(1)} &-\mat{T}^{(1)}\\
        -\mat{T}^{(1)} &\mat{T}^{(1)}
    \end{pmatrix} }_{\mat{R}^{(1)}} + \underbrace{\begin{pmatrix}
        \mat{T}^{(2)} &-\mat{T}^{(2)}\\
        -\mat{T}^{(2)} &\mat{T}^{(2)}
    \end{pmatrix}}_{\mat{R}^{(2)}}
\end{equation}
In the above, we have defined for $k \in \{0,1,2\}$
\[
\mat{T}^{(k)}_{ij} = \begin{cases}
    C_{ij} &\text{if, } d_\cH(i,j) = k\\
    0 &\text{if, } d_\cH(i,j) \neq k
\end{cases} \quad 
\]
The above means $\mat{T}^{(k)}$ contains only entries for pairs of nodes at distance $k$. 
By the subadditivity of the operator norm, we have
\begin{equation}\label{eq:norm_subadditivity}
\lammax(R) \leq \lammax(\mat{R}^{(0)}) + \lammax(\mat{R}^{(1)}) + \lammax(\mat{R}^{(2)})
\end{equation}
We now proceed to bound the operator norms of $\mat{R}^{(0)}, \mat{R}^{(1)}$ and $\mat{R}^{(2)}$, which will result in our final bound of the variance.
In order to bound the operator norm of these three matrices, we need to establish bounds for the entries $C_{(i,k),(j,l)}$ that are contained in these three matrices. 
These are established in Lemma~\mainref{lemma:individual-covariance-terms}*. 
With these bounds in hand, we are ready to analyze the finite sample variance of the estimator under the design.

For $\mat{R}^{(0)}$, we notice that all four sub-blocks are diagonal matrices. Thus, by simple linear algebra, the maximum eigenvalue of $\mat{R}^{(0)}$ can be bounded by the maximum element of these diagonal matrices.
Indeed, we can apply Lemma~\suppref{lemma:operator_norm_diagonal} with $\mat{D}^{(1)} = \mat{D}^{(2)} = \mat{T}^0$, $\mat{D}^{(3)}= -\mat{I}$, which gives
\[ 
\norm{\mat{R}^{(0)}} \leq 1 + \max\paren[\big]{\max_{i\in [n]}(|C_{(i,0),(i,0)}|)} \leq 1 + \constvar \lamH \enspace.
\]
In the last inequality we used Lemma~\mainref{lemma:individual-covariance-terms}*. 
% Thus, we have
% \begin{equation}\label{eq:term1_bound}
% \lambda_{\max}(\mat{R}^{(0)}) = O\lp(\max_{i \in [n]} |C_{ii}|\rp) = O(\lambda(H))
% \end{equation}
% In the last inequality, we used the property established in Lemma~\ref{lem:Cij_bound}. 
Notice that the bound in $C_{ii}$ comes from 
lower bounding the probabilities $\Pr{E_i^k}$ for all $i \in [n]$ and $k \in \{0,1\}$. Intuitively, this means that all units receive the 
desired exposures with high-enough probability, and this indeed shows up in the variance calculation.

Let us now handle correlations between nodes that are distance $1$ away in $\cH$. We notice that $\mat{T}^{(1)}$ is equal to the negative adjacency matrix $\adjH$ of the conflict graph $\mathcal{H}$, meaning $\mat{T}^{(1)} = - \adjH$. 
Thus, the eigenvalues of $\mat{T}^{(1)}$ are in 1-1 correspondence to the ones of $A_\cH$ with just a sign reversal. 
Some simple linear algebra on the block matrix can now be used to relate the maximum eigenvalue of $\mat{R}^{(1)}$ with that of $\mat{T}^{(1)}$.  
Indeed, we can apply Lemma~\suppref{lem:block_operator_norm} with $S = \mat{T}^{(1)}$, which will give us
\[
\lammax(\mat{R}^{(1)}) = 2 \lammax(- A_\mathcal{H}) \leq 2 \lamH \enspace.
\]
In the last inequality, we used the fact that $A_{\mathcal{H}}$ is an adjacency matrix, and its largest in absolute value eigenvalue is equal to its largest eigenvalue $\lamH$ \citep{Spielman2019Book}. 

Finally, let's bound the operator norm of $\mat{R}^{(2)}$, which contains pairs of nodes that are at distance $2$ away in $H$. We will try to relate the entries of $\mat{T}^{(2)}$ to the number of paths of length $2$ between two nodes. 
Indeed, by Lemma~\mainref{lemma:individual-covariance-terms}*, 
\[
\mat{T}^{(2)}_{ij} 
\leq \constcovar \frac{|\mathcal{N}_b^\pi(i) \cap \mathcal{N}_b^\pi(j)|}{\lamH} 
\leq \frac{|\mathcal{N}^\pi(i) \cap \mathcal{N}^\pi(j)|}{\lamH}
\enspace.
\]
Now, notice that for any pair of nodes $i,j$ that are at distance $2$ in $\cH$, the number of paths between $i$ and $j$ of length $2$ is exactly
$|\mathcal{N}^\pi(i) \cap \mathcal{N}^\pi(j)|$. 
On the other hand, we know that the matrix $\adjH^2$ contains for every pair $i,j$ the number of paths between $i$ and $j$ of length $2$. 
Thus, for any $i,j$, we have established the entrywise bound 
$$
\mat{T}^{(2)}_{ij} \leq \constcovar \frac{1}{\lamH} (\adjH^2)_{ij}
\enspace.
$$
Since both $\mat{T}^{(2)}$ and $\adjH^2$ are symmetric matrices of nonnegative entries, the Peron-Frobenius Theorem implies monotonicity of the maximum eigenvalue \citep{Spielman2019Book}, from which we conclude that
\begin{align*}
\lammax(\mat{T}^{(2)}) &\leq \constcovar \frac{\lammax(A_\mathcal{H}^2)}{\lamH} \\
&= \constcovar \frac{\lamH^2}{\lamH}\\
& = \constcovar \lamH
\end{align*}
We can now apply Lemma~\suppref{lem:block_operator_norm} with $S = \mat{T}^{(2)}$, which yields 
\[
\lammax(\mat{R}^{(2)}) \leq 2 \constcovar \lamH
\]

Combining the above bounds and using \eqref{eq:norm_subadditivity}, we obtain that whenever $\lamH \geq 2$, 
\[
\lammax(R) \leq \paren[\Big]{\constvar + 3 + 2 \constcovar}\lamH
\]
Observe that for any $r \in [1.8, 2.8]$, this constant is at most $12.5$, which is why we choose $r = 2$ in the statement of the algorithm.
Moreover, we can optimize the choice of $r > 0$ to obtain the best constant.
Standard calculations show that this expression is minimized when $r \approx 2.19$ and the bound we obtain its $\lammax(R) \leq \finalconst \cdot \lamH$. 
\end{proof}

\subsection{Consistency of Effect Estimator (Theorem~\mainref{thm:variance-analysis-asymptotic})}

\begin{reftheorem}{\mainref{thm:variance-analysis-asymptotic}}
	\varianceasymptotic
\end{reftheorem}

\begin{proof}
We suppress the dependence of the sequence of potential outcomes on the index $n$ for simplicity. 
The proof follows from applying the finite result of Theorem~\mainref{thm:variance-analysis-finite-sample}.
In particular, since $\eate$ is unbiased, we have 
\begin{align*} 
     \sqrt{ \frac{n}{\lamH} } \cdot \E[\Big]{ \paren[\big]{ \ate - \eate }^2 }^{1/2} &= \sqrt{ \frac{n}{\lamH} } \cdot \sqrt{\Var{\eate}}\\
     &\leq \finalconst \sqrt{\frac{1}{n} \sum_{i=1}^n y_i(e_0)^2 + \frac{1}{n} \sum_{i=1}^n y_i(e_1)^2}\\
	 & = \bigO{1}\,,
\end{align*}
where in the last step we used Assumption~\mainref{assumption:bounded-second-moment}. Taking limits as $n \to \infty$ finishes the proof. 
\end{proof}

\subsection{Slow rate example for Standard Horvitz--Thompson Estimator (Theorem~\mainref{thm:true_Horvitz_Thompson_counterexample})}

\begin{reftheorem}{\mainref{thm:true_Horvitz_Thompson_counterexample}}
	\truehtlowerbound
\end{reftheorem}

\begin{proof}
In the remainder of this Section, we set $r=1$ for simplicity in the Conflict Graph Design, since we are not interested in the constants but on rates of convergence.
We start by defining the sequence of graphs $G^{(n)} = (V_n , E_n)$. 
For every $n \in \Naturals$, we define $V_n =  \{1\} \cup K_n \cup R_n$, where $K_n$ contains $m_n = \sqrt{n}/ \log n$ nodes. Also, $R_n$ contains $n$ nodes. We enumerate the nodes $K_n = \{k_1,\ldots, k_{m_n}\}$ and $R_n = \{r_1, \ldots, r_{n}\}$. We assume without loss of generality that $n$ is a power of $2$, otherwise we can modify the construction by replacing $\sqrt{n}$ with $\lceil \sqrt{n} \rceil$ and $\log n$ with $\lceil \log n \rceil$, without affecting the asymptotics. 
Also, notice that overall, we have $n + \sqrt{n}/\log n + 1$ nodes instead of $n$. 
However, since this is $n(1+o(1))$, it will not change the asymptotics of the rates, so we keep the same normalization $1/n$ for both estimators $\eate$ and $\eatetrue$. 

\begin{figure}[t]
    \centering
    \resizebox{\textwidth}{!}{
        \input{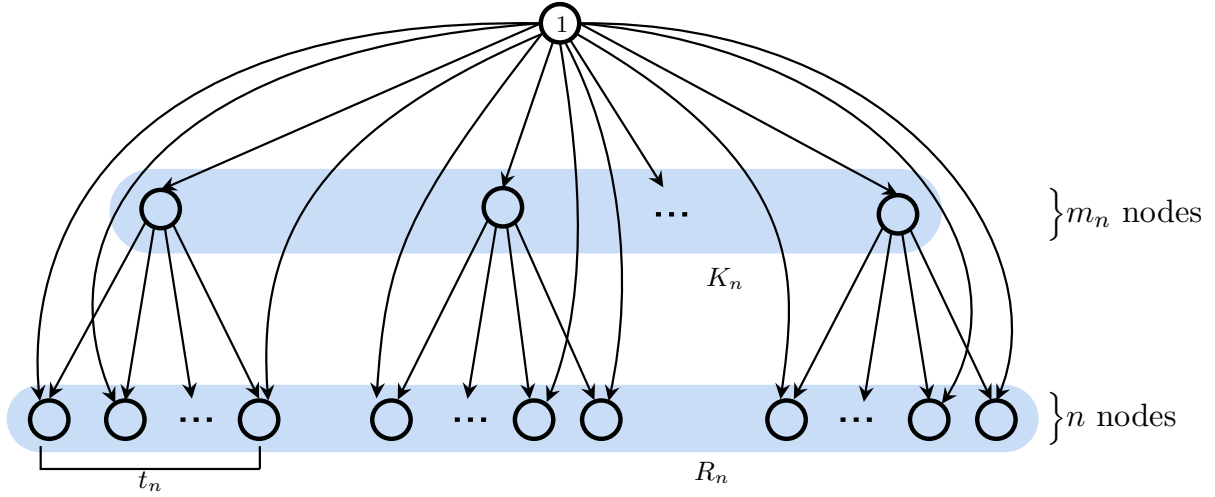}
    }
    \caption{
    	Graph $G^{(n)}$ that demonstrates a gap between the variance of the modified and the standard Horvitz-Thompson estimators.
    	Directions of edges indicate the eigenvalue ordering.
    }
    \label{fig:ht-bat-counterexample}
\end{figure}

We now describe the set of edges $E_n$. First of all, there is an edge $(1, r_i)$ between $1$ and each of the nodes $r_i \in R_n$, for $i = 1, \ldots, n$. 
Also, $R_n$ is divided in subsets of size $t_n = \sqrt{n} \log n = n / m_n$ each. We have a total of $m_n $ subsets , let $R_n^{(i)}$ be the $i$-th subset. We have an edge $(k_i,r_j)$ for any $i \in [m_n]$ and $j \in R_n^{(i)}$. 
We can see pictorially how $G^{(n)}$ would look like in Figure~\suppref{fig:ht-bat-counterexample}.

We focus on the Average Direct Treatment Effect. As we have seen in Section~\mainref{sec:conflict_graph}, in that case the conflict graph $\mathcal{H}$ will be $G^{(n)}$ itself, with added self loops for all nodes. Thus, we focus on analyzing the spectral properties of $G^{(n)}$ for the remainder of this section. We start by determining the eigenvalue ordering $\eigpi$ induced by the graph $G^{(n)}$. Since self loops only add an identity matrix to the adjacency, we can analyze the adjacency matrix $A_n$ of $G^{(n)}$ without self loops for simplicity.
Let $\lamH$ be the largest eigenvalue of $A_n$. To determine the eigenvector $\ev \in \Reals^n$ corresponding to $\lamH$, we first notice that due to symmetry, all nodes in $K_n$ will have the same value in the eigenvector and all nodes in $R_n$ will also have the same value. 
Let $p_1$ be the entry of $\ev$ for node $1$, $p_2$ be the entry for nodes in $K_n$ and $p_3$ the entry for nodes in $R_n$. 
Our goal is to solve the linear system
\[
A_n \ev = \lamH \ev
\enspace.
\]
Because of symmetry, this system reduces to the following $3$ equations

\begin{align*}
    n p_3 &= \lamH p_1 \\
    t_n p_3 &= \lamH p_2\\
    p_1 + p2 &= \lamH p_3 
    \enspace.
\end{align*}
Solving these gives us
\begin{align*}
    p_2 &= \frac{t_n}{n} p_1\\
    p_3 &= \frac{\lamH}{n} p_1\\
    \lamH p_1 &= \frac{n}{\lamH}\paren[\big]{1 + \frac{t_n}{n}} p_1
    \enspace.
\end{align*}
Since the graph is connected, $p_1 > 0$ and hence the third equation gives us $\lamH = \sqrt{n (1 + t/n)} = \Theta(\sqrt{n})$, since $t_n = o(n)$ by construction. Thus, $\lamH = o(n)$, so Assumption~\mainref{assumption:lam-bound} on the sublinear growth of the largest eigenvalue is satisfied. We also get $p_2 = (t_n / n)p_1 = (\log n / \sqrt{n}) p_1$ and $p_3 = (\lamH /n) p_1 = \Theta(1/\sqrt{n}) p_1$. Thus, for sufficiently large $n$ we have $p_1 > p_2 > p_3$. This means that in $\eigpi$ node $1$ is the earliest, followed by all nodes in $K_n$ and then all nodes on $R_n$. Thus, we have that for all $k_i \in K_n$ , $\mathcal{N}_b^\pi (k_i) = \emptyset$. Also, for each $r_j \in R_n^{(i)}$ we have  $\mathcal{N}_b^\pi (k_i) = \{k_i\}$. Thus, the children of $1$ are all nodes in $R_n$ and the children of each $k_i \in K_n$ are the nodes in $R_n^{(i)}$. 

Now, let us define the outcome sequence. In particular, we define the following outcomes
\begin{equation*}
y_i^{(n)}(e_0) = \left\{
    \begin{array}{ll}
          0 & i \notin K_n \\
          \sqrt{t_n} & i \in K_n\\
    \end{array} 
    \right.  
    \quad \quad y_i^{(n)}(e_1) = 0 , \forall i
    \enspace.
\end{equation*}
Intuitively, we only assign non-trivial outcomes on nodes in $K_n$, since we are interested in capturing the correlations between them. We see later that this is enough to make the variance of $\eatetrue$ blow up.
Notice that Assumption~\mainref{assumption:bounded-second-moment} is satisfied, since
\[
\frac{1}{n} \sum_{i \in K_n}(y_i^{(n)}(e_0))^2 = \frac{m_n t_n}{n } = 1
\enspace.
\]

We can thus apply Theorem~\mainref{thm:variance-analysis-asymptotic} for $\eate$ to conclude that 
$$
		\limsup_{n\to \infty} n^{1/4}\cdot \E[\Big]{ \paren[\big]{ \ate - \eate }^2 }^{1/2} < \infty \enspace. 
$$

We now focus on the performance of $\eatetrue$.
To do that, we focus on the nodes $k_i \in K_n$, $i = 1,\ldots, m_n$. 
We start by calculating the exposure probabilities for these nodes, i.e. the probability $\Pr{d_{k_i}(Z) = d_{k_i}(\contrastz{i}{0})}$.
Recall that the standard Horvitz--Thompson uses these proper exposure probabilities rather than the desired exposure probabilities $\Pr{E_{(i,k)}}$ used by the modified Horvitz--Thompson estimator.
Let's analyze the possible outcomes that would lead to $d_{k_i}(Z) = d_{k_i}(\contrastz{i}{0})$. This event can be written equivalently as $\{Z_{k_i} = 0, Z_{r_j} = 0, \forall r_j \in R_n^{(i)}\}$. 
We now analyze all possible assignments of the $U_i$ variables that lead to that exposure.
\begin{itemize}
    \item $U_{k_i} = e_0$. In that case, $k_i$ has the desired exposure of all control, and since it has no parents, it will receive this exposure. 
    \item $U_{k_i} = * \land U_1 \neq *$. This corresponds to the case where at least one of the parents of the nodes in $R_n^{(i)}$ has a desired exposure that is not $*$.  In that case, 
    the nodes in $R_n^{(i)}$ \emph{yield} to the desires of one of their parents, whose desired exposures require all the nodes in $R_n^{(i)}$ to remain in control.
    \item $U_{k_i} = U_1 = *, U_{r_j} \neq e_1 \forall r_j \in R_n^{(i)}$. This corresponds to the case where both parents do not have a desired exposure, meaning that they allow nodes in $R_n^{(i)}$ to receive their desired exposures. This means that if at least one node in $R_n^{(i)}$ has $U_{r_j} = e_1$, then it ``desires'' to receive the direct treatment exposure, leading to $Z_{r_j} = 1$. Thus, in order for $k_i$ to receive the all control exposure, this should not happen for any node in $R_n^{(i)}$.
\end{itemize}
The three events discussed above are pairwise disjoint. 
Thus, they help us write the exposure probabilities in the following nice form. 

\begin{align*}
\Pr{d_{k_i}(Z) = d_{k_i}(z_i^{(1)})} &= \Pr{U_{k_i} = e_0 } + \Pr{U_{k_i} = * , U_{1} \neq *} \\
&\quad +\Pr{U_{k_i} = U_1 = *, U_{r_j} \neq e_1 \forall r_j \in R_n^{(i)}}\\
 &= \frac{1}{2\lamH} + \paren[\Big]{1 - \frac{1}{\lamH}} \frac{1}{\lamH} + \paren[\Big]{1 - \frac{1}{\lamH}}^2 \paren[\Big]{1 - \frac{1}{2\lamH}}^{t_n} \\
 &= \Theta\lp(\frac{1}{\sqrt{n}}\rp) + \Theta\lp(\lp(1 - \frac{1}{\sqrt{n}}\rp)^{\sqrt{n} \log n}\rp) \\
 &= \Theta\lp(\frac{1}{\sqrt{n}}\rp) \enspace, 
\end{align*}
since the second term is $\bigO{1/n}$. 

The next step would be to calculate the covariance between two different units $k_i, k_s \in K_n$. 
In particular, the following term will appear in the expanded form of the variance of $\eatetrue$ 
\begin{align*}
D_{(k_i,0),(k_s,0)}&:= \Cov[\Bigg]{\frac{\indicator{d_{k_i}(Z) = d_{k_i}(\contrastz{i}{0})}}{\Pr[\Big]{d_{k_i}(Z) = d_{k_i}(\contrastz{k_i}{0})}}, \frac{\indicator{d_{k_s}(Z) = d_{k_s}(\contrastz{i}{0})}}{\Pr[\Big]{d_{k_s}(Z) = d_{k_s}(\contrastz{k_s}{0})}}} 
\\
&=\frac{\Pr[\Big]{d_{k_i}(Z) = d_{k_i}(\contrastz{k_i}{0}), d_{k_s}(Z) = d_{k_s}(\contrastz{k_s}{0})}}{\Pr[\Big]{d_{k_i}(Z) = d_{k_i}(\contrastz{k_i}{0})}\Pr[\Big]{d_{k_s}(Z) = d_{k_s}(\contrastz{k_s}{0})}} - 1
\enspace.
\end{align*}

To compute the probability in the numerator, let us analyze all ways in which the following event occurs
\[
    \lp\{d_{k_i}(Z) = d_{k_i}(\contrastz{k_i}{0}), d_{k_s}(Z) = d_{k_s}(\contrastz{k_s}{0}) \rp\} = \lp\{Z_{k_i} = Z_{k_s} = 0, Z_{r} =0 ,  \forall r \in R_n^{(i)} \cup R_n^{(s)}\rp\}
    \enspace.
\]
This event can be broken down in the following disjoint events. 

\begin{itemize}
    \item $U_{1} \neq *, U_{k_i} \neq e_1, U_{k_s} \neq e_1$. In that case, all children of $k_i,k_s$ will receive control in the final assignment, since they are ``yielding'' to the exposure of their parent $1$. Thus, this event happens with probability
    \[
    \frac{1}{\lamH} \paren[\Big]{1 - \frac{1}{2\lamH}}^2 = \Theta\lp(\frac{1}{\sqrt{n}}\rp) \enspace.
    \]
    \item $U_1 = *$. In that case, the events $d_{k_i}(Z) = d_{k_i}(\contrastz{k_i}{0})$ and $d_{k_s}(Z) = d_{k_s}(\contrastz{k_s}{0})$ are now independent. This event happens with probability
    \begin{align*}
    \Pr{U_1 = *} & \Pr[\big]{d_{k_i}(Z) = d_{k_i}(\contrastz{k_i}{0}) | U_1 = *} \Pr[\big]{d_{k_s}(Z) = d_{k_s}(\contrastz{k_s}{0}) | U_1 = *}  \\
    &= \paren[\Big]{1 - \frac{1}{\lamH}} \Theta\lp(\frac{1}{\sqrt{n}}\rp)  \Theta\lp(\frac{1}{\sqrt{n}}\rp) \\
    &= \Theta\lp(\frac{1}{n}\rp)
    \enspace.
    \end{align*}
    For these conditional probabilities we can use the same case analysis as for calculating the propensity scores above to conclude that each one is $\Theta(1/\sqrt{n})$. 
\end{itemize}
Thus, overall we have that 
\[
    \Pr[\Big]{d_{k_i}(Z) = d_{k_i}(\contrastz{k_i}{0}), d_{k_s}(Z) = d_{k_s}(\contrastz{k_s}{0})} = \Theta\lp(\frac{1}{\sqrt{n}}\rp)
    \enspace.
\]
Thus, 
\[
D_{(k_i,0),(k_s,0)} = \frac{\Theta(1/\sqrt{n})}{\Theta(1/\sqrt{n})\Theta(1/\sqrt{n})}- 1 = \Theta(\sqrt{n})
\enspace.
\]
This shows that the correlation between nodes $k_i,k_s$ will have a large contribution to the variance of $\eatetrue$. 

Indeed, for this choice of potential outcomes, we have for the variance of $\eatetrue$
\begin{align*}
\Var{\eatetrue} &\geq  \frac{1}{n^2} \sum_{k_i, k_s \in K_n, k_i \neq k_s}  y_{k_i}(e_0) y_{k_s}(e_0) D_{(k_i,0),(k_s,0)} \\
&= \frac{t_n {m_n \choose 2}}{n^2} \Theta(\sqrt{n}) \\
&= \Theta\lp(\frac{m_n}{\sqrt{n}}\rp) \\
&= \Theta\lp(\frac{1}{\log n}\rp)
\enspace.
\end{align*}
The first inequality follows by ignoring the individual variance terms of each unit, which are always nonnegative.
Using that the standard Horvitz--Thompson estimator is unbiased and taking the limit $n \to \infty$, we get
\[
\liminf_{n\to \infty} \sqrt{\log n} \cdot \E{ \paren{\eatetrue - \ate}^2 }^{1/2} > 0
\enspace.
\qedhere
\] 
\end{proof}

	\section{Analysis of Inferential Methods} 

 In this section, we provide proofs for results arising in Section~\mainref{sec:inference} regarding the inferential methods. 

% \subsection{Variance Estimation}
% In this Section, we have collected proofs of statements made in Section~\ref{sec:variance-estimation}. 

\subsection{Validity of Inference Assumptions under i.i.d. Outcomes}\label{sec:iid-outcomes}

In this Section, we prove a claim that was made in Section~\mainref{sec:inference}, namely that Assumptions~\mainref{assumption:bounded-fourth-moments} and \mainref{assumption:not-superefficient} both hold with with high probability if the outcomes are drawn from a distribution with bounded fourth moments.
To be more precise, let us introduce the following assumption about the generation of potential outcomes. 

\begin{assumption}\label{ass:iid-outcomes}
	Suppose that the potential outcomes $\setb{ y_i(e_0),y_i(e_1) }_{i=1}^n$ are sampled i.i.d. from a distribution $\cD$ over $\Reals^2$ with the following properties:
	\begin{itemize}
		\item \textbf{Mean Zero}: If $(X,Y) \sim \cD$ then $\E{X} = \E{Y} = 0$.
		\item \textbf{Fourth Moments}: If $(X,Y) \sim \cD$ then $\E{|X|^{4}}, \E{|Y|^4} < \infty$.
		\item \textbf{Not Perfectly Correlated}: If $(X,Y) \sim \cD$ then 
		\[
		\rho:= \frac{\Cov{X,Y}}{\sqrt{\Var{X}\Var{Y}}} < 1
		\enspace.
		\] 
	\end{itemize}
\end{assumption}

We give here a useful Lemma that bounds some mixed moments of the distribution $\cD$ by the fourth moments. 

\begin{lemma}\label{lem:moment_bound}
    Let $\cD$ be a distribution satisfying Assumption~\ref{ass:iid-outcomes}. 
    Define 
    $$R \triangleq \max\lp(\E{|X|^{4}}, \E{|Y|^4}\rp)\enspace,$$ 
    where $X,Y \sim \cD$.
    Then,
    \begin{itemize}
    	\item $\E{(X^2 - \E{X^2})^2} \leq R$
    	\item $\E{(Y^2 - \E{Y^2})^2} \leq R$
    	\item $\E{(XY - \E{XY})^2} \leq R$
    	\item $\E{X^2}\E{Y^2} \leq R$.
    \end{itemize}
\end{lemma}
\begin{proof}
    Since the variance is always bounded by the second moment, we have 
    \[
        \E{(X^2 - \E{X^2})^2} = \Var{X^2} \leq \E{X^4} \leq R \enspace.
    \]
    Similarly, we conclude that $\E{(Y^2 - \E{Y^2})^2} \leq R$. 
    Observe also that
    \[
        \E{(XY - \E{XY})^2} = \Var{XY} \leq \E{X^2Y^2} \leq \sqrt{\E{X^4}\E{Y^4}} \leq R \enspace,
    \]
    where in the second to last step we have used the Cauchy-Schwartz inequality. 
    By Jensen's inequality we have that $\E{X^2} \leq \sqrt{\E{X^4}}$ and $\E{Y^2} \leq \sqrt{\E{Y^4}}$.
    Using this, we obtain that
    \[
    \E{X^2}\E{Y^2} \leq \sqrt{ \E{X^4} } \cdot \sqrt{ \E{Y^4} } \leq R
    \enspace.
    \qedhere
    \]
\end{proof}

Next, we prove that if outcomes are sampled according to Assumption~\suppref{ass:iid-outcomes}, then with probability tending to $1$ as $n \to \infty$, Assumption~\mainref{assumption:bounded-fourth-moments} holds. 

\begin{proposition}
    Suppose Assumption~\ref{ass:iid-outcomes} holds.
    Then, there exists a constant $M >0$, such that for both $k \in \{0,1\}$
    \[
    \lim_{n \to \infty} \Pr[\Bigg]{\frac{1}{n} \sum_{i=1}^n y_i(e_k)^4 > M} = 0
    \]
\end{proposition}
\begin{proof}
Let $(X,Y)\sim \cD$. We will prove the claim for $k=0$, the proof for $k=1$ is identical.
By Assumption~\ref{ass:iid-outcomes}, denote $R = \E{X^4} < \infty$ and
 consider any constant $M > R$. Then, we have
    \begin{align*}
        \Pr[\Bigg]{\frac{1}{n} \sum_{i=1}^n y_i(e_k)^4 > M} 
        = \Pr[\Bigg]{ \frac{1}{n} \sum_{i=1}^n y_i(e_k)^4 -  \E[\Bigg]{ \frac{1}{n} \sum_{i=1}^n y_i(e_k)^4 }  > M - R}
        = o(1)\enspace,
    \end{align*}
    where the last line follows from the strong Law of Large Numbers \citep{durrett2019probability}. 
\end{proof}

Next, we show that if Assumption~\suppref{ass:iid-outcomes} holds, then the estimator is not superefficient with high probability. 
We emphasize that there are two levels of randomness. The first level comes from the randomness of the potential outcomes $y_i(e_k)$. 
Conditioned on the values of the potential outcomes, the second level of randomness comes from the design that assigns the randomized treatment vector. To distinguish between the two sources, we will use the notation $\Var{\eate|y}$ to denote the variance of $\eate$, after conditioning on the potential outcomes $y_i(e_k)$. Thus, this quantity is only a function of the potential outcomes, so the probability in the statement is with respect to the randomness of the potential outcomes. We use a subscript of $\cD$ to denote that. 
\begin{proposition}
    Suppose Assumption~\ref{ass:iid-outcomes} holds.
    Then, there exists a constant $M >0$, such that
    \[
    \lim_{n \to \infty}\Prsub[\Bigg]{\cD}{\Var{\eate|y} < M\frac{\lamH}{n}} = 0
    \]
\end{proposition}
\begin{proof}
    In the below calculations, the expectation is taken with respect to the randomness of the potential outcomes, as explained above. 
    Using \eqref{eq:var_expression}, we get that
    \begin{align*}
    &\Esub{\cD}{\Var{\eate|y}} \\
    &= \Esub[\Bigg]{\cD}{\frac{1}{n^2} \sum_{i=1}^n \sum_{j=1}^n \sum_{\substack{k \in \setb{0,1} \\ \ell \in \setb{0,1}}} (-1)^{k - \ell} y_i(e_k) y_j(e_\ell) \Cov[\Bigg]{ \frac{ \indicator{ E_{(i,k)} } }{ \Pr{ E_{(i,k)} } }  , \frac{ \indicator{ E_{(j,\ell)} } }{ \Pr{ E_{(j,\ell)} } } }}\\
    &= \frac{1}{n^2} \sum_{i=1}^n \sum_{k=0}^1 \E{y_i(e_k)^2} \Var[\Bigg]{ \frac{ \indicator{ E_{(i,k)} } }{ \Pr{ E_{(i,k)} } }  }\\
    &\quad\quad + \frac{2}{n^2} \sum_{i=1}^n \E{y_i(e_1) y_i(e_0)}  \cdot \Cov[\Bigg]{ \frac{ \indicator{ E_{(i,0)} } }{ \Pr{ E_{(i,0)} } }  , \frac{ \indicator{ E_{(i,1)} } }{ \Pr{ E_{(i,1)} } } }\\
    \intertext{
    	where we used the fact that $y_i(e_k)$ and $y_j(e_\ell)$ are independent zero mean if $i \neq j$.
    	We will first handle the variance terms.
    	Using Lemma~\suppref{lem:Cij_expression} together with $r \geq 1$, we obtain that
    }
    &\geq \frac{1}{n}  \sum_{i=1}^n \sum_{k \in \setb{0,1}} \E{y_i(e_k)^2} (2 \lamH - 1) \\
    &quad\quad+  \frac{2}{n^2} \sum_{i=1}^n \E{y_i(e_1) y_i(e_0)}  \cdot \Cov[\Bigg]{ \frac{ \indicator{ E_{(i,0)} } }{ \Pr{ E_{(i,0)} } }  , \frac{ \indicator{ E_{(i,1)} } }{ \Pr{ E_{(i,1)} } } }\\
    \intertext{
    	Next, we analyze the covariance terms. 
    	Recall that these events $E_{i,1}$ and $E_{i,0}$ are mutually exclusive.
    	Thus, the covariance terms are $-1$.
    	Using this, we obtain
    }
    &\geq \frac{1}{n} \sum_{i=1}^n  \paren[\Big]{\E{y_i(e_0)^2} (2 \lamH - 1) + \E{y_i(e_1)^2} (2 \lamH - 1) - 2 \E{y_i(e_1) y_i(e_0)}} \\
    \intertext{Using the definition of the correlation $\rho$, we have that}
    &= \frac{1}{n} \sum_{i=1}^n \paren[\Big]{\E{y_i(e_0)^2} (2 \lamH - 1) + \E{y_i(e_1)^2} (2 \lamH - 1) - 2 \rho \sqrt{\E{y_i(e_1)^2} \E{y_i(e_0)^2}} }\\
    \intertext{Using Cauchy-Schwartz to lower bound this third term and rearranging, we obtain that}
    &\geq \frac{1}{n} \sum_{i=1}^n \paren[\Big]{\E{y_i(e_0)^2} (2 \lamH - 1) + \E{y_i(e_1)^2} (2 \lamH - 1) - 2 \rho \paren{\E{y_i(e_1)^2} + \E{y_i(e_0)^2}} }\\
    &= \frac{\lamH}{n}  \sum_{i=1}^n \paren[\Bigg]{\E{y_i(e_0)^2}\paren[\Big]{2 - \frac{1 + \rho}{\lamH}} + \E{y_i(e_1)^2}\paren[\Big]{2 - \frac{1 + \rho}{\lamH}}}\\
    &\geq \frac{\lamH}{n} \sum_{i=1}^n \paren[\Bigg]{\E{y_i(e_0)^2}(1 - \rho) + \E{y_i(e_1)^2}(1 - \rho)} \enspace,
    \end{align*}
    where the last line using the fact that $\lamH \geq 1$.
    Overall, this yields the lower bound
    \begin{equation}\label{eq:exp_lower_bound}
        \Esub{\cD}{\Var{\eate|y}} \geq \frac{\lamH}{n} \paren[\Bigg]{\E{y_i(e_0)^2}(1 - \rho) + \E{y_i(e_1)^2}(1 - \rho)}
    \end{equation}
    Now, to apply the second moment method, let's calculate the variance of $\Var{\eate|y}$ with respect to the randomness of the outcomes. 
    \begin{align*}
        &\Varsub{\cD}{\Var{\eate}} \\
        &=
        \Varsub[\Bigg]{\cD}{\frac{1}{n^2} \sum_{i=1}^n \sum_{j=1}^n \sum_{\substack{k \in \setb{0,1} \\ \ell \in \setb{0,1}}} (-1)^{k - \ell} y_i(e_k) y_j(e_\ell) \Cov[\Bigg]{ \frac{ \indicator{ E_{(i,k)} } }{ \Pr{ E_{(i,k)} } }  , \frac{ \indicator{ E_{(j,\ell)} } }{ \Pr{ E_{(j,\ell)} } }  }}\\
        &= \mathbb{E}\lp[
        \lp(
            \frac{1}{n^2} \sum_{i=1}^n \sum_{k \in \{0,1\}} (y_i(e_k)^2 - \E{y_i(e_k)^2}) \Var[\Big]{\frac{ \indicator{ E_{(i,k)} } }{ \Pr{ E_{(i,k)} } }} \rp.\rp.\\
            &\lp.\lp. + \frac{1}{n^2}\sum_{i \neq j} \sum_{\substack{k \in \{0,1\}\\ \ell \in \{0,1\}} } (-1)^{k - \ell} y_i(e_k) y_j(e_\ell) \Cov[\Bigg]{ \frac{ \indicator{ E_{(i,k)} } }{ \Pr{ E_{(i,k)} } }  , \frac{ \indicator{ E_{(j,\ell)} } }{ \Pr{ E_{(j,\ell)} } } }\rp.\rp.\\
            &\lp.\lp. + \frac{2}{n^2}\sum_{i=1}^n \paren{y_i(e_0) y_i(e_1) - \E{y_i(e_0) y_i(e_1)}} \Cov[\Bigg]{ \frac{ \indicator{ E_{(i,0)} } }{ \Pr{ E_{(i,0)} } }  , \frac{ \indicator{ E_{(i,1)} } }{ \Pr{ E_{(i,1)} } } }
        \rp)^2    
        \rp]
    \end{align*}
    Now, we use the fact that for any random variables $X,Y,Z$, using Cauchy-Schwartz 
    \[
    \E{(X+Y+Z)^2} \leq 3 \E{X^2} + 3 \E{Y^2} + 3 \E{Z^2} \enspace.
    \]
    Using this, the variance can be bounded as 
\begin{align*}
    &\Varsub{\cD}{\Var{\eate}} \\
    &\leq 
    3\mathbb{E}\lp[
        \lp(
            \frac{1}{n^2} \sum_{i=1}^n \sum_{k \in \{0,1\}} (y_i(e_k)^2 - \E{y_i(e_k)^2}) \Var[\Big]{\frac{ \indicator{ E_{(i,k)} } }{ \Pr{ E_{(i,k)} } }} \rp)^2\rp]\\
            &+3\mathbb{E}\lp[\frac{1}{n^2}\lp(  \sum_{i \neq j} \sum_{\substack{k \in \{0,1\}\\ \ell \in \{0,1\}} } (-1)^{k - \ell} \paren{y_i(e_k) y_j(e_\ell) \rp.\rp.\\
            &\quad\quad\lp.\lp.- \E{y_i(e_k) y_j(e_\ell)}} \Cov[\Bigg]{ \frac{ \indicator{ E_{(i,k)} } }{ \Pr{ E_{(i,k)} } }  , \frac{ \indicator{ E_{(j,\ell)} } }{ \Pr{ E_{(j,\ell)} } }  }
        \rp)^2   \rp] \\
        &+ 
        3\mathbb{E}\lp[\frac{1}{n^2}\lp(  2\sum_{i=1}^n \paren{y_i(e_0) y_i(e_1) - \E{y_i(e_0) y_i(e_1)}} \Cov[\Bigg]{ \frac{ \indicator{ E_{(i,0)} } }{ \Pr{ E_{(i,0)} } }  , \frac{ \indicator{ E_{(i,1)} } }{ \Pr{ E_{(i,1)} } } }
        \rp)^2 \rp]
\end{align*}
Let's focus on the first term. As before, let $R = \max(\E{X^4}, \E{Y^4})$, where $(X,Y) \sim \cD$. 
For any random variables $X,Y$, we have $\E{(X+Y)^2} \leq 2\E{X^2} + 2 \E{Y^2}$. Applying this for the first term gives
\begin{align*}
    &\mathbb{E}\lp[
        \lp(
            \frac{1}{n^2} \sum_{i=1}^n \sum_{k \in \{0,1\}} (y_i(e_k)^2 - \E{y_i(e_k)^2}) \Var[\Big]{\frac{ \indicator{ E_{(i,k)} } }{ \Pr{ E_{(i,k)} } }} \rp)^2\rp] \\
            &\leq
            \frac{2}{n^4} \mathbb{E}\lp[
             \sum_{i=1}^n  \E[\big]{(y_i(e_0)^2 - \E{y_i(e_0)^2})^2} \Var[\Big]{\frac{ \indicator{ E_{(i,0)} } }{ \Pr{ E_{(i,0)} } }}^2 \rp]\\
            &\quad+ \frac{2}{n^4}\mathbb{E}\lp[
                \sum_{i=1}^n  \E[\big]{(y_i(e_1)^2 - \E{y_i(e_1)^2})^2} \Var[\Big]{\frac{ \indicator{ E_{(i,1)} } }{ \Pr{ E_{(i,1)} } }}^2 \rp]\\
            &=
            \frac{1}{n^4} 
             \sum_{i=1}^n \sum_{k \in \{0,1\}} \E[\big]{(y_i(e_k)^2 - \E{y_i(e_k)^2})^2} \Var[\Big]{\frac{ \indicator{ E_{(i,k)} } }{ \Pr{ E_{(i,k)} } }}^2 \\
             &\leq 
             \frac{2R}{n^4}  \sum_{i=1}^n \sum_{k \in \{0,1\}} \paren[\Bigg]{\frac{2r\lamH}{(1-1/(2r))^2}}^2 \\
             \intertext{where we have used Lemma~\mainref{lemma:individual-covariance-terms} to bound the variance of the indicators and Lemma~\suppref{lem:moment_bound} to bound the moments}
            &= \frac{4R}{n^3} \paren[\Bigg]{\frac{2r}{(1-1/(2r))^2}}^2 \lamH^2\\
            &= \bigO{\lamH^2/n^3}
\end{align*}
For the second term, again by independence we get
\begin{align*}
    &\frac{1}{n^4}\mathbb{E}\lp[\lp(  \sum_{i \neq j} \sum_{\substack{k \in \{0,1\}\\ \ell \in \{0,1\}} }  (-1)^{k - \ell} \paren{y_i(e_k) y_j(e_\ell)\rp.\rp.\\
    &\lp.\lp.\quad\quad - \E{y_i(e_k) y_j(e_\ell)}} \Cov[\Bigg]{ \frac{ \indicator{ E_{(i,k)} } }{ \Pr{ E_{(i,k)} } }  , \frac{ \indicator{ E_{(j,\ell)} } }{ \Pr{ E_{(j,\ell)} } }  }
        \rp)^2    
        \rp] \\
        &=  \frac{1}{n^4} \sum_{i \neq j} \sum_{\substack{k \in \setb{0,1} \\ \ell \in \setb{0,1}}} \E{y_i(e_k)^2} \E{y_j(e_\ell)^2} \Cov[\Bigg]{ \frac{ \indicator{ E_{(i,k)} } }{ \Pr{ E_{(i,k)} } }  , \frac{ \indicator{ E_{(j,\ell)} } }{ \Pr{ E_{(j,\ell)} } }  }^2\\
        &= \frac{1}{n^4} \sum_{\substack{i \neq j\\ d_\cH(i,j) \geq 2}} \sum_{\substack{k \in \setb{0,1} \\ \ell \in \setb{0,1}}} \E{y_i(e_k)^2} \E{y_j(e_\ell)^2} \Cov[\Bigg]{ \frac{ \indicator{ E_{(i,k)} } }{ \Pr{ E_{(i,k)} } }  , \frac{ \indicator{ E_{(j,\ell)} } }{ \Pr{ E_{(j,\ell)} } }  }^2 \\
        &\quad + \frac{1}{n^4} \sum_{\substack{i \neq j\\ d_\cH(i,j) \leq 1}} \sum_{\substack{k \in \setb{0,1} \\ \ell \in \setb{0,1}}} \E{y_i(e_k)^2} \E{y_j(e_\ell)^2} \Cov[\Bigg]{ \frac{ \indicator{ E_{(i,k)} } }{ \Pr{ E_{(i,k)} } }  , \frac{ \indicator{ E_{(j,\ell)} } }{ \Pr{ E_{(j,\ell)} } }  }^2\\
        &\leq C \frac{R}{n^4} \sum_{i,j: d_\cH(i,j) \geq 2}   \paren[\Bigg]{\frac{|\mathcal{N}_b(i) \cap \mathcal{N}_b(j)|}{\lamH}}^2
        + C \frac{R \cdot |E_\cH| }{n^4}
        \\
        \intertext{for some absolute constant $C$. 
        	To see this, observe that if $i,j$ are neighbors in $\cH$, their covariance is $-1$ by Lemma~\mainref{lemma:individual-covariance-terms}, which gives a total contribution of $|E_\cH|$ for these terms. 
        	If $i,j$ are at distance $2$, their covariance can be bounded by the number of common parents of $i,j$ as per Lemma~\mainref{lemma:individual-covariance-terms}.
        	We have also used Lemma~\suppref{lem:moment_bound} to bound the moments by $R$.
        	Moreover, we can further upper bound the above by}
        &\leq C \frac{R}{n^4} \sum_{i \neq j}   \frac{|\mathcal{N}_b(i) \cap \mathcal{N}_b(j)|}{\lamH} + C \frac{R \cdot \lamH n}{n^4}\\
        \intertext{where for the first term we have used using the importance ordering property $|\mathcal{N}_b(i) \cap \mathcal{N}_b(j)| \leq \lamH$ so that $|\mathcal{N}_b(i) \cap \mathcal{N}_b(j)| / \lamH \leq 1$.
        	For the second term, we use that that the average degree is always bounded by the largest eigenvalue, i.e. $n^{-1} |E_\cH| = d_{avg}(\cH) \leq \lamH$.
        	To bound the first term further, we use the same observation as in the proof of Theorem~\mainref{thm:variance-analysis-finite-sample} about paths of length $2$ and the squared adjacency matrix (see Section~\suppref{sec:finite-sample-proof} for more details). Overall, this yields that }
        &\leq 2C \frac{R\cdot \lamH}{n^3} \\
        & \leq \bigO{ \lamH / n^3 }\enspace,
\end{align*}

For the third term, we have by independence that
\begin{align*}
    &\frac{1}{n^4}\mathbb{E}\lp[\lp(  2\sum_{i=1}^n \paren{y_i(e_0) y_i(e_1) - \E{y_i(e_0) y_i(e_1)}} \Cov[\Bigg]{ \frac{ \indicator{ E_{(i,0)} } }{ \Pr{ E_{(i,0)} } }  , \frac{ \indicator{ E_{(i,1)} } }{ \Pr{ E_{(i,1)} } } }
        \rp)^2 \rp]\\
    &\qquad = 4 \sum_{i=1}^n \E{\paren{y_i(e_0) y_i(e_1) - \E{y_i(e_0) y_i(e_1)}}^2} \Cov[\Bigg]{ \frac{ \indicator{ E_{(i,0)} } }{ \Pr{ E_{(i,0)} } }  , \frac{ \indicator{ E_{(i,1)} } }{ \Pr{ E_{(i,1)} } } }^2\\
    &\qquad  \leq \frac{4R}{n^4} \sum_{i=1}^n \Cov[\Bigg]{ \frac{ \indicator{ E_{(i,0)} } }{ \Pr{ E_{(i,0)} } }  , \frac{ \indicator{ E_{(i,1)} } }{ \Pr{ E_{(i,1)} } } }^2\\
    &\qquad  = \frac{4R}{n^3}\\
    	\intertext{by Lemma~\suppref{lem:moment_bound}} \\
    &\qquad  = \bigO[\Bigg]{\frac{1}{n^3}}\\
    	\intertext{by Lemma~\suppref{lem:Cij_expression}.}
\end{align*}

Overall, the above analysis gives us that $\Varsub{\cD}{\Var{\eate|y}} = \bigO{\lamH^2/n^3} $. 
To complete the proof, we apply Chebyshev's inequality.
Define $Q = \E{y_i(e_0)^2}(1 - \rho) + \E{y_i(e_1)^2}(1 - \rho)$.
An earlier calculation showed that $ \Esub{\cD}{\Var{\eate|y}} \geq \frac{\lamH}{n} \cdot Q$.
By Assumption~\suppref{ass:iid-outcomes}, we have that the correlations are not perfect so that $Q > 0$.
Set $M = 2Q$ and observe that
\begin{align*}
    \Prsub[\Bigg]{\cD}{\Var{\eate|y} < \frac{M}{n}  \lamH} 
    &=  \Prsub[\Bigg]{\cD}{\Var{\eate|y} - \Esub{\cD}{\Var{\eate|y}} < \frac{M}{n}  \lamH - \Esub{\cD}{\Var{\eate|y}} } \\
    &\leq \Prsub[\Bigg]{\cD}{\Var{\eate|y} - \Esub{\cD}{\Var{\eate|y}} < \frac{\lamH}{n} \cdot \paren{ M - Q } }
    \intertext{where the last line follows from the lower bound on $\Esub{\cD}{\Var{\eate|y}}$.
    Now applying the choice of $M$ and Chebyshev's inequality, we have that}
    &= \Prsub[\Bigg]{\cD}{\Var{\eate|y} - \Esub{\cD}{\Var{\eate \mid y}} < \frac{\lamH}{n} \cdot Q } \\
    &\leq \Pr[\Bigg]{\abs[\Big]{\Var{\eate} - \Esub{\cD}{\Var{\eate \mid y}}} > \frac{\lamH}{n}  \cdot Q}\\
    &\leq \frac{\Varsub{\cD}{\Var{\eate}}}{\paren[\Big]{\frac{\lamH}{n} Q}^2} 
    \intertext{
    	Finally, using that $Q > 0$ together with the upper bound on $\Varsub{\cD}{\Var{\eate}}$, we have that
    }
    &= \bigO[\Bigg]{\frac{\lamH^2/n^3}{\lamH^2/n^2}}  \\
    &= \bigO{1/n} \enspace,
\end{align*}
which concludes the proof. 
\end{proof}

\subsection{Variance Bound is Conservative (Proposition~\mainref{prop:conservative-vb-is-tight})}

\begin{refproposition}{\mainref{prop:conservative-vb-is-tight}}
	\conservativevbistight
\end{refproposition}

\begin{proof}
    The fact that $\Var{\eate} \leq \vb$ follows from the proof of Theorem~\mainref{thm:variance-analysis-finite-sample}.
    In particular, from the discussion after equation\eqref{eq:variance-as-eigenvalue} it follows that
    \[
    \sup_{\vec{y}(e_0), \vec{y}(e_1)}\frac{n \Var{\eate}}{\frac{1}{n} \sum_{i=1}^n y_i(e_1)^2 + \frac{1}{n}\sum_{i=1}^n y_i(e_1)^2 } = \lamV
    \]
    Thus, when the potential outcomes are aligned with the leading eigenvector of $\varM$, we get equality in the above expression. 
\end{proof}

\subsection{Estimator $\evb$ is Unbiased (Proposition~\mainref{prop:unbiased-var-est})}
We start with the proof that the variance estimator $\evb$ is unbiased for estimating the bound $\vb$ on the variance.

\begin{refproposition}{\mainref{prop:unbiased-var-est}}
	\unbiasedvarest
\end{refproposition}

\begin{proof}
	For the first part, we use the observation that $Y_i^2 \indicator{E_{(i,k)}} = y_i(e_k)^2 \indicator{E_{(i,k)}} $, which was justified in the proof of Proposition~\mainref{prop:unbiased-estimator} and requires Assumption~\mainref{assumption:ani-model}. 
    \begin{align*}
    \E{\evb} &= \E[\Bigg]{\frac{\lamV}{n} \paren[\Big]{\frac{1}{n}\sum_{i=1}^n Y_i^2 \paren[\Big]{\frac{\indicator{E_{(i,1)}}}{\Pr{E_{(i,1)}}} + \frac{\indicator{E_{(i,1)}}}{\Pr{E_{(i,1)}}}}}}\\
	& = \frac{\lamV}{n} \paren[\Big]{\frac{1}{n}\sum_{i=1}^n \paren[\Big]{y_i^2(e_1) + y_i^2(e_0)} }\\
	&= \vb
    \end{align*}

	For the second part, it follows from the proof of Theorem~\mainref{thm:variance-analysis-finite-sample} that $\lambda(\varM) \leq \cleanconst \cdot \lamH$.
	Thus, using Assumption~\mainref{assumption:bounded-second-moment} we get $VB = \bigO{\lamH/n}$, which concludes the proof.
\end{proof}

\subsection{Consistency of $\evb$ (Proposition~\mainref{prop:var-est-consistency})}

We next prove that $\evb$ converges to $\vb$ at a sufficiently fast rate that will allow us to construct confidence intervals later. 

\begin{refproposition}{\mainref{prop:var-est-consistency}}
	\varestconsistency
\end{refproposition}

\begin{proof}
Let's calculate the variance of $\evb$. 
Using the exact same calculations as in calculating the variance of the modified Horvitz-Thompson estimator, we get 
\begin{align*}
	\Var[\Bigg]{\frac{n}{\lamH} \cdot \evb} 
	&=
	\Var[\Bigg]{ 
		\frac{n}{\lamH} \cdot 
		\frac{\lamV}{n} \bracket[\Bigg]{
			 \frac{1}{n}\sum_{i=1}^n Y_i^2 \braces[\Big]{
			 		\frac{\indicator{E_{(i,1)}}}{\Pr{E_{(i,1)}}}
			 		+ \frac{\indicator{E_{(i,0)}}}{\Pr{E_{(i,0)}}} 
		 		}  
		 	} 
		} \\
	&= \paren[\Big]{ \frac{\lamV}{\lamH} }^2 \cdot 
		\Var[\Big]{ 
			\frac{1}{n}\sum_{i=1}^n Y_i^2 \braces[\Big]{
			\frac{\indicator{E_{(i,1)}}}{\Pr{E_{(i,1)}}}
			+ \frac{\indicator{E_{(i,0)}}}{\Pr{E_{(i,0)}}} 
			}
		} \\
	&= \paren[\Big]{ \frac{\lamV}{\lamH} }^2 \cdot 
	\Var[\Big]{ 
		\frac{1}{n}\sum_{i=1}^n
			y_i(e_1)^2 \frac{\indicator{E_{(i,1)}}}{\Pr{E_{(i,1)}}}
			+ y_i(e_0)^2 \frac{\indicator{E_{(i,0)}}}{\Pr{E_{(i,0)}}} 
		} \\
	&= \paren[\Big]{ \frac{\lamV}{\lamH} }^2 \cdot 
	\frac{1}{n^2} \sum_{i=1}^n \sum_{j = 1}^n \sum_{\substack{k \in \setb{0,1} \\ \ell \in \setb{0,1}}} C_{(i,k),(j,l)} \cdot y_i^2(e_k) y_j^2(e_l)
	\enspace.
\end{align*}
By the proof of Theorem~\mainref{thm:variance-analysis-finite-sample} it follows that $\lamV \leq \cleanconst \lamH$ so that the first term in the expression is at most $\cleanconst^2$. 
Next, notice the similarity of the sum in the above expression and the expression for the variance of the effect estimator in \eqref{eq:var_expression}.
In fact, they differ only in the extra $(-1)^{k-l}$ factor that is appearing in \eqref{eq:var_expression} and the extra squares on the outcomes that are appearing in the expression above.
To make these two expressions appear exactly the same, imagine we define the following potential outcomes $y'$.
\[
y_i'(e_0) := y_i^2(e_0) \quad\quad y_i'(e_1) := - y_i^2(e_1)
\enspace.
\]
Then, notice that the variance of $\eate$ for these particular outcomes $y'$ is exactly the same as the expression above (if we disregard the constant factor that multiplies it). 
Moreover, for the $y'$ outcomes we have
\[
\frac{1}{n} \sum_{i=1}^n (y_i'(e_0))^2 + \frac{1}{n} \sum_{i=1}^n (y_i'(e_1))^2 = \frac{1}{n} \sum_{i=1}^n y_i^4(e_0) + \frac{1}{n} \sum_{i=1}^n y_i^4(e_1) = \bigO{1}\,,
\]
where the last step follows from Assumption~\mainref{assumption:bounded-fourth-moments}. Thus, we can apply Theorem~\mainref{thm:variance-analysis-finite-sample} for the $y'$ outcomes to get 
\begin{align*}
	\Var[\Bigg]{\frac{n}{\lamH}\evb}  \leq \cleanconst^2 \cdot \frac{\cleanconst \cdot \lamH}{n} \paren[\Big]{\frac{1}{n} \sum_{i=1}^n (y_i'(e_0))^2 + \frac{1}{n} \sum_{i=1}^n (y_i'(e_1))^2} = \bigO[\Big]{\frac{\lamH}{n}}
	\enspace.
\end{align*}
By Proposition~\mainref{prop:unbiased-var-est}, the variance estimator is unbiased for the conservative upper bound.
Thus, an application of Chebyshev's inequality now implies the desired result.
\end{proof}

Next, we prove that the above approximation is strong enough to imply convergence of the ratio $\vb/\evb$ to 1.

\begin{proposition} \label{prop:var-ratio-convergence}
	Under Assumptions~\mainref{assumption:ani-model}-\mainref{assumption:not-superefficient},
	the ratio of the variance estimator and the variance bound converges to 1 in probability: $\vb / \evb \xrightarrow{p} 1$.
\end{proposition}
\begin{proof}
It will be more convenient to prove the equivalent claim $\evb/ \vb \xrightarrow{p} 1$.
By Proposition~\mainref{prop:var-est-consistency} we get
\[
\abs[\Big]{\frac{\evb}{\vb} - 1} 
= \abs[\Big]{\frac{\evb - \vb}{\vb}} 
\leq \frac{\bigOp[\Big]{(\lamH/ n)^{3/2}}}{\Omega\lp(\lamH/n\rp)} 
\leq \bigOp{\sqrt{\lamH/n}} 
= \littleOp{1} 
\enspace,
\]
where the inequality uses Assumption~\mainref{assumption:not-superefficient} to lower bound the denominator and Proposition~\mainref{prop:var-est-consistency} to upper bound the denominator.
Finally, the second inequality uses Assumption~\mainref{assumption:lam-bound} which states that $\lamH = \littleO{n}$.
\end{proof}

\subsection{Chebyshev-Style Confidence Intervals (Theorem~\mainref{theorem:intervals-cover-chebyshev})}

We restate the Theorem for convenience.

\begin{reftheorem}{\mainref{theorem:intervals-cover-chebyshev}}
	\intervalschebyshev
\end{reftheorem}

\begin{proof}
	We will use Chebyshev's inequality. Indeed, let's fix some $\epsilon \in (0,1)$. By Chebyshev's inequality, together with the Proposition~\mainref{prop:conservative-vb-is-tight}, we have that
	\begin{align}\label{eq:decomposition}
	\Pr{ \ate \notin C_{\textrm{Cheb}}(\alpha) } &= 
	\Pr[\Bigg]{\abs{\eate - \ate} > \frac{\sqrt{\evb}}{\sqrt{\alpha}}, \abs[\Big]{\frac{\evb}{\vb} -1} \leq \epsilon}\\
	&\quad\quad +
	\Pr[\Bigg]{\abs{\eate - \ate} > \frac{\sqrt{\evb}}{\sqrt{\alpha}}, \abs[\Big]{\frac{\evb}{\vb} -1} > \epsilon}\nonumber\\
	&\leq \Pr[\Bigg]{\abs{\eate - \ate} > \frac{\sqrt{\vb}\sqrt{1-\epsilon}}{\sqrt{\alpha}}}+
	\Pr[\Bigg]{\abs{\eate - \ate} > \frac{\sqrt{\evb}}{\sqrt{\alpha}}, \abs[\Big]{\frac{\evb}{\vb} -1} > \epsilon}\nonumber\\
	&\leq \frac{\alpha \Var{\eate}}{(1 -\epsilon)\evb} + 
	\Pr[\Bigg]{\abs{\eate - \ate} > \frac{\sqrt{\evb}}{\sqrt{\alpha}}, \abs[\Big]{\frac{\evb}{\vb} -1} > \epsilon}
	\end{align}
	In the last step, we used Chebyshev's inequality. Now, using Proposition~\mainref{prop:conservative-vb-is-tight} we have that
	\[
		\frac{\alpha \Var{\eate}}{(1 -\epsilon)\evb} \leq \frac{\alpha}{1-\epsilon}
	\]
	Also, by Proposition~\suppref{prop:var-ratio-convergence}(which we can apply since Assumptions~\mainref{assumption:ani-model}-\mainref{assumption:not-superefficient} are satisfied) we obtain
	\[
		\limsup_{n \to \infty}\Pr[\Bigg]{\abs{\eate - \ate} > \frac{\sqrt{\evb}}{\sqrt{\alpha}}, \abs[\Big]{\frac{\evb}{\vb} -1} > \epsilon}
		\leq \limsup_{n \to \infty}\Pr[\Bigg]{\abs[\Big]{\frac{\evb}{\vb} -1} > \epsilon} = 0
	\]
	Overall, \eqref{eq:decomposition} implies that
	\[
	\limsup_{n\to \infty} \Pr{ \ate \notin C_{\textrm{Cheb}}(\alpha) } \leq \frac{\alpha}{1 - \epsilon}
	\]
	This holds for any $\epsilon \in (0,1)$. Taking $\epsilon \to 0$ finishes the claim. 
\end{proof}

\subsection{Central Limit Theorem (Theorem~\mainref{thm:clt})} \label{sec:supp-clt-proof}

To prove a Central Limit Theorem, we use Stein's method with dependency graphs, which is a common technique in the literature on causal inference under interference.
We will use the following result found in the survey of \citet{Ross2011Fundamentals}.

\begin{definition}[Definition 3.1 of \citet{Ross2011Fundamentals}]
	A collection of random variables $X_1 \dots X_n$ has \emph{dependency neighborhoods} $N(i) \subseteq \setb{1 \dots n}$ if $X_i$ is independent of $\setb{X_j}_{j \notin N(i)}$
\end{definition}

\begin{proposition}[Theorem of 3.5 \citet{Ross2011Fundamentals}] \label{prop:steins-method}
	Let $X_1 \dots X_n$ be random variables with $\E{ X_i^4 } < \infty$, $\E{X_i} = 0$, and $\sigma^2 = \Var{ \sum_{i=1}^n X_i }$, and define $W = \sum_{i=1}^n X_i / \sigma$.
	Suppose that $X_1 \dots X_n$ has dependency neighborhoods with $D = \max_{i \in [n]} \abs{ N(i) }$.
	Then for a standard normal random variable $Z$, 
	\[
	d_{W}(W,Z) \leq 
	\frac{ D^2 }{\sigma^3} \sum_{i=1}^n \E{ \abs{X_i}^3 }
	+ \sqrt{\frac{26}{\pi}} \frac{D^{3/2}}{\sigma^2}
	\sqrt{ \sum_{i=1}^n \E{X_i^4} }
	\enspace,
	\]
	where $d_{W}$ is the Wasserstein distance.
\end{proposition}

The proof of the following theorem relies on using these results applied to the modified Horvitz--Thompson estimator.

\begin{reftheorem}{\mainref{thm:clt}}
	\clttheorem
\end{reftheorem}

\begin{proof}
	We consider the case where the graph has at least one edge so that $\lamH \geq 2$, otherwise there is no interference and standard results may be applied.
	Recall that the Horvitz--Thompson estimator may be written as the average of individual effect estimators, 
	\[
	\eate = \frac{1}{n} \sum_{i=1}^n \eate_i
	\quadwhere
	\eate_i = Y_i \paren[\Big]{ \frac{\indicator{E_{(i,1)}}}{\Pr{E_{(i,1)}}} - \frac{\indicator{E_{(i,0)}}}{\Pr{E_{(i,0)}}} }
	\enspace.
	\]
	Under the arbitrary neighborhood interference model (Assumption~\mainref{assumption:ani-model}), we have that the individual effect estimators take the form
	\[
	\eate_i = y_i(e_1) \cdot \frac{\indicator{E_{(i,1)}}}{\Pr{E_{(i,1)}}}
	- y_i(e_0) \cdot \frac{\indicator{E_{(i,0)}}}{\Pr{E_{(i,0)}}}
	\enspace.
	\]
	For each subject $i \in [n]$, define the random variable $X_i$ to be the error of the individual effect estimator, i.e.
	\[
	X_i 
	= \eate_i - \ate_i 
	= y_i(e_1) \cdot \paren[\Bigg]{  \frac{\indicator{E_{(i,1)}}}{\Pr{E_{(i,1)}}} - 1}
	- y_i(e_0) \cdot \paren[\Bigg]{ \frac{\indicator{E_{(i,0)}}}{\Pr{E_{(i,0)}}} - 1}
	\enspace.
	\]
	Define $\sigma^2 = \Var{\sum_{i=1}^n X_i}$ and $W = \frac{1}{\sigma} \sum_{i=1}^n X_i$.
	By construction, we have that the standardized estimator is equal to $W$, i.e.
	\[
	\frac{\eate - \ate}{\sqrt{ \Var{\eate} }}
	= W
	\enspace.
	\]
	We now seek to show that the conditions of Proposition~\suppref{prop:steins-method} hold.
	There will be four main conditions that we need to check.
	
	First, we need to check that the variables $X_i$ have mean zero.
	To this end, observe that by Lemma~\mainref{lemma:prob-of-desired-exposure-events}, the probabilities $\Pr{E_{(i,k)}}$ are positive so that the individual effect estimators are unbiased for the individual effects, i.e. $\E{ \eate_i } = \ate_i$.
	This means that $\E{ X_i } = 0$.

	Second, we need to bound the third and fourth moments of the individual variables $X_i$.
	To this end, consider a positive integer $m$ and observe that
	\begin{align*}
		\E[\big]{ \abs{X_i}^m }
		&= \E[\Bigg]{ \abs[\Bigg]{
				y_i(e_1) \cdot \paren[\Bigg]{ \frac{\indicator{E_{(i,1)}}}{\Pr{E_{(i,1)}}} - 1}
				- y_i(e_0) \cdot \paren[\Bigg]{ \frac{\indicator{E_{(i,0)}}}{\Pr{E_{(i,0)}}} - 1}}^m 
			} \\
		&\leq 2^{m-1} \E[\Bigg]{ \abs[\Bigg]{y_i(e_1) \cdot \paren[\Bigg]{ \frac{\indicator{E_{(i,1)}}}{\Pr{E_{(i,1)}}} - 1}}^m }\\
		&\quad\quad+ 2^{m-1} \E[\Bigg]{ \abs[\Bigg]{y_i(e_0) \cdot \paren[\Bigg]{ \frac{\indicator{E_{(i,0)}}}{\Pr{E_{(i,0)}}} - 1}}^m } \\
		&\leq 2^{m-1} \paren[\Big]{ \abs{ y_i(e_1) }^m + \abs{ y_i(e_0) }^m  } \cdot \max_{k \in \setb{0,1}} \E[\Bigg]{ \abs[\Bigg]{  \frac{\indicator{E_{(i,k)}}}{\Pr{E_{(i,k)}}} - 1  }^m  }
	\end{align*}
	We can calculate this last term directly as
	\begin{align*}
		\E[\Bigg]{ \abs[\Bigg]{  \frac{\indicator{E_{(i,k)}}}{\Pr{E_{(i,k)}}} - 1  }^m  }
		&= (1 - \Pr{ E_{(i,k)} }) + \Pr{ E_{(i,k)} } \cdot \paren[\Big]{ \frac{1}{\Pr{E_{(i,k)}}}  - 1 }^m \\
		&\leq 1 + \frac{1}{\Pr{E_{(i,k)}}^{m-1}} \\
		&\leq \bigO{\lamH^{m-1}} \enspace,
	\end{align*}
	where the last inequality follows from the analysis that $1 / \Pr{ E_{(i,k)} } \leq \bigO{\lamH}$.
	Thus, we have that
	\[
	\E[\big]{ \abs{X_i}^m } \leq \paren[\Big]{ \abs{ y_i(e_1) }^m + \abs{ y_i(e_0) }^m  } \cdot \bigO{ \lamH^{m-1} }
	\enspace.
	\]
	
	Next, we derive a lower bound on $\sigma$ using Assumption~\mainref{assumption:not-superefficient}, which states that the variance of the modified Horvitz--Thompson estimator is bounded below as $\Var{\eate} \geq \bigOmega{ \lamH / n }$.
	To this end, observe that 
	\[
	\sigma^2 
	= \Var[\Big]{ \sum_{i=1}^n X_i }
	= \Var[\Big]{ \sum_{i=1}^n \eate_i - \ate_i } 
	= n^2 \Var[\Big]{ \frac{1}{n} \sum_{i=1}^n \eate_i } 
	= n^2 \Var{ \eate }
	\geq \bigOmega{n \cdot \lamH}
	\enspace.
	\]
	
	Finally, we discuss the dependency neighborhoods associated to these random variables.
	For a variable $i \in [n]$, the neighborhood $N(i)$ may be taken to be the 2-hop neighborhood in the conflict graph $\cH$, i.e.
	\[
	N(i) = \setb[\big]{ j \in [n] : \mathcal{N}(i) \cap \mathcal{N}(j) \neq \emptyset }
	\enspace.
	\]
	Let us verify that this indeed forms a dependency neighborhood.
	First, observe that each $X_i$ is a function of the independent random variables $U_\ell : \ell \in \mathcal{N}(i)$.
	Fix $i \in [n]$ and consider the collection of random variables $\setb{X_j}_{j \notin N(i)}$.
	Let us use the shorthand $R_i = [n] \setminus N(i)$.
	The random variables in this collection are functions of the independent random variables $U_k : k \in \mathcal{N}( R_i )$.
	Note that $\setb{ U_\ell : \ell \in \mathcal{N}(i) }$ and $\setb{ U_k : k \in \mathcal{N}(R_i) }$ are, by definition, disjoint sets of independent random variables.
	Thus, $X_i$ is independent from the collection of random variables $\setb{X_j}_{j \notin N(i)}$ so that $N(i)$ forms a valid dependency graph.
	Thus, the maximum degree of the dependency graph may be bounded as $D \leq d_{\max}(\cH)^2$.
	
	We are now ready to plug these results into Stein's method via dependency graphs.
	By directly applying Proposition~\suppref{prop:steins-method}, we obtain the bound on the Wasserstein distance:
	\begin{align*}
		d_{W}(W,Z) 
		&\leq \frac{ D^2 }{\sigma^3} \sum_{i=1}^n \E{ \abs{X_i}^3 }
		+ \sqrt{\frac{26}{\pi}} \frac{D^{3/2}}{\sigma^2}
		\sqrt{ \sum_{i=1}^n \E{X_i^4} } \\
			\intertext{by Proposition~\ref{prop:steins-method}}\\
		&\leq 
			\bigO[\Bigg]{ \frac{ d_{\max}(\cH)^4 \lamH^{2} }{\paren[\Big]{ n \cdot \lamH }^{3/2} } } 
			\cdot \sum_{i=1}^n \abs{y_i(e_1)}^3 + \abs{ y_i(e_0) }^3 \\
			&\qquad+ \bigO[\Bigg]{ \frac{d_{\max}(\cH)^{3} \lamH^{3/2} }{ n \cdot \lamH } } 
			\cdot \sqrt{\sum_{i=1}^n \abs{y_i(e_1)}^4 + \abs{ y_i(e_0) }^4} \\
				\text{by calculations above}\\
		&= \bigO[\Bigg]{ \frac{ d_{\max}(\cH)^4 \lamH^{1/2} }{ n^{1/2} } } \cdot \frac{1}{n} \sum_{i=1}^n \abs{y_i(e_1)}^3 + \abs{ y_i(e_0) }^3  \\
		&\qquad+ \bigO[\Bigg]{ \frac{d_{\max}(\cH)^{3} \lamH^{1/2} }{ n^{1/2} } } \cdot \sqrt{ \frac{1}{n}\sum_{i=1}^n \abs{y_i(e_1)}^4 + \abs{ y_i(e_0) }^4} \\
			\intertext{by rearranging}\\
		&= \bigO[\Bigg]{ \frac{ d_{\max}(\cH)^4 \lamH^{1/2} }{ n^{1/2} } }\\
			\intertext{by Assumption~\mainref{assumption:bounded-fourth-moments}} \\
		&= \bigO[\Bigg]{ \paren[\Big]{ \frac{d_{\max}(\cH)^9}{n} }^{1/2} }\\
			\intertext{by using $\lamH \leq d_{\max}(\cH)$.}
	\end{align*}
	Thus, the standardized estimator converges to a normal distribution in Wasserstein metric if $d_{\max}(\cH)^9 \lamH = \littleO{n}$.
	This implies convergence in distribution, which completes the proof.
\end{proof}

\subsection{Wald-Type Confidence Intervals (Corollary~\mainref{corollary:intervals-cover-wald})}

\begin{refcorollary}{\mainref{corollary:intervals-cover-wald}}
	\intervalswald
\end{refcorollary}

\begin{proof}
	We start again by fixing some $\epsilon \in (0,1)$. We can write
	\begin{align}
		&\Pr{ \ate \notin C_{\textrm{Wald}}(\alpha) } \\
		&= \Pr[\Bigg]{\abs{\eate - \ate} >  \Phi^{-1}(1 - \alpha/2)\cdot\sqrt{\evb}, \abs[\Big]{\frac{\evb}{\vb} -1} \leq \epsilon} \\
			&\qquad +
			\Pr[\Bigg]{\abs{\eate - \ate} >  \Phi^{-1}(1 - \alpha/2)\cdot \sqrt{\evb}, \abs[\Big]{\frac{\evb}{\vb} -1} > \epsilon}\nonumber\\
		&\leq \Pr[\Bigg]{\abs{\eate - \ate} >  \Phi^{-1}(1 - \alpha/2)\cdot \sqrt{\vb}\sqrt{1-\epsilon}} \\
			&\qquad +
			\Pr[\Bigg]{\abs{\eate - \ate} >  \Phi^{-1}(1 - \alpha/2) \cdot \sqrt{\evb}, \abs[\Big]{\frac{\evb}{\vb} -1} > \epsilon}\nonumber\\
		&\leq \Pr[\Bigg]{\abs{\eate - \ate} > \Phi^{-1}(1 - \alpha/2)\cdot \sqrt{\Var{\eate}}\sqrt{1-\epsilon} } \\
			&\qquad +
			\Pr[\Bigg]{\abs{\eate - \ate} > \Phi^{-1}(1 - \alpha/2) \cdot \sqrt{\vb}, \abs[\Big]{\frac{\evb}{\vb} -1} > \epsilon}\enspace. \label{eq:decomposition_wald}
	\end{align}
	The last step follows since by Proposition~\mainref{prop:conservative-vb-is-tight} we have $\Var{\eate} \leq \evb$.
	By Proposition~\suppref{prop:var-ratio-convergence} (which we can apply since Assumptions~\mainref{assumption:ani-model}-\mainref{assumption:not-superefficient} are satisfied) we obtain
	\[
	\limsup_{n \to \infty}\Pr[\Bigg]{\abs{\eate - \ate} > \Phi^{-1}(1 - \alpha/2) \cdot \sqrt{\vb}, \abs[\Big]{\frac{\evb}{\vb} -1} > \epsilon}
	\leq \limsup_{n \to \infty}\Pr[\Bigg]{\abs[\Big]{\frac{\evb}{\vb} -1} > \epsilon} = 0
	\enspace.
	\]
	Also, since Assumptions~\mainref{assumption:ani-model}-\mainref{assumption:not-superefficient} are satisfied and $d_{\textup{max}}(\cH) = o(n^{1/9})$, we can apply the result of Theorem~\mainref{thm:clt}, which yields
	
	\begin{align*}
		&\lim_{n \to \infty} \Pr[\Bigg]{\abs{\eate - \ate} > \Phi^{-1}(1 - \alpha/2)\cdot \sqrt{\Var{\eate}}\sqrt{1-\epsilon} }\\
		&\qquad= 
		\lim_{n \to \infty} \Pr[\Bigg]{\frac{\eate - \ate}{\sqrt{\Var{\eate}}} \notin \lp[- \Phi^{-1}(1 - \alpha/2)\cdot \sqrt{1-\epsilon},\Phi^{-1}(1 - \alpha/2)\cdot \sqrt{1-\epsilon}\rp]  }\\
		&\qquad= 2 \cdot \paren[\Bigg]{1-\Phi\paren[\Big]{\Phi^{-1}(1 - \alpha/2)\cdot\sqrt{1-\epsilon}}}
		\enspace.
	\end{align*}
	Thus, overall using \eqref{eq:decomposition_wald} we get 
	\[
	\limsup_{n\to \infty} \Pr{ \ate \notin C_{\textrm{Wald}}(\alpha) }  \leq 2 \cdot \paren[\Bigg]{1-\Phi\paren[\Big]{\Phi^{-1}(1 - \alpha/2)\cdot\sqrt{1-\epsilon}}}
	\enspace.
	\]
	Since this holds for every $\epsilon \in (0,1)$, the above bound also holds when taking $\epsilon \to 0$, which yields by continuity of $\Phi$
	\[
	2 \cdot \lim_{\epsilon \to 0} \paren[\Bigg]{1-\Phi\paren[\Big]{\Phi^{-1}(1 - \alpha/2)\cdot\sqrt{1-\epsilon}}} = 2 \cdot \paren[\Big]{1-\Phi\paren[\Big]{\Phi^{-1}(1 - \alpha/2)}} =  \alpha
	\enspace.
	\qedhere
	\]
\end{proof}

\subsection{Counterexample for Central Limit Theorem (Proposition~\mainref{prop:clt-counter-example})} \label{sec:supp-clt-counterexample}

In this section, we provide the explicit construction of the sequence of graphs where the standardized estimator fails to converge in distribution. 

\begin{refproposition}{\mainref{prop:clt-counter-example}}
	\cltcounter
\end{refproposition}

\begin{proof}
Our strategy will be to define two sequences of graphs $G_1^{(n)}, G_2^{(n)}$ and corresponding sequences of potential outcomes, so that $\eate$ has different limiting distributions across the two sequences. Then, we can construct a third sequence by alternating between these two, which will clearly not have a limiting distribution. Throughout this section, we will use the notation $X = o_P(1)$ to denote that $X$ converges to $0$ in probability. 

We start by describing the first sequence $G_1^{(n)} = ([n],E_n)$. This is a star graph on $n$ vertices, meaning that node $1$ is the center and there is an edge $(1,i)$ for all $i \in \{2,\ldots,n\}$. Since we focus on the Average Direct Treatment Effect, the conflict graph $\mathcal{H}$ is $G_1^{(n)}$ itself. Standard calculations show that the maximum eigenvalue $\lamH = \sqrt{n-1}$ and the eigenvalue ordering $\eigpi$ has $1$ as the most important node, followed by all the rest. Thus, all nodes $\geq 2$ have $1$ as a parent, and this is the only parent they have.

We consider the following sequence of outcomes
\[
y_i(e_0) = 0 , \forall i \in [n] \quad \quad y_i(e_1) = 1 \forall i \in [n]
\enspace.
\]
Clearly, the outcomes satisfy Assumption~\mainref{assumption:bounded-fourth-moments}.
Also, $\lamH = \Theta(\sqrt{n})$, as the premise of the proposition suggests.
We will calculate the limiting distribution of the standardized estimator $(\eate - \ate)/ \sqrt{\Var{\eate}}$. 
We start by calculating the variance, using the expressions for the covariances established in Lemma~\suppref{lem:Cij_expression}.
For simplicity, we consider the Conflict Graph Design with $r=1$. 

\begin{align*}
n^2 \Var{\eate} &= (2 \lamH - 1) + \sum_{i=2}^n \paren[\Bigg]{\frac{2\lamH}{1 - \frac{1}{\lamH}}-1}\\
&\quad\quad + \sum_{i=2}^n \paren{-1} + \sum_{j , k \geq 2, j \neq k} \paren[\Big]{(1-1/\lamH)^{-1} - 1}\\
&= 2 \sqrt{n-1} - 1 + 2 n \sqrt{n-1} -2n + \frac{(n-1)(n-2)}{\sqrt{n-1}} + o(n^{3/2})\\
&\asymp 3 n^{3/2} \enspace.
\end{align*}
In the above, the notation $a_n \asymp b_n$ means that $a_n/b_n \to 1$ as $n \to \infty$. 
Next, observe that the error of the estimator can be written as 
\begin{align*}
n (\eate - \ate) &= \paren[\Big]{\frac{\indicator{U_1 = e_1}}{\Pr{U_1 = e_1}}-1} + \sum_{i \geq 2} \paren[\Big]{\frac{\indicator{U_1 = *, U_i = e_1}}{\Pr{U_1 = *, U_i = e_1}}-1}\\
&= \paren[\Big]{\frac{\indicator{U_1 = e_1}}{\Pr{U_1 = e_1}}-1} + \indicator{U_1 = *}\sum_{i \geq 2} \paren[\Big]{\frac{\indicator{ U_i = e_1}}{\Pr{U_1 = *, U_i = e_1}}-1} \\
&\quad\quad+ \indicator{U_1 \neq *} (-(n-1))\\
&= \paren[\Big]{\frac{\indicator{U_1 = e_1}}{\Pr{U_1 = e_1}}-1} + \indicator{U_1 = *}\underbrace{\sum_{i \geq 2} \paren[\Big]{\frac{\indicator{ U_i = e_1}}{\Pr{U_1 = *, U_i = e_1}}-(1-1/\lamH)^{-1}}}_{S_n}  \\
&\quad + \indicator{U_1 = *} \underbrace{(n-1)\paren[\Big]{(1-1/\lamH)^{-1} - 1}}_{\asymp \sqrt{n}} + \underbrace{\indicator{U_1 \neq *} (-(n-1))}_{o_P(1)}
\enspace.
\end{align*}
In the above, the second term, which we have denoted by $S_n$, is a sum of $n-1$ independent zero mean random variables. The last term is $o_P(1)$ since $\Pr{U_1 \neq *} = 1/\lamH = 1/\sqrt{n-1} \to 0$ as $n \to \infty$. 
For $S_n$, we can use the standard Central Limit Theorem for independent variables, which yields 
\[
	\frac{S_n}{\sqrt{\Var{S_n}}} \xrightarrow{d} \mathcal{N}(0,1)
	\enspace.
\]
Next, we calculate $\Var{S_n}$.
Because it is a sum of independent random variables, we easily get
\[
	\Var{S_n} = (n-1)(1-1/\lamH)^{-2}(2\lamH - 1) \asymp 2 n^{3/2}
	\enspace.
\]
Thus, we have that 
\[
	\frac{n (\eate - \ate)}{\sqrt{\Var{S_n}}} 
	= \frac{S_n}{\sqrt{\Var{S_n}}} + O(n^{-1/4}) + o_P(1) 
	\xrightarrow{d} \mathcal{N}(0,1)
	\enspace.
\]
In the final step, we used Slutsky's Theorem.
However, notice that $\Var{S_n}$ and $\Var{\eate}$ are not asymptotically the same. 
Indeed, we have that 
\[
	\frac{\eate - \ate}{\sqrt{\Var{\eate}}} 
	= \frac{n(\eate - \ate)}{\sqrt{n^2 \Var{\eate}}} 
	= \frac{n (\eate - \ate)}{\sqrt{\Var{S_n}}} \frac{\sqrt{\Var{S_n}}}{\sqrt{n^2 \Var{\eate}}}
	\enspace.
\]
By the preceding calculations, 
\[
    \frac{\sqrt{\Var{S_n}}}{\sqrt{n^2 \Var{\eate}}} \to \sqrt{\frac{2}{3}} \enspace.
\]
Thus, again by using Slutsky's Theorem, we conclude that 
\[
    \frac{\eate - \ate}{\sqrt{\Var{\eate}}} \xrightarrow{d} \mathcal{N}(0,2/3) \enspace.
\]
Thus, the limiting distribution has variance strictly smaller than $1$. The reason for this ``loss of variance'' becomes clear once we examine the role of node $1$. It is influencing all nodes in the graph, meaning that it's fluctuations account for a constant fraction of the overall variance of the estimator. On the other hand, the terms that are affected by the fluctuations of $1$ converge to $0$ in probability and hence do not influence the limiting distribution. This is the part of the variance that was ``lost'' in the limit. 
Such phenomena seem to arise whenever there is a large discrepancy between the maximum eigenvalue and the largest degree in the graph. This is because nodes with large degree can affect the overall variance by a constant factor, but this effect is lost in the limiting distribution. 
Thus, it is clear that determining the limiting distribution is a significantly more subtle problem than point estimation of $\ate$. 

\begin{figure}[t]
    \centering
    \input{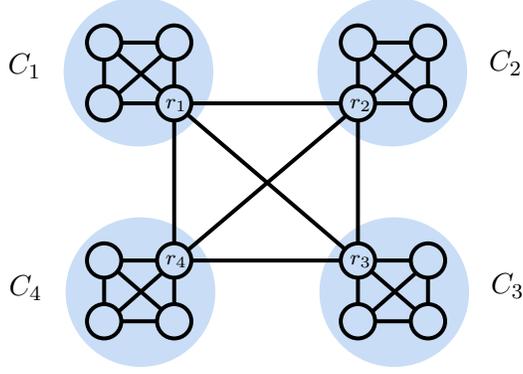}
    \caption{Graph $G_{16}^{(2)}$ used in the counterexample for the Central Limit Theorem. This graph has $4$ cliques of size $4$.}
    \label{fig:clt-counterexample}
\end{figure}

Going back to the proof, we construct the second sequence of graphs $G_n^{(2)}$. 
For simplicity, we will focus on the case where $\sqrt{n}$ is an integer. Otherwise, we replace $\sqrt{n}$ with $\lceil \sqrt{n}\rceil$. That way, we might end up with $n - \sqrt{n}$ vertices in the end, so we can add $O(\sqrt{n})$ independent vertices, which do not affect the limiting distribution, in order to have a total of $n$ vertices. 
Since nothing changes asymptotically, to avoid those complications we assume $\sqrt{n}$ is an integer from now on. 
    
To construct $G_2^{(n)}$, consider $\sqrt{n}$ copies of the complete graph on $\sqrt{n}$ vertices $K_{\sqrt{n}}$ and let $C_i$ be the $i$-th copy of $K_{\sqrt{n}}$. 
Let $r_i$ be a distinguished node in $C_i$. For all $i \neq j$, connect $r_i$ and $r_j$ by an edge. This is the topology of $G_n^{(2)}$ and Figure~\suppref{fig:clt-counterexample} contains an example for $n = 16$. 

First of all, notice that the maximum degree in this graph is $2 \sqrt{n}$ and the minimum degree is $\sqrt{n}$. Thus, for the maximum eigenvalue $\lamH$ of $G_2^{(n)}$, we have that $\lamH = \Theta(\sqrt{n})$, which again satisfies the premise of our Proposition. 
It's also easy to see that in the eigenvalue ordering $\eigpi$, the nodes $r_1,\ldots, r_{\sqrt{n}}$ come first in arbitrary order and then the rest of the variables follow. Thus, every node $r_i$ has the rest of the nodes in $C_i$ as children. 
We will calculate the limiting distribution of $(\eate - \ate) / \sqrt{\Var{\eate}}$ for this graph sequence under the Conflict Graph Design. 

Next, let us define the potential outcomes. We set 
\[
y_i(e_0) = 0 , \forall i \in [n] \quad \quad y_i(e_1) = 1, \forall i \in [n]
\enspace.
\]
It is clear that these outcomes satisfy Assumption~\mainref{assumption:bounded-fourth-moments}. 

We continue with some calculations involving the variance of $\eate$.   
In particular, denoting $p_i = \Pr{E_{(i,1)}}$ for every unit $i$, we have 
\begin{align*}
n^2 \Var{\eate} &= \Var[\Bigg]{\sum_{i=1}^{\sqrt{n}} \sum_{j \in C_i} \lp(\frac{\indicator{E_{(j,1)}}}{p_j} - 1\rp)}\\
&= 
\Var[\Bigg]{\sum_{i=1}^{\sqrt{n}} \sum_{j \in C_i, j \neq r_i} \lp(\frac{\indicator{E_{(j,1)}}}{p_j} - 1\rp) + \sum_{i=1}^{\sqrt{n}} \lp(\frac{\indicator{E_{(r_i,1)}}}{p_{r_i}} - 1\rp)} \\
&= \Var[\Bigg]{\sum_{i=1}^{\sqrt{n}} \sum_{j \in C_i, j \neq r_i} \lp(\frac{\indicator{E_{(j,1)}}}{p_j} - 1\rp)} + \Var[\Bigg]{\sum_{i=1}^{\sqrt{n}} \lp(\frac{\indicator{E_{(r_i,1)}}}{p_{r_i}} - 1\rp)}\\
& + \Cov[\Bigg]{\sum_{i=1}^{\sqrt{n}} \sum_{j \in C_i, j \neq r_i} \lp(\frac{\indicator{E_{(j,1)}}}{p_j} - 1\rp) , \sum_{i=1}^{\sqrt{n}} \lp(\frac{\indicator{E_{(r_i,1)}}}{p_{r_i}} - 1\rp)}
\enspace.
\end{align*}
Let us focus on the first term, as the remaining ones will be negligible as we will see in the sequel. 
Notice that since we have excluded $r_i$ from each sum, the sums corresponding to different clusters $C_i$ are independent. This is because for $j \in C_i$, with $j \neq r_i$, the random variable $\indicator{E_{(j,1)}}$ only depends on $U$ variables of $C_i$. Thus, we get
\begin{align*}
	\Var[\Bigg]{\sum_{i=1}^{\sqrt{n}} \sum_{j \in C_i, j \neq r_i} \lp(\frac{\indicator{E_{(j,1)}}}{p_j} - 1\rp)} &= 
	\sum_{i=1}^{\sqrt{n}} \Var[\Bigg]{\sum_{j \in C_i, j \neq r_i} \lp(\frac{\indicator{E_{(j,1)}}}{p_j} - 1\rp)} = 
	\sqrt{n} v \enspace,
\end{align*}
where
\[
v = v_i := \Var{\sum_{j \in C_i, j \neq r_i} \lp(\frac{\indicator{E_{(j,1)}}}{p_j} - 1\rp)}
\enspace.
\]
The preceding observation follows
by symmetry, since all $C_i$ blocks will have the same variance $v$. 
Note also that an easy calculation shows that $v = \Theta(n)$, since it is essentially a clique on $\sqrt{n}$ vertices. Thus, the first term is $\Omega(n^{3/2})$.
Now, let's show that the remaining terms are negligible. The second term is essentially the variance of a clique of $\sqrt{n}$ vertices,
where the vertices are the $r_i$ nodes. Thus, it can be upper bounded by $O(n)$. 
For the third term, notice that each $r_i$ is independent of all nodes in $C_j$ except when $j = i$. Thus, we can write

\begin{align*}
&\Covsym\lp(\sum_{i=1}^{\sqrt{n}} \sum_{j \in C_i, j \neq r_i} \lp(\frac{\indicator{E_{(j,1)}}}{p_j} - 1\rp) , \sum_{i=1}^{\sqrt{n}} \lp(\frac{\indicator{E_{(r_i,1)}}}{p_{r_i}} - 1\rp)\rp)  \\
&\qquad= \sum_{i=1}^{\sqrt{n}} \sum_{j \in C_i, j \neq r_i} \Covsym\lp(\frac{\indicator{E_{(r_i,1)}}}{p_{r_i}} - 1,\frac{\indicator{E_{(j,1)}}}{p_j} - 1\rp) \\
&\qquad= \sum_{i=1}^{\sqrt{n}} \sum_{j \in C_i, j \neq r_i} (-1) \\
&\qquad= - \sqrt{n}(\sqrt{n}-1)
\enspace.
\end{align*}
Overall, this implies that $n^2 \Var{\eate} = \Theta(n^{3/2})$ and in particular
\[
n^2 \Var{\eate}
= \sqrt{n} v \paren[\big]{ 1 + \littleO{1} }
\enspace.
\]
Thus, we can substitute in the normalization of the estimator $\sqrt{n}v$ instead of $\Var{\eate}$ without changing the limiting distribution, by Slutsky's Theorem. 
The reason we do this will become apparent when we analyze the estimator. 
Indeed, we can write 
\begin{equation}\label{eq:slutsky}
\frac{\eate - \ate}{\sqrt{\Var{\eate}}} 
= \frac{\sum_{i=1}^{\sqrt{n}} \overbrace{\sum_{j \in C_i, j \neq r_i} \lp(\frac{\indicator{E_{(j,1)}}}{p_j} - 1\rp)}^{L_i}  }{\sqrt{\sqrt{n}v}} \paren[\big]{ 1 + \littleO{1} } 
+ \frac{\sum_{i=1}^{\sqrt{n}} \lp(\frac{\indicator{E_{(r_i,1)}}}{p_{r_i}} - 1\rp)}{\sqrt{\sqrt{n}v}} \paren[\big]{ 1 + \littleO{1} }
\enspace.
\end{equation}
For the first term, as we explained earlier, variables $L_i$ are completely independent from each other, since we have excluded the $r_i$ nodes. Each one of them have variance $v$, so we have the correct normalization in the denominator. Thus, we can use the standard Central Limit Theorem for independent random variables. This guarantees that
\[
\frac{\sum_{i=1}^{\sqrt{n}} L_i }{\sqrt{\sqrt{n}v}} \xrightarrow{d} \mathcal{N}(0,1)
\enspace.
\]
For the second term, we now show it converges to $0$ in probability.
Indeed, we can easily bound
\[
	-\frac{\sqrt{n}}{\sqrt{\sqrt{n}v}} \leq \frac{\sum_{i=1}^{\sqrt{n}} \lp(\frac{\indicator{E_{(r_i,1)}}}{p_{r_i}} - 1\rp)}{\sqrt{\sqrt{n}v}} \leq \frac{\sum_{i=1}^{\sqrt{n}} \frac{\indicator{E_{(r_i,1)}}}{p_{r_i}} }{\sqrt{\sqrt{n}v}}
	\enspace.
\] 
For the left hand side, by the preceding observations it decays as $O(n^{-1/4})$. Thus, it suffices to show that the right hand sice converges to $0$ in probability. 
For the right hand side, we can write

\[
	\frac{\sum_{i=1}^{\sqrt{n}} \frac{\indicator{E_{(r_i,1)}}}{p_{r_i}} }{\sqrt{\sqrt{n}v}} \leq \frac{\sum_{i=1}^{\sqrt{n}} \frac{\indicator{u_{r_i} = 1}}{p_{r_i}} }{\sqrt{\sqrt{n}v}}
	\enspace.
\]
The right hand side is a sum of $\sqrt{n}$ independent scaled Bernoulli random variables with probability $1/\lamH = \Theta(1/\sqrt{n})$. Since $p_{r_i} = \Omega(1/\sqrt{n})$, it suffices to show convergence to $0$ in probability of the following quantity:
\[
\frac{\sum_{i=1}^{\sqrt{n}} \indicator{u_{r_i}= 1}}{n^{1/4}}
\enspace.
\] 
The numerator has mean and variance $\Theta(1)$, so by standard second-moment arguments, the fraction is $o_P(1)$. Thus, using \eqref{eq:slutsky} and applying Slutksy's Theorem once more, we have shown that for the sequence $G_2^{(n)}$
\[
	\frac{\eate - \ate}{\sqrt{\Var{\eate}}} \xrightarrow{d} \mathcal{N}(0,1) \enspace.
\]
To conclude the proof of the proposition, we can now construct a third sequence of graphs

\[
G_3^{(n)} = \left\{
	\begin{array}{ll}
  G_1^{(n)} & \text{ if $n$ is even} \\
  G_2^{(n)} & \text{ if $n$ is odd}
	\end{array} 
	\right. 
\]
The outcome functions are chosen similarly corresponding to the sequence for even and odd $n$. We now notice that the sequence of random variables $(\eate - \ate)/\sqrt{\Var{\eate}}$ does not have a limit in distribution. That is because the subsequence corresponding to even $n$ converges in distribution to $\mathcal{N}(0,2/3)$ and the subsequence corresponding to odd $n$ converges to $\mathcal{N}(0,1)$. 
This concludes the proof. 
\end{proof}

	\section{Analysis of Standard Horvitz-Thompson estimator (Theorem~\mainref{thm:true_HT_d_regular})}
\label{sec:supp-true-ht-d-regular-proof}

In this Section, we present an analysis of the standard Horvitz--Thompson estimator for the direct effect under the Conflict Graph Design under the stronger condition that $d_{\max}(G) = \bigTheta{ \lamG }$.
The proof technique used here is considerably more involved than for analyzing the modified Horvitz--Thompson estimator.

Because the theorem focuses on the direct treatment effect, the conflict graph $\cH$ is simply to $G$ with added self-cycles, as discussed in Section~\mainref{sec:conflict_graph}.
Thus, $\lamH = \lamG = \norm{\mat{A}_G} +1$, where as usual $\mat{A}_G$ is the adjacency matrix of $G$, so we will simply use $\lamG$ for the remainder of this discussion. 
Given the graph $G$, we refer to its maximum degree as $\dmax$.
The assumption in Theorem~\mainref{thm:true_HT_d_regular} then becomes $\dmax \leq C \cdot \lamG$.
Throughout this section, $C$ will be this specific constant. We will use notation such as $C',C''$ to denote constants that depend on $C$ but are not computed  explicitly for convenience. Note that $C',C''$ might refer to different constants in different places in the proof.
We will denote by $N^{(G)}(i),N_b^{(G)}(i), N_a^{(G)}(i)$ the set of parents and children of $i$ in $G$, respectively. As usual, we use the notation $\widetilde{N}^{(G)}(i) = N^{(G)}(i) \cup \{i\} $ for the extended neighborhood of a node $i$. 
We also abuse notation and denote for any subset $S \subseteq [n]$ the set of parents of nodes in $S$ to be $N_b^{(G)}(S) = \cup_{i \in S} N^{(G)}_b(i)$.
We will sometimes suppress the dependence of the neighborhood on the graph to avoid overloading notation, when it is clear from the context.  

\subsection{Proof of Theorem~\mainref{thm:true_HT_d_regular}}

\begin{reftheorem}{\mainref{thm:true_HT_d_regular}}
	\standardhtdregular
\end{reftheorem}

We make a few straightforward remarks about the Conflict Graph Design specialized to the case of the Direct Effect. 
First, recall that the direct effect has contrastive interventions $\zonei = \vec{a}_i$ and $\zzeroi = \vec{0}$, where $\vec{a}_i \in \Reals^n$ is the standard basis vector with $i$-th entry being $1$ and the rest $0$ and $\vec{0} \in \Reals^n$ is the all zeroes vector.
For convenience, we restate the Conflict Graph Design below, where we make a few cosmetic simplifications due to the fact that we are focusing specifically on the direct effect.
Indeed, we replaced Line~\mainref{algline:set-intervention} in the general description of Conflict Graph Design (Algorithm~\mainref{alg:conflict-graph-design}) with a more explicit description: if $U_i = e_0$, then $i$ and all of its neighbors in $G$ receive control. If $U_i = e_1$, then $i$ receives treatment and all of its neighbors  in $G$ receive control. 
We emphasize again that this is a difference only in presentation and the design remains unchanged.

\begin{algorithm}
    \DontPrintSemicolon
    \caption{\textsc{Conflict-Graph-Design-Direct-Effect}}\label{alg:conflict-graph-design-direct-effect}
    \KwIn{Importance ordering $\pi$, maximum eigenvalue $\lamG$, and  graph $G$.}
    \KwOut{Random intervention $Z \in \mathcal{Z} = \setb{0,1}^n$}
    Sample desired exposure variables $U_1, \dots, U_n$ independently and identically as
    $$
    U_i \gets
    \left\{
    \begin{array}{lr}
        e_1 & \text{with probability } \frac{1}{2 \lamG}\\
        e_0 & \text{with probability } \frac{1}{2 \lamG}\\
        * & \text{with probability } 1 - \frac{1}{ \lamG }
    \end{array}
    \right.
    $$
    Initialize the intervention vector $Z \gets \vec{0}$. \;
    \For{$i =1 \dots n$}{
        \If{$U_i \in \setb{e_1, e_0}$ and $U_j = *$ for all $j \in N_b(i)$}{
            \If{$U_i = e_1$}{
                $Z_i \gets 1$ and $Z_j \gets 0$ for all $j \in N(i)$ \label{algline:set-treatment-direct}
            }
            \Else{
                $Z_i \gets 0$ and $Z_j \gets 0$ for all $j \in N(i)$  \label{algline:set-control-direct}
            }
        }
    }
\end{algorithm}

We next state some useful results that will be helpful for the proof of Theorem~\mainref{thm:true_HT_d_regular}.
At the end of the Section we give the proof of Theorem~\mainref{thm:true_HT_d_regular}, based on these results. 

We start by showing that for the direct effect, the desired exposure event $E_{(i,1)}$ is precisely the event of true exposure.
Define the events $F_{(i,0)} = \{d_i(Z) = d_i(\vec{0})\}$ and $F_{(i,1)} = \{d_i(Z) = d_i(\vec{a}_i)\}$.
These are exactly the events appearing in the standard Horvitz--Thompson estimator, which we seek to analyze.

\begin{lemma} \label{lemma:e1-event-is-the-same}
	Let $Z$ be the output of the Conflict Graph Design. Then $ F_{(i,1)}$ holds if and only if $E_{(i,1)}$ holds.
\end{lemma}
\begin{proof}
	Clearly, if $E_{(i,1)}$ holds, then by Lemma~\mainref{lem:desired_exposures} $d_i(Z) = d_i(\vec{a}_i)$, so $ F_{(i,1)}$ holds, so it suffices to prove the converse. Suppose $F_{(i,1)}$ holds. This means in particular that $Z_i = 1$. In the execution of Algorithm~\mainref{alg:conflict-graph-design}, the vector $Z$ is initialized to $\vec{0}$. The only time when $Z_i$ can obtain the value $1$ is when it is $i$'s turn in the For loop and Line 7 is executed. The fact that Line $7$ is executed means that $U_i = e_i$ and $U_j = *$ for all $j \in N_b(i)$. Thus, $E_{(i,1)}$ holds and the proof is complete. 
\end{proof}

% When upper bounding the variance of $\eatetrue$, it will be convenient to bound correlations between subsets of nodes in the graph. 
In order to bound the variance of $\eatetrue$, we will need to upper bound the covariance terms between all pairs of units $i$ and $j$.
Lemma~\suppref{lemma:e1-event-is-the-same} shows that the correlation analysis in Theorem~\mainref{thm:variance-analysis-finite-sample} will be appropriate for analyzing the direct exposure $e_1$.
The correlation of the control exposure $e_0$ terms will be more involved they will depend in an intricate way on the relative topologies of their neighborhoods.

It will prove easier to abstract away these details and provide a general way of bounding correlations between subsets of nodes.
We now give some definitions of useful quantities for arbitrary subsets. 
For a subset of nodes $A \subseteq [n]$, we define the following event
\[
E_A := \{Z_k = 0, \forall k \in A\}
\enspace.
\]
For any two subsets of nodes $A , B \subseteq [n]$, define their \emph{correlation}
\begin{equation}\label{def:CAB}
C_{A,B} := \frac{\Prsub[\big]{G}{Z_k = 0 , \forall k \in A\cup B}}{\Prsub[\big]{G}{Z_k = 0 , \forall k \in A}\Prsub[\big]{G}{Z_k = 0 , \forall k \in  B}}-1 = \Covsub[\Bigg]{G}{\frac{\indicator{E_A}}{\Prsub{G}{E_A}},\frac{\indicator{E_B}}{\Prsub{G}{E_B}} }
\end{equation}
Note that $A$ and $B$ could potentially be overlapping.
For any two subsets $A,B$, we define the set of edges $F_{A,B}$ to be
\begin{align*}
&F_{A,B} \\
&:= \abs[\Bigg]{ 
	\setb[\Big]{(i,j) \in E_G: i,j \in A \cup B } \setminus \\
	&
	\paren[\Bigg]{\setb[\Big]{(i,j) \in E_G: i,j \in A \cap B } \cup \setb[\Big]{(i,j) \in E_G: i,j \in A \setminus B }  \cup \setb[\Big]{(i,j) \in E_G: i,j \in B \setminus A } }
} 
\enspace.
\end{align*}
In words, if we think of $A\setminus B, B\setminus A, A \cap B$ as a partition of $A \cup B$, then $F_{A,B}$ counts the number of edges where the endpoints lie in different subsets of the partition.
Moreover, for any two subsets $A,B \subseteq [n]$, define
\[
R_{A,B} := \setb[\big]{k \in V_G \setminus (A \cup B): \Na{k}{G} \cap A \neq \emptyset \text{ and } \Na{k}{G} \cap B \neq \emptyset}
\]
In other words, $R_{A,B}$ contains the nodes outside of $A\cup B$ that have children in both $A$ and $B$. 
Intuitively, edges in $F_{A,B}$ and nodes in $R_{A,B}$ induce correlations between subsets $A,B$ when we run the Conflict Graph Design. 

We now present a general theorem that bounds $C_{A,B}$ for arbitrary sets $A,B$, which we refer to as the \emph{correlation lemma}.
This result will be instantiated in the proof of Theorem~\mainref{thm:true_HT_d_regular} with $A$ and $B$ being the extended neighborhoods of two nodes in the graph. 

\begin{restatable}[Correlation Lemma]{theorem}{corrlemma} \label{thm:CAB_bound}
Let $G$ be a graph with maximum eigenvalue $\lamG$ satisfying $\dmax \leq C \cdot \lamG$ for some constant $C>0$. 
Let $A,B \subseteq [n]$ be two arbitrary subsets with $|A|,|B| \leq \dmax +1$ and let $I = A \cap B$. Then, there exists an absolute constant $C'$ that depends only on $C$, such that
\begin{equation}\label{eq:CAB_bound}
    |C_{A,B}| \leq C' \cdot\paren[\Bigg]{\frac{|I|}{\lamG} + \frac{|F_{A,B}|}{\lamG^2} + \frac{\sum_{k \in R_{A,B}} |N_a(k) \cap A|\cdot |N_a(k) \cap B|}{\lamG^3}}
    \enspace.
\end{equation}
\end{restatable}

Intuitively, the events $E_A,E_B$ depend on the assignments of the $U$ variables in a neighborhood of $A,B$.
These neighborhoods could be overlapping, which introduces a lot of dependence between the two events for the two subsets.
We will disentangle these two events by slowly transforming the local topology around $A$ and $B$, with the final goal of making $E_A$ and $E_B$ independent in the end. 
The key technical challenge will be to monitor how $C_{A,B}$ changes along this transformation.
This requires a detailed analysis of the probability distribution induced by our design.
The proof of Theorem~\suppref{thm:CAB_bound} is presented in Section~\suppref{sec:correlation-lemma-proof}, while the transformation process is described in Sections~\suppref{sec:PhaseI}, \suppref{sec:PhaseII} and \suppref{sec:PhaseIII}. 

Theorem~\suppref{thm:CAB_bound} will be instantiated in the proof of Theorem~\mainref{thm:true_HT_d_regular} with $A$ and $B$ being the extended neighborhoods of two nodes in the graph. It will thus be useful to provide an interpretation of the quantities $A \cap B, F_{A,B}, R_{A,B}$ in terms of more familiar combinatorial quantities in the graph $G$. 
In particular, let $\Path{i}{j}{s}$ denote the number of undirected walks of length $s$ that start in $i$ and end in $j$ (repetitions of edges are allowed). 
The following \emph{counting lemma} makes the interpretation quantitative. 

\begin{restatable}[Counting Lemma]{lemma}{countinglemma}\label{lem:counting}
    Let $G=(V,E)$ be a simple graph and $i,j \in V$ be two distinct vertices.  Then
    \begin{align}
    |\widetilde{N}(i) \cap \widetilde{N}(j)| &= 2 \Path{i}{j}{1} + \Path{i}{j}{2} \label{eq:intersection}\\
    \abs[\big]{F_{\widetilde{N}(i), \widetilde{N}(j)}} &\leq 2 \Path{i}{j}{2} + \Path{i}{j}{3} \label{eq:edges-between}\\
    \sum_{k \in R_{\widetilde{N}(i), \widetilde{N}(j)}}\abs[\big]{N_a(k) \widetilde{N}(i)}\abs[\big]{N_a(k) \cap \widetilde{N}(j)} 
    &\leq \Path{i}{j}{4} \label{eq:common_parents}
    \end{align}
\end{restatable}

The counting lemma (Lemma~\suppref{lem:counting} above) will enable us to interpret the correlation lemma (Theorem~\suppref{thm:CAB_bound}) in terms of walks in the graph $G$.
This will be a crucial ingredient in analyzing the correlation terms appearing in the standard Horvitz--Thompson estimator.
The proof of Lemma~\suppref{lem:counting} is deferred to Section~\suppref{sec:counting}. 

With these lemmas in hand, we are now ready to prove Theorem~\mainref{thm:true_HT_d_regular}.

\begin{proof}[Proof of Theorem~\mainref{thm:true_HT_d_regular}]

As discussed at the beginning of this section, $d_{\max}(\cH) = \bigTheta{\lamH}$ is equivalent to $d_{\max}(G) = \bigTheta{ \lamG }$ because we are considering the direct effect.
This means that there exists a constant $C > 0$ such that $d_{\max}(G) \leq C \lamG$.

Our goal will be to establish a finite sample result similar to the one in Theorem~\mainref{thm:variance-analysis-finite-sample}.
In particular, we will prove that there exists an absolute constant $K(C) > 0$ (which depends only on $C$) such that for arbitrary potential outcomes $y_i(e_0), y_i(e_1)$ 

\begin{equation}\label{eq:goal}
    \Var{\eatetrue} \leq \frac{K(C) \lamH}{n} \cdot \paren[\Big]{ \frac{1}{n} \sum_{i=1}^n y_i(e_1)^2 + \frac{1}{n} \sum_{i=1}^n y_i(e_0)^2 }
	\enspace.
\end{equation}
Since $\eatetrue$ is unbiased, the result of Theorem~\mainref{thm:true_HT_d_regular} then immediately follows using Assumption~\mainref{assumption:bounded-second-moment}, similarly to the proof of Theorem~\mainref{thm:variance-analysis-asymptotic}. 
We will analyze the Conflict Graph Design with sampling parameter $r=1$ for simplicity, since our focus is not on the optimal constant $K(C)$ that we can obtain.

Let's remember the definition of the standard Horvitz-Thompson estimator $\eatetrue$. 
\[
    \eate_{\textrm{standard}}
    = \frac{1}{n} \sum_{i=1}^n Y_i \cdot \paren[\Bigg]{ 
        \frac{\indicator[\big]{d_i(Z) = d_i(\zonei)}}{\Pr[\big]{d_i(Z) = d_i(\zonei)}} 
        -  
        \frac{\indicator[\big]{d_i(Z) = d_i(\zzeroi)}}{\Pr[\big]{d_i(Z) = d_i(\zzeroi)}} 
    }
    \enspace.
\]
We proceed by splitting $\eatetrue$ into the following two parts, one for estimating average potential outcome under the $e_1$ exposure and the other for estimating the average potential outcome under the $e_0$ exposure.
Lemma~\suppref{lemma:e1-event-is-the-same} implies that the exposure event is precisely the desired exposure event $E_{(i,1)}$, so that we can write this decomposition as
\[
\eatetrue =  \underbrace{\frac{1}{n} \sum_{i=1}^n Y_i \cdot 
   \frac{\indicator[\big]{E_{(i,1)}}}{\Pr[\big]{E_{(i,1)}}} }_{\eatetrue^{(1)}}
    -  
    \underbrace{\frac{1}{n} \sum_{i=1}^n Y_i \cdot \frac{\indicator[\big]{F_{(i,0)}}}{\Pr[\big]{F_{(i,0)}}} }_{\eatetrue^{(0)}}
    \enspace.
\]
We can bound the variance of $\eatetrue$ by the variances of the individual components as follows
\begin{align*}
\Var{\eatetrue} &= \Var{\eatetrue^{(1)}} + \Var{\eatetrue^{(0)}} - 2 \Cov{\eatetrue^{(1)}, \eatetrue^{(0)}}\\
&\leq \Var{\eatetrue^{(1)}} + \Var{\eatetrue^{(0)}} + 2 \sqrt{\Var{\eatetrue^{(1)}}\Var{\eatetrue^{(0)}}} \\
&\leq 2 \Var{\eatetrue^{(1)}} + 2 \Var{\eatetrue^{(0)}}\,,
\end{align*}
where in the first inequality we used Cauchy-Schwartz and in the second the AM-GM inequality. 
It thus suffices to bound the variance of each of the two components of $\eatetrue$. For $\eatetrue^{(1)}$, we observed that it is exactly the same as in the modified estimator $\eate$ that we analyzed in Theorem~\mainref{thm:variance-analysis-finite-sample}. In fact $\eatetrue^{(1)}$ is exactly the same as $\eate$ in the setting where the outcomes $y_i(e_0) = 0$. Thus, we can immediately apply Theorem~\mainref{thm:variance-analysis-finite-sample} to obtain 
\[
    \Var[\big]{\eatetrue^{(1)}}  \leq \frac{\cleanconst \lamH}{n} \cdot \paren[\Big]{ \frac{1}{n} \sum_{i=1}^n y_i(e_1)^2 }
    \enspace.
\]
Thus, we are left with upper bounding $\Var{\eatetrue^{(0)}}$. 

 Lemma~\mainref{lem:desired_exposures} implies that $E_{(i,0)} \subseteq F_{(i,0)}$. Unfortunately, equality does not hold in general.
 This is because there are many different paths that would lead to $F_{(i,0)}$ occurring at the end of execution.
 For example, suppose a unit $i$ has $U_i = *$ and all of its neighbors $j \in N(i)$ also satisfy $U_j = *$.
 Then, in the final treatment assignment $d_i(Z) = d_i(\vec{0})$, but clearly $E_{(i,0)}$ doesn't hold.
 The fact that the event $F_{(i,0)}$ may be significantly larger than $E_{(i,0)}$ is the main source of difficulty in analyzing $\eatetrue$.
 This is because these larger events introduce further correlations that lead to a significantly more complicated analysis.
 We will see that now units at distance $3$ or $4$ away could also be correlated, and we will bound their correlation in terms of the number of paths of a given length between them in the graph $G$. 

To start the analysis, let's write the variance in expanded form
\[
    n \cdot \Var{\eatetrue^{(0)}} = \frac{1}{n} \sum_{i=1}^n \sum_{j=1}^n y_i(e_0)y_j(e_0) \Cov[\Bigg]{ \frac{ \indicator{ F_{(i,0)} } }{ \Pr{ F_{(i,0)} } }  , \frac{ \indicator{ F_{(j,0)} } }{ \Pr{ F_{(j,0)} } }  }
    \enspace.
\]
We will use the notation 
\[
C_{ij} := \Cov[\Bigg]{ \frac{ \indicator{ F_{(i,0)} } }{ \Pr{ F_{(i,0)} } }  , \frac{ \indicator{ F_{(j,0)} } }{ \Pr{ F_{(j,0)} } }  }
\enspace.
\]
If $\mat{C} \in \Reals^{n \times n}$ is the matrix with entries $C_{ij}$, then we can write the variance as a quadratic form
\[
    n \cdot \Var{\eatetrue^{(0)}} = \frac{1}{n}\vec{y}(e_0)^\top \mat{C} \vec{y}(0) 
    \enspace.
\]
As discussed in the proof of Theorem~\mainref{thm:variance-analysis-finite-sample}, proving \eqref{eq:goal} is thus equivalent to showing that for some absolute constant $K(C)>0$ (which depends only on $C$) we have
\begin{equation}\label{eq:goal_lambda}
\norm{\mat{C}} \leq K(C) \cdot \lamH
\enspace.
\end{equation}
Thus, in the remainder of the proof, our goal will be to establish \eqref{eq:goal_lambda}. 

First, notice that if there are no edges in the graph, then the conflict graph only has self edges, which means that $\lamH = 1$.
In that case, as we also remarked in the proof of Theorem~\mainref{thm:variance-analysis-finite-sample}, all nodes are independent and $\norm{\mat{C}} = 1$, meaning that \eqref{eq:goal_lambda} follows trivially. 

In the remaining, let's assume that there is at least one edge in $G$, meaning $\lamH \geq 2$ so that $\lamG \geq 1$. 
As in the proof of Theorem~\mainref{thm:variance-analysis-finite-sample}, our strategy will be to upper bound $|C_{ij}|$ for all pairs of units $i,j$, based on their distance in the graph. However, the situation in Theorem~\mainref{thm:variance-analysis-finite-sample} was simpler, since we only had correlations up to nodes of distance $2$ from each other. We will see that now we have non-zero correlations up to distance $4$. In addition, even nodes that are at distance less than $4$ could exhibit correlations based on paths of length $3$ or $4$ between them. 
Thus, we need a more general way of bounding $C_{ij}$ than simply splitting $\mat{C}$ into parts according to the distance. This is provided by Theorem~\suppref{thm:CAB_bound}. Indeed, let's fix two nodes $i,j \in V$. Consider the set $A = \widetilde{N}(i), B= \widetilde{N}(j)$, where we recall that $\widetilde{N}(i) = N(i) \cup \setb{i}$.
Then, by Definition \eqref{def:CAB} we have that
\[
C_{ij} = C_{A,B}
\enspace.
\]
It is clear that $|\widetilde{N}(i)|, |\widetilde{N}(j)| \leq \dmax + 1 \leq C\lamG$ by assumption. Thus, we can apply Theorem~\suppref{thm:CAB_bound}, which says that there exists a constant $C'$ depending on $C$, such that 
\begin{multline}\label{eq:Cij-bound-general}
    |C_{ij}| \leq C' \cdot\lp(\frac{|\widetilde{N}(i) \cap \widetilde{N}(j)|}{\lamG} \rp.\\
   \lp.+ \frac{|F_{\widetilde{N}(i),\widetilde{N}(j)}|}{\lamG^2} + \frac{\sum_{k \in R_{\widetilde{N}(i),\widetilde{N}(j)}} |\Na{k}{G} \cap \widetilde{N}(i)|\cdot |\Na{k}{G} \cap \widetilde{N}(j)|}{\lamG^3}\rp)
\end{multline}
Let $P_{ij}^{(s)}$ be the number of undirected paths of length $s$ from $i$ to $j$ in $G$ (possibly with repeated vertices or edges). Using Lemma~\suppref{lem:counting} along with \eqref{eq:Cij-bound-general}, we conclude that
\begin{align}
    |C_{ij}| &\leq C' \cdot\paren[\Bigg]{\frac{2 \Path{i}{j}{1} + \Path{i}{j}{2}}{\lamG} + \frac{2 \Path{i}{j}{2} + \Path{i}{j}{3}}{\lamG^2} + \frac{\Path{i}{j}{4}}{\lamG^3}}\nonumber \\
    &\leq  C' \cdot\paren[\Bigg]{2 \cdot \frac{\Path{i}{j}{1} }{\lamG}+  3 \cdot\frac{\Path{i}{j}{2} }{\lamG} + \frac{ \Path{i}{j}{3}}{\lamG^2} + \frac{\Path{i}{j}{4}}{\lamG^3}}\nonumber \\
    &\leq 3C' \cdot \paren[\Bigg]{\frac{P^{(1)}_{ij}}{\lamG} + \frac{\Path{i}{j}{2}}{\lamG} +  \frac{\Path{i}{j}{3}}{\lamG^2} +  \frac{\Path{i}{j}{4}}{\lamG^3}} \label{eq:path_goal}
\end{align}
In the above, we used the fact that $\lamG \geq 2$. 

Now, note that by simple linear algebra, $\Path{i}{j}{s}$ is exactly the $(i,j)$-th entry in the matrix $\mat{A}_G^s$, where $\mat{A}_G$ is the adjacency matrix of $G$. Thus, \eqref{eq:path_goal} implies that 
\[
|C_{ij}| \leq 3C' \cdot \paren[\Bigg]{\frac{(\mat{A}_G)_{ij}}{\lamG} + \frac{(\mat{A}_G^2)_{ij}}{\lamG} +  \frac{(\mat{A}_G^3)_{ij}}{\lamG^2} +  \frac{(\mat{A}_G^4)_{ij}}{\lamG^3}}
\enspace.
\]
It is well known that the operator norm $\norm{\mat{C}}$ could only increase if we take absolute values on all entries of $\mat{C}$. It also follows from Peron-Frobenius theorem for matrices with non-negative entries that increasing the entries can only increase the operator norm \citep{Spielman2019Book}. These observations imply 
\[
\norm{\mat{C}} \leq 3C'\cdot \paren[\Bigg]{\frac{\norm{\mat{A}_G}}{\lamG} + \frac{\norm{\mat{A}_G^2}}{\lamG} +  \frac{\norm{\mat{A}_G^3}}{\lamG^2} +  \frac{\norm{\mat{A}_G^4}}{\lamG^3}}
\enspace.
\]
Moreover, it holds that $\norm{\mat{A}_G^s} = \lamG^s$.
Thus, setting $K(C) = 12 C'$, the above reduces to 
\[
    \norm{\mat{C}} \leq K(C) \cdot \lamG \leq K(C) \cdot \lamH \enspace,
\]
which is exactly what we wanted to prove in \eqref{eq:goal_lambda}. This concludes the proof.
\end{proof}

\subsection{Proof of Counting Lemma (Lemma~\suppref{lem:counting})}\label{sec:counting}

\countinglemma*

\begin{proof}
    In the following, if we want to denote a path with nodes $u_1,\ldotp,u_s$, we will write $u_1 \to \ldotp \to u_s$. We emphasize that the path is still undirected, so this notation is not referring to the direction of edges. 
    
    We first prove \eqref{eq:intersection}. We distinguish between two cases.
    \begin{itemize}
        \item \underline{$(i,j) \in E$.} In that case, $\widetilde{N}(i) \cap \widetilde{N}(j) = \{i,j\} \cup (N(i) \cap N(j))$. Notice that the set $N(i) \cap N(j)$ is in one to one correspondence with paths of length $2$ between $i$ and $j$ in $G$ (each path of length $2$ has the form $i \to k \to j$ with $k \in N(i) \cap N(j)$). Thus, in that case 
        \[
        \abs[\big]{\widetilde{N}(i) \cap \widetilde{N}(j)} = 2 + \Path{i}{j}{2}
        \enspace.
        \]
        \item \underline{$(i,j) \notin E$.} In that case, $\widetilde{N}(i) \cap \widetilde{N}(j) =  N(i) \cap N(j)$. As we explained in the previous case, the set $N(i) \cap N(j)$ is in one to one correspondence with paths of length $2$ between $i$ and $j$ in $G$. Thus, in that case
        \[
            \abs[\big]{\widetilde{N}(i) \cap \widetilde{N}(j)} = \Path{i}{j}{2}
            \enspace.
        \] 
    \end{itemize}
    Combining these two cases proves \eqref{eq:intersection}. 
    
    Moving on to \eqref{eq:edges-between}, let $\mathcal{P}_{ij}^{(s)}$ be the set of walks from $i$ to $j$ of length $s$. 
    By definition $F_{\widetilde{N}(i), \widetilde{N}(j)}$ contains the edges with endpoints belonging to different sets in the partition with 3 parts: the intersection of extended neighborhoods $\widetilde{N}(i) \cap \widetilde{N}(j)$, extended neighbors of $i$ who are not extended neighbors of $j$ $\widetilde{N}(i) \setminus \widetilde{N}(j)$, and extended neighbors of $j$ which are not extended neighbors of $i$ $\widetilde{N}(j) \setminus \widetilde{N}(i) $. 
    We will map each of these edges injectively to 
    $\mathcal{P}_{ij}^{(2)} \cup (\mathcal{P}_{ij}^{(2)})' \cup \mathcal{P}_{ij}^{(3)}$, where $(\mathcal{P}_{ij}^{(2)})'$ is a copy of $\mathcal{P}_{ij}^{(2)}$.
    We again distinguish between two cases depending on whether $(i,j) \in E$ and also examine edges $(u,v) \in F_{\widetilde{N}(i), \widetilde{N}(j)}$, depending on which two of the three subsets $u$ and $v$ belong to.  
    \begin{itemize}
        \item \underline{$(i,j) \in E$.} In this case, observe that the intersection of extended neighborhoods is $\widetilde{N}(i) \cap \widetilde{N}(j) = \{i,j\} \cup (N(i) \cap N(j))$, the extended neighbors of $i$ not $j$ are $\widetilde{N}(i) \setminus \widetilde{N}(j) = N(i) \setminus N(j)$, and the extended neighbors of $j$ not $i$ are $\widetilde{N}(j) \setminus \widetilde{N}(i) = N(j) \setminus N(i)$.
        \begin{itemize}
            \item \underline{$u \in \widetilde{N}(i) \cap \widetilde{N}(j)$ and $v \in \widetilde{N}(i) \setminus \widetilde{N}(j)$.} In that case, $ u \in  \{i,j\} \cup (N(i) \cap N(j)), v \in  N(i) \setminus N(j)$. If $u \in N(i) \cap N(j)$ and $v \in N(i) \setminus N(j)$, then $i \to u \to v \to j$ is a valid path of length $3$, so we map $(u,v)$ to that path. If $u = i$, then $i \to u \to i \to j$ is a valid path of length $3$, which has not been accounted in the previous cases, so we map $(u,v)$ to it. If $u = j$, then there are not edges between $u$ and $N(i) \setminus N(j)$ by definition.
            \item \underline{$u \in \widetilde{N}(i) \cap \widetilde{N}(j)$ and $v \in \widetilde{N}(j) \setminus \widetilde{N}(i)$.} This case is symmetric to the previous one, and $(u,v)$ will again be mapped to paths of length $3$ that have not appeared before. In particular, $u \in  \{i,j\} \cup (N(i) \cap N(j)),v \in N(j) \setminus N(i)$. If $u \in N(i) \cap N(j)$ and $v \in N(j) \setminus N(i)$, then $i \to u \to v \to j$ is a valid path of length $3$ that has not appeared before, so we map $(u,v)$ to that path. If $u = j$, then $i \to j \to v \to j$ is a valid path of length $3$, which has not been accounted in the previous cases, so we map $(u,v)$ to it. If $u = i$, then there are not edges between $u$ and $N(j) \setminus N(i)$ by definition
            \item \underline{$u \in \widetilde{N}(i) \setminus \widetilde{N}(j)$ and $v \in \widetilde{N}(j) \setminus \widetilde{N}(i)$.}
            In that case, we have $u \in N(i) \setminus N(j)$ and $v \in N(j) \setminus N(i)$. Thus, $(u,v)$ can be mapped to the path $i \to u \to v \to j$, which has not been accounted before. 
        \end{itemize}
        It is clear that the above mapping is injective to $\mathcal{P}_{ij}^{(3)}$, which implies in that case 
        \[
        \abs[\big]{F_{\widetilde{N}(i), \widetilde{N}(j)}} \leq \Path{i}{j}{3}
        \enspace.
        \]
        \item \underline{$(i,j) \notin E$.} In that case, the intersection of extended neighborhood is $\widetilde{N}(i) \cap \widetilde{N}(j) = N(i) \cap N(j)$, the extended neighbors of $i$ not $j$ are $\widetilde{N}(i) \setminus \widetilde{N}(j) = \{i\}\cup(N(i) \setminus N(j))$, and the extended neighbors of $j$ not $i$ are $\widetilde{N}(j) \setminus \widetilde{N}(i) = \{j\} \cup(N(j) \setminus N(i))$.
        \begin{itemize}
            \item \underline{$u \in \widetilde{N}(i) \cap \widetilde{N}(j)$ and $v \in \widetilde{N}(i) \setminus \widetilde{N}(j)$.}
            In that case, $u \in N(i) \cap N(j), v \in \{i\}\cup(N(i) \setminus N(j))$. If $v \in N(i) \setminus N(j)$, then we map $(u,v)$ to the path $i \to u \to v \to j$. If $v = i$, then $(u,v)$ is mapped to the path $i \to u \to j$, which belongs to $\mathcal{P}_{ij}^{(2)}$. 
            \item \underline{$u \in \widetilde{N}(i) \cap \widetilde{N}(j)$ and $v \in \widetilde{N}(j) \setminus \widetilde{N}(i)$.}
            In that case, $u \in N(i) \cap N(j), v \in \{j\}\cup(N(j) \setminus N(i))$. If $v \in N(j) \setminus N(i)$, then we map $(u,v)$ to the path $i \to u \to v \to j$. If $v = j$, then $(u,v)$ is mapped to the path $i \to u \to j$, which belongs to the copy $(\mathcal{P}_{ij}^{(2)})'$. 
            \item \underline{$u \in \widetilde{N}(i) \setminus \widetilde{N}(j)$ and $v \in \widetilde{N}(j) \setminus \widetilde{N}(i)$.}
            In that case, $u \in N(i) \setminus N(j), v \in \{j\}\cup(N(j) \setminus N(i))$. Then, we map $(u,v)$ to the path $i \to u \to v \to j$. 
        \end{itemize}
    	\eqref{eq:edges-between} now follows because the mapping was injective:
        \[
            \abs[\big]{F_{\widetilde{N}(i), \widetilde{N}(j)}} \leq 2 \cdot \Path{i}{j}{3} + \Path{i}{j}{3}
            \enspace.
        \]
    \end{itemize}

    Moving to \eqref{eq:common_parents}, we will map each $k \in R_{\widetilde{N}(i), \widetilde{N}(j)} $ to a subset of paths $\phi(k) \subseteq \mathcal{P}_{ij}^{(4)}$ such that 
    \begin{equation}\label{eq:mapping}
    |\phi(k)| = |N_a(k) \cap \widetilde{N}(i)|\cdot |N_a(k) \cap \widetilde{N}(j)|\enspace.
    \end{equation}
     We will also see that if $k \neq k'$ we have $\phi(k) \cap \phi(k') = \emptyset$. \eqref{eq:common_parents} would then follow trivially. 
    Indeed, let $k \in R_{\widetilde{N}(i), \widetilde{N}(j)} $. Let us consider all ordered pairs $(u,v)$ with $u \in N_a(k) \cap \widetilde{N}(i)$ and $v \in N_a(k) \cap \widetilde{N}(j)$. Notice that each pair defines a path of lenght $4$ as $i \to u \to k \to v \to j$. Moreover, since all pairs $(u,v)$ are distinct, all of these paths will be distinct. Thus, if we define $\phi(k)$ to be the set of all those paths defined for all such ordered pairs $(u,v)$, then clearly \eqref{eq:mapping} holds. 
    Moreover, notice that the third node in each path of $\phi(k)$ is $k$, hence $\phi(k) \cap \phi(k') = \emptyset$ whenever $k \neq k'$. This concludes the proof.  
\end{proof}

\subsection{Correlation Lemma: Preliminary Results for Interpolation}

In this Section we provide statements that will be useful for the proof of Theorem~\suppref{eq:CAB_bound}, which is the only remaining claim to establish Theorem~\mainref{thm:true_HT_d_regular}. 
As discussed earlier in this section, the correlation lemma is proved using a novel coupling argument where the original graph $G$ is transformed into a graph with simpler correlations.
A key aspect in the proof will be the \emph{interpolation between different graph topologies}.

It will be more convenient to view $G$ as a directed graph, where the direction of an edge is determined by the parent-child relationship between the two endpoints. In particular, if $(i,j)$ is an undirected edge, then the directed edge is $i \to j$ if and only if $i \in N_b^{(G)}(j)$. 
We will also view the interpolated graphs as being directed. 
Therefore, we now provide a definition of what it means to run the Conflict Graph Design on an arbitrary Directed Acyclic Graph (DAG).
Naturally, it corresponds to running Algorithm~\suppref{alg:conflict-graph-design-direct-effect} using a topological ordering induced by the DAG.

\begin{definition}
	Let $H= (V,E)$ be a DAG. Then, for each node $i \in V$ we define the set of \emph{parents} $N_b(i)$ to be 
	\[
	N_b^{(H)}(i) := \{j \in V: j \to i \in E\}
	\]
	Similarly, we define the set of \emph{children} of $i$ to be
	\[
	N_a^{(H)}(i) := \{j \in V: i \to j \in E\}
	\]
	Then, we say that $Z \in \setb{0,1}^n$ is the \emph{output of the Conflict Graph Design run on $H$}, if $Z$ is the output of Algorithm~\suppref{alg:conflict-graph-design-direct-effect}, run using the definition of $N_b^{(H)}(i)$ above.
\end{definition}

We can view running the Conflict Graph Design on $H$ as running Algorithm~\mainref{alg:conflict-graph-design} using any topological ordering $\pi$ of the nodes in $H$.
Notice that the probabilities of the $U_i$ variables still depend on $\lamG$, which is our original graph, and not on the eigenvalues of the directed graph $H$. 

Suppose we run the Conflict Graph Design on a DAG $H$. Whenever we refer to the probability of an event $E$ under this design, we will use $\Prsub{H}{E}$ to indicate the fact that Algorithm~\suppref{alg:conflict-graph-design-direct-effect} is run on $H$. Likewise, for the expectation of a function $f$, we will use $\Esub{H}{f}$. For every node $i$, we introduce the notations
\[
\alphaind{k}{H} := \indicator{U_k = e_1} \qquad \betaind{k}{H} := \indicator{U_k = *}
\]
to denote the random variables used in the execution of Algorithm~\suppref{alg:conflict-graph-design-direct-effect} for graph $H$. 

We begin our analysis by providing a simple expression for the event that a node receives treatment in the Conflict Graph Design. Specifically, we show that the only possibility for a node to receive treatment is if it has desired exposure $e_1$ and all of its parents have desired exposures $*$. The arguments are similar to the ones used for Lemma~\mainref{lem:desired_exposures}, where the event under which a unit receives the desired exposure is analyzed.

\begin{lemma}\label{lem:treatment-event-standard-ht}
Suppose we run the Conflict Graph Design on a directed graph $H$. Then, for any node $k$ we have that 
$$
\{Z_k = 1\} = \{U_k = e_1, U_{k'} = *, \forall k' \in \Nb{k}{H}\}\enspace.
$$
\end{lemma}
\begin{proof}
    First, observe that for a node $k$, the only way the assignment $Z_k = 1$ happens during execution is if Line~\suppref{algline:set-treatment-direct} is executed for the iteration corresponding to $k$, which means that the condition $U_k=e_1, U_{k'} = *, \forall k' \in \Nb{k}{H}$ is satisfied. Thus, 
    \[
        \{Z_k = 1\} \subseteq \{U_k = e_1, U_{k'} = *, \forall k' \in \Nb{k}{H}\}\enspace.
    \]
    To prove the reverse inclusion, suppose $U_k = e_1, U_{k'} = *, \forall k' \in \Nb{k}{H} $ holds. This is equivalent to saying that $E_{(i,k)}$ holds, thus by Lemma~\mainref{lem:desired_exposures}, we know that the assignment of $Z_k$ will not be altered in the subsequent iterations. This completes the proof. 
\end{proof}

% \subsection{A Useful Probability Expression}

The standard Horvitz-Thompson depends on the probability of receiving exposure, i.e. $\Pr{ d_i(Z) = e_k }$.
This exposure event $\setb{ d_i(Z) = e_k }$ is more complicated than the desired exposure event $E_{(i,k)}$ used in the construction of the modified Horvitz--Thompson estimator.
As such, it will be useful to obtain formulas for calculating these events.
One very useful such formula is given in the lemma below, and is based on rewriting the probability in terms of indicator random variables.
We remind the reader of the definitions 
\[
    \alphaind{k}{H} := \indicator{U_k = e_1} \qquad \betaind{k}{H} := \indicator{U_k = *}
\]

\begin{lemma}\label{lem:control-expression-direct-effect}
   Suppose we run the Algorithm~\suppref{alg:conflict-graph-design-direct-effect} for a DAG $H$. Let $S$ be an arbitrary subset of the nodes. 
   Then
   \[
   \Prsub[\big]{H}{Z_k = 0 , \forall k \in S} = \Esub[\Bigg]{H}{\prod_{k \in S} \paren[\Big]{1 - \alphaind{k}{H} \prod_{k ' \in N_b^{(H)}} \betaind{k'}{H}}}
   \]
\end{lemma}
 \begin{proof}
    By Lemma~\suppref{lem:treatment-event-standard-ht}, we can write 
    \[
     \indicator{Z_k = 1}= \indicator{U_k = e_1, U_{k'} = *, \forall k' \in N_b^{(H)}(i)} = \alphaind{k}{H} \prod_{k' \in N_b^{(H)}(i)} \betaind{k'}{H}
     \enspace.
     \]
     This allows us to write
      \begin{align*}
        \Prsub[\big]{H}{Z_k = 0 , \forall k \in S} &= \Esub[\Bigg]{H}{\prod_{k \in S}\paren[\Big]{1 - \indicator{Z_k = 1}}}\\
        &= \Esub[\Bigg]{H}{\prod_{k \in S} \paren[\Big]{1 - \alphaind{k}{H} \prod_{k ' \in N_b^{(H)}} \betaind{k}{H}}}
        \enspace.
        \qedhere
      \end{align*}
 \end{proof}

The next Lemma allows us to linearize product expressions, which we will very often encounter in the proof.
Its proof is a standard application of Taylor's theorem. 

\begin{lemma}\label{lem:linearization}
    Let $a_1,\ldotp,a_p \in Reals$ be real numbers such that 
    \[
    \sum_{k=1}^p |a_k| \leq C\enspace,
    \]
    where $C>0$. Then, 
    \[
        \abs[\Big]{\prod_{k=1}^p (1 + a_k) -1} \leq e^C \sum_{k=1}^p |a_k|
    \]
\end{lemma}
\begin{proof}
    Let's start by observing that for every $l \leq p$, using the inequality $1 + x \leq e^x$, we have that
    \begin{equation}\label{eq:aux_bound}
    \abs[\Bigg]{\prod_{k=1}^l (1 + a_k)} = \prod_{k=1}^l (1 + |a_k|) \leq e^{\sum_{k=1}^l |a_k|} \leq e^C
    \end{equation}
    Now, we can write inductively
    \begin{align*}
        \abs[\Bigg]{\prod_{k=1}^p (1 + a_k) -1} &= \abs[\Big]{\prod_{k=1}^{p} (1 + a_k) - \prod_{k=1}^{p-1} (1 + a_k) +\prod_{k=1}^{p-1} (1 + a_k)  -1} \\
        &\leq |a_p|\abs[\Bigg]{\prod_{k=1}^{p-1} (1 + a_k)} + \abs[\Bigg]{\prod_{k=1}^{p-1} (1 + a_k) +\prod_{k=1}^{p-1} (1 + a_k)  -1}\\
        \intertext{by triangle inequality}
        &\leq e^C |a_p| + \abs[\Bigg]{\prod_{k=1}^{p-1} (1 + a_k) +\prod_{k=1}^{p-1} (1 + a_k)  -1}\\
        \intertext{by \eqref{eq:aux_bound}}
        &\leq \ldots \leq e^C \sum_{k=1}^p |a_k|
        \enspace.
        \qedhere
    \end{align*}
\end{proof}

We now prove that the probability that a subset of nodes are untreated is lower bounded by a constant, as long as the size of this subset is $\bigO{\lamG}$. 

\begin{lemma}\label{lem:probability-lower-bound}
    Suppose we run the Conflict Graph Design on a DAG $H$ and let $S$ be a subset of nodes.
    Furthermore, assume $\lamG > 1$.
    Then, if $|S| \leq C \cdot \lamG$, then 
    \[
    \Prsub{H}{Z_k = 0, \forall k \in S} \geq e^{-C}
    \enspace.
    \]
\end{lemma}
\begin{proof}
    We observe that if all units in $S$ receive $*$ exposure, then they will all be untreated. Thus
    \begin{align*}
        \Prsub{H}{Z_k = 0, \forall k \in S}  \geq \Prsub{H}{U_k = *, \forall k \in S} = \paren[\Big]{1 - \frac{1}{\lamG}}^{|S|} \geq \paren[\Big]{1 - \frac{1}{\lamG}}^{C \cdot \lamG} 
        \enspace.
    \end{align*}
    In the last step, we used the assumption on the size of $S$. 
    Consider the function:
    \[
    f(\lambda):= \paren[\Big]{1 - \frac{1}{\lambda}}^{C\lambda}
    \enspace.
    \]
    We can show that $f$ is decreasing as long as $\lamG > 1$. Indeed,
    \[
    f'(\lambda) = C\paren[\Big]{\ln\paren[\Big]{1 - \frac{1}{\lambda}} + \frac{1}{\lambda - 1}} f(\lambda) < 0 
    \enspace,
    \]
    where we use the inequality $\ln(1 - 1/\lambda) + 1/(\lambda - 1) < 0$, for all $\lambda > 1$. 
    Thus, 
    \[
        \paren[\Big]{1 - \frac{1}{\lamG}}^{C\lamG}  \geq \lim_{\lambda \to \infty} f(\lambda) = e^{-C}
        \enspace.
        \qedhere
    \]
\end{proof}

We next present an important lemma that bounds the multiplicative difference between the probability of the same event, evaluated at two different directed graphs $H_1$ and $H_2$ that only differ locally. This situation will appear often once we start interpolating between different graphs that only differ locally in the proof of Theorem~\mainref{thm:true_HT_d_regular}.

\begin{proposition}\label{prop:interpolation-step}
    Let $H_1=(V,E_1),H_2= (V,E_2)$ be two directed graphs defined on the same set of vertices. Let $v \in V$ be a fixed node and $S\subseteq R \subseteq V$ be two arbitrary subsets, such that $v \notin S$ and $|R| \leq C \cdot \lamG$.
    Suppose further $H_1$ and $H_2$ differ only in edges that are incident to $v$ and in particular that $v$ has no edges with nodes in $S$ in $H_1$ but has directed edges to every node in $S$ in $H_2$, i.e. $E_2 = E_1 \cup \{k \to v:k \in S\}$.
    Then,
    \[
    \abs[\Bigg]{\frac{\Prsub{H_2}{Z_k = 0 ,\forall k \in R}}{\Prsub{H_1}{Z_k = 0 ,\forall k \in R}} - 1} \leq \frac{e^C}{2} \cdot \frac{|S|}{\lamG^2} 
    \enspace.
    \]
\end{proposition}
\begin{proof}
    We begin by writing the difference in the two probabilities according to Lemma~\suppref{lem:control-expression-direct-effect}
    \begin{align*}
        &\Prsub{H_2}{Z_k = 0 ,\forall k \in R} - \Prsub{H_1}{Z_k = 0 ,\forall k \in R}\\
        &\qquad = \Esub[\Bigg]{H_2}{\prod_{k \in R} \paren[\Big]{1 - \alphaind{k}{H_2} \prod_{k ' \in N_b^{(H_2)}} \betaind{k}{H_2}}}
        - \Esub[\Bigg]{H_1}{\prod_{k \in R} \paren[\Big]{1 - \alphaind{k}{H_1} \prod_{k ' \in N_b^{(H_1)}} \betaind{k}{H_1}}}
        \enspace.
    \end{align*}
    The crucial observation here is that we can \emph{couple} the two distributions of the treatment assignment vector for $H_1$ and $H_2$.
    We can do this simply by coupling the two executions of the Conflict Graph Design for $H_1$ and $H_2$ as follows: for all $k \in V$, we
    sample independently a common variable $U_k$ for both graphs and set    
\begin{align*}
\alphaind{k}{H_1} &= \alphaind{k}{H_2} := \alpha_k = \indicator{U_k = e_1}\\
\betaind{k}{H_1} &= \betaind{k}{H_2} := \beta_k = \indicator{U_k = *}
\enspace.
\end{align*}
We remark that while the random $U_1 \dots U_n$ are shared between $H_1$ and $H_2$, the resulting distribution of $Z$ will be different under $H_1$ and $H_2$ because the graphs, and thus topological orderings, are different.
This is a valid coupling as it preserves the marginal distribution of $Z$ under $H_1$ and $H_2$. 
We will denote expectation with respect to this coupling simply as $\mathbb{E}$, since now everything lies in the same probability space.
The only difference in sampling for $H_1$ and $H_2$ now is that in $H_2$ the nodes in $S$ have one extra parent, which is $v$. 
Armed with this observation, we can write

\begin{align*}
    &\Prsub{H_2}{Z_k = 0 ,\forall k \in R} - \Prsub{H_1}{Z_k = 0 ,\forall k \in R}\\
    &\qquad = \operatorname{E}\lp[\paren[\Bigg]{\prod_{k \in S}\paren[\Big]{1 - \alpha_k \beta_v \prod_{k' \in \Nb{k'}{H_1}}\beta_{k'}} - 
    \prod_{k \in S}\paren[\Big]{1 - \alpha_k  \prod_{k' \in \Nb{k'}{H_1}}\beta_{k'}}}\rp.\\
    &\lp.\qquad \times
    \prod_{k \in R\setminus S} \paren[\Big]{1 - \alpha_k \prod_{k ' \in N_b^{(H_1)}(k)} \beta_{k'}}\rp]
    \enspace.
\end{align*}
Now, some simple algebra, along with the fact that $\alpha_v \beta_v = 0$ always yields
\begin{align*}
    &\prod_{k \in S}\paren[\Big]{1 - \alpha_k \beta_v \prod_{k' \in \Nb{k}{H_1}}\beta_{k'}} -        
    \prod_{k \in S}\paren[\Big]{1 - \alpha_k  \prod_{k' \in \Nb{k}{H_1}}\beta_{k'}}
    \\ 
    &= 
    \paren[\Bigg]{\sum_{T \subseteq S} (-1)^{|T|} \paren[\Big]{\beta_{v}^{\indicator{T \neq \emptyset}}-1} \prod_{k \in T} \paren[\Big]{\alpha_k \prod_{k' \in \Nb{k}{H_1}} \beta_{k'}}}\\
    &=  
    \paren[\Bigg]{\sum_{T \subseteq S, T\neq \emptyset} (-1)^{|T|} \paren[\Big]{\beta_{v}-1} \prod_{k \in T} \paren[\Big]{\alpha_k \prod_{k' \in \Nb{k}{H_1}} \beta_{k'}}}\\
    &= \paren[\Big]{\beta_{v}-1}
    \sum_{T \subseteq S, T\neq \emptyset} (-1)^{|T|}  \prod_{k \in T} \paren[\Big]{\alpha_k \prod_{k' \in \Nb{k}{H_1}} \beta_{k'}}\\
    &= \paren[\Big]{1-\beta_{v}}
    \paren[\Bigg]{1-\prod_{k \in S} \paren[\Big]{1 - \alpha_k \prod_{k' \in \Nb{k}{H_1}} \beta_{k'}}}\\
    &= \indicator{U_v \neq *}\indicator{\cup_{k \in S} \{Z_k = 1\}} \enspace.
    \end{align*}
    The last step follows from the definition of the indicators $\alpha_k, \beta_k$. Substituting back to the expectation, we get
    \begin{align*}
        &\Prsub{H_2}{Z_k = 0 ,\forall k \in R} - \Prsub{H_1}{Z_k = 0 ,\forall k \in R} \\
        &\qquad=
        \E[\Bigg]{\paren[\Big]{1-\beta_{v}}
        \paren[\Bigg]{1-\prod_{k \in S} \paren[\Big]{1 - \alpha_k \prod_{k' \in \Nb{k}{H_1}} \beta_{k'}}}
        \prod_{k \in R\setminus S} \paren[\Big]{1 - \alpha_k \prod_{k ' \in N_b^{(H_1)}(k)} \beta_{k'}}}\\
        &\qquad= \Prsub[\Big]{H_1}{\{U_v \neq *\} \cap (\cup_{k \in S} \{Z_k = 1\}) \cap (\cap_{k \in R \setminus S}\{Z_k = 0\})}
        % &\qquad= \Pr{U_v \neq *} \frac{\Pr[\Big]{\{U_v \neq *\} \cap (\cup_{k \in S} \{Z_k = 1\}) \cap (\cap_{k \in R \setminus S}\{Z_k = 0\})}}{\Pr{U_v \neq *}}\\
        % &\qquad= \frac{1}{\lamG} \Pr[\Bigg]{ \braces[\Big]{ \cup_{k \in S} \{Z_k = 1\}} \cap \braces[\Big]{ \cap_{k \in R \setminus S} \setb{Z_k = 0} } \mid U_v \neq *}
        \enspace.
    \end{align*}
    The last step follows since in the expression of the probability we have excluded $v$ as a parent of the nodes in $S$, hence we follow the topology of $H_1$. 
    At this point, we can draw some conclusions. The first is that clearly $\Prsub{H_2}{Z_k = 0 ,\forall k \in R} \geq \Prsub{H_1}{Z_k = 0 ,\forall k \in R}$.
    This is to be expected because $H_2$ contains more edges so that it is more likely that nodes in $S$ will ``yield'' to the desires of their parent $v$ and thus receive control.
    We also get a nice expression for the difference in these probabilities. 
    By Lemma~\suppref{lem:treatment-event-standard-ht}, we conclude that $U_v= *$ affects the treatment assignment of itself and its children.
    % Indeed, conditioned on $U_v \neq *$, we have that $Z_k   = 0$ for $k \in \Na{v}{H_1}$ with probability one by Lemma~\ref{lem:treatment-event-standard-ht}. 
    % The value of $Z_v$ is not determined by $U_v \neq *$ alone, but we can always upper bound the probability by not considering node $v$, in case $v \in R \setminus S$. 
    Thus, we can write
    \begin{align*}
        &\Prsub[\Bigg]{H_1}{ \setb[\big]{U_v \neq *} \cap \braces[\Big]{ \cup_{k \in S} \{Z_k = 1\}} \cap \braces[\Big]{ \cap_{k \in R \setminus S} \setb{Z_k = 0} } } \\
        &\qquad\leq
        \Prsub[\Bigg]{H_1}{ \setb[\big]{U_v \neq *} \cap \braces[\Big]{ \cup_{k \in S} \{Z_k = 1\} } \cap \braces[\Big]{ \cap_{k \in R \setminus (S\cup \{v\})}\{Z_k = 0\}} }\\
        \intertext{The above follows by monotonicity since $R \setminus S \subseteq R \setminus (S\cup \{v\})$. Notice that it might be the case that $v$ does not belong to $R \setminus S$, but the conclusion still holds. 
        The only treatment assignments that variable $U_v$ affects other than $Z_v$ are $Z_k $ for $k \in \Na{v}{H_1}$. We can thus continue as
        }
        &\qquad\leq
        \Prsub[\Bigg]{H_1}{ \setb[\big]{U_v \neq *} \cap \braces[\Big]{ \cup_{k \in S} \{Z_k = 1\} } \cap \braces[\Big]{ \cap_{k \in R \setminus (S\cup \{v\}\cup \Na{v}{H_1})}\{Z_k = 0\}} }\\
        \intertext{which follows since, by Lemma~\suppref{lem:treatment-event-standard-ht}, $U_v \neq *$ implies that $Z_k = 0$ for all $k \in \Na{v}{H_1}$. Now notice that the treatment of units in $R \setminus \{\{v\} \cup \Na{v}{H_1}\}$ is independent of $U_v$. Hence we can continue as}
        &\qquad=
        \Pr{U_k \neq *}  \cdot      
        \Prsub[\Bigg]{H_1}{  \braces[\Big]{ \cup_{k \in S} \{Z_k = 1\} } \cap \braces[\Big]{ \cap_{k \in R \setminus (S\cup \{v\}\cup \Na{v}{H_1})}\{Z_k = 0\}} }\\
        &\qquad\leq \frac{1}{\lamG}\sum_{k \in S} \operatorname{Pr}_{H_1}\lp(\{U_k = e_1\} \cap \braces[\Big]{ \cap_{k' \in \Nb{k}{H_1}} \{U_{k'} = *\} } \cap \rp.\\
        &\qquad \lp. \braces[\Big]{ \cap_{k \in R \setminus (S\cup \Na{v}{H_1} \cup \{v\})}\{Z_k = 0\} }\rp)\\
        \intertext{where the above follows from a union bound. We get a further upper bound by evaluating the probability of the larger event, }
        &\qquad\leq \frac{1}{\lamG} \sum_{k \in S} \Pr{U_k = e_1}  \\
        &\qquad\leq \frac{|S|}{2\lamG^2} \enspace. 
    \end{align*}
    We now use Lemma~\suppref{lem:probability-lower-bound} for graph $H_1$ and set $S$, which has size at most $C \lamG$ units by assumption. Thus, we have
    \begin{align*}
    \Prsub{H_2}{Z_k = 0 ,\forall k \in R} &- \Prsub{H_1}{Z_k = 0 ,\forall k \in R} \\
    &\leq \frac{|S|}{2\lamG^2}
    	&\text{(above)}\\
    &\leq \frac{|S|}{2\lamG^2} \cdot \braces[\Big]{ e^C \Prsub{H_1}{Z_k = 0 ,\forall k \in R} }
    	&\text{(Lemma~\ref{lem:probability-lower-bound})} \\
    &= \frac{e^C}{2} \cdot \frac{|S|}{\lamG^2} \cdot \Prsub{H_1}{Z_k = 0 ,\forall k \in R} 
    \enspace.
    \end{align*}
	Moreover, we have that $\Prsub{H_2}{Z_k = 0 ,\forall k \in R} \geq \Prsub{H_1}{Z_k = 0 ,\forall k \in R}$.
	By rearranging terms and using this non-negativity, we attain the desired result:
    \[
        \abs[\Bigg]{\frac{\Prsub{H_2}{Z_k = 0 ,\forall k \in R}}{\Prsub{H_1}{Z_k = 0 ,\forall k \in R}} - 1} 
        \leq e^C \cdot \frac{|S|}{2\lamG^2} 
        \enspace.
        \qedhere
    \]
\end{proof}

\subsection{Correlation Lemma: Interpolation Phase I}\label{sec:PhaseI}
The next Sections will be devoted to establishing results that will aid with the proof of Theorem~\suppref{thm:CAB_bound}.
Intuitively, the events $E_A,E_B$ depend on the assignments of the $U$ variables in a neighborhood of $A,B$. These neighborhoods could be overlapping, which introduces a lot of dependence between the two events for the two subsets. We will disentangle these two events by slowly transforming the local topology around $A$ and $B$, with the final goal of making $E_A$ and $E_B$ independent in the end. 
The key technical challenge will be to monitor how $C_{A,B}$ changes along this transformation.

Our goal is to bound 
\[
\abs[\big]{ C_{A,B} }
= \abs[\Bigg]{\frac{\Prsub{G}{E_A \cap E_B}}{\Prsub{G}{E_A} \Prsub{G}{E_B}}-1}
\enspace.
\]
We would essentially like to know how close events $E_A, E_B$ are to being independent. To do that, we will slowly transform $G$ into different topologies, with the goal of keeping the marginal probabilities $\Pr{E_A}, \Pr{E_B}$ fixed, but slowly changing the probability of the intersection, so that in the final topology $\Pr{E_A \cap E_B} = \Pr{E_A} \cdot \Pr{E_B}$ holds.

\begin{figure}
	\centering
    \resizebox{\textwidth}{!}{
	    \input{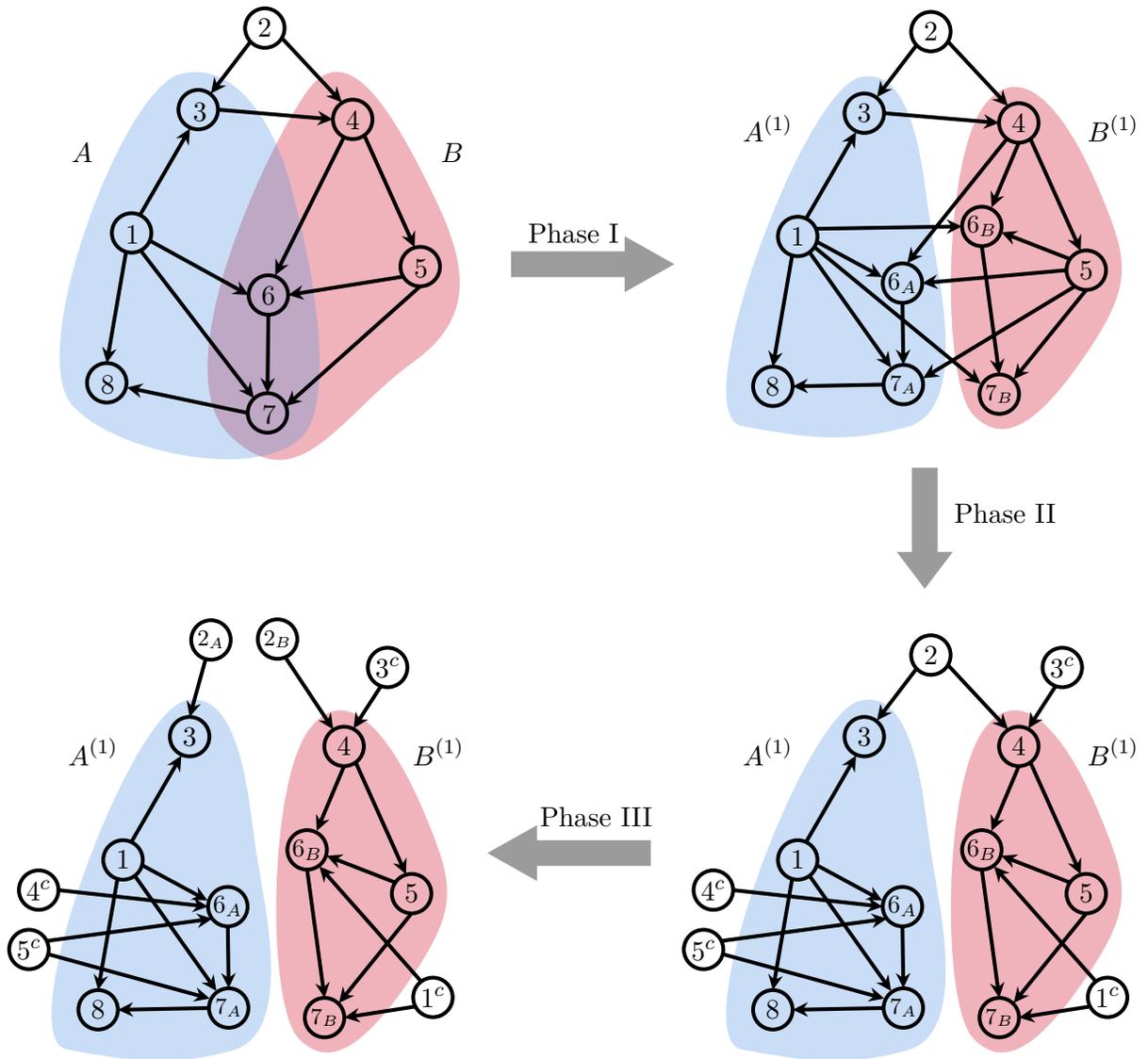}
    }
	\caption{\textbf{Example of Full Interpolation}: 
		The three phases of the interpolation process for two selected sets $A, B$ of an example graph. Note that the only phase that extends the sets $A, B$ is Phase I, when we get the sets $A^{(1)}, B^{(1)}$ that are used for the rest of the interpolation. We remark that we never need to explicitly construct these graphs; they are only used in the variance analysis for the standard Horvitz-Thompson estimator.}
	\label{fig:full-interpolation}
\end{figure}

The interpolation is carried out in three phases.
In this section, we describe Phase I of this interpolation process and focus on bounding the changes in the probability of $E_{A \cup B}$. 
Phase II and Phase III are described in Sections~\suppref{sec:PhaseII} and \suppref{sec:PhaseIII}, respectively.
In Figure~\suppref{fig:full-interpolation}, we present an illustration of a particular graph $G$ going through the three phases.

The first source of dependence between events $E_A$ and $E_B$ comes from the fact that their intersection of $A$ and $B$ might be nonempty.
The idea will be to compare the distribution of the Conflict Graph Design when run on the original graph $G$ versus a modification of $G$ where $A$ and $B$ no longer intersect each other. 
We call this modified graph $\Gone$ and its construction is described in pseudocode in Algorithm~\suppref{alg:fromGtoG1}. We note that we do not actually need to run this Algorithm and it is only useful to consider it for the sake of analysis. 

\begin{algorithm}[h]
    \DontPrintSemicolon
    \caption{\textsc{Phase I Interpolation: from $G$ to $\Gone$}}\label{alg:fromGtoG1}
    \KwIn{Graph $G=(V,E)$, ordering $\pi$, subsets $A,B \subseteq V$.}
    \KwOut{Graph $\Gone = (V_1,E_1)$ and sets $\Aone$ $\Bone$}
    Initialize vertices $V^{(1)} \gets V$ and edges $E^{(1)} \gets E$ \\
    Initialize sets $A_0 \gets A, B_0 \gets B$ and graph $G_0 = G$. \\
    Construct the intersection $I \gets A \cap B$ \\
    Let $\{k_1,\ldotp,k_{|I|}\}$ be the nodes of $I$ in decreasing order of $\pi$. \\
    \For{$i=1$ to |I|}{
    	\tcc{Step 1: Make two copies $k_{i,A}$ and $k_{i,B}$ of $k_i$.} 
        $V^{(1)} \gets (V^{(1)} \cup \{k_{i,A},k_{i,B}\}) \setminus \{k_i\}$\\
        $E^{(1)} \gets (E^{(1)} \cup \{k' \to k_{i,A}: k' \in \Nb{k_i}{G}\} \cup \{k' \to k_{i,B}: k' \in \Nb{k_i}{G}\}) \setminus \{k' \to k_i: k' \to k_i \in E\}$\\
        $E^{(1)} \gets E^{(1)} \cup \{k_{i,B} \to k': k' \in \Na{k_i}{G_{i-1}}\}$\\
        $A_i \gets \paren[\big]{ A_{i-1} \setminus \setb{k_i} } \cup \{k_{i,A}\}$\\
        $B_i \gets \paren[\big]{ B_{i-1} \setminus \setb{k_i} } \cup \{k_{i,B}\}$\\
        $G_i \gets (V^{(1)}, E^{(1)})$\\
        \tcc{Step 2: Connect $k_{i,B}$ to children of $k_i$ in $B_i$}
        $E^{(1)} \gets E^{(1)} \cup \{k_{i,A} \to k': k' \in \Na{k_i}{G_{i}} \cap A_i \}$\\
        $G_i' \gets (V^{(1)}, E^{(1)})$\\
        \tcc{Step 3: Disconnect $k_{i,A}$ from children of $k_i$ in $B_i$}
        $E^{(1)} \gets E^{(1)} \setminus \{k_{i,B} \to k': k' \in \Na{k_i}{G_{i}'} \cap A_i \}$\\
        $G_i'' \gets (V^{(1)}, E^{(1)})$\\
    }
\    $\Aone \gets A_{|I|}$\\
    $\Bone \gets B_{|I|}$
\end{algorithm}

\begin{figure}
    \centering
    \resizebox{\textwidth}{!}{
        \input{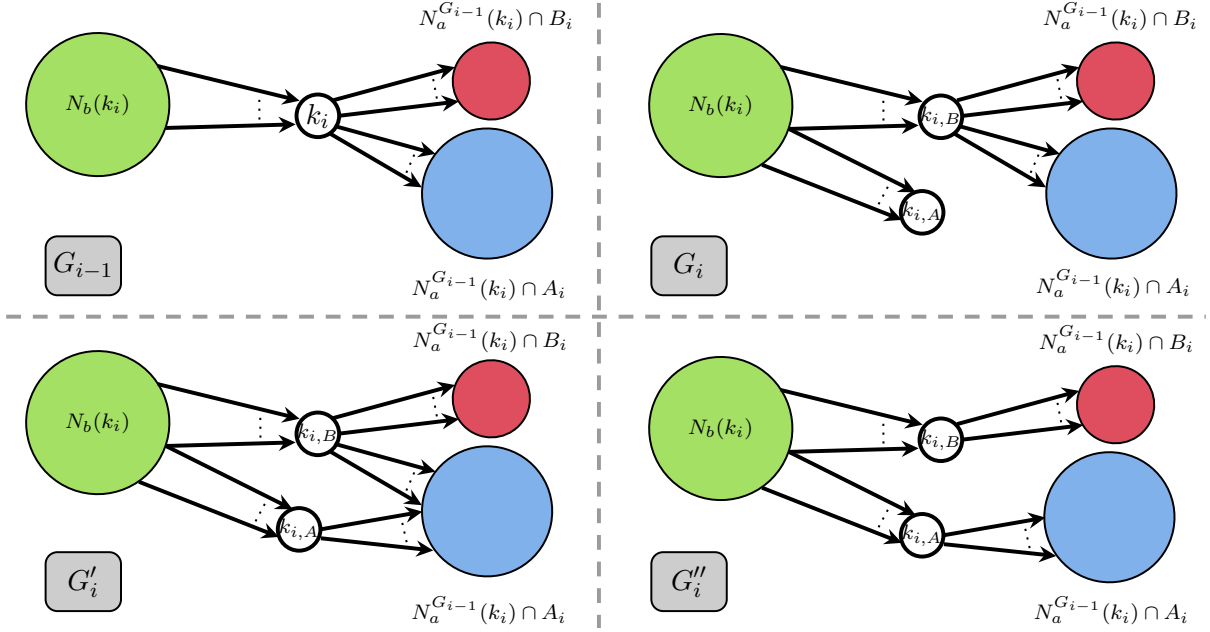}
    }
    \caption{Interpolation Phase I.}
    \label{fig:phase-1}
\end{figure}

See Figure~\suppref{fig:phase-1} for an illustration of this first phase.
The algorithm begins by constructing the intersection $A \cap B = I$.
The procedure then iterates over nodes $k_i \in I$ in the intersection, starting from the least important node, and creates two copies $k_{i,A}$ and $k_{i,B}$. The iteration for $k_i$ can be broken down in three smaller steps.  In the first step, copy $k_{i,B}$ is identified with $k_i$, while $k_{i,A}$ has the same parent set as $k$. We call this topology $G_{i} = (V_{k}, E_{k})$. 
In the second step, we add edges from $k_{A}$ to all units in $\Na{k}{G} \cap A$. We call this topology $G_{i}' = (V_{k}', E_{k}')$.  Finally, in the third step, we remove all edges from $k_{B}$ to $\Na{k}{G} \cap A$. We call this topology $G_{i}''=(V_{k}'', E_{k}'')$. 
This process continues until $I$ becomes empty. 
If we denote the copies $I_A,I_B$ of $I$, then
notice that at the end of Algorithm~\suppref{alg:fromGtoG1} $\Aone = \paren{ A \setminus I} \cup I_A$ and $\Bone = \paren{B \setminus I} \cup I_B$.
If the intersection is empty in the original graph (i.e. $A \cap B = I = \emptyset$) then the graph $G^{(1)}$ returned by Phase I is identical to the original input graph $G$.

An example of the first phase transformation is given in Figure~\suppref{fig:full-interpolation}.

The purpose of this transformation is to make the sets $\Aone$ and $\Bone$ disjoint in $\Gone$, while ensuring that the local topology of $\Aone$ and $\Bone$ in $\Gone$ matches that of $A$ and $B$ from $G$.
In fact, the following lemma shows that the marginal distributions of $\Aone$ and $\Bone$ in $\Gone$ match those of $A$ and $B$ in $G$. 

\begin{lemma}\label{lem:same_marginals_G1}
	Given the construction of the Phase I graph $\Gone$ in Algorithm~\suppref{alg:fromGtoG1} and the sets $\Aone$ and $\Bone$ above, the marginals are preserved:
	\[
	\Prsub{G}{E_A} = \Prsub{\Gone}{E_{\Aone}}
	\quadand
	\Prsub{G}{E_B} = \Prsub{\Gone}{E_{\Bone}}
	\enspace.
	\]
\end{lemma}
\begin{proof}
We prove the claim for $A$ and $\Aone$, as the situation for $B$ and $\Bone$ is identical. 
The proof essentially follows from observing that we can couple the execution of Algorithm~\suppref{alg:conflict-graph-design-direct-effect} for $G$ and $\Gone$ so that the outputs are the same for all variables in $A$ and $\Aone$, respectively. 
More specifically, we can find a one-to-one and onto mapping $\phi: V \to V_1$ by simply mapping every $k \in I$ to its corresponding copy $k_A \in I_A$, while the remaining nodes are mapped to themselves. 
We can check that by the construction of $\Gone$ (Algorithm~\suppref{alg:fromGtoG1}), that for any $k \in A$, the sets $\Nb{k}{G}$ and $\Nb{\phi(k)}{\Gone}$ are in one-to-one correspondence using $\phi$. 

To define the coupling, for each $k \in A$ and corresponding $\phi(k) \in \Aone$, we use the same variable $U_k$ to run Algorithm~\suppref{alg:conflict-graph-design-direct-effect}. In other words
\[
\alphaind{k}{G} = \alphaind{\phi(k)}{\Gone} = \indicator{U_k = e_1} \qquad \betaind{k}{G} = \betaind{\phi(k)}{\Gone} = \indicator{U_k = *}
\enspace.
\]
It is then clear that the output of Algorithm~\suppref{alg:conflict-graph-design-direct-effect} when run on $G$ and $\Gone$ will be exactly the same, restricted to the sets $A$ and $\Aone$ respectively. This concludes the argument. 
\end{proof}

Our goal will be to study how the probability $\Pr{E_A \cup E_B}$ changes when we interpolate from $G$ to $\Gone$. Since the ``analog'' of $A,B$ in $\Gone$ is $\Aone,\Bone$, we seek to control the quantity:
 \[
 \abs[\Bigg]{\frac{\Prsub{G}{E_{A \cup B}}}{\Prsub{\Gone}{E_{{\Aone} \cup \Bone}}}-1} = 
 \abs[\Bigg]{\frac{\Pp[G]{Z_k = 0 ,\forall k \in A \cup B}}{\Pp[\Gone]{Z_k = 0, \forall k \in \Aone \cup \Bone}} -1}
 \enspace.
 \]

 To analyze the above quantity, it will be useful to consider the smaller steps of the transformation 
 that are described in Algorithm~\suppref{alg:fromGtoG1}. 

 In particular, let $k_1,\ldotp k_{|I|}$ be the decreasing ordering of the nodes in $I$. For a node $k_i \in I$, we wish to analyze how the probability $\Prsub{G}{E_{A\cup B}}$ changes along the transition $G_{{i-1}}'' \to G_{i} \to G_{i}' \to G_{i}''$.
 In order to do this, we must define the ``analog'' of  $A \cup B$ in $G_{i}, G_{i}', G_{i}''$, since $A,B$ do not exist anymore in these graphs. 
 To this end, we need natural ``analogues'' of $A$ and $B$ in these graphs. These are given by the intermediate sets $A_i,B_i$. In particular, $A_i$ contains all copies of nodes $k_{1,A},\ldotp,k_{i,A}$ and the rest of the original nodes on $A$, similarly for $B_i$. 
Our goal will be to study the following three quantities
\[
\mathcal{P}_{i} := \Prsub{G_{i}}{E_{A_i \cup B_i}}
\quad
\mathcal{P}_{i}' := \Prsub{G_{i}'}{E_{A_i \cup B_i}}
\quad
\mathcal{P}_{i}'' := \Prsub{G_{i}''}{E_{A_i \cup B_i}}
\enspace.
\]
Our first Lemma establishes that $\mathcal{P}_{i-1}'', \mathcal{P}_{i}$ are multiplicatively close. 

\begin{lemma}\label{lem:phaseIfirst}
    For some constant $C'$ we have that for every $i$
    \[
        \abs[\Bigg]{\frac{\mathcal{P}_{i-1}''}{\mathcal{P}_{i}}-1} \leq \frac{C'}{\lamG}    
    \]
\end{lemma}
\begin{proof}
    We will couple the execution of Algorithm~\suppref{alg:conflict-graph-design-direct-effect} for $G$ and $G_i$. We can identify the $U_k$ variables used in Algorithm~\suppref{alg:conflict-graph-design-direct-effect} for all nodes in $G$ and $G_i$ using a one-to-one correspondence, except for node $k_{i,A}$, which only exists in $G_i$. Thus, in the coupling $k_{i,A}$ will have an independent variable $U_{k_{i,A}}$ assigned to it. Unit $k_i$ in $G$ corresponds to unit $k_{i,B}$ in $G_i$ and we use the variables $\alpha_{k_i}, \beta_{k_i}$ for both. 
    Thus, we have
       \begin{align*}
       &\mathcal{P}_{i-1}'' - \mathcal{P}_{i} \\
       &= 
       \operatorname{E}\lp[\prod_{k\in A \cup B}
       \paren[\Big]{1 - \alpha_k \prod_{k' \in N_b(k_i)} \beta_{k'} } - \rp.\\
       &\qquad \lp.
       \paren[\Big]{1 - \alpha_{k_{i,A}} \prod_{k' \in N_b(k_{i,A})} \beta_{k'}}\prod_{k\in A \cup B}
       \paren[\Big]{1 - \alpha_k \prod_{k' \in N_b(k_i)} \beta_{k'} }\rp]\\
       &= \E[\Bigg]{
           \alpha_{k_{i,A}} \prod_{k' \in N_b(k_{i,A})} \beta_{k'}    
       \prod_{k\in A \cup B}
       \paren[\Big]{1 - \alpha_k \prod_{k' \in N_b(k_i)} \beta_{k'} } }\\
       &\leq
       \E[\Bigg]{
           \alpha_{k_{i,A}}    
       \prod_{k\in A \cup B}
       \paren[\Big]{1 - \alpha_k \prod_{k' \in N_b(k_i)} \beta_{k'} } }\\
       &= \frac{1}{2\lamG} \E[\Bigg]{   
       \prod_{k\in A \cup B}
       \paren[\Big]{1 - \alpha_k \prod_{k' \in N_b(k_i)} \beta_{k'} } }\\
       &= \frac{1}{2\lamG} \mathcal{P}_{i} \enspace.
       \end{align*}
   In the second to last step, we used the fact that variable $\alpha_{k_{i,A}}$ does not affect any other nodes in the graph $G_i$, since $k_{i,A}$ has no children in $G_i$. 
    It is also clear from the above expression that $\mathcal{P}_{i-1}'' \geq \mathcal{P}_{i}$.
   Rearranging the above gives 
   \begin{equation}\label{eq:fromGtoGk0}
   \abs[\Bigg] {\frac{\Pp[G]{E_{A\cup B}}}{\mathcal{P}_i }-1} \leq \abs[\Bigg]{\frac{1}{1 - 1/(2\lamG)} - 1} \leq \frac{C'}{2\lamG}\,,
   \end{equation}
   where $C'$ is an absolute constant and  we have used that $\lamG \geq 2$.
\end{proof}
% Let's analyze each of these three quantities separately. 
% \chriscomment{this would be much more readable if you can organize into lemmas / proofs.}

Our second Lemma establishes that $\mathcal{P}_{i}, \mathcal{P}_{i}'$ are multiplicatively close.

\begin{lemma}\label{lem:phaseIsecond}
    For some constant $C'$ we have that for every $i$
    \[
        \abs[\Bigg]{\frac{\mathcal{P}_{i}}{\mathcal{P}_{i}'}-1} \leq \frac{C'}{\lamG}    
    \]
\end{lemma}

\begin{proof}
    In this step, we add edges from $k_{i,A}$ to all nodes in $N_a(k_i) \cap A_i$.
We can view this as an instance of Proposition~\suppref{prop:interpolation-step} where $H_1 = G_{i}, H_2 = G_{i}', R =A_i \cup B_i, S = N_a(k_i) \cap A, v = k_{i,A}$.
Observe that $S \subseteq R$ and that $|R| \leq 2C\lamG + 1$.
Hence, there exists constants $C'$ and $C''$ that only depend on $C$, such that
\[
\abs[\Bigg]{\frac{\mathcal{P}_i}{\mathcal{P}_i' }-1} 
\leq C' \frac{|S|}{\lamG^2}
\leq C' \cdot \frac{2C\lamG + 1}{\lamG^2}
\leq C'' \frac{1}{\lamG}
\enspace.
\qedhere
\]
\end{proof}

Our third Lemma establishes that $\mathcal{P}_{i}', \mathcal{P}_{i}''$ are multiplicatively close.

\begin{lemma}\label{lem:phaseIthird}
    For some constant $C'$ we have that for every $i$
    \[
        \abs[\Bigg]{\frac{\mathcal{P}_{i}'}{\mathcal{P}_{i}''}-1} \leq \frac{C'}{\lamG}   
        \enspace. 
    \]
\end{lemma}

\begin{proof}
    At this step, we remove the edges from $k_{i,B}$ to the set $N_a(k_i)\cap A$. 
Again, we can view this as an instance of Proposition~\suppref{prop:interpolation-step} where $H_1 = G_{i}'', H_2 = G_{i}', R =A_i \cup B_i, S = N_a(k_i) \cap A, v = k_{i,B}$.
Observe that $S \subseteq R$ and that $|R| \leq 2C\lamG + 1$.
Hence, by Proposition~\ref{prop:interpolation-step}, there exists constants $C'$ and $C''$ that only depends on $C$, such that
\[
    \abs[\Bigg]{\frac{\mathcal{P}_i'}{\mathcal{P}_i'' }-1} 
    \leq C' \frac{|S|}{\lamG^2} 
    \leq C' \cdot \frac{2C\lamG + 1}{\lamG^2}
    \leq C'' \frac{1}{\lamG}
    \enspace.
    \qedhere
\]
\end{proof}

Finally, we provide a Lemma that quantifies the change in probability between graphs $G$ and $\Gone$, using the previous three Lemmas as building blocks. The proof proceeds by essentially adding up all the errors accumulated over the individual steps in Algorithm~\suppref{alg:fromGtoG1}. 

\begin{lemma}\label{lem:fromGtoG1}
    There exists a constant $C'$ only depending on $C$, such that
    \begin{equation}
    \abs[\Bigg]{\frac{\Prsub{G}{E_{A\cup B}}}{\Prsub{G^{(1)}}{E_{A^{(1)} \cup B^{(1)}}}} - 1} \leq C' \frac{|I|}{\lamG}
    \enspace.
    \end{equation} 
\end{lemma}

\begin{proof}
    By definition, we have $G_{|I|}'' = \Gone$ and thus $\mathcal{P}_{|I|}'' = \Prsub{\Gone}{E_{\Aone\cup \Bone}}$. 
    Using the convention $\Prsub{G}{E_{A \cup B}} = G_0''$, we can write the telescopic product
    \begin{align*}
        \abs[\Bigg]{\frac{\Prsub{G}{E_{A\cup B}}}{\Prsub{G^{(1)}}{E_{A^{(1)} \cup B^{(1)}}}} - 1}  &= 
        \abs[\Bigg]{\prod_{i=1}^{|I|}\paren[\Big]{\frac{\mathcal{P}_{i-1}''}{\mathcal{P}_{i}} \cdot \frac{\mathcal{P}_{i}}{\mathcal{P}_{i}'} \cdot \frac{\mathcal{P}_{i}'}{\mathcal{P}_{i}''}} - 1} 
        \enspace.
    \end{align*}
    We can now use Lemma~\suppref{lem:linearization} with 
    \[
        a_{1,i} = \frac{\mathcal{P}_{i-1}''}{\mathcal{P}_{i}} - 1 \qquad a_{2,i} = \frac{\mathcal{P}_{i}}{\mathcal{P}_{i}'}-1 \qquad
        a_{3,i} = \frac{\mathcal{P}_{i}'}{\mathcal{P}_{i}''}   -1 
        \enspace.
    \]
    Indeed, by Lemmas~\suppref{lem:phaseIfirst},~\suppref{lem:phaseIsecond} and ~\suppref{lem:phaseIthird} we get that, for a constant $C'$ 
    \[
    \sum_{i=1}^{|I|} \paren{|a_{1,i}| + |a_{2,i}| + |a_{3,i}|} \leq C' \frac{|I|}{\lamG} \leq C' \frac{\dmax}{\lamG} \leq C' \cdot C
    \enspace.
    \]
    Thus, by Lemma~\suppref{lem:linearization} we get that for some constant $C'$
    \[
    \abs[\Bigg]{\frac{\Prsub{G}{E_{A\cup B}}}{\Prsub{G^{(1)}}{E_{A^{(1)} \cup B^{(1)}}}} - 1} 
    \leq C' \sum_{i=1}^{|I|} \paren{|a_{1,i}| + |a_{2,i}| + |a_{3,i}|}  
    \leq C'' \frac{|I|}{\lamG}
    \enspace.
    \qedhere
    \]
\end{proof}

% \textbf{Phase II: Removing Edges Between $A^{(1)}$, $B^{(1)}$}
\subsection{Correlation Lemma: Interpolation Phase II}\label{sec:PhaseII}
Phase I ensures that $A^{(1)}$ and $B^{(1)}$ are disjoint sets, i.e. $A^{(1)} \cap B^{(1)} = \emptyset$. The goal of Phase II will be to eliminate edges from $A^{(1)}$ to $B^{(1)}$. Intuitively, this will reduce correlations even further between the two sets. As always, our goal will be to preserve the marginal probabilities for $A^{(1)}$ and $B^{(1)}$. 
We now define the graph $G^{(2)} = (V_2,E_2)$ through an iterative procedure described in Algorithm~\suppref{alg:fromG1toG2}. Again, we emphasize that this is a procedure that we study for the sake of analysis, we don't need to actually run it. 

\begin{algorithm}[h]
    \DontPrintSemicolon
    \caption{\textsc{Phase II Interpolation: from $\Gone$ to $\Gtwo$}}\label{alg:fromG1toG2}
    \KwIn{Graph $G^{(1)}=(V^{(1)},E^{(1)})$, subsets $\Aone,\Bone \subseteq V_1$.}
    \KwOut{Graph $\Gtwo = (\Vtwo,\Etwo)$}
    $\Vtwo \gets \Vone, \Etwo \gets \Eone$\\
    Let $\Aone = \{k_1,\ldotp,k_{|\Aone|}\}, \Bone = \{l_1,\ldotp,l_{|\Bone|}\}$ be arbitrary ordering. \\
    \For{$i=1$ to $|\Aone|$}{
    	\tcc{Step 1: Create a copy $k_i^{(c)}$ of $k_i$ and connect it to children of $k_i$ in $\Bone$}
        $\Vtwo \gets \Vtwo \cup \{k_i^{(c)}\}$\\
        $\Etwo \gets \Etwo \cup \{k_i^{(c)} \to k': k' \in \Na{k_i}{\Gone} \cap \Bone\}$\\
        $G_i^{(1)} \gets (\Vtwo,\Etwo)$\\
        \tcc{Step 2: Remove edges from $k_i$ to its children in $\Bone$}
        $\Etwo \gets \Etwo \setminus \{k_i \to k': k' \in \Na{k_i}{\Gone} \cap \Bone\}$\\
        $\paren[\big]{G_i^{(1)}}' \gets (\Vtwo,\Etwo)$
    }
    \For{$i=1$ to $|\Bone|$}{
    	\tcc{Step 1: Create a copy of $l_i^{(c)}$ of $l_i$ and connect it to children of $l_i$ in $\Aone$}
        $\Vtwo \gets \Vtwo \cup \{l_i^{(c)}\}$\\
        $\Etwo \gets \Etwo \cup \{l_i^{(c)} \to l': l' \in \Na{l_i}{\Gone} \cap \Aone\}$\\
        $G_{|A|+i}^{(1)} \gets (\Vtwo,\Etwo)$\\
        \tcc{Step 2: Remove edges from $l_i$ to its children in $\Aone$}
        $\Etwo \gets \Etwo \setminus \{k_i \to l': l' \in \Na{l_i}{\Gone} \cap \Aone\}$\\
        $\paren[\big]{G_{|A|+i}^{(1)}}' \gets (\Vtwo,\Etwo)$
    }
\end{algorithm}

In every iteration of Phase II, we select a node $k_i$ that has children on the other subset. Without loss of generality, suppose we choose $k \in A^{(1)}$ with $N_a(k) \cap B^{(1)} \neq \emptyset$. 
We then create a copy $k^{(c)}$ and connect $\kc{k_i}$ as a parent to $\Na{k_i}{G^{(1)}} \cap B^{(1)} $.
We call this topology $G^{(1)}_i$.  
Then, we delete all edges from $k_i$ to $\Na{k_i}{G^{(1)}} \cap B^{(1)} $. 
We call the resulting topology $(G^{(1)}_{i})'$. 
Note that the sets $A^{(1)}, B^{(1)}$ remain the same, i.e. $\kc{k_i}$ is not added to them. 
Pictorially,the steps of a single iteration are shown in Figure~\suppref{fig:phase-2}.
An example of the phase II transformation is given in Figure~\suppref{fig:full-interpolation}.

\begin{figure}
    \centering
    \resizebox{\textwidth}{!}{
        \input{\figpath/standard-ht/interpolation-phase2.tex}
    }
    \caption{Interpolation Phase II.}
    \label{fig:phase-2}
\end{figure}

As in the previous phase, we argue that the marginals of the distribution on $A^{(1)}, B^{(1)}$ don't change from $G^{(1)}$ to $G^{(2)}$. 

\begin{lemma}\label{lem:same_marginals_G2}
Given the construction of the Phase II graph $\Gtwo$ in Algorithm~\suppref{alg:fromG1toG2} and the sets $\Aone$ and $\Bone$ above, the marginals are preserved:
\[
 \Prsub{G^{(1)}}{E_{A^{(1)}}} = \Prsub{G^{(2)}}{E_{A^{(1)}}}
 \quadand
\Prsub{G^{(1)}}{E_{B^{(1)}}} = \Prsub{G^{(2)}}{E_{B^{(1)}}}
\enspace.
\]
\end{lemma}
\begin{proof}
    We prove the claim for $A^{(1)}$, as the proof for $B^{(1)}$ is similar. 
    Similarly to the proof of Lemma~\suppref{lem:same_marginals_G1}, we couple the distributions of $\Aone$ when Algorithm~\suppref{alg:conflict-graph-design-direct-effect} is run for $\Gone$ and $\Gtwo$, respectively. In particular, we can define a mapping $\phi: V_1 \mapsto V_2$, where every node in $\Aone$ is mapped to itself and every node $k \in \Bone$ with $\Na{k}{\Gone} \cap \Aone \neq \emptyset$ is mapped to it's copy $\kzeroc$. 
    Then, we identify $k \in V_1$ and $\phi(k) \in V_2$ with the same random variable $U_k$ in the execution. 
    It is straightforward to check using Algorithm~\suppref{alg:fromG1toG2} that for every $k \in \Aone$, the sets $\Nb{k}{\Gone}$ and $\Nb{k}{\Gtwo}$ are in one-to-one correspondence using $\phi$. Thus, the output of the coupled executions will be the same for $\Gone$ and $\Gtwo$ and the proof is complete. 
\end{proof}

Our goal will be to study how $\Pr{E_{\Aone \cup \Bone}}$ changes from $(G_{i-1}^{(1)})'$ to $G^{(1)}_{i}$ and then to $(G^{(1)}_{i})'$. 
Thus, let us define the  quantities
\[
\mathcal{P}_{i} = \Prsub{G^{(1)}_{i}}{E_{\Aone \cup \Bone}} 
\quadand
\mathcal{P}_{i}' = \Prsub{(G^{(1)}_{i})'}{E_{\Aone \cup \Bone}} \enspace.
\]
The first lemma establishes that $\mathcal{P}_{i-1}', \mathcal{P}_i$ are multiplicatively close to each other. 

\begin{lemma}\label{lem:PhaseIIfirst}
    For every $0 \leq i \leq |A| + |B|$, the following holds
    \begin{itemize}
        \item If $i \leq |A|$, then 
        \[
            \abs[\Bigg]{\frac{\mathcal{P}_{i-1}'}{\mathcal{P}_{i}}-1} \leq  C' \frac{|\Na{k_i}{\Gone} \cap \Bone|}{\lamG^2}
        \]
        \item If $i > |A|$, then 
        \[
            \abs[\Bigg]{\frac{\mathcal{P}_{i-1}'}{\mathcal{P}_{i}}-1} \leq  C' \frac{|\Na{l_{i - |A|}}{\Gone} \cap \Aone|}{\lamG^2}
        \]
    \end{itemize}
\end{lemma}
\begin{proof}
    The case analysis on the indices simply corresponds to analysing the copies in $\Aone$ and $\Bone$, respectively. We focus on the case $i \leq |A|$ for simplicity, the proof is identical for the other case. 
    Fix $k_i \in \Aone$. The difference from $\mathcal{P}_{i-1}'$ to $\mathcal{P}_{i}$ is that we are adding an extra parent $\kc{k_i}$ for all nodes in $\Na{k_i}{\Gone} \cap \Bone$.
    Thus, we can apply Proposition~\suppref{prop:interpolation-step} with $H_1 = (G_{i-1}^{(1)})', H_2 = G^{(1)}_{i}, R = \Aone \cup \Bone, S = \Na{k_i}{\Gone} \cap \Bone, v = \kc{k_i}$. The assumption of Proposition~\suppref{prop:interpolation-step} is satisfied, since $|R| \leq 2C \lamG$. 
Thus, there exists a constant $C'$ such that
\[
 \abs[\Bigg]{\frac{\mathcal{P}_{i-1}'}{\mathcal{P}_{i}}-1} 
 \leq C' \frac{|S|}{\lamG^2} 
 = C'' \frac{|\Na{k_i}{\Gone} \cap \Bone|}{\lamG^2}
 \enspace.
 \qedhere
\]
\end{proof}

The second lemma establishes that $\mathcal{P}_{i}, \mathcal{P}_i'$ are multiplicatively close to each other. 

\begin{lemma}\label{lem:PhaseIIsecond}
    For every $0 \leq i \leq |A| + |B|$, the following holds
    \begin{itemize}
        \item If $i \leq |A|$, then 
        \[
            \abs[\Bigg]{\frac{\mathcal{P}_{i}}{\mathcal{P}_{i}'}-1} \leq  C' \frac{|\Na{k_i}{\Gone} \cap \Bone|}{\lamG^2}
        \]
        \item If $i > |A|$, then 
        \[
            \abs[\Bigg]{\frac{\mathcal{P}_{i}}{\mathcal{P}_{i}'}-1} \leq  C' \frac{|\Na{l_{i - |A|}}{\Gone} \cap \Aone|}{\lamG^2}
        \]
    \end{itemize}
\end{lemma}
\begin{proof}
    As in Lemma~\suppref{lem:PhaseIIfirst}, the case analysis on the indices simply corresponds to analysing the copies in $\Aone$ and $\Bone$, respectively. We focus on the case $i \leq |A|$ for simplicity, the proof is identical for the other case. 
    Fix $k_i \in \Aone$. The difference from $\mathcal{P}_{i}$ to $\mathcal{P}_{i}'$ is that we are removing the parent $k_i$ from all nodes in $\Na{k_i}{\Gone} \cap \Bone$.
    Thus, we can apply Proposition~\suppref{prop:interpolation-step} with $H_1 = (G_{i}^{(1)})', H_2 = G^{(1)}_{i}, R = \Aone \cup \Bone, S = \Na{k_i}{\Gone} \cap \Bone, v = k_i$. The assumption of Proposition~\suppref{prop:interpolation-step} is satisfied, since $|R| \leq 2C \lamG$. 
Thus, there exists a constant $C'$ such that
\[
\abs[\Bigg]{\frac{\mathcal{P}_{i}}{\mathcal{P}_{i}'}-1} 
\leq C' \frac{|S|}{\lamG^2} 
= C'' \frac{|\Na{k_i}{\Gone} \cap \Bone|}{\lamG^2}
\enspace.
\qedhere
\]
\end{proof}

We can now combine the previous two lemmas to bound the multiplicative change in the probability of event $E_{\Aone \cup \Bone}$ from $\Gone$ to $\Gtwo$. 

\begin{lemma}\label{lem:fromG1toG2}
    There exists a constant $C'$ depending only on $C$, such that
    \begin{equation}
        \abs[\Bigg]{\frac{\Prsub{\Gone}{E_{\Aone\cup \Bone}}}{\Prsub{\Gtwo}{E_{A^{(1)} \cup B^{(1)}}}} - 1} \leq C' \frac{F_{A,B}}{\lamG^2}
        \end{equation} 
\end{lemma}

\begin{proof}
    Notice that $G^{(1)}_{|A|+|B|} = \Gtwo$ by definition. By taking the convention $(\Gone = G_0^{(1)})'$, we can write the telescopic product
    \[
        \abs[\Bigg]{\frac{\Prsub{\Gone}{E_{\Aone\cup \Bone}}}{\Prsub{\Gtwo}{E_{A^{(1)} \cup B^{(1)}}}} - 1}  = 
        \abs[\Bigg]{\prod_{i=1}^{|A|+|B|}\paren[\Big]{\frac{\mathcal{P}_{i-1}'}{\mathcal{P}_{i}} \cdot \frac{\mathcal{P}_{i}}{\mathcal{P}_{i}'} } - 1} 
    \]
    We can now use Lemma~\suppref{lem:linearization} with 
    \[
        a_{1,i} = \frac{\mathcal{P}_{i-1}'}{\mathcal{P}_{i}} - 1 \qquad a_{2,i} = \frac{\mathcal{P}_{i}}{\mathcal{P}_{i}'}-1 \enspace.
    \]
    Indeed, by Lemmas~\suppref{lem:PhaseIIfirst},~\suppref{lem:PhaseIIsecond} we get that, for a constant $C'$ 
    \[
    \sum_{i=1}^{|I|} \paren{|a_{1,i}| + |a_{2,i}| } \leq C' \frac{\sum_{k \in A}|\Na{k}{\Gone} \cap \Bone| + \sum_{k \in B}|\Na{k}{\Gone} \cap \Aone|  }{\lamG^2} \leq C'' \enspace,
    \]
    where we used the fact that each node in $\Aone \cup \Bone$ has at most $\dmax \leq C \lamG$ children, and $|\Aone \cup \Bone| = \bigO{\lamG}$. 
    Thus, by Lemma~\suppref{lem:linearization} we get that for some constant $C'$
    \begin{align*}
        \abs[\Bigg]{\frac{\Prsub{G}{E_{A\cup B}}}{\Prsub{G^{(1)}}{E_{A^{(1)} \cup B^{(1)}}}} - 1} 
        &\leq C' \sum_{i=1}^{|I|} \paren{|a_{1,i}| + |a_{2,i}| + |a_{3,i}|}  \\
        &\leq C'' \frac{\sum_{k \in A}|\Na{k}{\Gone} \cap \Bone| + \sum_{k \in B}|\Na{k}{\Gone} \cap \Aone|  }{\lamG^2} \enspace.
    \end{align*}
    To prove the claim, it suffices to establish that
    \[
        \sum_{k \in A}|\Na{k}{\Gone} \cap \Bone| + \sum_{k \in B}|\Na{k}{\Gone} \cap \Aone|  = F_{A,B}
        \enspace.
    \]
    We show that the left hand side counts exactly the edges that contribute to $F_{A,B}$. Indeed, edges with both endpoints in 
    $A\setminus B$ or both endpoints in $B \setminus A$ will not be altered in Phase I and thus will not be counted in the left hand side on Phase II. Also, edges $(u,v)$ with both endpoints in $I$ in $G$ will be duplicated to two edges $(u_A,v_A)$ and $(u_B,v_B)$, which lie inside $\Aone,\Bone$ respectively and thus will not be counted in the left hand side. 
    The remaining types of edges, which have the two endpoints in different subsets of the partition $A \setminus B, B \setminus A, A \cap B$ can easily be shown to correspond to edges between $\Aone$ and $\Bone$ after Phase I. This concludes the proof. 
\end{proof}

% \textbf{Phase III: Removing Common Parents of $\Aone$ and $\Bone$}
\subsection{Correlation Lemma: Interpolation Phase III}\label{sec:PhaseIII}
Phase II removed all edges between $\Aone, \Bone$. The only remaining correlations between $\Aone,\Bone$ is now coming from common parents of nodes in $\Aone,\Bone$ that do not belong to $\Aone \cup \Bone$. 
Our goal in Phase III will be to remove these dependencies. 
We now describe the final topology $\Gthree$ that will be the end of interpolation. The topology of $\Gthree$ can be constructed iteratively from $\Gtwo$ by the process that is described in Algorithm~\ref{alg:fromG2toG3}. 

\begin{algorithm}
    \DontPrintSemicolon
    \caption{\textsc{Phase III Interpolation: from $\Gtwo$ to $\Gthree$}}\label{alg:fromG2toG3}
    \KwIn{Graph $G^{(2)}=(V^{(2)},E^{(2)})$, subsets $\Aone,\Bone \subseteq V^{(2)}$.}
    \KwOut{Graph $\Gthree = (\Vthree,\Ethree)$}
    $\Vthree \gets \Vtwo, \Ethree \gets \Etwo$\\
    $T_{A,B} \gets \{k \notin \Aone \cup \Bone : \Na{k}{\Gtwo} \cap \Aone \neq \emptyset, \Na{k}{\Gtwo} \cap \Bone \neq \emptyset \}$\\
    Let $T_{A,B} = \{k_1,\ldotp,k_{|R_{A,B}|}\}$ be arbitrary ordering.\\
    \For{$i=1$ to $|T_{A,B}|$}{
    	\tcc{Create two copies $k_{i,A}$ and $k_{i,B}$ of $k_i$}
        $\Vthree \gets (\Vthree \cup \{k_{i,A}, k_{i,B}\}) \setminus \{k_i\}$\\
        \tcc{Remove $k_i$ and its edges from $\Gtwo$}
        $\Ethree \gets \Ethree \setminus \{(k_i, k'): k' \in N^{\Gtwo}(k_i)\} $\\
        \tcc{Connect $k_{i,A}$ to the children of $k_i$ in $\Aone$ and $k_{i,B}$ to the children of $k_i$ in $\Bone$}
        $\Ethree \gets \Ethree \cup \{k_{i,A} \to k': k' \in \Na{k_i}{\Gtwo} \cap \Aone\} \cup \{k_{i,B} \to k': k' \in \Na{k_i}{\Gtwo} \cap \Bone\}$\\
        $G_i^{(2)} \gets (\Vthree, \Ethree)$
    }
\end{algorithm}

\begin{figure}
	\centering
    \resizebox{0.9 \textwidth}{!}{
	    \input{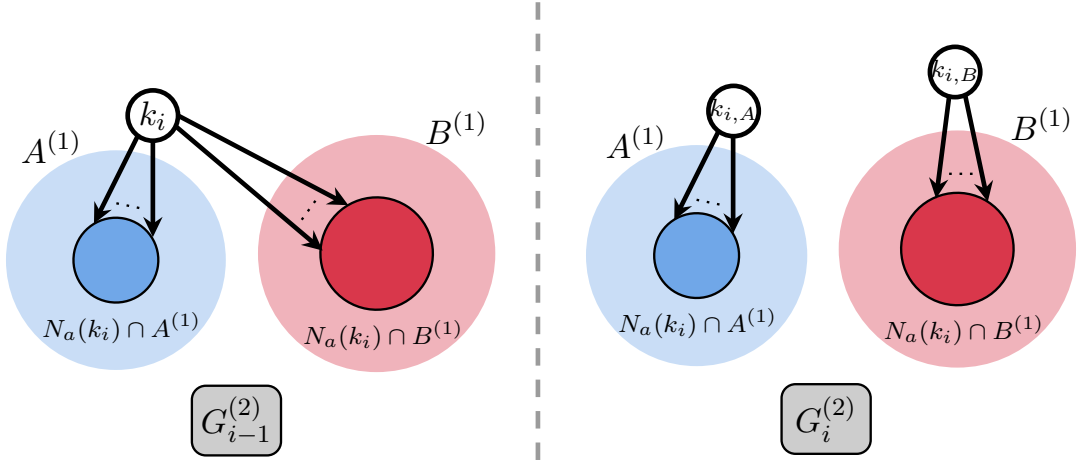}
    }
	\caption{Interpolation Phase III.}
	\label{fig:phase-3}
\end{figure}

Algorithm~\suppref{alg:fromG2toG3} selects one common parent $k_i \in T_{A,B}$ in each iteration. It creates two copies $k_{i,A}, k_{i,B}$, and connects $k_{i,A}$ to the children of $k_i$ in $\Aone$ and $k_{i,B}$ with the children of $k_i$ in $\Bone$. 
We call this topology $G^{(2)}_i$.
Pictorially, the steps of a single iteration are shown in Figure~\suppref{fig:phase-3}.
An example of the Phase III transformation is given in Figure~\suppref{fig:full-interpolation}.

As in the previous phases, we establish that the marginals of $\Aone$ and $\Bone$ do not change from $\Gtwo$ to $\Gthree$.
In addition, we claim that events $E_{\Aone}$ and $E_{\Bone}$ are independent in $\Gthree$:

\begin{lemma}\label{lem:same_marginals_G3}
    The events $E_A$ and $E_B$ are independent under the Conflict Graph Design given by the topological ordering of $\Gthree$:
    \[
    \Prsub[\Big]{\Gthree}{E_{\Aone \cup \Bone}} =  \Prsub[\Big]{G}{E_A} \cdot \Prsub[\Big]{G}{E_B }
    \enspace.
    \]
\end{lemma}
\begin{proof}
    We have that
    \begin{align*}
        \Prsub[\Big]{\Gthree}{E_{\Aone \cup \Bone}} &=
    \Esub[\Bigg]{\Gthree}{\underbrace{\prod_{k \in \Aone}\paren[\Big]{1 - \alpha_k\prod_{k' \in N_b(k)} \beta_{k'}} }_{Y_A}\cdot \underbrace{\prod_{k \in \Bone}\paren[\Big]{1 - \alpha_k\prod_{k' \in N_b(k)} \beta_{k'}}}_{Y_B} }
    \end{align*}
    Notice that all variables that appear in the expression of random variable $Y_A$ are independent from all variables that appear in $Y_B$. This follows from the property of $\Gthree$ that $\Aone \cap \Bone = \emptyset$ (because of Phase I), there are no edges between $\Aone$ and $\Bone$ (because of Phase II) and no-common parents between nodes in $\Aone$ and $\Bone$ (because of Phase III). Thus, the expectation can be broken into the product of expectations
     \begin{align*}
    \Prsub[\Big]{\Gthree}{E_{\Aone \cup \Bone}} &=
     \Esub[\Bigg]{\Gthree}{\prod_{k \in \Aone}\paren[\Big]{1 - \alpha_k\prod_{k' \in N_b(k)} \beta_{k'}} } \\
     &\qquad\cdot \Esub[\Bigg]{\Gthree}{\prod_{k \in \Bone}\paren[\Big]{1 - \alpha_k\prod_{k' \in N_b(k)} \beta_{k'}} }\\
     &=  \Prsub[\Big]{\Gthree}{E_{\Aone} } \cdot  \Prsub[\Big]{\Gthree}{E_{\Bone}}
     \end{align*}
    To finish the proof, by Lemmas~\suppref{lem:same_marginals_G1} and~\suppref{lem:same_marginals_G2}, it suffices to prove that the marginals are the same as in $\Gtwo$.  Let's focus on $\Aone$, since the proof for $\Bone$ is identical. The strategy will be the same as in the proof of Lemma~\suppref{lem:same_marginals_G2}. Let $\phi:V_2 \mapsto V_3$ be the mapping where every node $k \in T \subseteq V_2$ is mapped to its copy $k_{i,A} \in T_A$. The rest of $\phi$ is identity. Then, we can couple the executions of Algorithm~\suppref{alg:conflict-graph-design-direct-effect} when run on $\Gtwo$ and $\Gthree$ by using the same variable $U_k$ for $k$ and $\phi(k)$, for all $k \in V_1$. 
    It is straightforward to check using Algorithm~\suppref{alg:fromG2toG3} that for every $k \in \Aone$, the sets $\Nb{k}{\Gtwo}$ and $\Nb{\phi(k)}{\Gthree}$ are in one-to-one correspondence using $\phi$. Thus, the outputs of the two executions will be identical for $\Aone$ and the proof is complete. 
\end{proof}

As in the previous Phases, we analyze the change in $\Pr{E_{\Aone \cup \Bone}}$ in one iteration of Algorithm~\suppref{alg:fromG2toG3}.
To do that, again we introduce the relevant notation
\[
\mathcal{P}_{i} = \Prsub{G^{(2)}_{i}}{E_{\Aone \cup \Bone}}\enspace.
\]
We also use the convention $\mathcal{P}_{0} = \Prsub{G^{(2)}}{E_{\Aone \cup \Bone}}$. 
The precise statement is given in the following Lemma. 

\begin{lemma}\label{lem:Phase3change}
    For $i \leq |T_{A,B}|$, we have for some constant $C'$ that only depends on $C$ that
    \[
    \abs[\Bigg]{\frac{\mathcal{P}_{i-1}}{\mathcal{P}_{i}}-1}
    \leq C' \frac{\abs[\Big]{\Na{k_i}{\Gtwo} \cap \Aone} \cdot \abs[\Big]{\Na{k_i}{\Gtwo} \cap \Bone}}{\lamG^3}
    \enspace.
    \]
\end{lemma}
\begin{proof}
    From $\mathcal{P}_{i-1}$ to $\mathcal{P}_{i}$, we are removing $k_i$ and its edges and are adding edges from $k_{i,A}$ to all nodes in $\Na{k_i}{\Gtwo} \cap \Aone$ and from $k_{i,B}$ to all nodes in $\Na{k_i}{\Gtwo} \cap \Bone$. 
    Thus, we can couple the executions of the Conflict Graph design for $G^{(2)}_{i-1}$ and $G^{(2)}_i$ as follows: all nodes $k \neq k_i$ in $G^{(2)}_{i-1}$ are in one to one correspondence with all nodes $k \neq k_{i,A},k_{i,B}$ in $G^{(2)}_i$, so we can use the same $U_k$ random variable to sample them. For $k_i,k_{i,A}, k_{i,B}$ we use random variables $U_{k_i} = U_{k_{i,A}}$ and $ U_{k_{i,B}}$, which are sampled independently. 
    Thus, with this coupling we can write
    \begin{align*}
    &\mathcal{P}_{i-1} - \mathcal{P}_i \\ 
    &\qquad = \mathbb{E}\lp[\prod_{k \in \Na{k_i}{\Gtwo} \cap \Aone}\paren[\Big]{1 - \alpha_k \cdot\beta_{k_i}\prod_{k' \in \Nb{k}{G^{(2)}_{i-1}}\setminus\{k_i\}}\beta_{k'}}\rp. \\
    &\qquad\qquad \lp. \cdot \prod_{k \in \Na{k_i}{\Gtwo} \cap \Bone}\paren[\Big]{1 - \alpha_k \cdot \beta_{k_i} \prod_{k' \in \Nb{k}{G^{(2)}_{i-1}}\setminus \{k_i\}}\beta_{k'}} \rp.\\
    &\qquad\qquad\qquad \cdot\lp.\prod_{k \in \Aone \cup \Bone \setminus \Na{k_i}{\Gtwo}}\paren[\Big]{1 - \alpha_k \prod_{k' \in \Nb{k}{G^{(2)}_{i-1}}}\beta_{k'}}\rp]\\
    &\qquad - \mathbb{E}\lp[\prod_{k \in \Na{k_i}{\Gtwo} \cap \Aone}\paren[\Big]{1 - \alpha_k \cdot\beta_{k_{i,A}}\prod_{k' \in \Nb{k}{G^{(2)}_{i-1}}\setminus\{k_i\}}\beta_{k'}}\rp.\\
    &\qquad\qquad \lp. \cdot \prod_{k \in \Na{k_i}{\Gtwo} \cap \Bone}\paren[\Big]{1 - \alpha_k \cdot \beta_{k_{i,B}} \prod_{k' \in \Nb{k}{G^{(2)}_{i-1}}\setminus \{k_i\}}\beta_{k'}} \rp.\\
    &\qquad\qquad\qquad \cdot\lp.\prod_{k \in \Aone \cup \Bone \setminus \Na{k_i}{\Gtwo}}\paren[\Big]{1 - \alpha_k \prod_{k' \in \Nb{k}{G^{(2)}_{i-1}}}\beta_{k'}}\rp]
    \end{align*}
    Notice that the only difference in the above two expressions is that the first has the same factor $\beta_{k_i}$ for nodes in $\Na{k_i}{\Gtwo} \cap (\Aone \cup \Bone)$, but the second expression has a factor $\beta_{k_{i,A}}$ for nodes in $\Na{k_i}{\Gtwo} \cap \Aone $ and a factor $\beta_{k_{i,B}}$ for nodes in $\Na{k_i}{\Gtwo} \cap \Bone $.
    Some simple algebra, using the facts that $\beta_{k_i}^2 = \beta_{k_i}, \beta_{k_i} = \beta_{k_{i,A}}$ now shows that 
    \begin{align*}
        &\prod_{k \in \Na{k_i}{\Gtwo} \cap \Aone}\paren[\Big]{1 - \alpha_k \cdot\beta_{k_i}\prod_{k' \in \Nb{k}{G^{(2)}_{i-1}}\setminus\{k_i\}}\beta_{k'}}\\
        &\qquad \cdot \prod_{k \in \Na{k_i}{\Gtwo} \cap \Bone}\paren[\Big]{1 - \alpha_k \cdot \beta_{k_i} \prod_{k' \in \Nb{k}{G^{(2)}_{i-1}}\setminus \{k_i\}}\beta_{k'}} \\
        &\qquad- 
        \prod_{k \in \Na{k_i}{\Gtwo} \cap \Aone}\paren[\Big]{1 - \alpha_k \cdot\beta_{k_{i,A}}\prod_{k' \in \Nb{k}{G^{(2)}_{i-1}}\setminus\{k_i\}}\beta_{k'}} \\
        &\qquad \qquad \cdot \prod_{k \in \Na{k_i}{\Gtwo} \cap \Bone}\paren[\Big]{1 - \alpha_k \cdot \beta_{k_{i,B}} \prod_{k' \in \Nb{k}{G^{(2)}_{i-1}}\setminus \{k_i\}}\beta_{k'}}\\
        &= \paren[\big]{1- \beta_{k_{i,B}}} \\
        &\qquad \cdot 
        \paren[\Bigg]{1 - \prod_{k \in \Na{k_i}{\Gtwo} \cap \Aone}\paren[\Big]{1 - \alpha_k \cdot \beta_{k_i}\prod_{k' \in \Nb{k}{G^{(2)}_{i-1}}\setminus\{k_i\}}\beta_{k'}}} \\
        &\qquad\cdot 
        \paren[\Bigg]{1 - \prod_{k \in \Na{k_i}{\Gtwo} \cap \Bone}\paren[\Big]{1 - \alpha_k \cdot \beta_{k_i} \prod_{k' \in \Nb{k}{G^{(2)}_{i-1}}\setminus\{k_i\}}\beta_{k'}}} 
    \end{align*}
    Now, notice that the random variable $\paren[\big]{1- \beta_{k_{i,B}}}$ is independent of everything else in the expectation, so we can write
    \begin{align*}
        &\mathcal{P}_{i-1} - \mathcal{P}_i \\ 
        &\quad= \mathbb{E}\lp[\paren[\big]{1- \beta_{k_{i,B}}}
        \rp]\cdot\\
        &\qquad \cdot \mathbb{E}\lp[   \paren[\Bigg]{1 - \prod_{k \in \Na{k_i}{\Gtwo} \cap \Aone}\paren[\Big]{1 - \alpha_k \cdot \beta_{k_i} \prod_{k' \in \Nb{k}{G^{(2)}_{i-1}}\setminus\{k_i\}}\beta_{k'}}}\rp.\\
        &\qquad\qquad\lp. \cdot 
        \paren[\Bigg]{1 - \prod_{k \in \Na{k_i}{\Gtwo} \cap \Bone}\paren[\Big]{1 - \alpha_k \cdot \beta_{k_i}\cdot\prod_{k' \in \Nb{k}{G^{(2)}_{i-1}}\setminus\{k_i\}}\beta_{k'}}} \rp.\\
        &\qquad\qquad \qquad\lp. \cdot \prod_{k \in \Aone \cup \Bone \setminus \Na{k_i}{\Gtwo}}\paren[\Big]{1 - \alpha_k \prod_{k' \in \Nb{k}{G^{(2)}_{i-1}}}\beta_{k'}}
        \rp]\\
        &\qquad= \frac{1}{\lamG} \mathbb{P}_{G^{(2)}_{i-1}}\lp[\paren[\Big]{\cup_{k \in  \Na{k_i}{\Gtwo} \cap \Aone} \{Z_k = 1\}}\cap \paren[\Big]{\cup_{k \in  \Na{k_i}{\Gtwo} \cap \Bone} \{Z_k = 1\}}\rp.\\
        &\qquad\qquad\lp.
        \cap \paren[\Big]{\cup_{k \in  (\Aone \cup \Bone)\setminus \Na{k_i}{\Gtwo} } \{Z_k = 0\}}
        \rp] \\
        \intertext{by definition of the indicator events}
        &\qquad\leq \frac{1}{\lamG} \cdot\\
        &\qquad\sum_{\substack{k \in  \Na{k_i}{\Gtwo} \cap \Aone\\
        k' \in \Na{k_i}{\Gtwo} \cap \Bone}} \Prsub[\Bigg]{G^{(2)}_{i-1}}{ \{Z_k = 1\}\cap  \{Z_{k'} = 1\}
        \cap \paren[\Big]{\cup_{k \in  (\Aone \cup \Bone)\setminus \Na{k_i}{\Gtwo} } \{Z_k = 0\}}}\\
        \intertext{by union bound}
        &\qquad\leq \frac{1}{\lamG} \sum_{\substack{k \in  \Na{k_i}{\Gtwo} \cap \Aone\\
        k' \in \Na{k_i}{\Gtwo} \cap \Bone}} \Prsub[\Bigg]{G^{(2)}_{i-1}}{ \{U_k = 1\}\cap  \{U_{k'} = 1\}}\\
        \intertext{by monotonicity}
        &\qquad= \frac{\abs[\big]{\Na{k_i}{\Gtwo} \cap \Aone} \cdot \abs[\big]{\Na{k_i}{\Gtwo} \cap \Bone}}{4 \lamG^3}\\
        &\qquad\leq C' \frac{\abs[\big]{\Na{k_i}{\Gtwo} \cap \Aone} \cdot \abs[\big]{\Na{k_i}{\Gtwo} \cap \Bone}}{4 \lamG^3} \mathcal{P}_i 
        \enspace,
    \end{align*}
	where we applied Lemma~\suppref{lem:probability-lower-bound} to relate back to $\mathcal{P}_i$.
    Rearranging the above gives the desired inequality. 
\end{proof}

We are now ready to quantify the multiplicative difference of $\Prsub{\Gtwo}{E_{\Aone \cup \Bone}}$ and $\Prsub{\Gthree}{E_{\Aone \cup \Bone}}$.

\begin{lemma}\label{lem:fromG2toG3}
    There exists a constant $C'$ depending only on $C$, such that
    \begin{equation}
        \abs[\Bigg]{\frac{\Prsub{\Gtwo}{E_{\Aone\cup \Bone}}}{\Prsub{\Gthree}{E_{A^{(1)} \cup B^{(1)}}}} - 1} \leq C' \frac{\sum_{k \in R_{A,B}} |N_a(k) \cap A|\cdot |N_a(k) \cap B|}{\lamG^3}
        \enspace.
    \end{equation} 
\end{lemma}
\begin{proof}
    We can write the telescopic product
    \[
        \abs[\Bigg]{\frac{\Prsub{\Gtwo}{E_{\Aone\cup \Bone}}}{\Prsub{\Gthree}{E_{A^{(1)} \cup B^{(1)}}}} - 1}  = 
        \abs[\Bigg]{\prod_{i=1}^{|T_{A,B}|}\frac{\mathcal{P}_{i-1}}{\mathcal{P}_{i}}- 1} 
    \]
    We can now use Lemma~\suppref{lem:linearization} with 
    \[
        a_{i} = \frac{\mathcal{P}_{i-1}}{\mathcal{P}_{i}} - 1 \enspace.
    \]
    Indeed, by Lemma~\suppref{lem:Phase3change} we get that, for a constant $C'$ 
    \[
    \sum_{i=1}^{|I|} \paren{|a_{i}|  } \leq C' \frac{\sum_{k \in T_{A,B}}\abs[\big]{\Na{k_i}{\Gtwo} \cap \Aone} \cdot \abs[\big]{\Na{k_i}{\Gtwo} \cap \Bone} }{\lamG^3} \leq C'' \enspace,
    \]
    where we used the fact that each node in $\Aone \cup \Bone$ has at most $\lamG$ parents, and $|\Aone \cup \Bone| = \bigO{\lamG}$. 
    Thus, by Lemma~\suppref{lem:linearization} we get that for some constant $C'$
    \begin{align*}
        \abs[\Bigg]{\frac{\Prsub{\Gtwo}{E_{\Aone\cup \Bone}}}{\Prsub{G^{(3)}}{E_{A^{(1)} \cup B^{(1)}}}} - 1} 
        &\leq C' \sum_{i=1}^{|I|} \paren{|a_{1,i}| + |a_{2,i}| + |a_{3,i}|}  \\
        &\leq C'' \frac{\sum_{k \in T_{A,B}}\abs[\big]{\Na{k_i}{\Gtwo} \cap \Aone} \cdot \abs[\big]{\Na{k_i}{\Gtwo} \cap \Bone} }{\lamG^3} \enspace.
    \end{align*}
    The claim would follow if we can establish that
    \begin{equation}\label{eq:common-parents-G-G2}
        \sum_{k \in T_{A,B}}\abs[\big]{\Na{k_i}{\Gtwo} \cap \Aone} \cdot \abs[\big]{\Na{k_i}{\Gtwo} \cap \Bone}  = \sum_{k \in R_{A,B}} |N_a(k) \cap A|\cdot |N_a(k) \cap B|
    \end{equation}
    Indeed, it is easy to check that for any $k \notin \Aone \cup \Bone$, $\Na{k}{G} \cap \Aone$ is in one-to-one corresponcence with $\Na{k}{\Gtwo}$. This is the case because children of $k$ that do not belong in the intersection $I = \Aone \cap \Bone$ have not been altered, and if $k \to k'$ was an edge in $G$ with $k' \in I$, then $k \to k'_A$ will be an edge in $\Gtwo$ and vice versa. Similar claims hold for $\Bone$. Thus, the sets $R_{A,B}$ and $T_{A,B}$ are the same and \eqref{eq:common-parents-G-G2} holds. 
\end{proof}

% \textbf{Phase III: Removing Common Parents of $\Aone$ and $\Bone$}
\subsection{Proof of Correlation Lemma (Theorem~\suppref{thm:CAB_bound})} \label{sec:correlation-lemma-proof}

We now provide a proof of the Correlation Lemma (Theorem~\suppref{thm:CAB_bound}).
For convenience, we restate the theorem here.

\corrlemma*

\begin{proof}

We now combine the results of interpolation of the probability from the three different phases.
Indeed, using Lemmas~\suppref{lem:fromGtoG1}, \suppref{lem:fromG1toG2} and \suppref{lem:fromG2toG3}, together with Lemma~\suppref{lem:same_marginals_G3} and Lemma~\suppref{lem:linearization}, we get that
\begin{align*}
    C_{A,B} &= \abs[\Bigg]{\frac{\Prsub{G}{E_{A\cup B}}}{\Prsub{G}{E_{A}}\Prsub{G}{E_{B}}}-1} \\
    &= \abs[\Bigg]{\frac{\Prsub{G}{E_{A\cup B}}}{\Prsub{\Gone}{E_{\Aone \cup \Bone}}} \cdot
    \frac{\Prsub{\Gone}{E_{\Aone\cup \Bone}}}{\Prsub{\Gtwo}{E_{\Aone \cup \Bone}}} \cdot
    \frac{\Prsub{\Gtwo}{E_{\Aone\cup \Bone}}}{\Prsub{\Gthree}{E_{\Aone \cup \Bone}}}
    -1} \\
    &\leq
    C'  \cdot\paren[\Bigg]{\frac{|I|}{\lamG} + \frac{|F_{A,B}|}{\lamG^2} + \frac{\sum_{k \in R_{A,B}} |N_a(k) \cap A|\cdot |N_a(k) \cap B|}{\lamG^3}} \enspace,
\end{align*}
and the proof is complete. 
\end{proof}

\end{document}